\pdfoutput=1 %

  \documentclass[fleqn,3p,authoryear]{elsarticle}

\usepackage[utf8]{inputenc}
\usepackage[T1]{fontenc}
\usepackage[ngerman,USenglish]{babel}
\usepackage{lineno}
\usepackage[english=american]{csquotes}
\usepackage{setspace}
\usepackage{etoolbox}
\usepackage{needspace}

  \usepackage{float}

\usepackage{amssymb,amsmath,amsfonts,mathtools,dsfont,pifont,bm,mleftright}

\usepackage{enumitem}

\usepackage{xspace}

\usepackage[table-align-text-post=false, retain-unity-mantissa=false, exponent-product=\cdot, detect-all, table-number-alignment=center]{siunitx}
\DeclareSIUnit\molar{\mole\per\cubic\metre}
\DeclareSIUnit\Molar{M}

\biboptions{sort&compress,semicolon}

\usepackage{booktabs}
  \usepackage{longtable}
  \usepackage{threeparttablex}
\usepackage{multicol,multirow}
\usepackage{array}
\newcolumntype{x}[1]{>{\centering\arraybackslash\hspace{0pt}}p{#1}}

\usepackage[dvipsnames]{xcolor}
\usepackage{tikz}
\usepackage{tikz-3dplot}
\usetikzlibrary{shapes.geometric,calc,intersections,angles,positioning}
\usetikzlibrary{arrows,arrows.meta,decorations,backgrounds,patterns,matrix,shapes,fit,shadows,plotmarks,spy}

\tikzset{
	stdArrowFixed/.style={arrows={-Stealth[round,length=4.8pt,width'=3.6pt,inset'=1.56pt,line width=0.4pt]}},
	stdArrow/.style={>={Stealth[round]}},
    bulkVolume/.style={fill={rgb,255:red,225; green,233; blue,242}},
	intVelocity/.style={draw=blue},
	bndAvgVolEE/.style={draw=red},
	bndAvgVolBead/.style={draw=blue},
	swArrow/.style={shorten <=3pt,shorten >=3pt,stdArrow},
	swComp/.style={draw=black,thick,inner sep=0.25cm},
	swLib/.style={swComp,rounded corners=0.25cm},
	swExternal/.style={swComp,densely dashed,rounded corners=0.25cm},
	swClass/.style={swComp,rounded corners=0.25cm},
	swAbstractClass/.style={swComp,rounded corners=0.25cm,font=\itshape},
	swIface/.style={swComp},
	swExe/.style={swComp},
	swFile/.style={shape=file,swComp},
	swLine/.style={thick},
	swRuns/.style={->,swArrow,swLine,dashed},
	swUses/.style={->,swArrow,swLine},
	swOwns/.style={<-,swArrow,swLine,>=Diamond,shorten <=0pt,shorten >=0pt},
	swHas/.style={<-,swArrow,swLine,>=Diamond[open],shorten <=0pt,shorten >=0pt},
	swKnows/.style={->,swArrow,swLine,dashed,shorten <=0pt,shorten >=0pt},
	swUsesClass/.style={->,swArrow,swLine,shorten <=0pt,shorten >=0pt},
	swImplements/.style={->,swArrow,swLine,>=Triangle[open],shorten <=0pt,shorten >=0pt},
	swExported/.style={densely dotted}
}

\tikzset{
	stateVecCell/.style = {rectangle, draw,outer sep=0pt,node distance=0pt,minimum height=0.6cm},
	blockDesc/.style = {node distance=0.3cm,font=\footnotesize},
	bracedescAbove/.style = {black,yshift=18pt,font=\footnotesize},
	bracedescBelow/.style = {black,yshift=-18pt,font=\footnotesize},
	ellipsisSize12N/.style={column 2/.style={nodes={minimum width=1cm}},column 5/.style={nodes={minimum width=1cm}},column 7/.style={nodes={minimum width=1cm}},column 9/.style={nodes={minimum width=1cm}}},
	arrayMatrixBase/.style={matrix of nodes,column sep=-\pgflinewidth,nodes in empty cells,matrix anchor=west,inner sep=0cm,outer sep=0cm},
	stateBlockInlet/.style={fill=cyan!10},
	stateBlockBulk/.style={fill=red!15},
	stateBlockParticle/.style={fill=blue!15},
	stateBlockFlux/.style={fill=yellow!15},
}

\usepackage{glossaries}

\usepackage[font=small]{caption}
\usepackage{subcaption}
\newsavebox{\figHelperBox}

\PassOptionsToPackage{hyphens}{url}\usepackage[hidelinks,breaklinks=true,hypertexnames=false]{hyperref}
\usepackage{amsthm}
\usepackage[capitalise, nameinlink]{cleveref} %

\usepackage{orcidlink}

\newcommand\swapifbranches[3]{#1{#3}{#2}}
\makeatletter
\catcode`\_=11\relax
\catcode`\:=11\relax
\patchcmd{\DeclarePairedDelimiter}{\@ifstar}{\swapifbranches\@ifstar}{}{}
\catcode`\:=12\relax
\catcode`\_=8\relax
\makeatother

\DeclarePairedDelimiter{\parensFences}{\lparen}{\rparen}

\DeclarePairedDelimiter{\curlyFences}{\{}{\}}
\DeclarePairedDelimiter{\abs}{\lvert}{\rvert}
\DeclarePairedDelimiterXPP{\norm}[1]{}{\lVert}{\rVert}{}{#1}
\DeclarePairedDelimiterXPP{\bigO}[1]{\mathrm{O}}{(}{)}{}{#1}

\let\Re=\relax
\DeclarePairedDelimiterXPP{\Re}[1]{\mathrm{Re}}{(}{)}{}{#1}
\let\Im=\relax
\DeclarePairedDelimiterXPP{\Im}[1]{\mathrm{Im}}{(}{)}{}{#1}

\newcommand{\supp}{\operatorname*{supp}\parensFences}

\newcommand{\rank}{\operatorname{rank}\parensFences}

\newcommand{\spann}{\operatorname{span}\curlyFences}

\newcommand{\numrows}{\operatorname{rows}\parensFences}

\newcommand{\vol}{\operatorname{vol}\parensFences}
\newcommand{\surfarea}{\omega\parensFences}

\newcommand{\diag}{\operatorname{diag}\parensFences}
\newcommand{\card}{\abs}
\newcommand{\closure}[1]{\bar{#1}}

\let\div=\relax
\newcommand{\divOp}{\nabla\cdot}
\newcommand{\div}{\nabla\cdot\parensFences}
\DeclarePairedDelimiter{\avg}{\langle}{\rangle}
\newcommand{\avgspace}{\!\!}
\newcommand{\vecnabla}{\bm{\nabla}}

\newcommand{\mat}[1]{\bm{#1}}
\DeclarePairedDelimiterXPP{\submat}[3]{#1}{[}{]}{}{\;\! #2,#3 \;\!}

\renewcommand{\vec}[1]{\bm{#1}}
\DeclarePairedDelimiterXPP{\subvec}[2]{#1}{[}{]}{}{\;\! #2 \;\!}

\newcommand{\all}{\mathrm{\text{\textendash}}}

\newcommand{\op}[1]{\mathcal{#1}}
\newcommand{\vecop}[1]{\mathcal{#1}}

\newcommand{\mset}[1]{\mathcal{#1}}

\newcommand{\Reals}{\mathds{R}}
\newcommand{\Naturals}{\mathds{N}}
\newcommand{\Integers}{\mathds{Z}}

\makeatletter
\newcommand{\overbar}[2][1.0]{{%
  \mspace{#1mu}%
  \overline{\mspace{-#1mu}\overbar@check#2\relax}%
}}
\newcommand\overbar@check[1]{%
  #1\@ifnextchar_{\overbar@sub}{}%
}
\newcommand{\overbar@sub}[2]{%
  _{#2}\mspace{-2mu}\aftergroup\overbar@compensate
}
\newcommand{\overbar@compensate}{\mspace{2mu}}
\makeatother

\renewcommand{\l}[0]{\mathrm{\ell}}
\renewcommand{\b}[0]{\mathrm{b}}
\newcommand{\p}[0]{\mathrm{p}}
\newcommand{\s}[0]{\mathrm{s}}
\newcommand{\crossPhase}[0]{\mathrm{c}}
\newcommand{\ps}[0]{\mathrm{par}}

\newcommand{\Rho}{P}

\newcommand{\T}[0]{\mathrm{T}}

\DeclarePairedDelimiterXPP{\laplaceT}[1]{\op{L}}{(}{)}{}{#1}
\DeclarePairedDelimiterXPP{\hankelT}[1]{\op{J}}{(}{)}{}{#1}

\newcommand*{\rom}[1]{\expandafter\@slowromancap\romannumeral #1@}

\newtheoremstyle{plain}
  {\topsep}   %
  {\topsep}   %
  {\itshape}  %
  {0pt}       %
  {\bfseries} %
  {.}         %
  {5pt plus 1pt minus 1pt} %
  {\thmname{#1}\thmnumber{ #2}\thmnote{ (#3)}}

\theoremstyle{definition} 
  \newtheorem{theorem}{Theorem}[section]
  
  \newtheorem{example}[theorem]{Example}

\usepackage{xparse}

\newtheorem{innerassumption}{Assumption}
\newtheorem{innerremark}{Remark}
\newtheorem{innerdefinition}{Definition}

\NewDocumentCommand{\DeclareManualTheorem}{mm}{%
  \NewDocumentEnvironment{#1}{m o}
  {%
    \expandafter\renewcommand\csname theinner#1\endcsname{##1}%
    \IfNoValueTF{##2}
      {\csname inner#1\endcsname}
      {\csname inner#1\endcsname[##2]}%
  }
  {\csname endinner#1\endcsname}
}

\DeclareManualTheorem{model}{Model}

\NewDocumentEnvironment{assumption}{o o}
{%
  \IfNoValueTF{#1}
    {%
      \IfNoValueTF{#2}
        {\innerassumption}
        {\innerassumption[#2]}%
    }
    {%
      \renewcommand{\theinnerassumption}{#1}%
      \IfNoValueTF{#2}
        {\innerassumption}
        {\innerassumption[#2]}%
    }
}
{\endinnerassumption}

\NewDocumentEnvironment{remark}{o o}
{%
  \IfNoValueTF{#1}
    {%
      \IfNoValueTF{#2}
        {\innerremark}
        {\innerremark[#2]}%
    }
    {%
      \renewcommand{\theinnerremark}{#1}%
      \IfNoValueTF{#2}
        {\innerremark}
        {\innerremark[#2]}%
    }
}
{\endinnerremark}

\NewDocumentEnvironment{definition}{o o}
{%
  \IfNoValueTF{#1}
    {%
      \IfNoValueTF{#2}
        {\innerdefinition}
        {\innerdefinition[#2]}%
    }
    {%
      \renewcommand{\theinnerdefinition}{#1}%
      \IfNoValueTF{#2}
        {\innerdefinition}
        {\innerdefinition[#2]}%
    }
}
{\endinnerdefinition}

\crefname{innermodel}{Model}{Models}
\Crefname{innermodel}{Model}{Models}
\crefname{innerassumption}{Assumption}{Assumptions}
\Crefname{innerassumption}{Assumption}{Assumptions}
\crefname{innerremark}{Remark}{Remarks}
\Crefname{innerremark}{Remark}{Remarks}
\Crefname{innerdefinition}{Definition}{Definitions}
\crefname{remark}{Remark}{Remarks}

  \newtheorem{suppdefinition}[theorem]{Definition}
  \newtheorem{suppassumption}[theorem]{Assumption}
  \newtheorem{suppremark}[theorem]{Remark}

  \crefname{suppdefinition}{Definition}{Definitions}
  \Crefname{suppdefinition}{Definition}{Definitions}
  \crefname{suppassumption}{Assumption}{Assumptions}
  \Crefname{suppassumption}{Assumption}{Assumptions}
  \crefname{suppremark}{Remark}{Remarks}
  \Crefname{suppremark}{Remark}{Remarks}

  \makeatletter
  \def\ps@pprintTitle{%
    \let\@oddhead\@empty
    \let\@evenhead\@empty
    \let\@oddfoot\@empty
    \let\@evenfoot\@empty}
  \makeatother

\begin{document}

\newacronym{hrm}{HRM}{High-Resolution Model}
\newacronym{rom}{ROM}{Reduced Order Model}
\newacronym{grm}{GRM}{General Rate Model}
\newacronym{2dgrm}{2DGRM}{Two-dimensional General Rate Model}
\newacronym{lrmp}{LRMP}{Lumped Rate Model with Pores}
\newacronym{lrm}{LRM}{Lumped Rate Model without Pores}
\newacronym{edm}{EDM}{Equilibrium Dispersive Model}
\newacronym{pfr}{PFR}{Plug Flow Reactor}
\newacronym{ipfr}{IPFR}{Ideal Plug Flow Reactor}
\newacronym{dpfr}{DPFR}{Dispersive Plug Flow Reactor}
\newacronym{str}{STR}{Stirred Tank Reactor}
\newacronym{bstr}{BSTR}{Batch Stirred Tank Reactor}
\newacronym{cstr}{CSTR}{Continuous Stirred Tank Reactor}
\newacronym{sma}{SMA}{Steric Mass Action}
\newacronym{zrm}{ZRM}{Zonal Rate Model}

\newacronym{pde}{PDE}{Partial Differential Equation}
\newacronym{pdae}{PDAE}{Partial Differential-Algebraic Equation}
\newacronym{ode}{ODE}{Ordinary Differential Equation}

\newacronym{rev}{REV}{Representative Elementary Volume}
\newacronym{fv}{FV}{Finite Volume}

\newacronym{lbm}{LBM}{Lattice Boltzmann Method}
\newacronym{pnm}{PNM}{Pore Network Models}

\newacronym{psd}{PSD}{Particle Size Distribution}
\newacronym{ptd}{PTD}{Particle Type Distribution}
\newacronym{sec}{SEC}{Size-Exclusion Chromatography}

\newacronym{ldf}{LDF}{Linear Driving Force}

\newacronym{smb}{SMB}{Simulated Moving Bed}

\newacronym{gui}{GUI}{Graphical User Interface}

  \newacronym{ipf}{IPF}{Ideal Plug Flows}
  \newacronym{dpf}{DPF}{Dispersive Plug Flows}
  \newacronym{cst}{CST}{Continuous Stirred Tanks}

\begin{frontmatter}

\title{A unified and composable framework for modeling packed-bed liquid chromatography}

\author[addrUoC,addrFZJ]{Jan Michael Breuer\,\orcidlink{0000-0002-1999-2439}}
\author[addrFZJ]{Samuel Leweke\,\orcidlink{0000-0001-9471-4511}}
\author[addrFZJ]{Johannes Schmölder\,\orcidlink{0000-0003-0446-7209}}
\author[addrFZJ,addrRWTH]{Antonia Berger\,\orcidlink{0009-0002-0207-9042}}
\author[addrUoC,addrCDS]{Gregor~J.~Gassner\,\orcidlink{0000-0002-1752-1158}}
\author[addrFZJ,addrRWTH]{Eric von~Lieres\,\orcidlink{0000-0002-0309-8408}\corref{cor1}}
\ead{e.von.lieres@fz-juelich.de}

\cortext[cor1]{Corresponding author.}

\address[addrUoC]{Department of Mathematics and Computer Science, University of Cologne, 50931 Cologne, Germany}
\address[addrFZJ]{Forschungszentrum J\"ulich, IBG-1: Biotechnology, 52425 J\"ulich, Germany}
\address[addrRWTH]{RWTH Aachen University, Institute of Fluid Process Engineering, 52074 Aachen, Germany}
\address[addrCDS]{Center for Data and Simulation Science, University of Cologne, 50931 Cologne, Germany}

\begin{abstract}

Packed-bed chromatography is a cornerstone of bioseparation, yet its modeling landscape remains fragmented by numerous model variants, inconsistent terminology, and limited structural clarity. While prior work has categorized and contrasted individual variants, our approach shifts the focus to unification. We present a unified, modular, and composable framework that integrates diverse mechanistic packed-bed chromatography models into a coherent structure. Starting from a high-resolution three-dimensional model, the framework systematically derives a hierarchy of reduced-order models through formal model order reduction. It encompasses a broad range of bulk transport phenomena, intraparticle mass transfer processes, binding mechanisms, and chemical reactions. Formulated orthogonally, these modules are independently meaningful and can be systematically combined into mathematically consistent models tailored to specific applications. The framework recovers and generalizes existing models, establishes a consistent taxonomy, and enables novel model combinations and extensions, including alternative column and particle geometries, particle-size and particle-type distributions, component-specific pore accessibility, rapid-equilibrium binding and reaction formulations, and size-exclusion with simultaneous binding. The framework is accessible through CADET-Equations, an interactive web application for model assembly and automatic generation of governing equations in LaTeX, promoting standardization and transparency. Its design principles are increasingly reflected throughout the CADET ecosystem, whose modular architecture supports the majority of the proposed model family. The generality and composable design of the framework provide a foundation for extending chromatography modeling software toward integrated bioseparation processes and advanced computational biotechnology applications.
\end{abstract}


\end{frontmatter}

\tableofcontents

\section{Introduction}

In packed-bed chromatography, there is no universal deterministic model, but a multitude of mechanistic models that describe the underlying physicochemical processes with varying levels of detail and diverse features~\mbox{(see e.g.}~\citealp{Felinger2004,Guiochon2006,Seidel-Morgenstern2020-Chap6,Gu2015,Schmidt-Traub2020,Nicoud2015}).
Although some of these models may be equivalent under specific conditions or share certain assumptions, physical mechanisms, or mathematical formulations, they often originate from distinct conceptual frameworks or practical considerations.
These considerations include various mass transfer phenomena, rate limiting kinetic processes, and rapid-equilibrium assumptions, as well as porous media effects.
As a result, the mathematical relationships and conceptual connections between different models are often unclear.
Furthermore, inconsistencies in the naming conventions, notation, and descriptive terminology add to the overall complexity of the modeling landscape and can create a misleading impression of ambiguity among these models.

This lack of clarity and consistency across the field of chromatography modeling poses significant challenges for researchers and practitioners as it hinders the understanding, comparison, and application of chromatography models.
For instance, models may be applied without a clear understanding of their underlying assumptions, which can lead to conceptual imprecision, complicate the selection of appropriate models, or result in the development and use of mathematically ill-defined model formulations.
One prominent example is the ambiguous use of a ``Lumped Kinetic Model'', which may refer either to finite-rate binding \citep{Guiochon2006} or to finite-rate film diffusion \citep{Schmidt-Traub2020}, illustrating how inconsistent terminology across the field can obscure the underlying mechanisms.
Several studies have examined these issues for specific model variants and parameter settings~\mbox{(see, e.g.,}~\citealp{GOLSHANSHIRAZI19921,Qamar20012019,KATSOULAS2023464345,CARTA2026466781}). For instance, \citet{KATSOULAS2023464345} derived an \gls{edm} as a limiting case of a pore diffusion formulation, showing that the effective dispersion parameters differ between the two models, despite their superficially similar mathematical structure. 
\cite{CARTA2026466781} compared four model variants for predicting gradient elution under linear binding conditions, showing mathematical relationships and identifying empirical correlations by analyzing moments of predicted elution profiles.
Nevertheless, such insights remain specific to the model variants considered and scattered throughout the literature, and a comprehensive unifying framework remains to be established.

All models are based on simplifications and are in that sense inherently ``wrong''~\citep{GeorgeBoxModelsAreWrong}. Nevertheless, many engineering problems can be mathematically characterized by a limited set of recurring mathematical descriptions. The variety of transport models found in the liquid chromatography literature reflects differences in physical interpretation, modeling assumptions, and level of resolution, rather than fundamentally distinct governing equations.
As a result, calibrating different models to the same experimental data often yields substantially different parameter estimates.
Band broadening, for instance, may be attributed to different transport phenomena, each justifiable at a different level of abstraction, since simpler models often combine multiple effects into a single parameter.
Even when a model is theoretically appropriate, calibration can still be limited by sparse or noisy data, and some parameters may not be identifiable.
More detailed models, while physically more precise, increase computational cost and the risk of overfitting.
Thus, model selection requires a careful balance between physical fidelity, computational demands, and robustness in calibration.
Addressing this challenge calls for a modeling framework that facilitates systematic model selection across a wide range of model configurations and levels of resolution.

Comprehensive works ~\mbox{(see e.g.}~\citealp{Felinger2004,Guiochon2006,Seidel-Morgenstern2020-Chap6,Gu2015,Schmidt-Traub2020,Nicoud2015,Leweke_840314}) contributed to the development and categorization of model variants in chromatography. These studies rightly emphasize the distinctions between different modeling approaches, offering deep insight into the assumptions, formulations, and application domains of each variant. Our approach builds on this foundation but shifts the perspective: rather than focusing on the differences and boundaries between models, we aim to establish a unified, modular and composable framework that systematically integrates these variants. This enables a coherent view of chromatography modeling, where model elements can be selected, combined, or refined as needed.
The proposed modeling framework provides a theoretical basis for modeling arbitrary and interconnected systems, as required in advanced computational biotechnology applications.

We present a modeling framework for packed-bed chromatography that is (1) unified, encompassing a wide range of established models from the literature as well as novel generalizations and combinations, (2) modular, unifying various features and operational modes of packed-bed chromatography in a flexible, systematic manner, and (3) composable, allowing these building blocks to be systematically combined into mathematically consistent model variants.
The resulting framework equips researchers and practitioners with a comprehensive modeling toolbox that facilitates the selection of the most appropriate model variant for a given application.
Furthermore, we introduce a unified taxonomy with distinctive and descriptive model names that clearly reflect the underlying equations and assumptions, while also providing a coherent notation and terminology.

Our modeling framework decomposes chromatography models into four orthogonal modules:
Transport in the bulk domain of the column, transport into and within the particle pores, binding mechanisms, and chemical reactions.
Orthogonality here implies that (1) each module is independent of the others and can function as a meaningful standalone model, and (2) the modules can also be systematically combined to construct comprehensive chromatography models.

We begin with a high-resolution three-dimensional representation of the column and derive a hierarchy of \glspl{rom} for transport in both the bulk and particle domains.
To achieve this, we perform model order reduction by progressively introducing simplifying assumptions.
Subsequently, we discuss model variants that modify key system characteristics, such as particle and column geometry, differing time scales of mass transfer processes, or flow direction, which necessitate distinct mathematical formulations.
Further, we show how binding and reaction models are incorporated into our modeling framework.
Throughout this publication, we lay the mathematical foundation for formulating models that are known from the literature, as well as novel combinations of their components, generalizations, and variants.

We demonstrate the utility of our modeling framework by showcasing how prominent models from the literature are represented in the generalized framework as important special cases.
Moreover, we propose a unified phenomenological taxonomy.
The modeling framework is accessible through \texttt{CADET-Equations}, an interactive web application that enables users to configure models via a \gls{gui} and download standardized, mathematically consistent governing equations in LaTeX format. The tool improves the transparency of \texttt{CADET-Core} simulations by allowing users to upload simulation files and generate the corresponding governing equations. In addition, \texttt{CADET-Equations} serves as an educational resource by enabling users to interactively explore how individual modeling choices affect the governing equations and how the resulting models are related.
\texttt{CADET-Equations} is hosted as part of the CADET documentation (\url{https://cadet.github.io}), and the source code is publicly available \citep{breuer_2026_18339287}.

The presented framework reflects the modeling philosophy developed by the CADET team with its principles increasingly integrated throughout the CADET software stack.
Recent developments in \texttt{CADET-Core} \citep{Leweke2025,leweke_2025_17132359, Leweke2018}, including a series of version 6.0.0 pre-releases \citep{breuer_2026}, have introduced agnostic implementations of bulk- and particle-transport models and their spatial discretizations, support for reactions under rapid-equilibrium assumptions, and alternative column geometries, among other features.
Together, these developments advance \texttt{CADET-Core} toward the unified and composable modeling framework presented here.

\section{Hierarchical Family of Transport Models}
\label{sec:a_hierarchical_family_of_transport_models}

In this section, we show that the various transport models for packed-bed chromatography presented in the literature can be understood as part of a hierarchy derived through systematic simplifications of a high-resolution reference model.
For conciseness, we derive this model family for cylindrical axial-flow columns. Other geometries for which the same hierarchy can be derived are discussed in \cref{sec:variant_column_device}.
In this context, we refer to solute-stationary phase interactions as binding.
A detailed overview of different categories of binding models, including explicit formulations of representative binding models, is given in Section~\ref{supp:sec:binding_models}.

To streamline the mathematical derivations throughout this work, all quantities and parameters are assumed to be sufficiently smooth and integrable, as required. We further assume that all mathematical limits arising in the derivations exist and may be interchanged with other limits or differential operators when necessary.

In packed-bed liquid chromatography, a cylindrical column is filled with densely packed, typically spherical particles that may vary in size.
This packing creates several distinct domains to be considered:
the void space between the particles, known as the interstitial or bulk domain $\Omega^\mathrm{\b}$, and the particle domain $\Omega^\ps$.
The particle domain is further divided into the particle pore phase $\Omega^\p$ and the particle solid phase $\Omega^\s$.
The corresponding bulk $c^\b$, pore $c^\p$, and solid $c^\s$ concentrations are the primary quantities of interest.
The derivation of the model family starts from a high-resolution consideration of the chromatography column that fully resolves the packed-bed geometry.
Reduced-order models are obtained by successively introducing simplifying assumptions and applying their corresponding mathematical implications.
This reduction procedure is carried out independently for the bulk (\cref{sec:bulk_domain_transport}) and particle (\cref{sec:intra_particle_transport}) domains.

\subsection{High-Resolution Model}

A \gls{hrm} of packed-bed liquid chromatography describes two main aspects:
the velocity field in the bulk domain, which is determined by the geometry of the packed bed, and the mass transfer of solute within and across the bulk, pore and solid domains.
At the highest level of spatial resolution, a model would explicitly resolve the geometry of both the packed bed and the pores of each particle, as described in \cref{sec:pore-resolved-particles}.
Such fully pore-resolved column models are currently not computationally tractable.
The highest-resolution models that have been applied to complete chromatographic beds resolve the individual particles and the bulk flow around them while representing the pore geometry of each particle through a homogenized description, as described in \cref{sec:homogenized-particles}. 
For completeness, we briefly discuss pore-scale modeling approaches and their current limitations for packed-bed liquid chromatography.

\subsubsection{Pore-Resolved Particles}
\label{sec:pore-resolved-particles}
A pore-scale high-resolution model explicitly resolves the pore geometry of the particles and directly solves the governing mass transfer equations within this domain. At the molecular scale, interactions between solutes and the pore walls influence diffusion and binding, particularly in narrow mesopores. Recent advances in multiscale simulation have demonstrated the feasibility of integrating representative pore-scale molecular dynamics simulations with reconstructed packed-bed geometries \citep{TALLAREK2024,TALLAREK2025}. 

Another approach uses \gls{pnm}, which represent the porous medium as a simplified network of pore bodies connected by throats. \gls{pnm} have been used to reconstruct pore networks from high-resolution imaging and simulate intraparticle transport, including reactive transport processes \citep{FATHIGANJEHLOU2024a,FATHIGANJEHLOU2024b}. While these approaches substantially reduce the computational burden compared with fully pore-resolved column simulations, they still rely on detailed pore-scale information and remain computationally demanding for realistic chromatographic columns.
Nevertheless, their application to complete chromatographic columns remains strongly limited by the geometric complexity of the porous structure and the associated computational cost. Consequently, pore-resolved models are not considered further in this work.

\subsubsection{Homogenized Particles (Model HR)}
\label{sec:homogenized-particles}

\begin{figure}[!htbp]
	\centering
	\begin{tikzpicture}
	\node[cylinder,draw=black,thick,aspect=1.25,minimum height=4cm,minimum width=1.75cm,shape border rotate=0,] (column) {};
	\draw[dashed]
		let \p1 = ($ (column.after bottom) - (column.before bottom) $),
			\n1 = {0.5*veclen(\x1,\y1)-0.5\pgflinewidth},
			\p2 = ($ (column.bottom) - (column.after bottom)!.5!(column.before bottom) $),
			\n2 = {veclen(\x2,\y2)-0.5\pgflinewidth}
		in
			([yshift=-0.5\pgflinewidth] column.before bottom) arc [start angle=-90, end angle=90, y radius=\n1, x radius=\n2];

	\draw[<-,very thick,stdArrow] ($(column.bottom) - (0.25,0)$) -- +(-0.75,0);
	\draw[->,very thick,stdArrow] ($(column.top) + (0.25,0)$) -- +(0.75,0) node[right] {Flow};

	\draw[] ($(column.before top |- column.top) + (0,0.2)$) -- +(0.75, 0.6) node[right] {$\Gamma^{\mathrm{out}}$};
	\draw[] ($(column.before bottom |- column.bottom) + (0,0.2)$) -- +(-0.75, 0.6) node[left] {$\Gamma^{\mathrm{in}}$};
	\draw[] ($(column.south) - (1,0)$) -- +(0,-0.25) node[below] {$\Gamma^{\mathrm{wall}}$};

	\fill[bulkVolume]
		let \p1 = ($ (column.after bottom) - (column.before bottom) $),
			\n1 = {0.5*veclen(\x1,\y1)-\pgflinewidth},
			\p2 = ($ (column.bottom) - (column.after bottom)!.5!(column.before bottom) $),
			\n2 = {veclen(\x2,\y2)-\pgflinewidth}
		in
			(column.center) circle [y radius=\n1, x radius=\n2];

	\draw[dashed]
		let \p1 = ($ (column.after bottom) - (column.before bottom) $),
			\n1 = {0.5*veclen(\x1,\y1)-\pgflinewidth},
			\p2 = ($ (column.bottom) - (column.after bottom)!.5!(column.before bottom) $),
			\n2 = {veclen(\x2,\y2)-\pgflinewidth}
		in
			([yshift=-0.5\pgflinewidth] column.south) arc [start angle=-90, end angle=90, y radius=\n1, x radius=\n2];
	\draw[]
		let \p1 = ($ (column.after bottom) - (column.before bottom) $),
			\n1 = {0.5*veclen(\x1,\y1)-\pgflinewidth},
			\p2 = ($ (column.bottom) - (column.after bottom)!.5!(column.before bottom) $),
			\n2 = {veclen(\x2,\y2)-\pgflinewidth}
		in
			([yshift=-0.5\pgflinewidth] column.north) arc [start angle=90, end angle=270, y radius=\n1, x radius=\n2];

	\draw[->,stdArrow] ($(column.south) - (0,0.25)$) -- ++(0.0, -0.75) -- ++(6.5,0) node[midway,above] {Cross-section} -- ++(0, 0.25) coordinate (crossectbottom);

	\node[shape=coordinate] (crosssectmid) at ($(crossectbottom) + (0,1.75)$) {};
	\filldraw[thick,draw=black,bulkVolume] (crosssectmid) circle[radius=1.5];

	\draw[rotate around={60:(crosssectmid)}] ($(crosssectmid) + (0,1.5)$) -- +(0,0.4) node[above] {$\Gamma^{\mathrm{wall}}$};

	\node at ($(crosssectmid) + (-0.55, -0.2)$) {$\Omega^{\b}$};

	\node[shape=coordinate] (bead1) at ($(crosssectmid) + (0.75, 0.5)$) {};
	\filldraw[draw=black,fill=white] (bead1) circle[radius=0.5];
	\node[shape=coordinate] (bead2) at ($(crosssectmid) + (.25, -0.85)$) {};
	\filldraw[draw=black,fill=white] (bead2) circle[radius=0.45];
	\node[shape=coordinate] (bead3) at ($(crosssectmid) + (-.45, 0.45)$) {};
	\filldraw[draw=black,fill=white] (bead3) circle[radius=0.4];

	\draw (bead1) -- +(-0.65,1.2) coordinate (beadInner);
	\draw (bead3) -- (beadInner);
	\draw (bead2) -- (beadInner);
	\node[above] at (beadInner) {$\Omega^{\mathrm{par}}$};

	\node[shape=coordinate] (beadSurf) at ($(crosssectmid) + (1.6,-0.9)$) {};
	\node[right] at (beadSurf) {$\Gamma^{\mathrm{par}}$};
	\draw ($(bead1)!0.5cm!(beadSurf)$) -- (beadSurf);
	\draw ($(bead2)!0.45cm!(beadSurf)$) -- (beadSurf);
	\draw ($(bead3)!0.4cm!(beadSurf)$) -- (beadSurf);
\end{tikzpicture}
	\caption{Geometry of the high-resolution cylindrical axial flow model with homogenized particles.}\label{fig:model-hrm-geometry}
\end{figure}
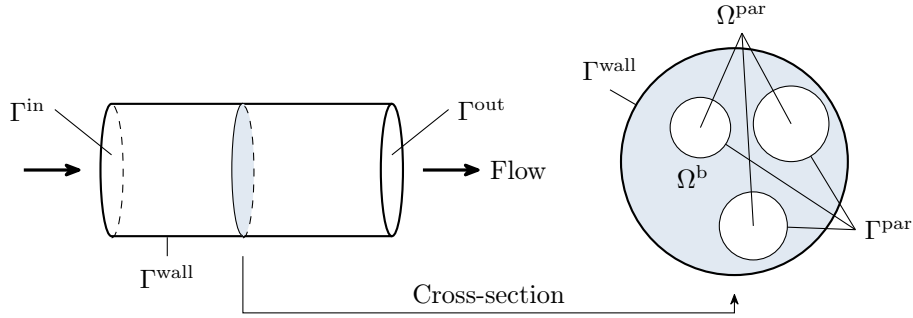

The \gls{hrm} presented in this work is based on the following customary assumptions, which also serve as the starting point for deriving reduced-order models.
\Needspace{6\baselineskip}
\begin{assumption}[(HR)]\label{asmpt:model-hrm}
We assume that \citep{Gu2015,Guiochon2006,Seidel-Morgenstern2020-Chap6}
\begin{multicols}{2}
\begin{itemize}\setlength{\itemsep}{0\baselineskip}
    \item the packed bed is confined within a rigid, perfectly cylindrical column of radius $\Rho^\mathrm{c} > 0$, with fluid entering and exiting through the inlet and outlet faces;
    \item fluid density and viscosity are constant in time and space (in particular, the fluid is incompressible);
    \item fluid flow is laminar;
    \item fluid inside the particles is stagnant (i.e., no convective flow in the particle pores);
    \item particles are perfectly spherical, rigid, and immobile;
    \item particles are homogeneous and have uniform porosity, and the intraparticle pores are equally accessible to all components;
    \item diffusion is Fickian and independent of component concentrations;
    \item solute concentrations are sufficiently low such that changes in solution volume caused by solute transport and binding can be neglected;
    \item solvent is not adsorbed.
\end{itemize}
\end{multicols}
\end{assumption}
A detailed discussion of most points in \cref{asmpt:model-hrm} is provided in Section~2.1.2 of \cite{Guiochon2006}, and selected aspects are highlighted in the Supplement~\ref{supp:sec:hrm}.
Following \cite{Puttmann2014}, these assumptions lead to a system comprising the Stokes equations for fluid flow, a convection-diffusion equation for mass transfer in the bulk domain, and diffusion equations for mass transfer within the particles.
The geometry of this model is illustrated in \cref{fig:model-hrm-geometry}.

\begin{model}{(HR)}[High-Resolution Model]\label{model:column-highres3D}
	The flow field $\vec{u}\colon \Omega^{\b} \to \Reals^3$ in the bulk column domain $\Omega^{\b}$ is governed by the Stokes equations \citep{Spurk2008}
	\begin{subequations}\label{eq:model-hrm-flow}
	\begin{alignat}{2}
		\vecnabla \cdot \mat{\sigma} &= \vec{0} & & \qquad \text{in } \Omega^{\b}, \label{eq:model-hrm-flow-momentum} \\
		\nabla \cdot \vec{u} &= 0 & & \qquad \text{in } \Omega^{\b} \label{eq:model-hrm-flow-incompressible}
	\end{alignat}
	\end{subequations}
	with boundary conditions
	\begin{subequations}\label{eq:model-hrm-flowbc}
	\begin{alignat}{2}
		\vec{u} &= \vec{u}^{\mathrm{in}} & & \qquad \text{on } \Gamma^{\mathrm{in}}, \label{eq:model-hrm-flow-inbc} \\
		\mat{\sigma} \vec{n} &= \vec{0} & & \qquad \text{on } \Gamma^{\mathrm{out}}, \label{eq:model-hrm-flow-outbc} \\
		\vec{u} &= \vec{0} & & \qquad \text{on } \Gamma^{\mathrm{wall}} \cup \Gamma^{\mathrm{par}} \label{eq:model-hrm-noslipbc},
	\end{alignat}
	\end{subequations}
	\Needspace{9\baselineskip}
	where $\vec{n}$ is the outward-facing unit normal, the fluid stress tensor is given by
	\begin{equation}\label{eq:model-hrm-flow-stress}
		\mat{\sigma}\colon \Omega^{\b} \to \Reals^{3 \times 3}, \qquad \vec{x} \mapsto -\mu \left(\left(\nabla \vec{u}(\vec{x})\right) + \left(\nabla \vec{u}(\vec{x})\right)^{\T} \right) + p(\vec{x}) \mat{I} 
	\end{equation}
	with dynamic viscosity $\mu > 0$, $p\colon \Omega^{\b} \to \Reals$ is the pressure, $\mat{I}$ is the identity matrix, and $\partial \Omega^{\b} = \Gamma^{\mathrm{in}} \cup \Gamma^{\mathrm{out}} \cup \Gamma^{\mathrm{wall}} \cup \Gamma^{\mathrm{par}}$.

    A no-slip boundary condition is imposed on the column wall and outer particle surfaces.
    We prescribe an inhomogeneous Dirichlet condition at the inlet $\Gamma^{\mathrm{in}}$ using a given inlet velocity profile $\vec{u}^{\mathrm{in}}\colon \Gamma^{\mathrm{in}} \to \Reals^3$ and a homogeneous Neumann condition (vanishing normal stress) at the outlet $\Gamma^{\mathrm{out}}$ \citep{Danckwerts1953}.
    The transport of $N^{\mathrm{c}}$ components in the bulk domain $\Omega^{\b}$ is governed by solute mass balance equations accounting for convection by the flow field $\vec{u}$ and isotropic molecular diffusion:
	\begin{subequations}
	\begin{alignat}{2}
		\frac{\partial c^{\b}_i}{\partial t} + \div{\vec{u} c^{\b}_i - D^{\b}_i \nabla c^{\b}_i} &= 0 & &\qquad \text{in } (0, T^{\mathrm{end}}) \times \Omega^{\b}, \label{eq:model-hrm-bulk} \\
		\left(\vec{u} c^{\b}_i - D^{\b}_i \nabla c^{\b}_i\right) \cdot \vec{n} &= \left(\vec{n} \cdot \vec{u} \right) c^{\mathrm{in}}_{i} & & \qquad \text{on } (0, T^{\mathrm{end}}) \times \Gamma^{\mathrm{in}}, \label{eq:model-hrm-inbc} \\
		-D^{\b}_i \nabla c^{\b}_i \cdot \vec{n} &= 0 & & \qquad \text{on } (0, T^{\mathrm{end}}) \times \left( \Gamma^{\mathrm{wall}} \cup \Gamma^{\mathrm{out}} \right), \label{eq:model-hrm-outbc} \\
		\left. c_i^{\b} \right|_{t = 0} &= c^{\b,\mathrm{init}}_{i} & & \qquad \text{in } \Omega^{\b}, \label{eq:model-hrm-initbc}
	\end{alignat}\label{eq:model-hrm-mtbulk}
	\end{subequations}
	where $c^{\b}_i\colon (0, T^{\mathrm{end}}) \times \Omega^{\b} \to \Reals$, $(t,\vec{x}) \mapsto c^{\b}_i(t, \vec{x})$, is the concentration of component $i \in \{1, \dots, N^{\mathrm{c}}\}$ in the bulk domain, $\vec{u}$ is the flow field from \cref{eq:model-hrm-flow,eq:model-hrm-flowbc}, $D^{\b}_i \geq 0$ is the molecular diffusion coefficient, $c^{\mathrm{in}}_{i}\colon (0, T^{\mathrm{end}}) \times \Gamma^{\mathrm{in}} \to \Reals$ is a given inlet concentration profile, and $c^{\b,\mathrm{init}}_{i} \colon \Omega^{\b} \to \Reals$ is a given initial concentration profile in the bulk domain.

    Inside the porous particles, solutes are transported solely by diffusion, while exchange between the pore and solid phases is governed by binding kinetics. The corresponding solute mass balance equations are defined on $(0, T^{\mathrm{end}}) \times \Omega^{\mathrm{par}}$ and for each component $i\in\{1,\dots,N^\mathrm{c}\}$ and associated bound state $k\in\Naturals^{\leq N^{\mathrm{b}}_i}$, where $N^{\mathrm{b}}_i\geq 0$ denotes the number of bound states of component $i$ as
	\begin{subequations}\label{eq:model-hrm-particle}
	\begin{alignat}{2}
		\frac{\partial c^{\p}_i}{\partial t} &= \div{D^{\p}_i \nabla c^{\p}_i} - \frac{1 - \varepsilon^{\p}}{\varepsilon^{\p}} \sum_{k=1}^{N^{\mathrm{b}}_i} f^{\mathrm{bind}}_{i,k}\mleft( \vec{c}^{\p}, \vec{c}^{\s} \mright), \label{eq:model-hrm-particleliquid} \\
        \frac{\partial c^{\s}_{i,k}}{\partial t} &= \div{D^{\s}_{i,k} \nabla c^{\s}_{i,k}} + f^{\mathrm{bind}}_{i,k}\mleft( \vec{c}^{\p}, \vec{c}^{\s} \mright). \label{eq:model-hrm-particlesolid}
	\end{alignat}
	\end{subequations}
	where $c^{\p}_i, c^{\s}_{i,k}\colon (0, T^{\mathrm{end}}) \times \Omega^{\mathrm{par}} \to \Reals$ are the concentrations in the particle pore and solid phases, respectively, $D^{\p}_i \geq 0$ and $D^{\s}_{i,k} \geq 0$ are the effective pore and surface diffusion coefficients, $\varepsilon^{\p} \in (0,1]$ is the particle porosity, and $f^{\mathrm{bind}}_{i,k}(\vec{c}^{\p}, \vec{c}^{\s})$ is the binding rate, which can depend on the concentrations of all liquid and solid-phase components.
    The boundary condition is given by
    \begin{alignat}{2}
        0 &= - D^{\s}_{i,k} \nabla c^{\s}_{i,k} \cdot \vec{n} & & \qquad \text{on } (0, T^{\mathrm{end}}) \times \Gamma^{\mathrm{par}}. \label{eq:model-hrm-particlebcsolid}
    \end{alignat}
	Initial conditions are given by 
	\begin{subequations}
	\begin{alignat}{2}
		\left. c_i^{\p} \right|_{t = 0} &= c^{\p,\mathrm{init}}_{i} & & \qquad \text{in } \Omega^{\mathrm{par}}, \label{eq:model-hrm-particlebcliquidinit} \\
		\left. c_{i,k}^{\s} \right|_{t = 0} &= c^{\s,\mathrm{init}}_{k} & & \qquad \text{in } \Omega^{\mathrm{par}}, \label{eq:model-hrm-particlebcsolidinit}
	\end{alignat}\label{eq:model-hrm-mtparticleinit}
	\end{subequations}
	where $c^{\p,\mathrm{init}}_{i}, c^{\s,\mathrm{init}}_{k} \colon \Omega^{\mathrm{par}} \to \Reals$ are prescribed initial concentration profiles.
	The particles are surrounded by a thin stagnant fluid film resulting from the no-slip condition \cref{eq:model-hrm-noslipbc}. This film introduces an external mass transfer resistance.
	Additional resistance arises from the finite size and accessibility of the pore entrances.
	\Needspace{9\baselineskip}
	The combined effect of these transport resistances is modeled by a film mass transfer coefficient $k^{\mathrm{f}}_{i} \geq 0$, hereafter referred to as the film diffusion coefficient, and Robin boundary conditions
	\begin{subequations}\label{eq:model-hrm-mtcoupling}
	\begin{alignat}{2}
		\left( \vec{u} c^{\b}_i - D^{\b}_i \nabla c^{\b}_i \right) \cdot \vec{n}^{\b} &= k^{\mathrm{f}}_{i} \left( c^{\b}_i - c^{\p}_i \right) & & \qquad \text{on } (0, T^{\mathrm{end}}) \times \Gamma^{\mathrm{par}}, \label{eq:model-hrm-bulkparticlebc} \\
		-\varepsilon^{\p} D^{\p}_i \nabla c^{\p}_i \cdot \vec{n}^{\p} &= k^{\mathrm{f}}_{i} \left( c^{\p}_i - c^{\b}_i \right) & & \qquad \text{on } (0, T^{\mathrm{end}}) \times \Gamma^{\mathrm{par}}, \label{eq:model-hrm-particlebulkbc}
	\end{alignat}
	\end{subequations}
	where $\vec{n}^{\b}$ is the unit normal vector pointing from $\Omega^{\b}$ into $\Omega^{\mathrm{par}}$ and $\vec{n}^{\p} = -\vec{n}^{\b}$.
\end{model}

\Needspace{6\baselineskip}
\begin{remark}[(HRa)]
For mathematical soundness, we assume that the bulk domain $\Omega^{\b}$ and the (joint) particle domain $\Omega^{\mathrm{par}}$ are bounded open subsets of $\Reals^3$ with piecewise smooth boundaries.
We also consider a finite end time $T^{\mathrm{end}} > 0$, which may be chosen arbitrarily.
Naturally, we assume that $\vec{u} \cdot \vec{n} < 0$ at the inlet cross section $\Gamma^{\mathrm{in}}$ and $\vec{u} \cdot \vec{n} > 0$ at the outlet $\Gamma^{\mathrm{out}}$, where $\vec{n}$ denotes the outward-facing unit normal and $\vec{u}$ is the fluid velocity field.
\end{remark}

\begin{remark}[(HRb)]\label{rem:model-highres}
	\Cref{model:column-highres3D} differs from the model published by \cite{Puttmann2014} in several aspects:
	\begin{itemize}\setlength{\itemsep}{0\baselineskip}
		\item We apply Danckwerts boundary conditions in \cref{eq:model-hrm-inbc,eq:model-hrm-outbc}, which are introduced by \cite{Danckwerts1953}.
			In contrast, \cite{Puttmann2014} employ a Dirichlet inlet condition instead of \cref{eq:model-hrm-inbc} and the identical outlet condition \cref{eq:model-hrm-outbc}.
		\item In the mass transfer boundary condition \cref{eq:model-hrm-bulkparticlebc} for $\Gamma^{\mathrm{par}}$, we use the full flux $\vec{u} c^{\b}_i - D^{\b}_i \nabla c^{\b}_i$ instead of just $-D^{\b}_i \nabla c^{\b}_i$.
			This is equivalent since $\vec{u} = \vec{0}$ on $\Gamma^{\mathrm{par}}$ due to \cref{eq:model-hrm-noslipbc}.
		\item Our model takes surface diffusion into account.
			Contrary to \citet[Section~6.2.6]{Seidel-Morgenstern2020-Chap6}, boundary condition \cref{eq:model-hrm-particlebcsolid} prevents a direct flux between the bulk domain and the particle solid phase.
			The absence of this flux can be problematic when surface diffusion dominates, particularly in the absence of a true parallel diffusion model \citep{Ma1996}.
	\end{itemize}
	In addition, we observe the following points:
	\begin{itemize}\setlength{\itemsep}{0\baselineskip}
		\item The bulk-particle boundary condition \cref{eq:model-hrm-bulkparticlebc} does not enforce continuity of solute concentration at the interface (i.e., in general, $c^{\b}_i \neq c^{\p}_i$ on $\Gamma^{\mathrm{par}}$).
			This effect vanishes when film diffusion does not limit mass transfer, that is, in the limit $k^{\mathrm{f}}_{i} \to \infty$.
		\item Enforcing $\vec{u} = \vec{0}$ on $\Gamma^{\mathrm{par}}$ gives rise to a stagnant film around the particles.
			The mass transfer equations remain invariant if the condition of zero normal velocity $\vec{u} \cdot \vec{n} = 0$ is imposed on $\Gamma^{\mathrm{par}}$ instead.
		\item The above formulation also allows for spatially varying diffusion coefficients $D^{\b}_i$, $D^{\p}_i$, and $D^{\s}_i$ as well as spatially varying film diffusion coefficients $k^{\mathrm{f}}_{i}$.
			Furthermore, we could replace the isotropic diffusivities by anisotropic ones using tensor fields for $D^{\b}_i$, $D^{\p}_i$, and $D^{\s}_i$.
		\item Non-ideal fluid distribution in the frits \citep{Yuan1999} can be modeled by appropriate profiles of $\vec{u}^{\mathrm{in}}$ and $c^{\mathrm{in}}_{i}$.
			This acts as an additional source of radial inhomogeneity besides the structure of the spherical packing.
		\item Even though the particles are assumed to be homogeneous, the concentration distribution inside a particle is generally not symmetric.
			This may, for example, be caused by locally non-uniform solute concentrations in the bulk domain, in particular near contact points between neighboring particles.
            Asymmetry inside a particle can also arise because concentration fronts in the bulk domain reach the upstream side of a particle slightly earlier than its downstream side \citep{Puttmann2014}.
		\item The model is not restricted to monodisperse packings; that is, the particles are not required to have the same radius.
			A particle radius distribution can be used when generating the spherical packing, which, for this model, is encoded in the geometry $\Omega^{\b}$ and $\Omega^{\mathrm{par}}$.
			The required geometry can either be generated in~silico, for example using methods described in \cite{Boccardo2015} and \cite{Wachs2012}, or obtained experimentally from a real packed column, for example by X-ray tomography \citep{Johnson2017,Astrath2007}.
		\item The particles are not required to be spherical. 
			Their shape can be arbitrary and multiple shapes may also be mixed within the same packing.
			The difficulty of generating a realistic packing and constructing a suitable mesh for discretization varies with particle shape, though.
		\item Core-shell particles can be modeled by placing a solid core inside the particle, which serves as an internal domain boundary.
			On this boundary, we apply no-flux Neumann boundary conditions
			\begin{equation*}
				-\varepsilon^{\p} D^{\p}_i \nabla c^{\p}_i \cdot \vec{n} = 0, \qquad 
				-\left(1 - \varepsilon^{\p}\right) D^{\s}_{i,k} \nabla c^{\s}_{i,k} \cdot \vec{n} = 0.
			\end{equation*}
		\item The assumption of a stationary flow field $\vec{u}$ in \cref{asmpt:model-hrm} can be relaxed by accounting for the fluid density $\rho$ and replacing \cref{eq:model-hrm-flow-momentum} with
			\begin{equation*}
				\rho \frac{\partial \vec{u}}{\partial t} + \vecnabla \cdot \mat{\sigma} = \vec{0} \qquad \text{in } \Omega^{\b}.
			\end{equation*}
	\item An alternative approach, briefly mentioned by \cite{Puttmann2015} and implemented by \cite{Gerontas2013}, is to replace the Stokes equations in \cref{eq:model-hrm-flow-momentum} with the (stationary) Navier-Stokes equations.
	\end{itemize}
\end{remark}

Further model variants, including the limiting cases of infinite (non-limiting) film diffusion and infinite (non-limiting) binding kinetics, are discussed in Section~\ref{supp:sec:models-hrm-variants}.

\subsection{Bulk Transport}
\label{sec:bulk_domain_transport}

Resolving microscale phenomena described by the \gls{hrm} is often impractical due to the associated data requirements and computational cost.
Homogenized macroscale models are therefore used in practice. In the following, we derive these models from the \gls{hrm} using formal model-order reduction.

To modularize the reduced-order modeling framework, we treat mass transport in the bulk domain (\cref{sec:bulk_domain_transport}) and the particle domain (\cref{sec:intra_particle_transport}) as separate, though physically coupled, submodels.
Each submodel receives boundary data either from the complementary submodel or from external sources.
This modular approach allows mass transport in the two domains to be simulated independently, provided the necessary interface data are available.

In this section, we consider mass transport in the bulk domain.
The particle liquid-phase concentration $c^{\p}_i$ at the particle surface can be defined by a particle model (\cref{sec:intra_particle_transport}) or supplied externally as boundary data.
Model variants for alternative column geometries and rapid-equilibrium assumptions exist for all transport models presented in this section and are thus described together in \cref{sec:model_variants}.

\subsubsection{3D Bulk Model (Model B3)}
\label{sec:3D_continuum_bulk_transport_model}

\begin{assumption}[(B3)]\label{asmpt:model-3D}
    In addition to \cref{asmpt:model-hrm}, we assume that
    \begin{itemize}\setlength{\itemsep}{0\baselineskip}
        \item The particles can be treated as a continuum inside the column (i.e., both bulk and particle phases are present at every point in the column);
        \item All solute concentrations ($c^\b, c^\p, c^\s$) are spatially constant over each particle surface;
        \item Particle heterogeneity can be represented by a \gls{ptd} consisting of a finite number $N^{\mathrm{p}}$ of particle types with different radii, diffusion coefficients, binding models and/or parameters;
        \item The fluid flows only in the axial direction of the column;
    \end{itemize}
\end{assumption}
The first point of \cref{asmpt:model-3D} describes a homogenization of the bulk and particle phases.
Instead of resolving the packed-bed geometry, the bulk domain is treated as a continuum, where every point in the column represents a mixture of both phases.
Consequently, spatially constant concentrations at the particle surfaces are required to define a consistent interface between the bulk and particle phases.
The third point results in the discretization of particle properties into a finite number of particle types.

Having reduced the problem, the unresolved structural features and transport phenomena must be accounted for in the \gls{rom} \citep[Section~2.1.2]{Guiochon2006}:
\begin{itemize}\setlength{\itemsep}{0\baselineskip}
	\item Random packing inhomogeneities (e.g., local particle agglomeration) that can hinder the flow,
	\item Wall-induced packing inhomogeneities that result in local variations in packing density and void fraction,
	\item Hydrodynamic dispersion arising from multiple flow paths (often referred to as eddy diffusion, although no eddies occur in the fluid-dynamical sense),
	\item Reduced effective molecular diffusion due to bed tortuosity (e.g., twists in the channels between particles).
\end{itemize}
Of these four, the wall effect stands out because it is confined to a region extending only a few particle diameters from the column wall \citep{Bruns2013}, whereas the other effects occur throughout the column.
The effects are additive~\citep[Section~2.3.4]{Seidel-Morgenstern2020-Chap2} and all contribute to the total band broadening.

Contrary to \cref{model:column-highres3D}, we introduce a coordinate system for the following models, which is also customary in the literature.
Considering a column of length $L^\mathrm{b} > 0$ and radius $\Rho^{\mathrm{c}} > 0$, we choose cylindrical coordinates (see~\ref{supp:sec:vectorcalculus-cylindrical}) for the bulk domain with axial coordinate $z \in [0, L^\mathrm{b}]$, radial coordinate $\rho \in [0, \Rho^{\mathrm{c}}]$, angular coordinate $\varphi \in [0, 2 \pi )$, and hence $\Omega^\b \equiv (0, L^\mathrm{b}) \times (0, \Rho^{\mathrm{c}}) \times [0, 2 \pi )$.

The effects on flow (and, consequently, on mass transport) in the bulk domain are accounted for by axial, radial, and angular dispersion with coefficients $D^{\mathrm{ax}}_{i}, D^{\mathrm{rad}}_{i}, D^{\mathrm{ang}}_{i} \geq 0$, respectively.
The effects of volume averaging and packing inhomogeneities are taken into account by a bulk porosity coefficient $\varepsilon^{\mathrm{b}} \in (0,1]$.
To account for spatial inhomogeneities, we allow the four coefficients, as well as the superficial velocity $u$, to depend on the coordinates.
That is, we have $D^{\mathrm{ax}}_{i}, D^{\mathrm{rad}}_{i}, D^{\mathrm{ang}}_{i} \colon \Omega^\b \to [0, \infty)$, $\varepsilon^{\mathrm{b}}\colon \Omega^\b \to (0,1]$, and $u\colon (0, \Rho^{\mathrm{c}}) \times [0, 2 \pi ) \to [0, \infty)$, where $u$ is assumed to be independent of the axial coordinate $z$ due to the incompressibility of the fluid (see \cref{asmpt:model-hrm} and \cref{eq:model-3d-incompress} below).
In the following derivation, we first apply volume averaging to the spatially resolved model and then exploit \cref{asmpt:model-3D} to simplify and reduce \cref{model:column-highres3D}.

We consider a discrete distribution of $N^{\mathrm{p}} \in \Naturals$ particle types with volume fractions $d_j \colon \Omega^\b \to [0, 1]$, $j \in \{1, \dots, N^{\mathrm{p}}\}$.
The volume fraction $d_j$ describes the proportion of the total particle volume occupied by type $j$.
Naturally, we require
\begin{equation*}
	\sum_{j=1}^{N^{\mathrm{p}}} d_j(z, \rho, \phi) = 1, \qquad \forall\ (z,\rho,\phi) \in \Omega^\b.
\end{equation*}
Throughout, $N^{\mathrm{p}} = 0$ denotes a particle-free domain: all sums over particle types are then empty, so the above normalization and every film diffusion term vanish identically.
If particles differ only in size, the types correspond to bins of a discrete \gls{psd}, with bin edges $\{R_{j\pm \frac{1}{2}}\}_{j=1}^{N^p}$ where $R_{\frac{1}{2}}=0, R_{N^p+\frac{1}{2}}=\infty$. Then, the volume fraction of type $j$ is given by
\begin{equation*}
    \int_{R_{j-\frac{1}{2}}}^{R_{j+\frac{1}{2}}} f(R; z,\rho,\phi) \mathrm{d}R = d_j(z,\rho,\phi),
\end{equation*}
where $f(R; z,\rho,\phi)\colon \Reals^+ \times \Omega^\b \to \Reals^+$ is the volume-based particle size density function.

We begin the derivation of the continuum bulk transport model by applying volume averaging theory, following the framework of \cite{Whitaker1999} (see Section~\ref{supp:sec:volaverages}).
Each point $\vec{y} \in \Omega^{\b} \cup \Omega^{\mathrm{par}}$ of the column domain is associated with a representative (i.e., containing parts of both bulk and particle phases), simply connected open averaging volume $V(\vec{y}) \subset \Omega^{\b} \cup \Omega^{\mathrm{par}}$ with piecewise smooth closed boundary $\partial V(\vec{y})$ and centroid $\vec{y}$.
This volume $V(\vec{y})$ is referred to as the \gls{rev} and its size may vary depending on the position $\vec{y}$ (e.g., it can be smaller near the column wall).
Since we focus on a single fixed point $\vec{y}$, we suppress this dependence and simply write $V$.
Conceptually, $V$ can be thought of as an open ball of some radius, which would be a reasonable choice, but it could also be an open cube or any other (convenient) shape.
Taking the volume average $\avg{\cdot}$ over $V$ of \cref{eq:model-hrm-bulk} yields
\begin{equation}
	\frac{1}{\vol{V}}\int_{V^{\b}} \frac{\partial c^{\b}_i}{\partial t}(t, \vec{x}) \,\mathrm{d}\vec{x} +  \frac{1}{\vol{V}} \int_{V^{\b}} \div{\vec{u}(\vec{x}) c^{\b}_i(t, \vec{x}) - D^{\b}_i(\vec{x}) \nabla c^{\b}_i(t, \vec{x})} \,\mathrm{d}\vec{x} = 0. \label{eq:model-3d-volavg1}
\end{equation}
Using Definition~\ref{supp:def:VolumeAverages}, we rewrite \cref{eq:model-3d-volavg1} as
\begin{equation}
	\avg{\frac{\partial c^{\b}_i}{\partial t}} + \avg{ \div{\vec{u} c^{\b}_i - D^{\b}_i \nabla c^{\b}_i} } = 0. \label{eq:model-3d-volavg2}
\end{equation}
Since the particles are assumed to be fixed in space, we interchange differentiation and integration of the first term and arrive at
\begin{equation}
	\frac{\partial \avg{ c^{\b}_i }}{\partial t} + \avg{ \div{\vec{u} c^{\b}_i - D^{\b}_i \nabla c^{\b}_i} } = 0. \label{eq:model-3d-volavg3}
\end{equation}

\begin{figure}[!htbp]
	\centering
	\begin{tikzpicture}

	\fill[bulkVolume,name path=v] (0.1,0.1) circle[radius=2cm];

	\begin{scope}
		\clip (3.2,3.4) rectangle (-2.55,-3.1);
		\filldraw[thick,draw=black,fill=white,name path=c1] (2,0) circle[radius=1cm] coordinate (C1);
		\filldraw[very thick,bndAvgVolBead,fill=white,name path=c2] (-0.25,-0.75) circle[radius=0.8cm] coordinate (C2);
		\filldraw[thick,draw=black,fill=white,name path=c3] (-1.45,0.65) circle[radius=0.9cm] coordinate (C3);
		\filldraw[thick,draw=black,fill=white,name path=c4] (0.5,2) circle[radius=1.2cm] coordinate (C4);
		\filldraw[thick,draw=black,fill=white,name path=c5] (1.1,-2) circle[radius=0.9cm] coordinate (C5);

		\filldraw[thick,draw=black,fill=white] (-2,2.9) circle[radius=1.2cm];
		\filldraw[thick,draw=black,fill=white] (-1.9,-2.8) circle[radius=1.4cm];
		\filldraw[thick,draw=black,fill=white] (3.5,-3.8) circle[radius=1.8cm];
		\filldraw[thick,draw=black,fill=white] (2.9,2.4) circle[radius=0.9cm];

		\path[name intersections={of=v and c1, name=i}] coordinate (v1-1) at (i-1);
		\path[name intersections={of=v and c1, name=i}] coordinate (v1-2) at (i-2);
		\path[name intersections={of=v and c3, name=i}] coordinate (v3-1) at (i-1);
		\path[name intersections={of=v and c3, name=i}] coordinate (v3-2) at (i-2);
		\path[name intersections={of=v and c4, name=i}] coordinate (v4-1) at (i-1);
		\path[name intersections={of=v and c4, name=i}] coordinate (v4-2) at (i-2);
		\path[name intersections={of=v and c5, name=i}] coordinate (v5-1) at (i-1);
		\path[name intersections={of=v and c5, name=i}] coordinate (v5-2) at (i-2);

		\path pic[draw,bndAvgVolBead,very thick,angle radius=1cm] {angle=v1-1--C1--v1-2};
		\path pic[draw,bndAvgVolBead,very thick,angle radius=0.9cm] {angle=v3-2--C3--v3-1};
		\path pic[draw,bndAvgVolBead,very thick,angle radius=1.2cm] {angle=v4-2--C4--v4-1};
		\path pic[draw,bndAvgVolBead,very thick,angle radius=0.9cm] {angle=v5-1--C5--v5-2};

		\draw[dashed,very thick] (0.1,0.1) circle[radius=2cm];

		\path coordinate (V) at (0.1,0.1);
		\path pic[draw,bndAvgVolEE,very thick,angle radius=2cm] {angle=v4-2--V--v3-1};
		\path pic[draw,bndAvgVolEE,very thick,angle radius=2cm] {angle=v1-1--V--v4-1};
		\path pic[draw,bndAvgVolEE,very thick,angle radius=2cm] {angle=v5-1--V--v1-2};
		\path pic[draw,bndAvgVolEE,very thick,angle radius=2cm] {angle=v3-2--V--v5-2};
	\end{scope}

	\draw[dashed] (3.2,3.4) rectangle (-2.55,-3.1);

	\filldraw[black] (0.1,0.1) circle[radius=0.05cm] node[above,left,yshift=0.15cm,xshift=0.05cm] (y) {$y$};

	\node[xshift=0.5cm,yshift=0.3cm] at (y.east) {$V_{\mathrm{int}}$};

	\node (GammaBead) at (0.35,1.65) {$\Gamma_{\mathrm{bb}}$};
	\draw[thick,bndAvgVolBead] (GammaBead.south west) -- +(-0.25,-0.25);

	\node (GammaBe) at (-2.2,-0.9) {$\Gamma_{\mathrm{be}}$};
	\draw[thick,bndAvgVolEE] (GammaBe.east) -- +(0.19,0); 

	\node[right] at (2.1,0.1) {$V(y)$};
\end{tikzpicture}
	\caption[Representative elementary volume with boundaries]{Representative elementary volume $V(\vec{y})$ centered at $\vec{y}$ showing the bulk-particle interface $\Gamma^{\mathrm{bb}}$, bulk-phase portion $V^{\b}$, and entry and exit boundaries $\Gamma^{\mathrm{be}}$} \label{fig:model-3d-volavg}
\end{figure}
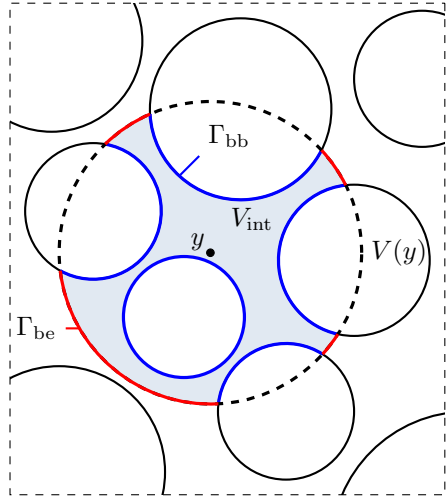

In order to apply the spatial averaging theorem (see Theorem~\ref{supp:thm:SpatialAveraging}) to the second term, we define the bulk volume portion of $V$ by $V^{\b} \coloneqq V \cap \Omega^{\b}$.
Furthermore, we split up the boundary $\partial V^{\b}$ into the disjoint parts $\Gamma^{\mathrm{be}} \coloneqq \partial V^{\b} \setminus (\partial V^{\b} \cap \Gamma^{\mathrm{par}})$, which are the entries and exits to the rest of the bulk domain, and $\Gamma^{\mathrm{bb}} \coloneqq \partial V^{\b} \cap \Gamma^{\mathrm{par}}$, which are the particle surfaces contained in $V$. 
The situation is depicted in \cref{fig:model-3d-volavg}, where $V(\vec{y})$ is an open ball.

Applying the spatial averaging theorem and using the boundary condition (\ref{eq:model-hrm-bulkparticlebc}) on $\Gamma^{\mathrm{bb}}$, we obtain 
\begingroup \allowdisplaybreaks[0]
\begin{align} \allowdisplaybreaks[0]
    &\avg{ \div{\vec{u} c^{\b}_i - D^{\b}_i \nabla c^{\b}_i} }(\vec{y}) \nonumber \\
    =\ & \nabla_{\vec{y}} \cdot \avg{ \vec{u} c^{\b}_i - D^{\b}_i \nabla c^{\b}_i }(\vec{y}) + \frac{1}{\vol{V(\vec{y})}} \int_{\Gamma^{\mathrm{bb}}(\vec{y})} \left( \vec{u}(\vec{x}) c^{\b}_i(t,\vec{x}) - D^{\b}_i(\vec{x}) \nabla c^{\b}_i(t,\vec{x}) \right) \cdot \vec{n}(\vec{x}) \,\mathrm{d}\omega(\vec{x})
    \nonumber \\
    =\ & \nabla_{\vec{y}} \cdot \avg{ \vec{u} c^{\b}_i - D^{\b}_i \nabla c^{\b}_i }(\vec{y}) + \frac{1}{\vol{V(\vec{y})}} \int_{\Gamma^{\mathrm{bb}}(\vec{y})} k^{\mathrm{f}}_{i} \left(c^{\b}_i(t,\vec{x}) - c^{\p}_i(t,\vec{x}) \right) \,\mathrm{d}\omega(\vec{x}). \label{eq:model-3d-bulk}
\end{align}
\endgroup
For the sake of clarity, we have added the subscript $\vec{y}$ to indicate that the gradient or divergence of an averaged quantity is taken with respect to the position $\vec{y}$ of the averaging volume $V(\vec{y})$ (as detailed in Theorem~\ref{supp:thm:SpatialAveraging}).
We split the average in the argument of the divergence operator and apply the spatial averaging theorem to the gradient $\nabla c^{\b}_i$:
\begin{align}
	&\nabla_{\vec{y}} \cdot \avg{ \vec{u} c^{\b}_i - D^{\b}_i \nabla c^{\b}_i }(\vec{y}) \nonumber \\
	=\ & \nabla_{\vec{y}} \cdot \left( \avg{\vec{u} c^{\b}_i }(\vec{y}) - \avg{D^{\b}_i \nabla c^{\b}_i }(\vec{y}) \right) \nonumber \\
	=\ & \nabla_{\vec{y}} \cdot \left( \vec{u}(\vec{y}) \avg{ c^{\b}_i }(\vec{y}) - D^{\b}_i(\vec{y}) \nabla_{\vec{y}} \avg{ c^{\b}_i }(\vec{y}) - \frac{D^{\b}_i}{\vol{V(\vec{y})}} \int_{\Gamma^{\mathrm{bb}}(\vec{y})} c^{\b}_i(t,\vec{x}) \vec{n}(\vec{x}) \,\mathrm{d}\omega(\vec{x}) \right), \label{eq:model-3d-volavg4}
\end{align}
where we have assumed that $V$ is small enough such that the superficial velocity $\vec{u}$ and the diffusivity $D^{\b}_i$ are approximately constant in $V$.
The surface integral on the right-hand side of \cref{eq:model-hrm-bulk} is rewritten as
\begin{equation}
    \frac{1}{\vol{V(\vec{y})}} \int_{\Gamma^{\mathrm{bb}}(\vec{y})} k^{\mathrm{f}}_{i}(\vec{x}) \left(c^{\b}_i(t,\vec{x}) - c^{\p}_i(t,\vec{x}) \right) \,\mathrm{d}\omega(\vec{x}) = \frac{\surfarea{\Gamma^{\mathrm{bb}}(\vec{y})}}{\vol{V(\vec{y})}} \avg{k^{\mathrm{f}}_{i} \left( c^{\b}_i - c^{\p}_i \right) }_{\!\omega}(\vec{y}), \label{eq:model-3d-volavg5}
\end{equation}
where $\omega$ denotes the surface measure and
\begin{equation*}
    \avg{k^{\mathrm{f}}_{i} \left( c^{\b}_i - c^{\p}_i \right) }_{\!\omega}(\vec{y}) \coloneqq \frac{1}{\surfarea{\Gamma^{\mathrm{bb}}(\vec{y})}} \int_{\Gamma^{\mathrm{bb}}(\vec{y})} k^{\mathrm{f}}_{i}(\vec{x}) \left( c^{\b}_i(t,\vec{x}) - c^{\p}_i(t,\vec{x})\right) \,\mathrm{d}\omega(\vec{x})
\end{equation*}
is the surface average.

Up to this point, we have computed (phase) averages with respect to the full volume of $V$.
However, we are interested in the (intrinsic) averages with respect to the volume of $V^{\b} \subset V$ (see Definition~\ref{supp:def:VolumeAverages}).
The conversion is achieved by inserting the relations
\begin{equation*}
    \varepsilon^{\mathrm{m}}(\vec{y}) \coloneqq \frac{\vol{V^{\b}(\vec{y})}}{\vol{V(\vec{y})}}, \qquad \avg{ c^{\b}_i }^{\varepsilon}(\vec{y}) \coloneqq \frac{1}{\vol{V^{\b}(\vec{y})}} \int_{V^{\b}(\vec{y})} c^{\b}_i(t,\vec{x}) \,\mathrm{d}\vec{x}, \qquad \avg{ c^{\b}_i } = \varepsilon^{\mathrm{m}} \avg{ c^{\b}_i }^{\varepsilon},
\end{equation*}
into the combination of \cref{eq:model-3d-volavg3,eq:model-3d-volavg4,eq:model-3d-volavg5}, where $\varepsilon^{\mathrm{m}}$ is the theoretical local porosity:\enlargethispage{1\baselineskip}
\begin{multline}\label{eq:model-3d-volavg6}
    \varepsilon^{\mathrm{m}} \frac{\partial \avg{ c^{\b}_i }^{\varepsilon}}{\partial t} + \nabla_{\vec{y}} \cdot \Biggl( \vphantom{\int_{\Gamma^{\mathrm{bb}}(\vec{y})}} \vec{u}(\vec{y}) \varepsilon^{\mathrm{m}}(\vec{y}) \avg{ c^{\b}_i }^{\varepsilon}(\vec{y}) - D^{\b}_i(\vec{y}) \nabla_{\vec{y}} \left( \varepsilon^{\mathrm{m}}(\vec{y}) \avg{ c^{\b}_i }^{\varepsilon}(\vec{y}) \right) 
    \\
     \qquad \qquad  - \frac{D^{\b}_i(\vec{y})}{\vol{V(\vec{y})}} \int_{\Gamma^{\mathrm{bb}}(\vec{y})} c^{\b}_i(t,\vec{x}) \vec{n}(\vec{x}) \,\mathrm{d}\omega(\vec{x}) \Biggr)
    + \frac{\surfarea{\Gamma^{\mathrm{bb}}}}{\vol{V}} \avg{ k^{\mathrm{f}}_{i} \left( c^{\b}_i - c^{\p}_i \right) }_{\!\omega} = 0. 
\end{multline}
Since the porosity $\varepsilon^{\mathrm{m}}$ also depends on the position $\vec{y}$ of $V$, we can use the product rule to expand the gradient in the argument of the divergence operator:
\begin{multline}
	\varepsilon^{\mathrm{m}} \frac{\partial \avg{ c^{\b}_i }^{\varepsilon}}{\partial t} + \nabla_{\vec{y}} \cdot \Biggl( \vphantom{\int_{\Gamma^{\mathrm{bb}}(\vec{y})}} \vec{u}(\vec{y}) \varepsilon^{\mathrm{m}}(\vec{y}) \avg{ c^{\b}_i }^{\varepsilon}(\vec{y}) - D^{\b}_i(\vec{y}) \varepsilon^{\mathrm{m}}(\vec{y}) \nabla_{\vec{y}} \avg{ c^{\b}_i }^{\varepsilon}(\vec{y})  \\
	  \qquad \qquad - D^{\b}_i(\vec{y}) \avg{ c^{\b}_i }^{\varepsilon}(\vec{y}) \nabla_{\vec{y}} \varepsilon^{\mathrm{m}}(\vec{y}) - \frac{D^{\b}_i(\vec{y})}{\vol{V(\vec{y})}} \int_{\Gamma^{\mathrm{bb}}(\vec{y})} c^{\b}_i(t,\vec{x}) \vec{n}(\vec{x}) \,\mathrm{d}\omega(\vec{x}) \Biggr)\\
	+ \frac{\surfarea{\Gamma^{\mathrm{bb}}}}{\vol{V}} \avg{ k^{\mathrm{f}}_{i} \left( c^{\b}_i - c^{\p}_i \right) }_{\!\omega} = 0. \label{eq:model-3d-volavg7}
\end{multline}
\citet[Sec.~1.3--1.4]{Whitaker1999} argues that the expression
\begin{equation*}
    D^{\b}_i(\vec{y}) \left( \avg{ c^{\b}_i }^{\varepsilon}(\vec{y}) \nabla_{\vec{y}} \varepsilon^{\mathrm{m}}(\vec{y}) + \frac{1}{\vol{V(\vec{y})}} \int_{\Gamma^{\mathrm{bb}}(\vec{y})} c^{\b}_i(t,\vec{x}) \vec{n}(\vec{x}) \,\mathrm{d}\omega(\vec{x}) \right)
\end{equation*}
in the argument of the divergence operator in \cref{eq:model-3d-volavg7} produces higher order contributions that can be lumped together with $D^{\b}_i$ into an effective diffusivity tensor $\mat{D}^{\b,\mathrm{eff}}_{i}$, which depends on the position $\vec{y}$ and the shape of $V$.
Summarizing, we arrive at the volume averaged form
\begin{multline}
    \varepsilon^{\mathrm{m}} \frac{\partial \avg{ c^{\b}_i }^{\varepsilon}}{\partial t} + \nabla_{\vec{y}} \cdot \left( \varepsilon^{\mathrm{m}}(\vec{y}) \vec{u}(\vec{y}) \avg{ c^{\b}_i }^{\varepsilon}(\vec{y}) - \varepsilon^{\mathrm{m}}(\vec{y}) \mat{D}^{\b,\mathrm{eff}}_{i}(\vec{y}) \nabla_{\vec{y}} \avg{ c^{\b}_i }^{\varepsilon}(\vec{y}) \right)
    \\
    + \left( 1 - \varepsilon^{\mathrm{m}} \right) \frac{\surfarea{\Gamma^{\mathrm{bb}}}}{\vol{V^{\mathrm{par}}}} \avg{ k^{\mathrm{f}}_{i} \left( c^{\b}_i - c^{\p}_i \right) }_{\!\omega} = 0. \label{eq:model-3d-volavg8}
\end{multline}

\begin{remark}[(B3a)]
    \citet[Sec.~1.4.5]{Whitaker1999} showed that film diffusion at the bulk-particle interface gives rise to an additional convective term
    \begin{equation*}
        \nabla_{\vec{y}} \cdot \left(\varepsilon^{\mathrm{m}}(\vec{y}) \vec{u}^{\mathrm{fd}}_{i}(\vec{y})\, \avg{ c^{\b}_i }^{\varepsilon}(\vec{y})\right)
    \end{equation*}
    on the left-hand side of \cref{eq:model-3d-volavg8}.
    The associated velocity field $\vec{u}_{\mathrm{fd},i}$ depends on the position $\vec{y}$ and is proportional to $k^{\mathrm{f}}_{i}$.
    For purely diffusional problems, this contribution is small and can be neglected~\citep[Sec.~1.4.5]{Whitaker1999}.
    In systems with significant convective transport, however, the term can become relevant, as reported by \cite{Paine1983}.
    Although chromatography models are usually convection dominated, we neglect this contribution.
    Since the film diffusion coefficient $k^{\mathrm{f}}_{i}$ is typically several orders of magnitude smaller than the (average) superficial velocity $\vec{u}$, the induced convective transport is correspondingly small. Neglecting this term therefore constitutes a leading-order approximation and is justified under the conditions considered here.
\end{remark}

We now apply \cref{asmpt:model-3D}.
This allows us to take volume averages with respect to arbitrarily small volumes $V$ (i.e., in the limit $\vol{V} \to 0$ with $V$ concentrating on the point $\vec{y}$), under the assumption of a locally homogeneous continuum.
In conclusion, we can simplify the notation by dropping the volume averages in \cref{eq:model-3d-volavg8} and replacing the local porosity $\varepsilon^{\mathrm{m}}$ with the macroscale porosity $\varepsilon^{\mathrm{b}}$.
This results in
\begin{equation}
	\varepsilon^{\mathrm{b}} \frac{\partial c^{\b}_i}{\partial t} + \nabla_{\vec{y}} \cdot \left(\varepsilon^{\mathrm{b}} \vec{u} c^{\b}_i - \varepsilon^{\mathrm{b}} \mat{D}^{\b,\mathrm{eff}}_{i} \nabla c^{\b}_i \right)
	+ \left( 1 - \varepsilon^{\mathrm{b}} \right) \frac{\surfarea{\Gamma^{\mathrm{bb}}}}{\vol{V^{\mathrm{par}}}} \avg{ k^{\mathrm{f}}_{i} \left( c^{\b}_i - c^{\p}_i \right) }_{\!\omega} = 0, \label{eq:model-3d-volavg9}
\end{equation}
where $c_i^{\b}, \avg*{ k^{\mathrm{f}}_{i} ( c^{\b}_i - c^{\p}_i) }_{\omega} \colon (0, T^{\mathrm{end}}) \times (0, \Rho^{\mathrm{c}}) \times (0, L^\mathrm{b}) \times [0, 2\pi ) \to \Reals$ are now functions of time $t$, radial position $\rho$, axial position $z$, and angular position $\varphi$.

In the continuum of bulk and particle phases, we consider a finite number of particle types $N^{\mathrm{p}}$ indexed by $j \in \{1, \dots, N^{\mathrm{p}}\}$.
Thus, in order to compute the total surface area that particles of size $R^{\mathrm{p}}_{j}$ contribute to $\Gamma^{\mathrm{bb}}$, we compute the number of particles and multiply this by the surface area of a sphere with the respective size:
\begin{equation}
    \label{eq:model-3d-surface-measure-sphere}
    \surfarea{\Gamma^{\mathrm{bb}}} = \sum_{j=1}^{N^{\mathrm{p}}} \frac{\vol{V^{\mathrm{par}}} d_j}{\frac{4}{3}\pi \left(R^{\mathrm{p}}_{j}\right)^3} 4\pi \left(R^{\mathrm{p}}_{j}\right)^2 = \vol{V^{\mathrm{par}}} \sum_{j=1}^{N^{\mathrm{p}}} \frac{3d_j}{R^{\mathrm{p}}_{j}}.
\end{equation}
Substituting this into the last term of \cref{eq:model-3d-volavg9} and assuming
\begin{equation*}
	\mat{D}^{\b,\mathrm{eff}}_{i}(\vec{y}) = \mat{D}_{\mathrm{disp},i} \coloneqq D^{\mathrm{ax}}_{i} \vec{e}_z \otimes \vec{e}_z + D^{\mathrm{rad}}_{i} \vec{e}_{\rho} \otimes \vec{e}_{\rho} + D^{\mathrm{ang}}_{i} \vec{e}_{\varphi} \otimes \vec{e}_{\varphi},
\end{equation*}
we obtain the counterpart of the bulk-domain \gls{pde} \eqref{eq:model-hrm-bulk}:
For all $(\rho, z, \varphi) \in (0, \Rho^{\mathrm{c}}) \times (0, L^\mathrm{b})  \times [0, 2\pi)$ and $t \in (0, T^{\mathrm{end}})$, it holds that
\begin{equation}
	\varepsilon^{\mathrm{b}} \frac{\partial c^{\b}_i}{\partial t} + \nabla_{y} \cdot \left(\varepsilon^{\mathrm{b}} \vec{u} c^{\b}_i - \varepsilon^{\mathrm{b}} \mat{D}_{\mathrm{disp},i} \nabla c^{\b}_i \right) + \left(1 - \varepsilon^{\mathrm{b}} \right) \sum_{j=1}^{N^{\mathrm{p}}} \frac{3 d_j}{R^{\mathrm{p}}_{j}} \avg{ k^{\mathrm{f}}_{j,i} \left( c^{\b}_i - c^{\p}_{j,i} \right) }_{\!\omega} = 0. \label{eq:model-3d-bulk-presimple}
\end{equation}

In order to simplify \cref{eq:model-3d-bulk-presimple}, we express the flow field as $\vec{u} = u(\rho, z, \varphi) \vec{e}_z$, where $\vec{e}_z$ denotes the axial unit vector.
Transforming the incompressibility condition \cref{eq:model-hrm-flow-incompressible} into cylindrical coordinates yields
\begin{equation}
	0 = \nabla \cdot \vec{u} = \frac{1}{\rho} \frac{\partial }{\partial \rho} \left( \rho \vec{u} \cdot \vec{e}_{\rho} \right) + \frac{\partial}{\partial z}\left( \vec{u} \cdot \vec{e}_{z} \right) + \frac{1}{\rho} \frac{\partial }{\partial \varphi} \left( \vec{u} \cdot \vec{e}_{\varphi} \right) = 0 + \frac{\partial u}{\partial z}(\rho, z) + 0 = \frac{\partial u}{\partial z}(\rho, z, \varphi). \label{eq:model-3d-incompress}
\end{equation}
This means that $\vec{u}$ and, hence, $u$ are constant in the axial direction for each value of $\rho$ and $\varphi$.
Consequently, we can write $\vec{u} = u(\rho, \varphi) \vec{e}_z$.

Using the preparations above, we split the linear divergence operator in \cref{eq:model-3d-bulk-presimple} and use vector calculus identities (see Section~\ref{supp:sec:vectorcalculus}) to simplify the first term:
\begin{equation}
    \label{eq:model-3d-bulk-convection}
	\div{\varepsilon^{\mathrm{b}} \vec{u} c^{\b}_i} =
    \varepsilon^{\mathrm{b}} c^{\b}_i \left(\nabla \cdot \vec{u}\right) + \left( \nabla \left(\varepsilon^{\mathrm{b}} c^{\b}_i \right) \right) \cdot \vec{u}
    = 0 + \left( \nabla \left( \varepsilon^{\mathrm{b}} c^{\b}_i \right) \right) \cdot u(\rho, \varphi) \vec{e}_z
    = u(\rho, \varphi) \frac{\partial \left( \varepsilon^{\mathrm{b}} c^{\b}_i \right)}{\partial z}.
\end{equation}
For the second term, we obtain
\begingroup
\medmuskip=1mu plus 1mu minus 1mu
\allowdisplaybreaks
\begin{align}
	\div{\varepsilon^{\mathrm{b}} \mat{D}_{\mathrm{disp},i} \nabla c^{\b}_i}
	&= \div{\varepsilon^{\mathrm{b}} \mat{D}_{\mathrm{disp},i} \left( \frac{\partial c^{\b}_i}{\partial \rho} \vec{e}_{\rho} + \frac{\partial c^{\b}_i}{\partial z} \vec{e}_{z} + \frac{1}{\rho} \frac{\partial c^{\b}_i}{\partial \varphi} \vec{e}_{\varphi} \right)} \nonumber \\
	&= \div{\varepsilon^{\mathrm{b}} \left( D^{\mathrm{ax}}_{i} \vec{e}_z \otimes \vec{e}_z + D^{\mathrm{rad}}_{i} \vec{e}_{\rho} \otimes \vec{e}_{\rho} + D^{\mathrm{ang}}_{i} \vec{e}_\varphi \otimes \vec{e}_\varphi \right) \left( \frac{\partial c^{\b}_i}{\partial \rho} \vec{e}_{\rho} + \frac{\partial c^{\b}_i}{\partial z} \vec{e}_{z} + \frac{1}{\rho} \frac{\partial c^{\b}_i}{\partial \varphi} \vec{e}_{\varphi} \right) } \nonumber \\
	&= \div{\varepsilon^{\mathrm{b}} D^{\mathrm{rad}}_{i}  \frac{\partial c^{\b}_i}{\partial \rho} \vec{e}_{\rho} + \varepsilon^{\mathrm{b}} D^{\mathrm{ax}}_{i} \frac{\partial c^{\b}_i}{\partial z} \vec{e}_{z} + \frac{\varepsilon^{\mathrm{b}} D^{\mathrm{ang}}_{i}}{\rho} \frac{\partial c^{\b}_i}{\partial \varphi} \vec{e}_{\varphi} } \nonumber \\
	&= \frac{1}{\rho} \frac{\partial}{\partial \rho} \left( \rho \varepsilon^{\mathrm{b}} D^{\mathrm{rad}}_{i}  \frac{\partial c^{\b}_i}{\partial \rho} \right) + \frac{\partial}{\partial z} \left( \varepsilon^{\mathrm{b}} D^{\mathrm{ax}}_{i} \frac{\partial c^{\b}_i}{\partial z} \right) + \frac{1}{\rho^2} \frac{\partial}{\partial \varphi} \left( \varepsilon^{\mathrm{b}} D^{\mathrm{ang}}_{i}  \frac{\partial c^{\b}_i}{\partial \varphi} \right). \label{eq:model-3d-bulk-dispersion}
\end{align}
\endgroup
Inserting \cref{eq:model-3d-bulk-convection,eq:model-3d-bulk-dispersion} into \cref{eq:model-3d-bulk-presimple} and rearranging terms, we arrive at the bulk \gls{pde}
\begin{multline}
	\varepsilon^{\mathrm{b}} \frac{\partial c^{\b}_i}{\partial t} = -u \frac{\partial \left( \varepsilon^{\mathrm{b}} c^{\b}_i \right)}{\partial z} + \frac{\partial}{\partial z} \left( \varepsilon^{\mathrm{b}} D^{\mathrm{ax}}_{i} \frac{\partial c^{\b}_i}{\partial z} \right) + \frac{1}{\rho} \frac{\partial}{\partial \rho} \left( \rho \varepsilon^{\mathrm{b}} D^{\mathrm{rad}}_{i} \frac{\partial c^{\b}_i}{\partial \rho} \right) + \frac{1}{\rho^2} \frac{\partial}{\partial \varphi} \left( \varepsilon^{\mathrm{b}} D^{\mathrm{ang}}_{i} \frac{\partial c^{\b}_i}{\partial \varphi} \right) \\
	 - \left(1 - \varepsilon^{\mathrm{b}} \right) \sum_{j=1}^{N^{\mathrm{p}}} \frac{3 d_j}{R^{\mathrm{p}}_{j}} \avg{ k^{\mathrm{f}}_{j,i} \left( c^{\b}_i - c^{\p}_{j,i} \right) }_{\!\omega} \qquad \text{in } (0, T^{\mathrm{end}}) \times (0, \Rho^{\mathrm{c}}) \times (0, L^\mathrm{b}) \times [0, 2\pi). \label{eq:model-3d-bulk-oom-almost}
\end{multline}

The Danckwerts boundary conditions \cref{eq:model-hrm-inbc,eq:model-hrm-outbc} simplify to
\begin{alignat*}{2}
	\left.\left( u \varepsilon^{\mathrm{b}} c^{\b}_i - \varepsilon^{\mathrm{b}} D^{\mathrm{ax}}_{i} \frac{\partial c^{\b}_i}{\partial z} \right)\right|_{z=0} &= u \left.\varepsilon^{\mathrm{b}}\right|_{z=0} c^{\mathrm{in}}_{i} & &\qquad\text{on } (0, T^{\mathrm{end}}) \times (0, \Rho^{\mathrm{c}}) \times [0, 2\pi), \\
	- \left( \varepsilon^{\mathrm{b}} D^{\mathrm{ax}}_{i} \left. \frac{\partial c^{\b}_i}{\partial z} \right) \right|_{z=L^\mathrm{b}} &= 0 & &\qquad\text{on } (0, T^{\mathrm{end}}) \times (0, \Rho^{\mathrm{c}}) \times [0, 2\pi).
\end{alignat*}
In addition, we keep the no-flow condition \cref{eq:model-hrm-outbc} at $\rho = \Rho^{\mathrm{c}}$ and add a symmetry condition at $\rho = 0$:
\begin{alignat*}{2}
	-\left( \varepsilon^{\mathrm{b}} D^{\mathrm{rad}}_{i} \left. \frac{\partial c^{\b}_i}{\partial \rho} \right) \right|_{\rho=\Rho^{\mathrm{c}}} &= 0 & &\qquad\text{on } (0, T^{\mathrm{end}}) \times (0, L^\mathrm{b}) \times [0, 2\pi), \\
	-\left( \varepsilon^{\mathrm{b}} D^{\mathrm{rad}}_{i} \left. \frac{\partial c^{\b}_i}{\partial \rho} \right) \right|_{\rho=0} &= 0 & &\qquad\text{on } (0, T^{\mathrm{end}}) \times (0, L^\mathrm{b}) \times [0, 2\pi).
\end{alignat*}
Since $\varepsilon^{\mathrm{b}} > 0$ by assumption, we can cancel this factor from all four equations.

Theoretically, the continuum model described here could be combined with a 3D particle model, i.e.~\cref{eq:model-hrm-particle}.
However, the bulk concentrations $c^\mathrm{b}_i$ in the particle boundary condition~\eqref{eq:model-hrm-mtcoupling} are uniform over the particle surface due to the continuum representation of the column domain according to \cref{asmpt:model-3D}.
Thus, variations in particle concentrations in directions other than the radial direction can only arise from the initial conditions.
Therefore, all continuum models derived in this section are complemented by reduced particle models that assume concentration variations only in the radial direction, i.e.~those presented in \cref{sec:intra_particle_transport}.
In this case, we can further simplify the particle surface area average in~\cref{eq:model-3d-bulk-oom-almost} because the concentrations $c_i^{\b}$ and $c_{j,i}^{\p}$ are uniform over the particle surfaces.
Factoring the concentration difference out of the integral in this average, we obtain
\begin{equation*}
	\begin{split}
		\avg{k^{\mathrm{f}}_{i} \left( c^{\b}_i - c^{\p}_i \right) }_{\!\omega}(\vec{y}) &= \left( c^{\b}_i(t,\rho,z,\varphi) - c^{\p}_i\mleft(t,\rho,z,\varphi,R^{\mathrm{p}}\mright)\right) \frac{1}{\surfarea{\Gamma^{\mathrm{bb}}(\vec{y})}} \int_{\Gamma^{\mathrm{bb}}(\vec{y})} k^{\mathrm{f}}_{i}(\vec{x}) \,\mathrm{d}\omega(\vec{x}) \\
		&= \left( c^{\b}_i(t,\rho,z,\varphi) - c^{\p}_i\mleft(t,\rho,z,\varphi,R^{\mathrm{p}}\mright)\right) \avg{k^{\mathrm{f}}_{j,i}}_{\!\omega}(\vec{y}).
	\end{split}
\end{equation*}
By overloading notation, we identify $\avg{k^{\mathrm{f}}_{j,i}}_{\!\omega}(\vec{y})$ with the position-dependent function $k^{\mathrm{f}}_{j,i}(\rho, z, \varphi)$ and obtain the final bulk \gls{pde} 
\begin{multline}
	\varepsilon^{\mathrm{b}} \frac{\partial c^{\b}_i}{\partial t}
    =
    -u \frac{\partial \left( \varepsilon^{\mathrm{b}} c^{\b}_i \right)}{\partial z} +
    \frac{\partial}{\partial z} \left( \varepsilon^{\mathrm{b}} D^{\mathrm{ax}}_{i} \frac{\partial c^{\b}_i}{\partial z} \right)
    + \frac{1}{\rho} \frac{\partial}{\partial \rho} \left( \rho \varepsilon^{\mathrm{b}} D^{\mathrm{rad}}_{i} \frac{\partial c^{\b}_i}{\partial \rho} \right)
    + \frac{1}{\rho^2} \frac{\partial}{\partial \varphi} \left( \varepsilon^{\mathrm{b}} D^{\mathrm{ang}}_{i} \frac{\partial c^{\b}_i}{\partial \varphi} \right) \\
	- \left(1 - \varepsilon^{\mathrm{b}} \right) \sum_{j=1}^{N^{\mathrm{p}}} \frac{3 d_j}{R^{\mathrm{p}}_{j}} k^{\mathrm{f}}_{j,i} \left( c^{\b}_i - \left.c^{\p}_{j,i} \right|_{r = R^{\mathrm{p}}_{j}} \right) \qquad \text{in } (0, T^{\mathrm{end}}) \times \Omega^\b. \label{eq:model-3d-bulk-oom}
\end{multline}

In summary, we arrive at the following bulk transport model, which combines and generalizes the models of \cite[Section~2.1.6]{Guiochon2006}, \cite{Qamar2017}, and \cite{Carta2003}.

\begin{model}{(B3)}[3D Continuum Bulk Model]\label{model:3D_continuum_bulk_transport_model}
	Consider a cylindrical column packed with $N^{\mathrm{p}}\geq 0$ different particle types, indexed by $j \in \{1, \dots, N^{\mathrm{p}}\}$, and distributed according to the volume fractions $d_j \colon \Omega^\b \to [0, 1]$.
	For every $(\rho,z,\varphi) \in \Omega^\b$, the volume fractions satisfy
	\begin{equation*}
		\sum_{j=1}^{N^{\mathrm{p}}} d_j(\rho, z, \varphi) = 1.
	\end{equation*}
	In the bulk domain, mass transport is governed by the convection-diffusion-reaction equation
	\begin{multline}
		\varepsilon^{\mathrm{b}} \frac{\partial c^{\b}_i}{\partial t} =
        -u \frac{\partial \left( \varepsilon^{\mathrm{b}} c^{\b}_i \right)}{\partial z}
        + \frac{\partial}{\partial z} \left( \varepsilon^{\mathrm{b}} D^{\mathrm{ax}}_{i} \frac{\partial c^{\b}_i}{\partial z} \right) 
        + \frac{1}{\rho} \frac{\partial}{\partial \rho} \left( \rho \varepsilon^{\mathrm{b}} D^{\mathrm{rad}}_{i}  \frac{\partial c^{\b}_i}{\partial \rho} \right) 
        + \frac{1}{\rho^2} \frac{\partial}{\partial \varphi} \left( \varepsilon^{\mathrm{b}} D^{\mathrm{ang}}_{i}  \frac{\partial c^{\b}_i}{\partial \varphi} \right) \\
		- \left(1 - \varepsilon^{\mathrm{b}} \right) \sum_{j=1}^{N^{\mathrm{p}}} \frac{3 d_j}{R^{\mathrm{p}}_{j}} k^{\mathrm{f}}_{j,i} \left(c^{\b}_i - \left. c^{\p}_{j,i} \right|_{r = R^{\mathrm{p}}_{j}} \right) \qquad \text{in } (0, T^{\mathrm{end}}) \times \Omega^\b \label{eq:model-3d-bulk-pde}
	\end{multline}
	with boundary conditions
	\begingroup
	\allowdisplaybreaks
	\begin{subequations}\label{eq:model-3d-bulk-bc}
	\begin{alignat}{2}
		u c^{\mathrm{in}}_{i} &= \left.\left( u c^{\b}_i - D^{\mathrm{ax}}_{i} \frac{\partial c^{\b}_i}{\partial z} \right)\right|_{z=0} & &\qquad\text{on } (0, T^{\mathrm{end}}) \times (0, \Rho^{\mathrm{c}}) \times [0,2\pi), \label{eq:model-3d-bulk-bc-inlet} \\
		0 &= - \left(D^{\mathrm{ax}}_{i} \left. \frac{\partial c^{\b}_i}{\partial z} \right) \right|_{z=L^\mathrm{b}} & &\qquad\text{on } (0, T^{\mathrm{end}}) \times (0, \Rho^{\mathrm{c}}) \times [0,2\pi), \\
		0 &= - \left(D^{\mathrm{rad}}_{i} \left. \frac{\partial c^{\b}_i}{\partial \rho} \right) \right|_{\rho=\Rho^{\mathrm{c}}} & &\qquad\text{on } (0, T^{\mathrm{end}}) \times (0, L^\mathrm{b}) \times [0,2\pi), \label{eq:model-3d-bulk-bc-outer} \\
		0 &= - \left(D^{\mathrm{rad}}_{i} \left. \frac{\partial c^{\b}_i}{\partial \rho} \right) \right|_{\rho=0} & &\qquad\text{on } (0, T^{\mathrm{end}}) \times (0, L^\mathrm{b}) \times [0,2\pi), \label{eq:model-3d-bulk-bc-center} \\
		0 &= c^{\b}_i \Big|_{\varphi=0} - c^{\b}_i \Big|_{\varphi=2\pi} & &\qquad\text{on } (0, T^{\mathrm{end}}) \times (0, L^\mathrm{b}) \times (0, \Rho^{\mathrm{c}}), \label{eq:model-3d-bulk-bc-periodic} \\
		0 &= \frac{\partial c^{\b}_i}{\partial \varphi} \bigg|_{\varphi=0} - \frac{\partial c^{\b}_i}{\partial \varphi} \bigg|_{\varphi=2\pi} & &\qquad\text{on } (0, T^{\mathrm{end}}) \times (0, L^\mathrm{b}) \times (0, \Rho^{\mathrm{c}}). \label{eq:model-3d-bulk-bc-periodic-flux}
	\end{alignat}
	\end{subequations}
	\endgroup
	The model can be complemented by $N^{\p}\geq 0$ particle models, as defined in \cref{sec:intra_particle_transport}, which provide the particle liquid-phase concentrations $c^{\p}_{j,i} \Big|_{r = R^{\mathrm{p}}_{j}}$ at the particle surfaces.
	The initial conditions at $t = 0$ are given by
	\begin{subequations}\label{eq:model-3d-bulk-ic}
	\begin{alignat}{2}
		\left. c^{\b}_i \right|_{t = 0} &= c^{\b,\mathrm{init}}_{i} & & \qquad\text{in } \Omega^\b, \\
		\left. c^{\p}_{j,i} \right|_{t = 0, r = (0, R^{\mathrm{p}}_{j})} &= c^{\p,\mathrm{init}}_{j,i} & & \qquad\text{in } \Omega^\b. \label{eq:model-3d-bulk-ic-solid}
	\end{alignat}
	\end{subequations}
\end{model}

\begin{remark}[(B3b)]\enlargethispage{1\baselineskip}
\begin{itemize}\setlength{\itemsep}{0\baselineskip}
We conclude with the following remarks.
    \item The interstitial velocity depends on $\varepsilon^\b$, whereas $u$ denotes the superficial velocity;
    \item By using \cref{asmpt:model-3D} after volume averaging, we have discarded information on the microscale geometry that is hidden in the higher order contributions lumped into $\mat{D}^{\b,\mathrm{eff}}_{i}$.
        In general, $\mat{D}^{\b,\mathrm{eff}}_{i}$ might not even be symmetric (see \cite[Sec.~3.3.3]{Whitaker1999}). We nevertheless replace it with the symmetric tensor $\mat{D}^{\b}_{\mathrm{disp},i}$.
    \item In the volume averaging process, we assumed a constant flow field $\vec{u}$ which is a crude approximation on the microscale that may be justified on the macroscale in the light of \cref{asmpt:model-3D}.
        Another approach would be to employ the volume-averaged velocity $\avg{\vec{u}}$, which is still divergence-free.
    \item \cref{asmpt:model-hrm} includes that $u \geq 0$.
        If the flow is reversed, only the boundary conditions at $z = 0$ and $z = L^\mathrm{b}$ in \cref{eq:model-3d-bulk-bc} have to be interchanged.
\end{itemize}
\end{remark}

\subsubsection{2D Bulk Model (Model B2)}
\label{sec:2D_continuum_bulk_transport_model}

To further reduce the complexity of \cref{model:3D_continuum_bulk_transport_model}, we eliminate the angular coordinate in the bulk domain.
The resulting two-dimensional representation of the column can be used, e.g., to investigate wall effects and polydisperse packings \citep{Brhane2020,Uche2020,Qamar2017,RAO2025134409}.
\begin{assumption}[(B2)]\label{asmpt:model-2D}
	In addition to \cref{asmpt:model-hrm,asmpt:model-3D}, we assume that the column is radially symmetric; that is, solute concentrations and model parameters depend only on the axial and radial coordinates.
\end{assumption}
This assumption removes all dependence on the angular coordinate $\varphi$ from the model and we define $\Omega^\b \equiv (0, R_c) \times (0, L^\mathrm{b})$.
As in the derivation of the three-dimensional continuum model from the \gls{hrm}, we apply volume averaging, here in the circumferential direction, to derive the corresponding two-dimensional model.
To this end, we define, for an integrable function $f\colon (0, R_c) \times (0, L^\mathrm{b}) \times [0, 2\pi) \to \Reals$, the angular average
\begin{equation*}
	\avg{f}_\varphi \coloneqq \frac{1}{2\pi} \int_0^{2\pi} f(\rho, z, \varphi)\, \mathrm{d}\varphi.
\end{equation*}

Applying angular averaging to the bulk-domain \gls{pde} \eqref{eq:model-3d-bulk-pde} yields
\begin{align}
    \avg{\varepsilon^{\mathrm{b}} \frac{\partial c^{\b}_i}{\partial t}}_{\avgspace\varphi}
    = &-\avg{u \frac{\partial \left( \varepsilon^{\mathrm{b}} c^{\b}_i \right)}{\partial z}}_{\avgspace\varphi}
    + \avg{\frac{\partial}{\partial z} \left( \varepsilon^{\mathrm{b}} D^{\mathrm{ax}}_{i} \frac{\partial c^{\b}_i}{\partial z} \right)}_{\avgspace\varphi}
    + \avg{\frac{1}{\rho} \frac{\partial}{\partial \rho} \left( \rho \varepsilon^{\mathrm{b}} D^{\mathrm{rad}}_{i}  \frac{\partial c^{\b}_i}{\partial \rho} \right)}_{\avgspace\varphi}
    \nonumber \\
    &+ \avg{\frac{1}{\rho^2} \frac{\partial}{\partial \varphi} \left( \varepsilon^{\mathrm{b}} D^{\mathrm{ang}}_{i} \frac{\partial c^{\b}_i}{\partial \varphi} \right)}_{\avgspace\varphi}
    \nonumber \\
    &- \left(1 - \varepsilon^{\mathrm{b}} \right) \sum_{j=1}^{N^{\mathrm{p}}} \frac{3}{R^{\mathrm{p}}_{j}} \avg{ d_j k^{\mathrm{f}}_{j,i} \left( c^{\b}_i - \left. c^{\p}_{j,i} \right|_{r = R^{\mathrm{p}}_{j}} \right) }_{\avgspace\varphi} \qquad \text{in } (0, T^{\mathrm{end}}) \times \Omega^\b. \label{eq:model-2d-bulk-pde-radavg}
\end{align}

Due to the periodic boundary conditions~\cref{eq:model-3d-bulk-bc-periodic,eq:model-3d-bulk-bc-periodic-flux}, the average of the angular dispersion term vanishes:
\begin{align*}
    \avg{\frac{1}{\rho^2} \frac{\partial}{\partial \varphi} \left( \varepsilon^{\mathrm{b}} D^{\mathrm{ang}}_{i}  \frac{\partial c^{\b}_i}{\partial \varphi} \right)}_{\avgspace\varphi}
    &= \frac{1}{2 \pi \rho^2} \int_0^{2\pi} \frac{\partial}{\partial \varphi} \left( \varepsilon^{\mathrm{b}} D^{\mathrm{ang}}_{i} \frac{\partial c^{\b}_i}{\partial \varphi} \right) \,\mathrm{d}\varphi \\
    &= \frac{1}{2 \pi \rho^2} \left[ \varepsilon^{\mathrm{b}} D^{\mathrm{ang}}_{i} \frac{\partial c^{\b}_i}{\partial \varphi} \right]_{\varphi = 0}^{\varphi = 2\pi} = 0.
\end{align*}

Next, we can move the parameters, which are now independent of the angular coordinate, out of the differential and averaging operators and interchange differentiation and averaging to arrive at
\begin{align}
    \varepsilon^{\mathrm{b}} \frac{\partial \avg{c^{\b}_i}_{\avgspace\varphi}}{\partial t}
    = &-u \frac{\partial \left( \varepsilon^{\mathrm{b}} \avg{c^{\b}_i}_{\avgspace\varphi} \right)}{\partial z}
    + \frac{\partial}{\partial z} \left( \varepsilon^{\mathrm{b}} D^{\mathrm{ax}}_{i} \frac{\partial \avg{c^{\b}_i}_{\avgspace\varphi}}{\partial z} \right)
    + \frac{1}{\rho} \frac{\partial}{\partial \rho} \left( \rho \varepsilon^{\mathrm{b}} D^{\mathrm{rad}}_{i}  \frac{\partial \avg{c^{\b}_i}_{\avgspace\varphi}}{\partial \rho} \right)
    \nonumber \\
    &- \left(1 - \varepsilon^{\mathrm{b}} \right) \sum_{j=1}^{N^{\mathrm{p}}} \frac{3 d_j}{R^{\mathrm{p}}_{j}} k^{\mathrm{f}}_{j,i} \left( \avg{c^{\b}_i}_{\avgspace\varphi} - \left. \avg{c^{\p}_{j,i}}_{\avgspace\varphi} \right|_{r = R^{\mathrm{p}}_{j}} \right) \qquad \text{in } (0, T^{\mathrm{end}}) \times \Omega^\b. \label{eq:model-2d-bulk-pde-radavg2}
\end{align}

Similarly, we apply the averaging process to the boundary conditions \cref{eq:model-3d-bulk-bc}, which yields~\cref{eq:model-2d-bulk-bc}.
If a particle model (\cref{sec:intra_particle_transport}) is used ($N^{\p} \geq 1$), we apply the same procedure by averaging the \gls{pde} over the angular coordinate and then interchanging the averaging and differentiation operations.

Finally, the initial conditions \cref{eq:model-3d-bulk-ic} at $t = 0$ are given by
\begin{subequations}\label[equation]{eq:model-2d-bulk-ic-radavg}
	\begin{alignat}{2}
		\left. \avg{c^{\b}_i}_{\avgspace\varphi} \right|_{t = 0} &= \avg{c^{\b,\mathrm{init}}_{i}}_{\avgspace\varphi} & & \qquad\text{in } \Omega^\b, \\
		\left. \avg{c^{\p}_{j,i}}_{\avgspace\varphi} \right|_{t = 0, r = R^{\mathrm{p}}_{j}} &= \avg{c^{\p,\mathrm{init}}_{j,i}}_{\avgspace\varphi} & & \qquad\text{in } \Omega^\b. \label{eq:model-2d-bulk-ic-solid}
	\end{alignat}
\end{subequations}

Having derived the model, we simplify the notation by dropping the angular averages.
That is, we overload notation to identify $\cramped{\avg*{c^{\b}_i}_{\varphi}}$ with the function $c^{\b}_i\colon (0, T^{\mathrm{end}}) \times \Omega^\b \to \Reals$ and analogously for the other averages.
This way, we have effectively removed the angular coordinate $\varphi$ from the model.

\begin{model}{(B2)}[2D Continuum Bulk Model]\label{model:2D_continuum_bulk_transport_model}
Consider a cylindrical column packed with $N^{\mathrm{p}} \geq 0$ different particle types, indexed by $j \in \{1, \dots, N^{\mathrm{p}}\}$, and distributed according to the volume fractions $d_j \colon \Omega^\b \to [0,1]$, which satisfy
\begin{equation*}
    \sum_{j=1}^{N^{\mathrm{p}}} d_j(\rho, z) = 1.
\end{equation*}
In the bulk domain, mass transport is governed by the convection-diffusion-reaction equation
\begin{multline}
    \varepsilon^{\mathrm{b}} \frac{\partial c^{\b}_i}{\partial t} = -u \frac{\partial \left( \varepsilon^{\mathrm{b}} c^{\b}_i \right)}{\partial z} + \frac{\partial}{\partial z} \left( \varepsilon^{\mathrm{b}} D^{\mathrm{ax}}_{i} \frac{\partial c^{\b}_i}{\partial z} \right) + \frac{1}{\rho} \frac{\partial}{\partial \rho} \left( \rho \varepsilon^{\mathrm{b}} D^{\mathrm{rad}}_{i}  \frac{\partial c^{\b}_i}{\partial \rho} \right) \\
    - \left(1 - \varepsilon^{\mathrm{b}} \right) \sum_{j=1}^{N^{\mathrm{p}}} \frac{3 d_j}{R^{\mathrm{p}}_{j}} k^{\mathrm{f}}_{j,i} \left(c^{\b}_i - \left. c^{\p}_{j,i} \right|_{r = R^{\mathrm{p}}_{j}} \right) \qquad \text{in } (0, T^{\mathrm{end}}) \times (0, \Rho^{\mathrm{c}}) \times (0, L^\mathrm{b}) \label{eq:model-2d-bulk-pde}
\end{multline}
with boundary conditions
\begingroup
\allowdisplaybreaks[0]
\begin{subequations}\label{eq:model-2d-bulk-bc}
\begin{alignat}{2}
    u c^{\mathrm{in}}_{i} &= \left.\left( u c^{\b}_i - D^{\mathrm{ax}}_{i} \frac{\partial c^{\b}_i}{\partial z} \right)\right|_{z=0} & &\qquad\text{on } (0, T^{\mathrm{end}}) \times (0, \Rho^{\mathrm{c}}), \label{eq:model-2d-bulk-bc-inlet} \\
    0 &= - \left(D^{\mathrm{ax}}_{i} \left. \frac{\partial c^{\b}_i}{\partial z} \right) \right|_{z=L^\mathrm{b}} & &\qquad\text{on } (0, T^{\mathrm{end}}) \times (0, \Rho^{\mathrm{c}}), \\
    0 &= - \left(D^{\mathrm{rad}}_{i} \left. \frac{\partial c^{\b}_i}{\partial \rho} \right) \right|_{\rho=\Rho^{\mathrm{c}}} & &\qquad\text{on } (0, T^{\mathrm{end}}) \times (0, L^\mathrm{b}), \label{eq:model-2d-bulk-bc-outer} \\
    0 &= - \left(D^{\mathrm{rad}}_{i} \left. \frac{\partial c^{\b}_i}{\partial \rho} \right) \right|_{\rho=0} & &\qquad\text{on } (0, T^{\mathrm{end}}) \times (0, L^\mathrm{b}). \label{eq:model-2d-bulk-bc-center}
\end{alignat}
\end{subequations}
\endgroup

The model is complemented by $N^{\p}\geq 0$ particle models as defined in \cref{sec:intra_particle_transport}, which provide the liquid-phase concentrations $c^{\p}_{j,i} \Big|_{r = R^{\mathrm{p}}_{j}}$ at the particle surfaces.

The initial conditions at $t = 0$ are given by
\begin{subequations}\label{eq:model-2d-bulk-ic}
\begin{alignat}{2}
    \left. c^{\b}_i \right|_{t = 0} &= c^{\b,\mathrm{init}}_{i} & & \qquad\text{in } \Omega^\b, \\
    \left. c^{\p}_{j,i} \right|_{t = 0, r = R^{\mathrm{p}}_{j}} &= c^{\p,\mathrm{init}}_{j,i} & & \qquad\text{in } \Omega^\b.
\end{alignat}
\end{subequations}
\end{model}

\Cref{model:2D_continuum_bulk_transport_model} is more general than formulations found in the literature \mbox{(see, e.g.,}~\citealp{Qamar2017,Uche2019,Brhane2020,RAO2025134409}), as it accommodates multiple particle types and supports spatially varying model parameters. We note, however, that homogeneous radial zones in which all model parameters are constant may be advantageous, either to facilitate the application of numerical schemes or because of limitations in the available data.

\subsubsection{1D Bulk Model (Model B1)}
\label{sec:1D_continuum_bulk_transport_model-1d}

The vast majority of chromatography models considered in the literature resolve only the axial flow direction of the column.
The model equations are derived under the following customary assumptions.
\begin{assumption}[(B1)]\label{asmpt:model-1D}
In addition to \cref{asmpt:model-3D,asmpt:model-2D}, we assume that the column is radially homogeneous; that is, solute concentrations and model parameters depend only on the axial coordinate.
\end{assumption}
This assumption removes all dependence on the radial coordinate $\rho$ from the model, and we define $\Omega^\b \equiv (0, L^\mathrm{b})$.
As previously, we apply volume averaging, here over radial cross-sections, to derive the corresponding one-dimensional model.
To this end, we define, for an integrable function $f\colon (0, \Rho^{\mathrm{c}}) \to \Reals$, the radial average
\begin{equation*}
	\avg{f}_\rho \coloneqq \frac{2}{(\Rho^{\mathrm{c}})^2} \int_0^{{\Rho^{\mathrm{c}}}} f(\rho) \rho \,\mathrm{d}\rho.
\end{equation*}

Applying radial averaging to the bulk-domain \gls{pde} \eqref{eq:model-2d-bulk-pde} yields
\begin{multline}
	\avg{\varepsilon^{\mathrm{b}} \frac{\partial c^{\b}_i}{\partial t}}_{\avgspace\rho} = -\avg{u \frac{\partial \left( \varepsilon^{\mathrm{b}} c^{\b}_i \right)}{\partial z}}_{\avgspace\rho} + \avg{\frac{\partial}{\partial z} \left( \varepsilon^{\mathrm{b}} D^{\mathrm{ax}}_{i} \frac{\partial c^{\b}_i}{\partial z} \right)}_{\avgspace\rho} + \avg{\frac{1}{\rho} \frac{\partial}{\partial \rho} \left( \rho \varepsilon^{\mathrm{b}} D^{\mathrm{rad}}_{i}  \frac{\partial c^{\b}_i}{\partial \rho} \right)}_{\avgspace\rho} \\
	- \left(1 - \varepsilon^{\mathrm{b}} \right) \sum_{j=1}^{N^{\mathrm{p}}} \frac{3}{R^{\mathrm{p}}_{j}} \avg{d_j k^{\mathrm{f}}_{j,i} \left( c^{\b}_i - \left. c^{\p}_{j,i} \right|_{r = R^{\mathrm{p}}_{j}} \right) }_{\avgspace\rho} \qquad \text{in } (0, T^{\mathrm{end}}) \times (0, L^\mathrm{b}). \label{eq:model-1d-bulk-pde-radavg}
\end{multline}

Due to the Neumann boundary conditions \cref{eq:model-2d-bulk-bc-outer,eq:model-2d-bulk-bc-center}, the average of the radial dispersion term vanishes: 
\begin{align*}
		\avg{\frac{1}{\rho} \frac{\partial}{\partial \rho} \left( \rho \varepsilon^{\mathrm{b}} D^{\mathrm{rad}}_{i}  \frac{\partial c^{\b}_i}{\partial \rho} \right)}_{\avgspace\rho} &= \frac{2}{(\Rho^{\mathrm{c}})^2} \int_0^{{\Rho^{\mathrm{c}}}} \frac{1}{\rho} \frac{\partial}{\partial \rho} \left( \rho \varepsilon^{\mathrm{b}}(\rho, z) D^{\mathrm{rad}}_{i}(\rho, z)  \frac{\partial c^{\b}_i}{\partial \rho}(\rho, z) \right) \rho \,\mathrm{d}\rho \\
		&= \frac{2}{(\Rho^{\mathrm{c}})^2} \left[ \rho \varepsilon^{\mathrm{b}}(\rho, z) D^{\mathrm{rad}}_{i}(\rho, z) \frac{\partial c^{\b}_i}{\partial \rho}(\rho, z) \right]_{\rho = 0}^{\rho = \Rho^{\mathrm{c}}} = 0.
\end{align*}
\cref{asmpt:model-1D} allows us to move the model parameters out of the averaging operators.
The convective term is closed by the approximation $\avg{u c^{\b}_i}_{\!\rho} \approx \avg{u}_{\!\rho} \avg{c^{\b}_i}_{\!\rho}$, whose residual, the correlation between the radial deviations of velocity and concentration, is an additional source of axial band broadening and is lumped into the axial dispersion coefficient $D^{\mathrm{ax}}_{i}$, as done for the unresolved effects in \cref{sec:3D_continuum_bulk_transport_model}.
Interchanging averaging with differentiation, we obtain
\begin{equation*}
		\varepsilon^{\mathrm{b}} \frac{\partial \avg{c^{\b}_i}_{\avgspace\rho}}{\partial t}
        = - \avg{u}_{\avgspace\rho} \frac{\partial \varepsilon^{\mathrm{b}} \avg{c^{\b}_i}_{\avgspace\rho} }{\partial z}
        + \frac{\partial}{\partial z} \left( \varepsilon^{\mathrm{b}} D^{\mathrm{ax}}_{i} \frac{\partial \avg{c^{\b}_i}_{\avgspace\rho}}{\partial z} \right) 
        - \left(1 - \varepsilon^{\mathrm{b}} \right) \sum_{j=1}^{N^{\mathrm{p}}} \frac{3 d_j}{R^{\mathrm{p}}_{j}} k^{\mathrm{f}}_{j,i} \left(\avg{ c^{\b}_i}_{\!\rho} - \left. \avg{c^{\p}_{j,i}}_{\!\rho} \right|_{r = R^{\mathrm{p}}_{j}}\right). \label{eq:model-1d-bulk-pde-radavg2}
\end{equation*}

Similarly, we apply the averaging process to the boundary conditions \cref{eq:model-2d-bulk-bc}, which yields
\begin{subequations}
\begin{alignat}{2}
	\avg{u}_{\!\rho} \avg{c^{\mathrm{in}}_{i}}_{\!\rho} &= \left.\left( \avg{u}_{\!\rho} \avg{c^{\b}_i}_{\!\rho} - D^{\mathrm{ax}}_{i} \frac{\partial \avg{c^{\b}_i}_{\!\rho}}{\partial z} \right)\right|_{z=0} & &\qquad\text{on } (0, T^{\mathrm{end}}), \\
	0 &= - D^{\mathrm{ax}}_{i} \left. \frac{\partial \avg{c^{\b}_i}_{\!\rho}}{\partial z} \right|_{z=L^\mathrm{b}} & &\qquad\text{on } (0, T^{\mathrm{end}}).
\end{alignat}\label[equation]{eq:model-2d-bulk-bc-radavg}
\end{subequations}

The particle equations are reduced to the lower-dimensional spatial domain by the same procedure, namely by averaging the \gls{pde} in the radial direction and again interchanging the averaging and differentiation operations.

Finally, the initial conditions \cref{eq:model-2d-bulk-ic} at $t = 0$ are given by
\begin{subequations}
\begin{alignat}{2}
	\left. \avg{c^{\b}_i}_{\!\rho} \right|_{t = 0} &= \avg{c^{\b,\mathrm{init}}_{i}}_{\!\rho} & & \qquad\text{in } (0, L^\mathrm{b}), \\
	\left. \avg{c^{\p}_{j,i}}_{\!\rho} \right|_{t = 0} &= \avg{c^{\p,\mathrm{init}}_{j,i}}_{\!\rho} & & \qquad\text{in } (0, L^\mathrm{b}) \times (0, R^{\mathrm{p}}_{j}).
\end{alignat}\label[equation]{eq:model-1d-bulk-ic-radavg}
\end{subequations}

Having derived the model, we simplify the notation by dropping the radial averages.
That is, we overload the notation by identifying $\cramped{\avg*{c^{\b}_i}_{\rho}}$ with the function $c^{\b}_i\colon (0, T^{\mathrm{end}}) \times (0, L^\mathrm{b}) \to \Reals$ and analogously for the other averaged quantities.
This way, we have effectively removed the radial coordinate $\rho$ from the model.

\begin{model}{(B1)}[1D Continuum Bulk Model]\label{model:1D_continuum_bulk_transport_model}
	Consider a cylindrical column packed with $N^{\mathrm{p}} \geq 0$ different particle types, indexed by $j \in \{1, \dots, N^{\mathrm{p}}\}$, and distributed according to the volume fractions $d_j \colon \Omega^\b \to [0,1]$, which satisfy
	\begin{equation*}
		\sum_{j=1}^{N^{\mathrm{p}}} d_j(z) = 1.
	\end{equation*}
	In the bulk domain, mass transport is governed by the convection-diffusion-reaction equation
	\begin{align}
		\varepsilon^{\mathrm{b}} \frac{\partial c^{\b}_i}{\partial t}
        = - u \frac{\partial (\varepsilon^{\mathrm{b}} c^{\b}_i)}{\partial z}
        + \frac{\partial}{\partial z} \left( \varepsilon^{\mathrm{b}} D^{\mathrm{ax}}_{i} \frac{\partial c^{\b}_i}{\partial z} \right)
        - \left(1 - \varepsilon^{\mathrm{b}} \right) \sum_{j=1}^{N^{\mathrm{p}}} \frac{3 d_j}{R^{\mathrm{p}}_{j}} k^{\mathrm{f}}_{j,i} \left(c^{\b}_i - \left. c^{\p}_{j,i} \right|_{r = R^{\mathrm{p}}_{j}} \right) \label{eq:model-1d-bulk-pde}
	\end{align}
	in $(0, T^{\mathrm{end}}) \times (0, L^\mathrm{b})$ with Danckwerts boundary conditions
	\begin{subequations}\label{eq:model-1d-bulk-bc}
	\begin{alignat}{2}
		u c^{\mathrm{in}}_{i} &= \left.\left( u c^{\b}_i - D^{\mathrm{ax}}_{i} \frac{\partial c^{\b}_i}{\partial z} \right)\right|_{z=0} & &\qquad\text{on } (0, T^{\mathrm{end}}), \\
		0 &= - D^{\mathrm{ax}}_{i} \left. \frac{\partial c^{\b}_i}{\partial z} \right|_{z=L^\mathrm{b}} & &\qquad\text{on } (0, T^{\mathrm{end}}).
	\end{alignat}
	\end{subequations}

	The model can be complemented by $N^{\p}$ particle models (see ~\cref{sec:intra_particle_transport}), which provide the liquid-phase concentrations $c^{\p}_{j,i} \Big|_{r = R^{\mathrm{p}}_{j}}$ at the particle surfaces.

	The initial conditions at $t = 0$ are given by
	\begin{subequations}
	\begin{alignat}{2}
		\left. c^{\b}_i \right|_{t = 0} &= c^{\b,\mathrm{init}}_{i} & & \qquad\text{in } (0, L^\mathrm{b}), \\
		\left. c^{\p}_{j,i} \right|_{t = 0, r = R^{\mathrm{p}}_{j}} &= c^{\p,\mathrm{init}}_{j,i} & & \qquad\text{in } (0, L^\mathrm{b}).
	\end{alignat}\label{eq:model-1d-bulk-ic}
	\end{subequations}
\end{model}
Note that most models in the literature do not consider multiple particle types.
Furthermore, they typically assume axially constant parameters, reducing the bulk-domain mass transport model to
\begin{align}
    \frac{\partial c^{\b}_i}{\partial t}
    = - u \frac{\partial c^{\b}_i}{\partial z}
    + D^{\mathrm{ax}}_{i} \frac{\partial^2 c^{\b}_i}{\partial z^2}
    - \frac{1 - \varepsilon^{\mathrm{b}}}{\varepsilon^\mathrm{b}} \sum_{j=1}^{N^{\mathrm{p}}} \frac{3 d_j}{R^{\mathrm{p}}_{j}} k^{\mathrm{f}}_{j,i} \left(c^{\b}_i - \left. c^{\p}_{j,i} \right|_{r = R^{\mathrm{p}}_{j}} \right), \label{eq:model-1d-bulk-pde-const-params}
\end{align}
which can be substituted for Eq.~\eqref{eq:model-1d-bulk-pde}.

\begin{remark}[(B1a)]
	Another approach to deriving \Cref{model:1D_continuum_bulk_transport_model} is to formulate the mass balances for a set of control volumes directly based on \cref{asmpt:model-1D}.
	The differential equations follow by shrinking the control volumes to a point.
	This approach is used, for example, in the derivations of the \gls{grm} by \citet[Sec.~6.2]{Seidel-Morgenstern2020-Chap6} and \citet[Sec.~2.1]{Guiochon2006}.
\end{remark}

\begin{remark}[(B1b)]
    Model parameters may be assumed constant or piecewise constant in space, allowing them to be taken outside differential operators and thereby reducing model complexity.
    For example, under the assumptions of constant porosity and axial dispersion, \cref{eq:model-1d-bulk-pde} reduces to the commonly used form
        \begin{equation}
        \label{eq:model-1d-bulk-pde-constParams}
        \frac{\partial c^{\b}_i}{\partial t}
        = - u \frac{\partial c^{\b}_i}{\partial z}
        + D^{\mathrm{ax}}_{i} \frac{\partial^2 c^{\b}_i}{\partial z^2}
        - \frac{\left(1 - \varepsilon^{\mathrm{b}} \right)}{\varepsilon^{\mathrm{b}}} \sum_{j=1}^{N^{\mathrm{p}}} \frac{3 d_j}{R^{\mathrm{p}}_{j}} k^{\mathrm{f}}_{j,i} \left(c^{\b}_i - \left. c^{\p}_{j,i} \right|_{r = R^{\mathrm{p}}_{j}} \right).
    \end{equation}
    Such parameter simplifications are often motivated by limited data availability or necessitated by the numerical discretization.
    For instance, \cite{RAO2025134409} assumed a piecewise constant radial dispersion coefficient in the \gls{2dgrm} (i.e.\ \cref{eq:model-2d-bulk-pde} coupled with \cref{eq:model_particle_1D}) before discretizing the resulting system in space using a \gls{fv} method.
\end{remark}

\subsubsection{0D Bulk Model (Model B0)}
\label{sec:models-cstr}
We consider mass transport in the bulk domain under the assumption of a perfectly mixed bulk phase, resulting in a 0D bulk transport model.
In combination with the particle transport models of \cref{sec:intra_particle_transport}, this model describes well-mixed particle suspensions, such as batch uptake experiments.
Without a particle transport model, it represents well-mixed vessels, including \glspl{bstr}, which operate without flow during the process, and \glspl{cstr}, which operate under continuous inflow and outflow.
The model can also represent filters in which unequal inflow and outflow rates are used to reduce the volume and thereby concentrate the retained solutes.
It can also be combined with the reaction models introduced in \cref{sec:models-reaction} and other submodels to describe, for instance, crystallization \citep{ZHANG2024108612,ZHANG2025108860}, fermentation \citep{LANZRATH2026109690}, and a wide range of other chemical and biotechnological processes.
Beyond these applications, the model is not restricted to representing the bulk domain of a column or \gls{str}.
It can be used more generally to model lumped system volumes with backmixing, including tubing, valves, detectors, and frits.
In these applications, the fluid is represented as a single well-mixed compartment \cite[Sec.~6.3.2.3]{Seidel-Morgenstern2020-Chap6}.

\begin{remark}[(B0)]
Representing only the packed bed of a chromatography column is usually not sufficient to model the complete process.
Additional system volumes contribute to band broadening and mixing \citep[Sec.~4.2.4]{Carta2010} and can significantly affect process performance, particularly in cyclic processes such as \gls{smb} chromatography \citep{Grosfils2010,Katsuo2009}, but also in single-column processes \citep{Schweiger2018,Schultze-Jena2017}.
System volumes are commonly modeled using (potentially reactive) \glspl{str} and (ideal or dispersive) \glspl{pfr} \citep{Shekhawat2019,Kumar2015}.
Within the present framework, these models emerge as special cases by neglecting certain transport phenomena or omitting the particles (see~\cref{table:taxonomy_primary} in \cref{sec:taxonomy}).
\end{remark}

To derive the model equations, we make the following customary assumptions.
\begin{assumption}[(B0)]\label{asmpt:model-0D}
We assume that
\begin{itemize}\setlength{\itemsep}{0\baselineskip}
    \item The modeled compartment is spatially homogeneous; that is, solute concentrations and model parameters are spatially uniform;
    \item The compartment contains $N^{\p} \geq 0$ particle types with radii $R^{\mathrm{p}}_{j} > 0$;
    \item The total volume $V\colon (0, T^{\mathrm{end}}) \to \Reals^{\geq 0}$ is variable and equals the sum of the bulk and particle volumes, that is, $V = V^\b + V^\ps$. The total particle volume $V^\ps \in \Reals^{\geq 0}$ remains constant; therefore, only the bulk volume $V^\b\colon (0, T^{\mathrm{end}}) \to \Reals^{\geq 0}$ and the porosity $\varepsilon\colon (0, T^{\mathrm{end}}) \to [0,1]$ may vary;
    \item The device is fed with a solution at volumetric flow rate $Q^{\mathrm{in}} \geq 0$ and solute concentrations $c^{\mathrm{in}}_{i}\colon (0, T^{\mathrm{end}}) \to \Reals$. Solution is withdrawn at volumetric flow rate $Q^{\mathrm{out}} \geq 0$, while solvent is withdrawn through a filter at volumetric flow rate $Q^{\mathrm{filter}} \geq 0$.
\end{itemize}
\end{assumption}
From mass balances, we derive the following compartment model with variable volume.
\begin{model}{(B0)}[0D Bulk Compartment Model]\label{model:tank}
Consider a spatially homogeneous compartment with liquid volume $V^b$ and solute concentrations $c^b_i\colon (0, T^{\mathrm{end}}) \to \Reals$, governed by
\begin{subequations}\label{eq:model-cstr}
\begin{align}
    \frac{\mathrm{d}}{\mathrm{d} t} \left( V^\b c^\b_i \right)
    &=
    Q^{\mathrm{in}} c^{\b,\mathrm{in}}_{i} - Q^{\mathrm{out}} c^\b_i
    - V^\ps \sum_{j=1}^{N^{\p}} \frac{3 d_j}{R^{\mathrm{p}}} k^{\mathrm{f}}_{j,i} \left(c^{\b}_i - \left. c^{\p}_{j,i} \right|_{r = R^{\mathrm{p}}_{j}} \right),
    \label{eq:cstr_mass_balance} \\
    \frac{\mathrm{d}V^\b}{\mathrm{d}t} &= Q^{\mathrm{in}} - Q^{\mathrm{out}} - Q^{\mathrm{filter}}
    \label{eq:model-cstr-vol}
\end{align}
\end{subequations}
with initial conditions
\begin{equation*}
    \left.V^\b\right|_{t=0} = V^{\b,\mathrm{init}}, \qquad \left.c^\b_i\right|_{t=0} = c^{\b,\mathrm{init}}_{i}.
\end{equation*}
In the presence of particles, i.e., $N^{\p}\geq 1$, the liquid-phase concentration at the particle surfaces \\ $\left. c^{\p}_{j,i} \right|_{r = R^{\mathrm{p}}_{j}}\colon (0, T^{\mathrm{end}}) \to \Reals$ is defined by the particle models in Section~\ref{sec:intra_particle_transport}.
\end{model}

\subsection{Intraparticle Transport}
\label{sec:intra_particle_transport}
After deriving a \gls{rom} hierarchy for the bulk transport \gls{pde} as a modular framework component, we now turn to the particle domain.
The particle transport models presented in this section can be coupled with any of the bulk transport models introduced in the previous section.
When coupled with a bulk transport model, the particles are modeled as a continuum throughout the column (i.e., both the bulk and particle phases are present at every point; see \cref{asmpt:model-3D}), and the particle equations are defined on the bulk domain $\Omega^\b$.
Accordingly, the bulk solute concentration at the particle surface~$c^{\b}_i$, required for the particle boundary condition, is provided by the coupled bulk transport model with $\varepsilon^\mathrm{b}\in(0,1)$.
Alternatively, the particle models may be used as standalone models, in which case~$\Omega^\b \equiv \{ * \}$ is the singleton set and~$c^{\b}_i$ is prescribed by external boundary data.

\subsubsection{1D Particle Model (Model P1)}
\label{sec:1D_particle}
We proceed directly with the formulation of a one-dimensional continuum particle model. 
Due to \cref{asmpt:model-3D}, the particle liquid-phase concentration is spatially uniform over the particle surface.
Consequently, no meaningful two-dimensional \gls{rom} can be derived from \cref{model:column-highres3D}.
The one-dimensional particle model is obtained under the following assumption:
\begin{assumption}[(P1)]\label{asmpt:model-1D-particle}
	In addition to \cref{asmpt:model-hrm}, we assume that the particle liquid-phase concentration depends only on the radial coordinate.
\end{assumption}
To couple the one-dimensional particle model with any of the continuum bulk models of~\cref{sec:bulk_domain_transport}, we overload notation to identify the spatial averages of the particle liquid and solid concentrations $c^p_{i}, c^s_{i}\colon (0, T^{\mathrm{end}}) \times \Omega^{\b} \times \Omega_\mathrm{par} \to \Reals$ with the corresponding functions on the bulk domain $\Omega^{\b}$.
We simplify the particle-domain \glspl{pde} \eqref{eq:model-hrm-particleliquid} and \eqref{eq:model-hrm-particlesolid} by employing the radial symmetry assumption and exploiting the representation of the differential operators in spherical coordinates (see Section~\ref{supp:sec:vectorcalculus-spherical}):
In $(0,T^{\mathrm{end}}] \times \Omega^{\b} \times (R^{\mathrm{c}}, R^{\mathrm{p}})$, it holds that
\begin{subequations}
\begin{align}
    \frac{\partial c^{\p}_{i}}{\partial t}
    &= \div{D^{\p}_{i} \nabla c^{\p}_{i}}
    - \frac{1 - \varepsilon^{\p}}{\varepsilon^{\p}} \sum_{k=1}^{N^{\mathrm{b}}_i} f^{\mathrm{bind}}_{i,k}\mleft( \vec{c}^{\p}, \vec{c}^{\s} \mright) \nonumber
    \\
    &= \frac{1}{r^2} \frac{\partial }{\partial r} \left( r^2 D_{i}^{\p} \frac{\partial c^{\p}_{i}}{\partial r} \right)
    - \frac{1 - \varepsilon^{\p}}{\varepsilon^{\p}} \sum_{k=1}^{N^{\mathrm{b}}_i} f^{\mathrm{bind}}_{i,k}\mleft( \vec{c}^{\p}, \vec{c}^{\s} \mright), \label{eq:model-2d-particleliquid-oom}
    \\
    \frac{\partial c^{\s}_{i,k}}{\partial t}
    &=
    \div{D^{\s}_{i,k} \nabla c^{\s}_{i,k}}
    + f^{\mathrm{bind}}_{i,k}\mleft( \vec{c}^{\p}, \vec{c}^{\s} \mright)
    \nonumber \\
    &=
    \frac{1}{r^2} \frac{\partial }{\partial r} \left( r^2 D_{i}^{\s} \frac{\partial c^{\s}_{i,k}}{\partial r} \right)
    + f^{\mathrm{bind}}_{i,k}\mleft( \vec{c}^{\p}, \vec{c}^{\s} \mright). \label{eq:model-2d-particlesolid-oom}
\end{align}
\end{subequations}
Here, we allow for arbitrary binding models $f^{\mathrm{bind}}$, diffusion coefficients $D_{i}^{\p}, D_{i}^{\s}~\geq~0$, and porosity $\varepsilon^{\p} \in (0,1)$.
Due to symmetry, we obtain the boundary conditions at $r = R^{\mathrm{c}}$:
\begin{alignat*}{2}
    -\left( \left. D^{\p}_{i} \frac{\partial c^{\p}_{i}}{\partial r} \right) \right|_{r=R^{\mathrm{c}}} &= 0 & &\qquad\text{on } (0,T^{\mathrm{end}}] \times \Omega^{\b},
    \\
    - \left( \left. D^{\s}_{i,k} \frac{\partial c^{\s}_{i,k}}{\partial r} \right) \right|_{r=R^{\mathrm{c}}} &= 0 & &\qquad\text{on } (0,T^{\mathrm{end}}] \times \Omega^{\b}.
\end{alignat*}
The complementary boundary condition at $r = R^{\mathrm{p}}$, which couples the particle domain with the bulk domain, follows from translating \cref{eq:model-hrm-particlebulkbc} to the present \gls{rom}.
Since the left-hand side of \cref{eq:model-hrm-particlebulkbc} is uniform over the particle surface, we obtain the total flux through the surface as
\begin{equation*}
\begin{split}
    &\int_{0}^{\pi} \int_{0}^{2\pi}  -\varepsilon^{\p} D^{\p}_{i}\mleft( R^{\mathrm{p}} \mright) \left( \nabla c^{\p}_{i}\mleft(t,\rho,z, R^{\mathrm{p}} \mright) \cdot \vec{n} \right) \left(R^{\mathrm{p}}\right)^2 \sin(\vartheta) \,\mathrm{d}\varphi\, \mathrm{d}\vartheta \\
    = & \int_{0}^{\pi} \int_{0}^{2\pi} -\varepsilon^{\p} D^{\p}_{i}\mleft( R^{\mathrm{p}} \mright)  \frac{\partial c^{\p}_{i}}{\partial r}\mleft(t,\rho,z, R^{\mathrm{p}} \mright) \left(R^{\mathrm{p}}\right)^2 \sin(\vartheta) \,\mathrm{d}\varphi\, \mathrm{d}\vartheta \\
    = & -4\pi \left(R^{\mathrm{p}}\right)^2 \varepsilon^{\p} \left( \left. D^{\p}_{i} \frac{\partial c^{\p}_{i}}{\partial r}\right) \right|_{r = R^{\mathrm{p}}}.
\end{split}
\end{equation*}
Multiplying the total surface flux by the number of particles per unit volume yields the total flux into the particles per unit volume:
\begin{equation}
    -4\pi \left(R^{\mathrm{p}}\right)^2 \varepsilon^{\p} \left(D^{\p}_{i} \left. \frac{\partial c^{\p}_{i}}{\partial r}\right) \right|_{r = R^{\mathrm{p}}} \frac{\left(1 - \varepsilon^{\mathrm{b}} \right) d}{\frac{4}{3}\pi \left(R^{\mathrm{p}}\right)^3} = -\left(1 - \varepsilon^{\mathrm{b}} \right) \frac{3 d}{R^{\mathrm{p}}} \varepsilon^{\p} \left(D^{\p}_{i} \left. \frac{\partial c^{\p}_{i}}{\partial r}\right) \right|_{r = R^{\mathrm{p}}}. \label{eq:model-2d-totalflux_parj}
\end{equation}
Similarly, for the right-hand side of \cref{eq:model-hrm-particlebulkbc} we get
\begin{equation*}
    \int_{0}^{\pi} \int_{0}^{2\pi} k^{\mathrm{f}}_{i}(\rho,z) \left( c^{\p}_{i}\mleft(t,\rho,z,R^{\mathrm{p}}\mright) - c^{\b}_i(t,\rho,z) \right) \left(R^{\mathrm{p}}\right)^2 \sin(\vartheta) \,\mathrm{d}\varphi\, \mathrm{d}\vartheta  =  4\pi \left(R^{\mathrm{p}}\right)^2 k^{\mathrm{f}}_{i} \left( \left. c^{\p}_{i} \right|_{r = R^{\mathrm{p}}} - c^{\b}_i \right),
\end{equation*}
which we also multiply by the total particle surface per unit volume:
\begin{equation}
    \frac{\left(1 - \varepsilon^{\mathrm{b}} \right) d}{\frac{4}{3}\pi \left(R^{\mathrm{p}}\right)^3} 4\pi \left(R^{\mathrm{p}}\right)^2 k^{\mathrm{f}}_{i} \left( \left. c^{\p}_{i} \right|_{r = R^{\mathrm{p}}} - c^{\b}_i \right) = \left(1 - \varepsilon^{\mathrm{b}} \right) \frac{3 d}{R^{\mathrm{p}}} k^{\mathrm{f}}_{i} \left( \left. c^{\p}_{i} \right|_{r = R^{\mathrm{p}}} - c^{\b}_i \right). \label{eq:model-2d-filmdiff-parj}
\end{equation}
Equating \cref{eq:model-2d-totalflux_parj} and \cref{eq:model-2d-filmdiff-parj} finally yields
\begin{alignat}{2}
    \left( 1 - \varepsilon^{\mathrm{b}} \right) k^{\mathrm{f}}_{i} \left( c^{\b}_i - \left. c^{\p}_{i} \right|_{r = R^{\mathrm{p}}} \right)
    &= \left( 1 - \varepsilon^{\mathrm{b}} \right) \varepsilon^{\p} \left(D^{\p}_{i} \left. \frac{\partial c^{\p}_{i}}{\partial r} \right)\right|_{r = R^{\mathrm{p}}}, \label{eq:model-1Dparticle-bulk-bc}
\end{alignat}
on $(0,T^{\mathrm{end}}] \times \Omega^{\b}$.
For the solid-phase concentration $c^{\s}_{i,k}$, we directly translate \cref{eq:model-hrm-particlebcsolid} to the new setting:
\begin{equation*}
    \left(1 - \varepsilon^{\mathrm{b}} \right) \left(1 - \varepsilon^{\p} \right) \left( \left. D^{\s}_{i,k} \frac{\partial c^{\s}_{i,k}}{\partial r} \right) \right|_{r = R^{\mathrm{p}}} = 0.
\end{equation*}

In summary, we arrive at the following particle model
\begin{model}{(P1)}[1D Continuum Particle Model]\label{model:particle_1D}
    Consider spherical particles of radius $R^\mathrm{p}>0$ with a potential solid core of radius $0 \leq R^\mathrm{c} < R^\mathrm{p}$.
    The mass transfer in the particle is governed by the following diffusion-reaction equations in $(0, T^{\mathrm{end}}) \times \Omega^{\b} \times (R^{\mathrm{c}}, R^\mathrm{p})$ and for all components $i\in\{1,\dots,N^\mathrm{c}\}$ and their respective bound states $k\in\{1,\dots,N^{\mathrm{b}}_i\}$:
\begin{subequations}
\label{eq:model_particle_1D}
\begin{align}    
    \frac{\partial c^{\p}_{i}}{\partial t}
    &=
    \frac{1}{r^2} \frac{\partial }{\partial r} \left( r^2 D_{i}^{\p} \frac{\partial c^{\p}_{i}}{\partial r} \right)
    - \frac{1 - \varepsilon^{\p}}{\varepsilon^{\p}} \sum_{k=1}^{N^{\mathrm{b}}_i} f^{\mathrm{bind}}_{i,k}\mleft( \vec{c}^{\p}, \vec{c}^{\s} \mright), \label{eq:model_particle_1D_liquid}
    \\
    \frac{\partial c^{\s}_{i,k}}{\partial t}
    &=
    \frac{1}{r^2} \frac{\partial }{\partial r} \left( r^2 D_{i}^{\s} \frac{\partial c^{\s}_{i,k}}{\partial r} \right)
    + f^{\mathrm{bind}}_{i,k}\mleft( \vec{c}^{\p}, \vec{c}^{\s} \mright).
    \label{eq:model_particle_1D_solid}
\end{align}
\end{subequations}

The boundary conditions are given on $(0, T^{\mathrm{end}})$ by
\begin{subequations}
\label{eq:model_particle_1D_BC}
\begin{align}
    - \varepsilon^\mathrm{p} \left. \left( D^{\p}_{i} \frac{\partial c^{\p}_{i}}{\partial r} \right) \right|_{r=R^{\mathrm{c}}}
    &= 0,
    \label{eq:model_particle_1D_BC_liquid_0} \\
    -\left( 1 - \varepsilon^\mathrm{p} \right) \left. \left( D^{\s}_{i,k} \frac{\partial c^{\s}_{i,k}}{\partial r} \right) \right|_{r=R^{\mathrm{c}}}
    &= 0,
    \label{eq:model_particle_1D_BC_solid_0}
    \\
    \varepsilon^\mathrm{p} \left(D^{\p}_{i} \left. \frac{\partial c^{\p}_{i}}{\partial r} \right)\right|_{r = R^{\mathrm{p}}}
    &=
    k^{\mathrm{f}}_{i} \left( c^{\b}_i - \left. c^{\p}_{i} \right|_{r = R^{\mathrm{p}}} \right)
    \label{eq:model_particle_1D_BC_liquid_Rp}
    \\
    \left( 1 - \varepsilon^\mathrm{p} \right) \left( \left. D^{\s}_{i,k} \frac{\partial c^{\s}_{i,k}}{\partial r} \right) \right|_{r = R^{\mathrm{p}}}
    &= 0. \label{eq:model_particle_1D_BC_solid_Rp}
\end{align}
\end{subequations}

The model is complemented by the bulk concentration $c^{\b}_{i}$ (or externally supplied data) and the initial conditions at $t = 0$ given by
\begin{subequations}
\begin{alignat}{2}
    \left. c^{\p}_{i} \right|_{t = 0} &= c^{\p,\mathrm{init}}_{i} & & \qquad\text{in } (R^{\mathrm{c}}, R^{\mathrm{p}}),\\
    \left. c^{\s}_{i,k} \right|_{t = 0} &= c^{\s,\mathrm{init}}_{i,k} & & \qquad\text{in } (R^{\mathrm{c}}, R^{\mathrm{p}}).
\end{alignat}\label{eq:model-particle1D-ic}
\end{subequations}

\end{model}

\subsubsection{0D Particle Model (Model P0)}
\label{sec:particle_0D}

In many applications, intraparticle transport is sufficiently fast such that concentration gradients inside the particles can be neglected. This assumption gives rise to a lumped particle model in which the particle liquid- and solid-phase concentrations are spatially uniform.

\begin{assumption}[(P0)]\label{asmpt:model-0D-particle}
	In addition to \cref{asmpt:model-hrm,asmpt:model-1D-particle}, we assume that pore and surface diffusion are infinitely fast; that is, we consider the limits $D_{i}^{\p} \to \infty$ and $D_{i}^{\s} \to \infty$.
\end{assumption}

Taking these limits reduces the particle transport equations from \glspl{pde} to \glspl{ode}.
To eliminate the pore diffusion term, we integrate \cref{eq:model_particle_1D_liquid} over the accessible particle domain $R^{\mathrm{c}} < r < R^{\mathrm{p}}$, assuming the concentrations $c^{\p}_{i}$ and their derivatives are sufficiently smooth and integrable.
Since no quantity depends on the angular coordinates, we obtain
\begin{multline*}
    4\pi \int_{R^{\mathrm{c}}}^{R^{\mathrm{p}}} \frac{\partial c^{\p}_{i}}{\partial t}(t,z,r) r^2 \,\mathrm{d}r 
    =
    4\pi \int_{R^{\mathrm{c}}}^{R^{\mathrm{p}}} \frac{1}{r^2} \frac{\partial }{\partial r} \left( r^2 D_{i}^{\p} \frac{\partial c^{\p}_{i}}{\partial r}(t,z,r) \right) r^2 \,\mathrm{d}r
    \\
    - 4\pi \frac{1 - \varepsilon^{\p}}{\varepsilon^{\p}} \int_{R^{\mathrm{c}}}^{R^{\mathrm{p}}} \sum_{k=1}^{N^{\mathrm{b}}_i} f^{\mathrm{bind}}_{i,k}\mleft( \vec{c}^{\p}(t,z,r), \vec{c}^{\s}(t,z,r) \mright) r^2 \,\mathrm{d}r.
\end{multline*}
Employing the boundary conditions \cref{eq:model_particle_1D_BC_liquid_Rp,eq:model_particle_1D_BC_liquid_0}, the first integral on the right-hand side simplifies to
\begin{align*}
    \left[ r^2 D_{i}^{\p} \frac{\partial c^{\p}_{i}}{\partial r}(t,z,r) \right]_{R^{\mathrm{c}}}^{R^{\mathrm{p}}} 
    &=
    \left(R^{\mathrm{p}}\right)^2 D_{i}^{\p} \frac{\partial c^{\p}_{i}}{\partial r}\mleft(t,z,R^{\mathrm{p}} \mright) - \left(R^{\mathrm{c}}\right)^2 D_{i}^{\p} \frac{\partial c^{\p}_{i}}{\partial r}\mleft(t,z,R^{\mathrm{c}} \mright)
    \\
    &= \frac{\left(R^{\mathrm{p}}\right)^2}{\varepsilon^{\p}} k^{\mathrm{f}}_{i} \left( c^{\b}_i(t,z) - c^{\p}_{i}\mleft(t,z,R^{\mathrm{p}} \mright) \right) - 0.
\end{align*}

Next, we show that the liquid-phase concentration $c^{\p}_{i}$ is spatially uniform, that is, independent of the radial coordinate.
Rearranging \cref{eq:model_particle_1D_liquid}, using that $D_{i}^{\p}$ is constant, dividing by $D_{i}^{\p}$, and taking the limit $D_{i}^{\p} \to \infty$, we obtain
\begin{equation*}
    \frac{1}{r^2} \frac{\partial }{\partial r} \left( r^2 \frac{\partial c^{\p}_{i}}{\partial r} \right)
    =
    \frac{1}{D_{i}^{\p}} \left[ \frac{\partial c^{\p}_{i}}{\partial t} + \frac{1 - \varepsilon^{\p}}{\varepsilon^{\p}} \sum_{k=1}^{N^{\mathrm{b}}_i} f^{\mathrm{bind}}_{i,k}\mleft( \vec{c}^{\p}, \vec{c}^{\s} \mright) \right] \to 0.
\end{equation*}
Since $1 / r^2 \neq 0$, we conclude that $\partial \left( r^2 \frac{\partial c^{\p}_{i}}{\partial r} \right) / \partial r = 0$, which implies that $r^2 \partial c^{\p}_{i} / \partial r$ is constant.
The no-flux boundary condition \cref{eq:model_particle_1D_BC_liquid_0} at $r = R^{\mathrm{c}}$ fixes this constant to zero.
Hence, $\partial c^{\p}_{i} / \partial r = 0$, i.e. $c^{\p}_{i}$ is spatially constant on $\Omega^\mathrm{par}$.
Applying the same argument to \cref{eq:model_particle_1D_solid} and its boundary condition \cref{eq:model_particle_1D_BC_solid_0}, we deduce that $c^{\s}_{i,k}$ is spatially constant on $\Omega^\mathrm{par}$.
Because $c^{\p}_{i}, c^{\s}_{i,k} \colon (0,T^{\mathrm{end}}] \times \Omega^\b \to \Reals$ no longer depend on the radial position $r$, we arrive at the \glspl{ode}
\begin{align*}
    4\pi \int_{R^{\mathrm{c}}}^{R^{\mathrm{p}}} \frac{\partial c^{\p}_{i}}{\partial t}(t,z) r^2 \,\mathrm{d}r
    & = 
    4\pi \frac{\left(R^{\mathrm{p}}\right)^2}{\varepsilon^{\p}} k^{\mathrm{f}}_{i} \left( c^{\b}_i(t,z) - c^{\p}_{i}(t,z) \right)
    \\
    &- 4\pi  \frac{1 - \varepsilon^{\p}}{\varepsilon^{\p}} \int_{R^{\mathrm{c}}}^{R^{\mathrm{p}}} \sum_{k=1}^{N^{\mathrm{b}}_i} f^{\mathrm{bind}}_{i,k}\mleft( \vec{c}^{\p}(t,z), \vec{c}^{\s}(t,z) \mright) r^2 \,\mathrm{d}r, 
    \\
    4\pi \int_{R^{\mathrm{c}}}^{R^{\mathrm{p}}} \frac{\partial c^{\s}_{i,k}}{\partial t}(t,z) r^2 \,\mathrm{d}r
    &=
    4\pi \int_{R^{\mathrm{c}}}^{R^{\mathrm{p}}} f^{\mathrm{bind}}_{i,k}\mleft( \vec{c}^{\p}(t,z), \vec{c}^{\s}(t,z) \mright) r^2 \,\mathrm{d}r.
\end{align*}
The binding rate $f^{\mathrm{bind}}_{i,k}( \vec{c}^{\p}, \vec{c}^{\s} )$ is now spatially uniform due to its constant arguments.
We compute the integrals and divide the equation by the particle volume, $4 \pi \left(\left(R^{\mathrm{p}}\right)^3 - \left(R^{\mathrm{c}}\right)^3 \right) / 3$:
\begin{align*}
    \frac{\partial c^{\p}_{i}}{\partial t}
    &=
    \frac{a^{\mathrm{p}}}{\varepsilon^{\p}} k^{\mathrm{f}}_{i} \left( c^{\b}_i - c^{\p}_{i} \right) - \frac{1- \varepsilon^{\p}}{\varepsilon^{\p}} \sum_{k=1}^{N^{\mathrm{b}}_i} f^{\mathrm{bind}}_{i,k}\mleft( \vec{c}^{\p}, \vec{c}^{\s} \mright),
    \\
    \frac{\partial c^{\s}_{i,k}}{\partial t}
    &=
    f^{\mathrm{bind}}_{i,k}\mleft( \vec{c}^{\p}, \vec{c}^{\s} \mright),
\end{align*}
where
\begin{equation}
    a^{\mathrm{p}} \coloneqq \frac{3 \left(R^{\mathrm{p}}\right)^2}{\left(R^{\mathrm{p}}\right)^3 - \left(R^{\mathrm{c}}\right)^3} \label{eq:model-particle0D-surface-to-volume}
\end{equation}
is the ratio of the outer (spherical) particle surface to volume, which reduces to $3 / R^{\mathrm{p}}$ for particles without a solid core.

The reduced particle transport equations are summarized in the following model.
\begin{model}{(P0)}[0D Lumped Particle Model]\label{model:particle_0D}
    Consider a single spherical particle of radius $R^\mathrm{p}>0$ with a potential solid core of radius $0 \leq R^\mathrm{c} < R^\mathrm{p}$ and surface-to-volume ratio $a^{\mathrm{p}}$ as defined in \cref{eq:model-particle0D-surface-to-volume}. Mass transport is governed by the following \glspl{ode} in $(0, T^{\mathrm{end}}) \times \Omega^{\b}$ and for all components $i\in\{1,\dots,N^\mathrm{c}\}$ and their respective bound states $k\in\{1,\dots,N^{\mathrm{b}}_i\}$:
\begin{subequations}
\label{eq:model_particle_0D}
\begin{align}
    \frac{\partial c^{\p}_{i}}{\partial t}
    &=
    \frac{a^{\mathrm{p}}}{\varepsilon^{\p}} k^{\mathrm{f}}_{i} \left( c^{\b}_i - c^{\p}_{i} \right)
    -\frac{1 - \varepsilon^{\p}}{\varepsilon^{\p}} \sum_{k=1}^{N^{\mathrm{b}}_i} f^{\mathrm{bind}}_{i,k}\mleft( \vec{c}^{\p}, \vec{c}^{\s} \mright), \label{eq:model_particle_0D_liquid}
    \\
    \frac{\partial c^{\s}_{i,k}}{\partial t}
    &=
    f^{\mathrm{bind}}_{i,k}\mleft( \vec{c}^{\p}, \vec{c}^{\s} \mright). \label{eq:model_particle_0D_solid}
\end{align}
\end{subequations}

    The model is complemented by the bulk concentration $c^{\b}_{i}$ (or externally supplied data) and the initial conditions at $t = 0$ given by
	\begin{equation}
		\left. c^{\p}_{i} \right|_{t = 0} = c^{\p,\mathrm{init}}_{i}, \qquad \left. c^{\s}_{i,k} \right|_{t = 0} = c^{\s,\mathrm{init}}_{i,k}  \qquad\text{in } \Omega^{\b}. \label{eq:model-0Dparticle_init}
	\end{equation}
\end{model}

\section{Model Variants and Extensions}
\label{sec:model_variants}

In this section, we derive several model variants that can be integrated into the transport models introduced in \cref{sec:a_hierarchical_family_of_transport_models} in an orthogonal manner, i.e., each variant modifies only a single mechanism without affecting the remaining modeling choices.
For brevity, we focus on variants that are based on a one-dimensional representation of the column coupled with $N^{\p}\geq 1$ particle types.
Variants of the \gls{hrm} require separate treatment and are therefore discussed in Section~\ref{supp:sec:models-hrm-variants}.

The only non-orthogonal combination of model variants arises when rapid equilibrium is assumed simultaneously for binding~(\cref{sec:binding_equilibrium_kinetics}) and film diffusion~(\cref{sec:rapid-equilibrium_film_diffusion}) in the presence of surface diffusion.
In this case, two non-equivalent modeling approaches are possible.
The appropriate choice depends on whether surface diffusion is assumed to significantly influence molecular access to the particle pores, as detailed in Section~\ref{supp:sec:model-reqBindFilmDiff-notorthogonal}.

\subsection{Geometric Variants}

\subsubsection{Alternative Column Geometries}
\label{sec:variant_column_device}
\begin{figure}[!htbp]
	\centering
	\begin{subfigure}{0.75\textwidth}
		\centering
		\resizebox{\textwidth}{!}{\begin{tikzpicture}

    \def\Douter{4.0}                    %
    \def\Router{0.5*\Douter}            %
    \def\Dinner{1.0}                    %
    \def\Rinner{0.5*\Dinner}            %
    \def\Bedlength{\Router-\Rinner}     %
    \def\Height{4.0}                    %
    \def\yradouter{0.5}                 %
    \pgfmathsetmacro{\yradinner}{(\Dinner/\Douter)*\yradouter}      %

    \coordinate (leftBottom) at (0,0);
    \coordinate (leftTop) at (0,\Height);
    \coordinate (rightBottom) at (\Douter,0);
    \coordinate (rightTop) at (\Douter,\Height);
    \coordinate (innerLeftBottom) at (\Bedlength,0);
    \coordinate (innerLeftTop) at (\Bedlength,\Height);
    \coordinate (innerRightBottom) at (\Bedlength+\Dinner,0);
    \coordinate (innerRightTop) at (\Bedlength+\Dinner,\Height);
    \coordinate (midHeight) at (0,0.5*\Height);

    \draw[solid,thick] (leftTop) arc[x radius=\Router,y radius=\yradouter,start angle=180,end angle=540];
    \draw[solid,thick] (innerLeftTop) arc[x radius=\Rinner,y radius=\yradinner,start angle=180,end angle=540];
    \draw[solid,thick] (leftBottom) arc[x radius=\Router,y radius=\yradouter,start angle=180,end angle=360];
    \draw[dashed,thick] (rightBottom) arc[x radius=\Router,y radius=\yradouter,start angle=0,end angle=180];
    \draw[dashed,thick] (innerLeftBottom) arc[x radius=\Rinner,y radius=\yradinner,start angle=180,end angle=540];

    \fill[bulkVolume, even odd rule]
        ($(midHeight)+(0,0)$) 
        arc[x radius=\Router, y radius=\yradouter, start angle=180, end angle=540]
        -- cycle
        ($(midHeight)+(\Bedlength,0)$)
        arc[x radius=\Rinner, y radius=\yradinner, start angle=180, end angle=540]
        -- cycle;
    \draw[thick,solid] ($(midHeight)$) arc[x radius=\Router, y radius=\yradouter, start angle=180, end angle=360];
    \draw[thick,dashed] ($(midHeight)+(\Douter,0)$) arc[x radius=\Router, y radius=\yradouter, start angle=0, end angle=180];
    \draw[thick,dashed] ($(midHeight)+(\Bedlength,0)$) arc[x radius=\Rinner, y radius=\yradinner, start angle=180, end angle=360];
    \draw[thick,dashed] ($(midHeight)+(\Router+\Rinner,0)$) arc[x radius=\Rinner, y radius=\yradinner, start angle=0, end angle=180];

    \draw[thick,solid] (leftBottom) -- (leftTop);
    \draw[thick,solid] (rightBottom) -- (rightTop);
    \draw[thick,dashed] (innerLeftBottom) -- (innerLeftTop);
    \draw[thick,dashed] (innerRightBottom) -- (innerRightTop);

    \draw[<-,very thick,stdArrow] (\Bedlength,0.5*\Height) -- +(-0.75,0);
    \draw[<-,very thick,stdArrow] (-0.2,0.5*\Height) -- +(-0.75,0.0) node[above] {Flow};
    \draw[<-,very thick,stdArrow] (\Bedlength+\Dinner,0.5*\Height) -- +(0.75,0);
    \draw[<-,very thick,stdArrow] (\Douter + 0.2,0.5*\Height) -- +(0.75,0.0) node[above] {Flow};

    \draw[] (\Bedlength,0.25*\Height) -- +(-1.75, -0.25) node[left] {$\Gamma^{\mathrm{out}}$};

    \draw[] (\Douter,0.25*\Height) -- +(0.75, -0.25) node[right] {$\Gamma^{\mathrm{in}}$};

    \draw[] ($(leftTop)+(0.75,0.0)$) -- +(-1.0,0.25) node[left] {$\Gamma^{\mathrm{wall}}$};

    \draw[] ($(rightBottom)+(-0.75,0.0)$) -- +(1.0,-0.25) node[right] {$\Gamma^{\mathrm{wall}}$};

	\draw[->,stdArrow] (\Router+0.75*\Bedlength,0.5*\Height+0.25) -- ++(0.0, 1.0) -- ++(4.5,0) node[pos=0.65,above] {Cross-section} -- ++(0, -0.25) coordinate (crossecttop);

	\node[shape=coordinate] (crosssectmid) at ($(crossecttop) + (0.5,-1.75)$) {};
	\filldraw[thick,draw=black,bulkVolume] (crosssectmid) circle[radius=1.75];

	\draw ($(crosssectmid) + (1.25,1.25)$) -- +(0.5,0.25) node[right] {$\Gamma^{\mathrm{in}}$};
    \draw ($(crosssectmid) - (0.25,0.25)$) -- +(-1.75,-1.25) node[above] {$\Gamma^{\mathrm{out}}$};

	\node at ($(crosssectmid) + (-1.0, -0.35)$) {$\Omega^{\b}$};

	\node[shape=coordinate] (innerCylinder) at ($(crosssectmid)$) {};
	\filldraw[draw=black,fill=white,thick,solid] (innerCylinder) circle[radius=0.5];

	\node[shape=coordinate] (bead1) at ($(crosssectmid) + (1.0, 0.5)$) {};
	\filldraw[draw=black,fill=white] (bead1) circle[radius=0.35];
	\node[shape=coordinate] (bead2) at ($(crosssectmid) + (-0.125, -1.125)$) {};
	\filldraw[draw=black,fill=white] (bead2) circle[radius=0.45];
	\node[shape=coordinate] (bead3) at ($(crosssectmid) + (-1.0, 0.5)$) {};
	\filldraw[draw=black,fill=white] (bead3) circle[radius=0.4];

	\draw (bead1) -- +(-0.75,1.5) coordinate (beadInner);
	\draw (bead3) -- (beadInner);
	\draw (bead2) -- (beadInner);
	\node[above] at (beadInner) {$\Omega^{\mathrm{par}}$};

	\node[shape=coordinate] (beadSurf) at ($(crosssectmid) + (2.0,-0.5)$) {};
	\node[right] at (beadSurf) {$\Gamma^{\mathrm{par}}$};
	\draw ($(bead1)!0.35cm!(beadSurf)$) -- (beadSurf);
	\draw ($(bead2)!0.45cm!(beadSurf)$) -- (beadSurf);
	\draw ($(bead3)!0.4cm!(beadSurf)$) -- (beadSurf);

\end{tikzpicture}}
		\caption{Schematic of an annular column with radial flow (left) and cross-section of corresponding packed bed (right; not to scale, $R^\mathrm{p} \ll \Rho^\mathrm{c}$).}
		\label{fig:radial-flow-device-geometries}
	\end{subfigure}
	\vspace{1em}
	\begin{subfigure}{0.75\textwidth}
		\centering
		\resizebox{\textwidth}{!}{\begin{tikzpicture}

    \def\Rleft{2}      %
    \def\Rright{1.2}   %
    \def\Rmiddle{1.6}   %
    \def\Length{4}     %
    \def\xrad{0.5}     %

    \coordinate (leftCenter) at (0,0);
    \coordinate (rightCenter) at (\Length,0);
    \coordinate (leftTop) at (0,\Rleft);
    \coordinate (leftBottom) at (0,-\Rleft);
    \coordinate (rightTop) at (\Length,\Rright);
    \coordinate (rightBottom) at (\Length,-\Rright);
    \coordinate (midCenter) at ($(leftCenter)!0.5!(rightCenter)$);
    \coordinate (midTop) at ($(midCenter)+(0,\Rmiddle)$);

    \draw[thick] (leftTop) -- (rightTop);
    \draw[thick] (leftBottom) -- (rightBottom);

    \draw[dashed,thick] (leftTop) arc[x radius=\xrad,y radius=\Rleft,start angle=90,end angle=-90];
    \draw[solid,thick] (leftTop) arc[x radius=\xrad,y radius=\Rleft,start angle=90,end angle=270];

    \draw[dashed,thick] (midTop)
    arc[x radius=\xrad, y radius=\Rmiddle, start angle=90, end angle=-90];
    \draw[solid,thick] (midTop)
    arc[x radius=\xrad, y radius=\Rmiddle, start angle=90, end angle=270];
    \fill[bulkVolume] (midCenter) ellipse [x radius=\xrad, y radius=\Rmiddle];

    \draw[solid,thick] (rightTop)arc[x radius=\xrad,y radius=\Rright,start angle=90,end angle=450];

    \draw[<-,very thick,stdArrow] ($(leftCenter) - (0.75,0)$) -- +(-0.75,0);
    \draw[->,very thick,stdArrow] ($(rightCenter) + (0.75,0)$) -- +(0.75,0) node[right] {Flow};

    \draw[] ($(leftCenter)+(0.0,0.75)$) -- +(-1.0, 0.75) node[left] {$\Gamma^{\mathrm{in}}$};

    \draw[] ($(leftCenter)+(4.0,0.5)$) -- +(1.0, 0.75) node[right] {$\Gamma^{\mathrm{out}}$};

    \draw[] ($(leftCenter)!0.5!(rightCenter)+(-0.75,\Rleft-0.25)$) -- +(-0.25,0.75) node[left] {$\Gamma^{\mathrm{wall}}$};

	\draw[->,stdArrow] ($(midCenter) - (0,\Rmiddle-0.5)$) -- ++(0.0, -0.75) -- ++(7.0,0) node[midway,above] {Cross-section} -- ++(0, 0.25) coordinate (crossectbottom);

	\node[shape=coordinate] (crosssectmid) at ($(crossectbottom) + (0.125,1.75)$) {};
	\filldraw[thick,draw=black,bulkVolume] (crosssectmid) circle[radius=1.5];

	\draw[rotate around={60:(crosssectmid)}] ($(crosssectmid) + (0,1.5)$) -- +(0,0.4) node[above] {$\Gamma^{\mathrm{wall}}$};

	\node at ($(crosssectmid) + (-0.55, -0.2)$) {$\Omega^{\b}$};

	\node[shape=coordinate] (bead1) at ($(crosssectmid) + (0.75, 0.5)$) {};
	\filldraw[draw=black,fill=white] (bead1) circle[radius=0.5];
	\node[shape=coordinate] (bead2) at ($(crosssectmid) + (.25, -0.85)$) {};
	\filldraw[draw=black,fill=white] (bead2) circle[radius=0.45];
	\node[shape=coordinate] (bead3) at ($(crosssectmid) + (-.45, 0.45)$) {};
	\filldraw[draw=black,fill=white] (bead3) circle[radius=0.4];

	\draw (bead1) -- +(-0.65,1.2) coordinate (beadInner);
	\draw (bead3) -- (beadInner);
	\draw (bead2) -- (beadInner);
	\node[above] at (beadInner) {$\Omega^{\mathrm{par}}$};

	\node[shape=coordinate] (beadSurf) at ($(crosssectmid) + (1.6,-0.9)$) {};
	\node[right] at (beadSurf) {$\Gamma^{\mathrm{par}}$};
	\draw ($(bead1)!0.5cm!(beadSurf)$) -- (beadSurf);
	\draw ($(bead2)!0.45cm!(beadSurf)$) -- (beadSurf);
	\draw ($(bead3)!0.4cm!(beadSurf)$) -- (beadSurf);

\end{tikzpicture}}
		\caption{Schematic of a conical frustum column with axial flow (left) and cross-section of corresponding packed bed (right; not to scale, $R^\mathrm{p} \ll \Rho^\mathrm{c}$).}
		\label{fig:frustum-device-geometries}
	\end{subfigure}
	\caption{Alternative column geometries and corresponding packed-bed cross-sections.}
	\label{fig:column-geometries}
\end{figure}
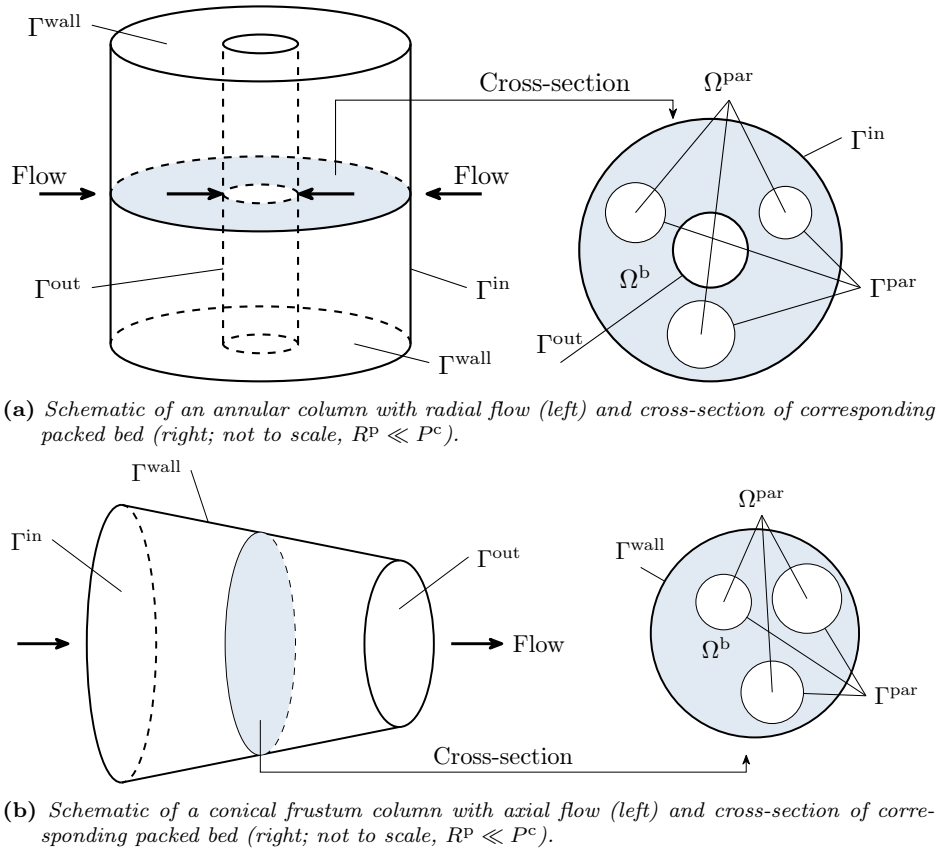

Cylindrical columns with axial flow are the most widely used in chromatography. However, alternative column geometries, such as annular columns with radial flow, are employed to meet specific performance objectives or to satisfy particular operating conditions, for example by reducing the pressure drop in large-scale columns.
The derivation presented in \cref{sec:bulk_domain_transport} can be extended to these geometries by adopting an appropriate coordinate system.
Rather than reiterating the derivation, we directly present the resulting one-dimensional governing equations for mass transport in the bulk domain of annular columns with radial flow and frustum columns with axial flow, both shown in \cref{fig:column-geometries}.

In annular columns with radial flow (\cref{fig:radial-flow-device-geometries}), we denote the radial coordinate by $\rho \in (\Rho^\mathrm{inner}, \Rho^\mathrm{outer})$, where $0 < \Rho^\mathrm{inner} < \Rho^\mathrm{outer}$ are the inner and outer radii of the annular bed, corresponding to the outlet and inlet, respectively.
The cross-sectional area normal to the flow varies with radial position according to $A(\rho) = 2\pi \rho H^\mathrm{b}$, where $H^\mathrm{b}$ is the axial bed height.
The superficial velocity is therefore given by
\begin{align}
u(\rho) = \frac{Q}{A(\rho)} = \frac{Q}{2\pi \rho H^\mathrm{b}},
\end{align}
and the dispersion coefficient $D_{i}(\rho)$ in the flow direction may depend on the local superficial velocity and, consequently, on the radial position.
The bulk transport equation can then be written as
\begin{multline}
\varepsilon^{\mathrm{b}} \frac{\partial c^{\b}_i}{\partial t}
= -\frac{Q}{2\pi H^\mathrm{b} \rho} \frac{\partial \left( \varepsilon^{\mathrm{b}} c^{\b}_i \right)}{\partial \rho}
+ \frac{1}{\rho} \frac{\partial}{\partial \rho} \left( \rho \varepsilon^{\mathrm{b}} D_{i} \frac{\partial c^{\b}_i}{\partial \rho} \right) \\
- \left(1 - \varepsilon^{\mathrm{b}} \right) \sum_{j=1}^{N^{\mathrm{p}}} \frac{3 d_j}{R^{\mathrm{p}}_{j}} k^{\mathrm{f}}_{j,i} \left(c^{\b}_i - \left. c^{\p}_{j,i} \right|_{r = R^{\mathrm{p}}_{j}} \right)
\quad \text{in } (0, T^\mathrm{end}) \times (\Rho^\mathrm{inner}, \Rho^\mathrm{outer}).
\end{multline}

In frustum columns with axial flow (\cref{fig:frustum-device-geometries}), we denote the axial coordinate by $z \in (0, L^\mathrm{b})$, measured from the inlet at the bottom of the frustum to the outlet at the top, where $L^\mathrm{b}>0$ is the axial bed height.
The cross-sectional area normal to the flow is given by
\begin{align*}
A(z) = \pi \rho(z)^2,
\end{align*}
where the radius varies linearly with the axial coordinate according to
\begin{align*}
\rho(z) = \Rho^\mathrm{0} + \frac{\Rho^\mathrm{L^\mathrm{b}} - \Rho^\mathrm{0}}{L^\mathrm{b}} z,
\end{align*}
with $\Rho^\mathrm{0}>\Rho^\mathrm{L^\mathrm{b}}>0$ denoting the radii at the bottom ($z=0$) and top ($z=L^\mathrm{b}$) of the frustum, respectively. 
In a conical frustum, the inclined wall geometry generally leads to a velocity field that varies over the cross section.
We therefore introduce the cross-sectionally averaged (superficial) velocity
\begin{align}
 \overline{u}(z) \coloneqq \frac{1}{A(z)} \int_{A(z)} \vec{u} \cdot \vec{n}\,\mathrm{d}A.
\end{align}
By incompressibility, the volumetric flow rate is constant, and hence
\begin{align}
\overline{u}(z) = \frac{Q}{A(z)} = \frac{Q}{\pi \rho(z)^2}.
\end{align}
The bulk transport equation can then be written as
\begin{multline}
\varepsilon^{\mathrm{b}} \frac{\partial c^{\b}_i}{\partial t}
= -\frac{Q}{\pi \rho(z)^2} \frac{\partial \left( \varepsilon^{\mathrm{b}} c^{\b}_i \right)}{\partial z}
+ \frac{1}{\rho(z)^2} \frac{\partial}{\partial z} \left( \rho(z)^2 \varepsilon^{\mathrm{b}} D_{i} \frac{\partial c^{\b}_i}{\partial z} \right)
\\
- \left(1 - \varepsilon^{\mathrm{b}} \right) \sum_{j=1}^{N^{\mathrm{p}}} \frac{3 d_j}{R^{\mathrm{p}}_{j}} k^{\mathrm{f}}_{j,i} \left(c^{\b}_i - \left. c^{\p}_{j,i} \right|_{r = R^{\mathrm{p}}_{j}} \right)
\quad \text{in } (0, T^\mathrm{end}) \times (0, L^\mathrm{b})
\end{multline}
where $D_{i}(z)$ is the dispersion coefficient.

\subsubsection{Alternative Particle Geometries}
Chromatography particles are generally not perfectly spherical, although they are usually approximated as such \cite[Section~3.1]{Gu2015}.
Bed compression caused by hydrodynamic or mechanical forces can lead to slight particle deformation.
However, simulations by \cite{Dorn2016} indicate that this deformation has only a minor effect.
\cite{EGHBALMANESH2024119396} reported differences in the flow field and pressure drop for spherical, cylindrical, and slab-shaped particles.
Overall, the literature reports mixed findings, and practical applications of non-spherical particle geometries remain niche.

In this section, we introduce two alternative particle geometries: cylindrical and slab-shaped particles.
We only summarize the derivation, as the order reduction procedure described in \cref{sec:1D_particle} can be applied analogously using these particle geometries.
These particle models are one-dimensional. Hence, film mass transfer is considered only in the direction of the resolved coordinate, while film mass transfer through the two end faces of the cylindrical particle and the four side faces of the slab-shaped particle is neglected.
This approximation is justified for cylindrical and slab-shaped particles with large aspect ratios, for which the neglected surfaces contribute negligibly to the overall film mass transfer. In particular, cylindrical particles represent long fibers, whereas slab-shaped particles represent thin flakes.

The alternative geometries induce two differences from the previously presented models: the coordinate system of the particle transport equations and the surface-to-volume ratio in the film diffusion term of the bulk transport equation.
The starting point of the derivation is a coordinate-free three-dimensional particle equation, similar to \cref{eq:model-hrm-particle}, defined on an arbitrary domain $\Omega^\mathrm{par}\subset\Reals^3$ rather than a spherical domain.
For brevity, we only consider the particle liquid equation; the solid equation is derived analogously.
As conceptually established before, we employ the symmetry \cref{asmpt:model-1D-particle}; the required representations of the differential operators are summarized in Section~\ref{supp:sec:vectorcalculus-cylindrical}.
For cylindrical particles, we exploit the representation of the differential operators in cylindrical coordinates and get
\begin{subequations}
\label{eq:model_particle_geometries}
\begin{align}    
    \frac{\partial c^{\p}_{i}}{\partial t}
    &=
    \frac{1}{r} \frac{\partial }{\partial r} \left( r D_{i}^{\p} \frac{\partial c^{\p}_{i}}{\partial r} \right)
    - \frac{1 - \varepsilon^{\p}}{\varepsilon^{\p}} \sum_{k=1}^{N^{\mathrm{b}}_i} f^{\mathrm{bind}}_{i,k}\mleft( \vec{c}^{\p}, \vec{c}^{\s} \mright). \label{eq:model_particle_cylinder_liquid}
    \intertext{For slab-shaped particles, we instead exploit the representation of the differential operators in Cartesian coordinates and get}
    \frac{\partial c^{\p}_{i}}{\partial t}
    &=
    \frac{\partial }{\partial r} \left( D_{i}^{\p} \frac{\partial c^{\p}_{i}}{\partial r} \right)
    - \frac{1 - \varepsilon^{\p}}{\varepsilon^{\p}} \sum_{k=1}^{N^{\mathrm{b}}_i} f^{\mathrm{bind}}_{i,k}\mleft( \vec{c}^{\p}, \vec{c}^{\s} \mright). \label{eq:model_particle_slab_liquid}
\end{align}
\end{subequations}
Similarly to the derivation for spherical particles, we obtain the boundary condition \cref{eq:model-1Dparticle-bulk-bc}.

Next, we consider the film diffusion term in the bulk transport model.
Following the procedure from \cref{sec:3D_continuum_bulk_transport_model}, we obtain the following surface measures for cylindrical and slab-shaped particles, analogous to \cref{eq:model-3d-surface-measure-sphere}:
\begin{subequations}
\label{eq:model-par-geom-surface-measure}
\begin{align}
    \label{eq:model-surface-measure-cylinder}
    \surfarea{\Gamma^{\mathrm{bb}}}
    &= \sum_{j=1}^{N^{\mathrm{p}}}
       \frac{\vol{V^{\mathrm{par}}} d_j}
            {\pi (R^{\mathrm{p}}_j)^2 H^\mathrm{cyl}_j}
       \, 2\pi R^{\mathrm{p}}_j H^\mathrm{cyl}_j
     = \vol{V^{\mathrm{par}}}
       \sum_{j=1}^{N^{\mathrm{p}}} \frac{2 d_j}{R^{\mathrm{p}}_j},
    \\
    \label{eq:model-surface-measure-slab}
    \surfarea{\Gamma^{\mathrm{bb}}}
    &= \sum_{j=1}^{N^{\mathrm{p}}}
       \frac{\vol{V^{\mathrm{par}}} d_j}{A^\mathrm{slab}_j 2 R^{\mathrm{p}}_j}
       \, 2 A^\mathrm{slab}_j
     = \vol{V^{\mathrm{par}}}
       \sum_{j=1}^{N^{\mathrm{p}}} \frac{d_j}{R^{\mathrm{p}}_j}.
\end{align}
\end{subequations}
Here, $H^\mathrm{cyl}_j>0$ denotes the height of cylindrical particle $j$.
Furthermore, $A^\mathrm{slab}_j>0$ denotes the face area of slab-shaped particle $j$, and $R^{\mathrm{p}}_j>0$ denotes half the slab thickness in the resolved coordinate direction.
Compared to spherical particles, only the surface-to-volume ratio in the film diffusion term of the bulk transport equation changes, resulting in geometry-dependent factors of $1$, $2$, and $3$ multiplying $d_j / R^{\mathrm{p}}_j$ for slabs, cylinders, and spheres, respectively.

\subsubsection{Component-Specific Pore Accessibility (Model PC)}
\label{sec:component_specific_pore_accessibility}
{
\captionsetup[subfigure]{labelformat=empty}
\begin{figure}[!ht]
	\centering
	\begin{subfigure}{0.55\textwidth}
		\centering
		\includegraphics[width=\textwidth]{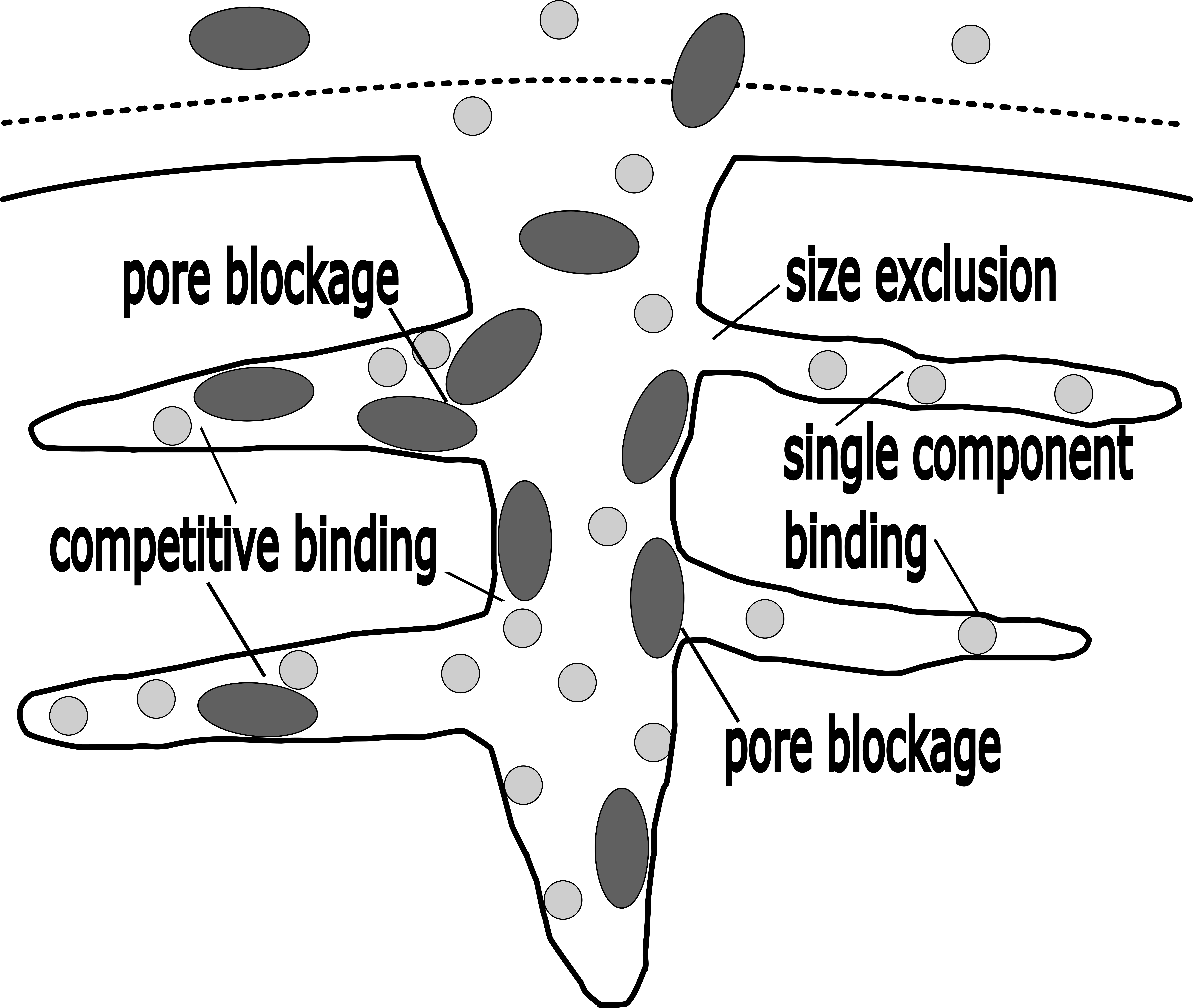}
		\caption{\textbf{(a)} Molecular-scale phenomena \mbox{(adapted from}~\citealp[figure 4.14]{worch2021adsorption}).}
	\end{subfigure}
    \hfill
    \begin{subfigure}{0.4\textwidth}
		\centering
		\includegraphics[width=\textwidth]{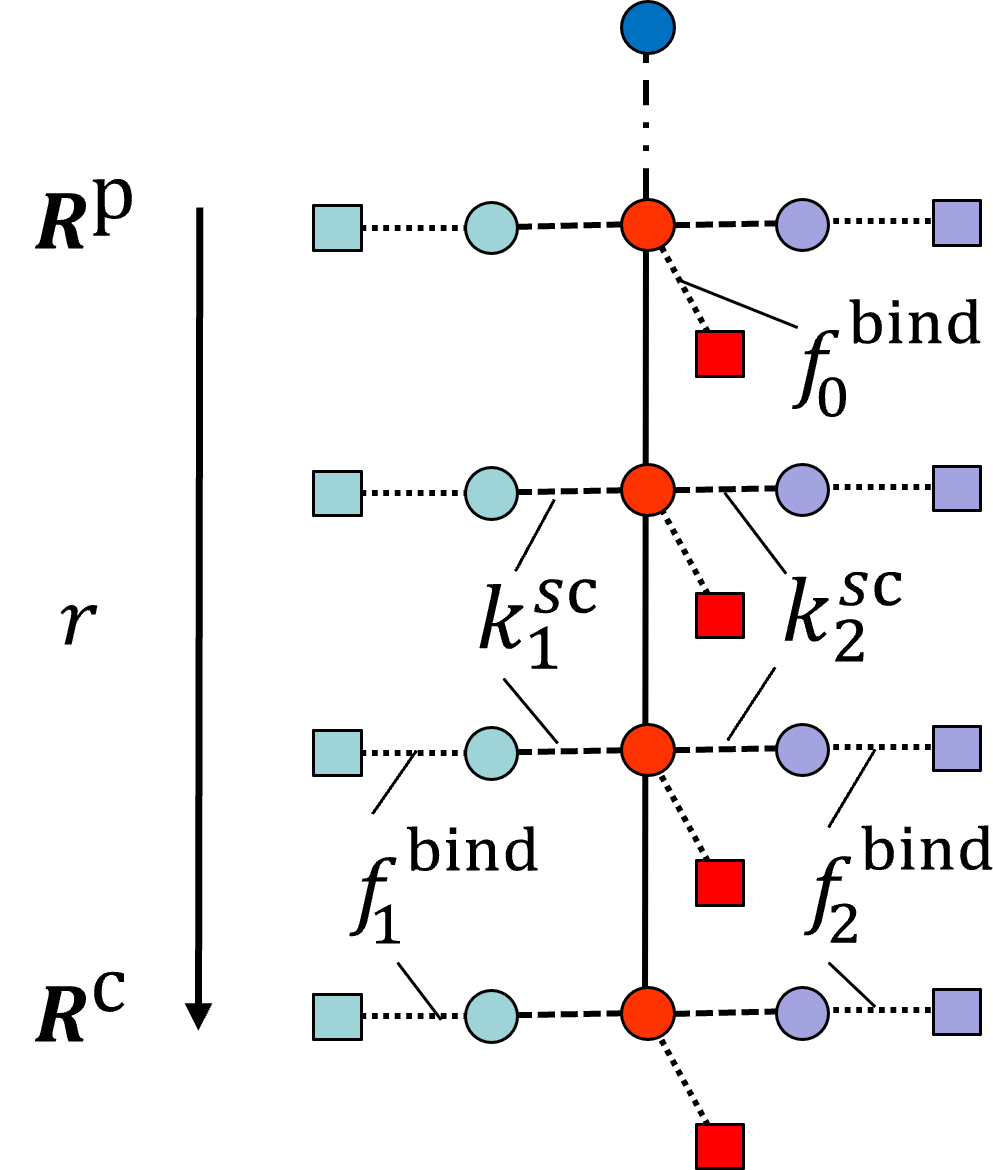}
		\caption{\textbf{(c.1)} Discretization and domain decomposition.}
	\end{subfigure}
	\hfill
	\begin{subfigure}{0.55\textwidth}
		\centering
		\includegraphics[width=\textwidth]{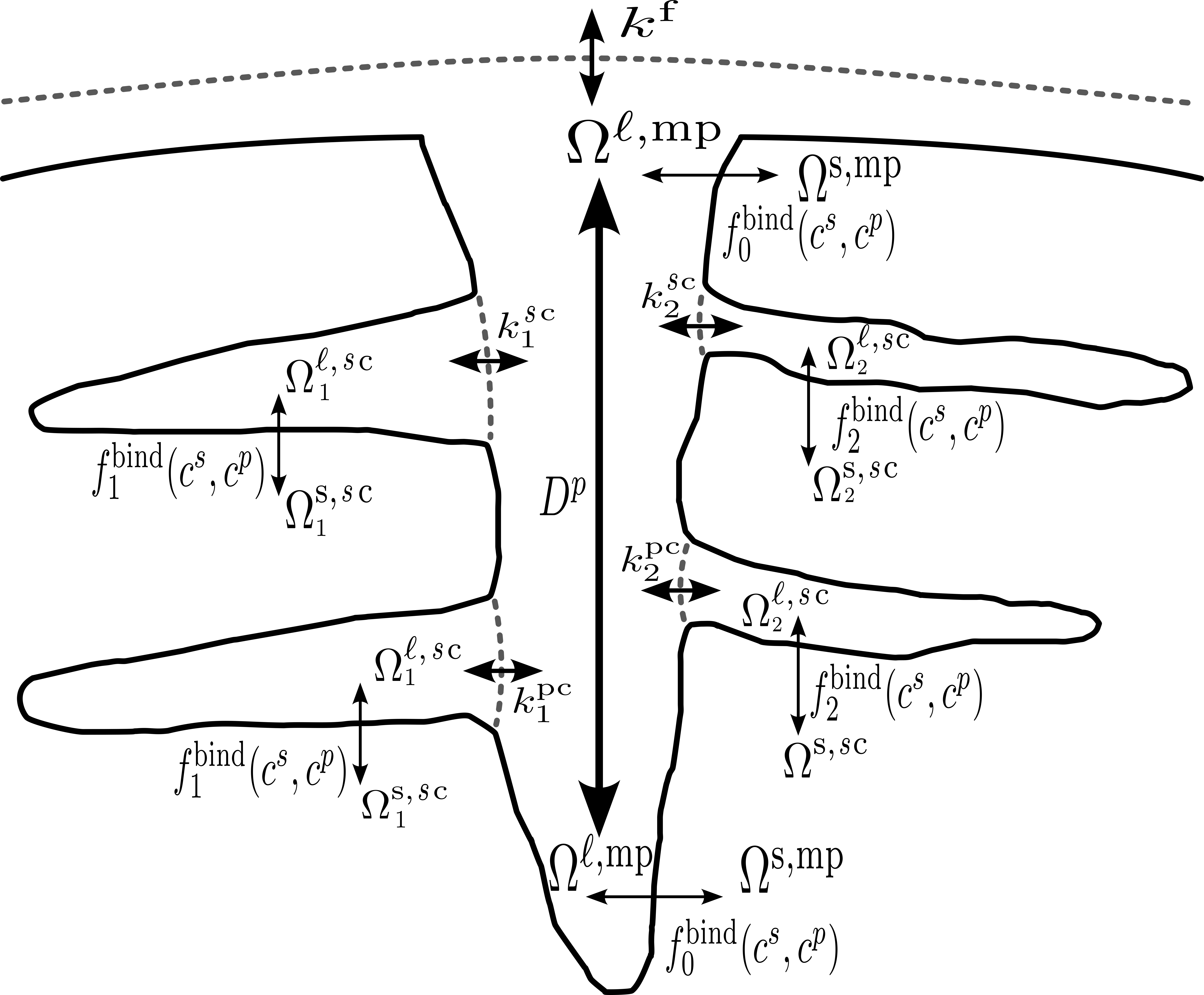}
		\caption{\textbf{(b)} Control volumes and mass transfer.}
	\end{subfigure}
    \hfill
    \begin{subfigure}{0.4\textwidth}
		\centering
		\includegraphics[width=\textwidth]{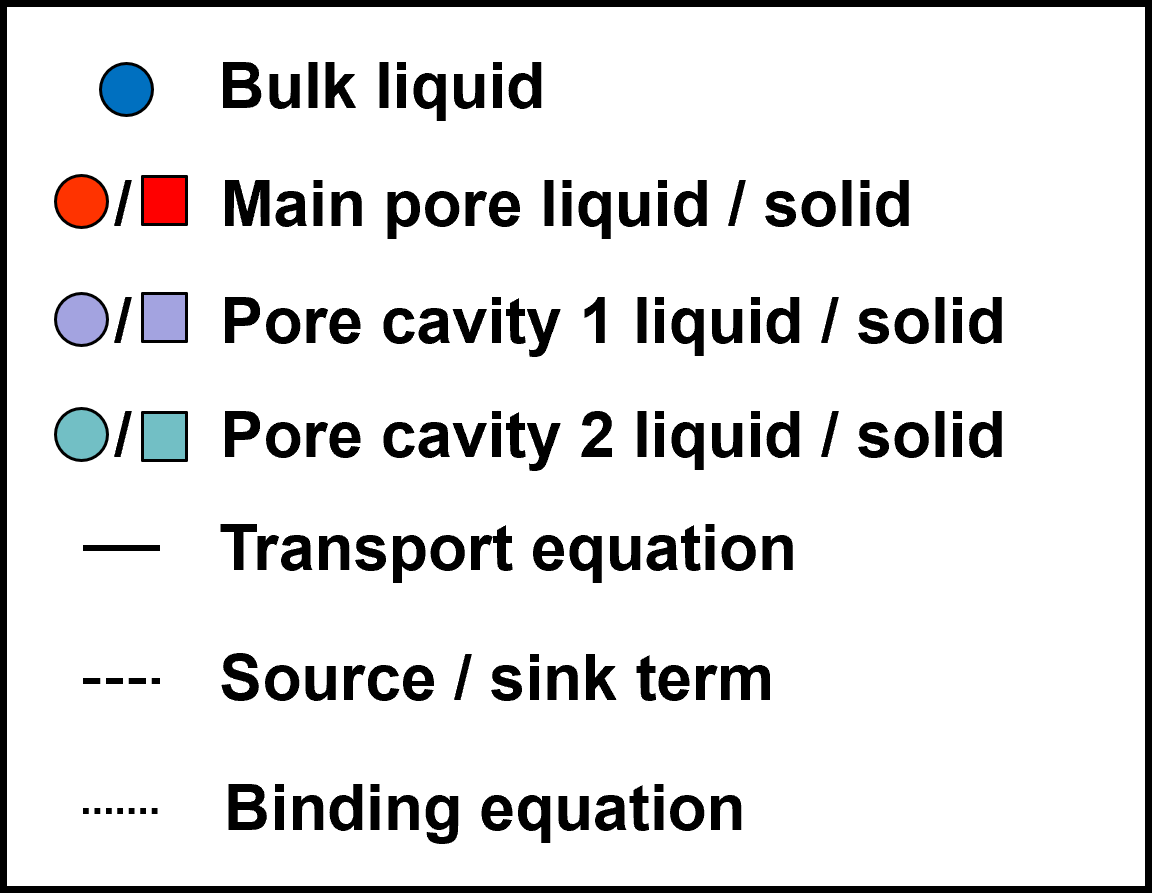}
		\caption{\textbf{(c.2)} Discretization and domain decomposition (legend).}
	\end{subfigure}
	\caption{Schematic of the particle model with two side-cavity types, illustrating progression from a) molecular-scale phenomena through b) continuum description based on control volumes and mass transfer to c) discretized domain suitable for numerical implementation.}
	\label{fig:side-cavity-model}
\end{figure}
}

\gls{sec} exploits differences in the hydrodynamic size of chemical species to achieve separation.
Within the established modeling framework, size-exclusion effects can be approximated via component-specific film and pore diffusion coefficients ($k^\mathrm{f}_{j,i}, D^\p_{j,i}$), as demonstrated, for example, by \cite{ALTERN2024464717} and \cite{Ojala2015}.

A more direct and effective approach to modeling size-exclusion introduces a component-specific pore accessibility factor $\mathrm{F}^\mathrm{acc}_{j,i}\in(0, 1]$, which defines the component-specific particle porosities
\begin{align}
    \label{eq:comp-specific-porosity}
    \varepsilon^{\mathrm{p}}_{j,i} \coloneqq \mathrm{F}^\mathrm{acc}_{j,i} \varepsilon^{\mathrm{p}}_j.
\end{align}
Although $\mathrm{F}^\mathrm{acc}_{j,i}=0$ is mathematically excluded, complete exclusion of a component from the particle domain can still be modeled by setting $k^\mathrm{f}_{j,i}=0$.
An important caveat of modeling size-exclusion using \cref{eq:comp-specific-porosity} is that the particle liquid and solid phases, as well as binding-related quantities such as the saturation capacity, are then defined with respect to different accessible pore volumes for each component.
Consequently, this approach is incompatible with binding models, which require these quantities to be defined with respect to the same pore volume for all components.

This limitation is acceptable as long as \gls{sec} is modeled as a non-binding separation mechanism \mbox{(as done, e.g., by}~\citealp{Gu2015}).
However, \cite{ALTERN2024464717} highlighted that this gap limits currently available modeling tools.
In their study of size-exclusion resins, where separation is governed by both steric exclusion and binding, they circumvented this limitation using a reduced-order modeling strategy that mimics the effect of component-specific accessible pore volumes through component-specific pore diffusion.
More broadly, the literature contains only a limited number of highly specialized approaches that simultaneously account for binding and size exclusion.
Another example of combined binding and size exclusion is the extended \gls{sma} model proposed by \cite{zhao_pore-blocking_2024}, which incorporates pore-blocking effects and thereby introduces variable pore accessibility.

In the spirit of our unified approach, we extend the particle models presented in \cref{sec:intra_particle_transport} to account for the combined effects of pore blocking, size exclusion, and competitive binding while remaining agnostic to other model components.
The proposed model extends \cref{model:particle_1D} by partitioning the particle pore phase into a main pore network with multiple side-cavities branching off from it, as illustrated in \cref{fig:side-cavity-model}.
This approach provides a pragmatic balance between model complexity and numerical tractability.
Within the main pore network, solutes are transported along the radial particle coordinate by pore diffusion.
The side-cavities do not contribute to transport along the radial particle coordinate and can be accessed only via interfacial mass exchange with the main pore network.
Hereafter, we refer to this interfacial mass exchange as cavity exchange.

\begin{remark}[(PCa)]
Although cavity exchange is conceptually analogous to film diffusion, the cavity exchange coefficients $k^\mathrm{sc}_{m,i}$ have units of $\frac{1}{s}$, rather than $\frac{m}{s}$. In contrast to the film diffusion coefficient $k^\mathrm{f}$, which describes mass transfer across the particle surface, the interfaces between the main pore network and the side cavities are not resolved geometrically. Instead, each side-cavity type represents a homogenized ensemble of cavities with an unresolved total interfacial area. Consequently, the cavity exchange process cannot be associated  with a unique physical interface area, and $k^\mathrm{sc}_{m,i}$ is defined directly as a homogenized first-order volumetric mass-transfer coefficient that lumps the effects of the unresolved interfacial area and the local mass-transfer resistance.
\end{remark}

\begin{assumption}[(PC)]\label{asmpt:model-side-cavities}
In addition to \cref{asmpt:model-hrm} -- except for the assumption of uniform particle porosity -- we assume that
\begin{itemize}\setlength{\itemsep}{0\baselineskip}
    \item The particle pore phase is partitioned into a main pore network $\Omega^{\mathrm{mp}}$ and a finite number of side-cavity types $\Omega^{\mathrm{sc}}_m$ with $m\in\{1,\dots,N^\mathrm{sc}\}$.
    \item The side-cavities are connected to the main pore network via interfacial mass transfer with cavity exchange coefficients $k^\mathrm{sc}_{m,i}\geq 0$.
    \item Pore diffusion along the radial particle coordinate occurs only within the main pore network.
    \item There is no direct transport between side-cavities, i.e., the side-cavities are connected only via the main pore network.
    \item The liquid-to-solid volume ratio may differ between side-cavity types.
    \item Solutes may be blocked from entering a side-cavity by other solutes occupying either the side-cavity or its entrance.
\end{itemize}
\end{assumption}

Following Assumption~\ref{asmpt:model-side-cavities}, the pore phase is partitioned into mutually disjoint domains
\begin{equation}
    \label{eq:pore_size_particle_volume_partition}
    \Omega^\mathrm{p} \equiv
    \Omega^{\mathrm{mp}} \cup \bigcup_{m=1}^{N^{\mathrm{sc}}} \Omega^{\mathrm{sc}}_m.
\end{equation}
The porosities $\varepsilon^{\mathrm{mp}},\varepsilon^{\mathrm{sc}}_m\in(0,1)$ of the main pore network and side-cavity types satisfy
\begin{equation} \label{eq:pore_size_distribution_volfrac_porosities}
    \varepsilon^\mathrm{p} = d^{\mathrm{mp}} \varepsilon^\mathrm{mp} + \sum_{m=1}^{N^{\mathrm{sc}}} d^{\mathrm{sc}}_m \varepsilon^{\mathrm{sc}}_m,
    \quad \text{with} \quad d^{\mathrm{mp}} + \sum_{m=1}^{N^{\mathrm{sc}}} d^{\mathrm{sc}}_m = 1,
\end{equation}
where $d^{\mathrm{mp}},d^{\mathrm{sc}}_m\in(0, 1)$ denote the volume fractions of the main pore network and side-cavity types, respectively.
Using this notation, we obtain the following one-dimensional particle transport model with side-cavities.

\begin{model}{(PC)}[1D Continuum Particle Model with Side-Cavities]\label{model:particle_side_cavities}
Consider a single spherical particle of radius $R^\mathrm{p}>0$ and a discrete partition of the pore phase into a main pore network and $N^{\mathrm{sc}}$ side-cavity types according to~\cref{eq:pore_size_particle_volume_partition,eq:pore_size_distribution_volfrac_porosities}.
Here, the pore and solid-phase concentrations $c^{\p}_{m,i}$ and $c^{\s}_{m,i,k}$, together with all associated functions and parameters, carry an additional index $m$, where $m=0$ denotes the main pore network and $m\in\{1,\dots,N^{\mathrm{sc}}\}$ denotes the side-cavity types.
Mass transport in the main pore network is governed by the following \glspl{pde} in $(0, T^{\mathrm{end}}) \times \Omega^{\b} \times (0, R^\mathrm{p})$ and for all components $i\in\{1,\dots,N^\mathrm{c}\}$ and their respective bound states $k\in\{1,\dots,N^{\mathrm{b}}_i\}$:
\begin{subequations}
\label{eq:model_particle_side_cavities}
\begin{align}    
    \frac{\partial c^{\p}_{0, i}}{\partial t}
    &=
    \frac{1}{r^2} \frac{\partial }{\partial r} \left( r^2 D_{i}^{\p} \frac{\partial c^{\p}_{0, i}}{\partial r} \right)
    - \frac{1 - \varepsilon^{\mathrm{mp}}}{\varepsilon^{\mathrm{mp}}} \sum_{k=1}^{N^{\mathrm{b}}_i} f^{\mathrm{bind}}_{0,i,k}\mleft( \vec{c}^{\p}_0, \vec{c}^{\s}_0 \mright)
     - \sum_{m=1}^{N^{\mathrm{sc}}} \frac{d^{\mathrm{sc}}_{m} \varepsilon^{\mathrm{sc}}_m}{d^{\mathrm{mp}} \varepsilon^{\mathrm{mp}}} k^{\mathrm{sc}}_{m, i} \left( c^{\p}_{0,i} - c^{\p}_{m, i} \right), \label{eq:model_particle_side_cavities_liquid}
    \\
    \frac{\partial c^{\s}_{0, i, k}}{\partial t}
    &=
    \frac{1}{r^2} \frac{\partial }{\partial r} \left( r^2 D_{i,k}^{\s} \frac{\partial c^{\s}_{0, i, k}}{\partial r} \right) + f^{\mathrm{bind}}_{0,i,k}\mleft( \vec{c}^{\p}_0, \vec{c}^{\s}_0 \mright).
    \label{eq:model_particle_side_cavities_solid}
\end{align}
\end{subequations}

For each side-cavity type, the following mass balances hold:
\begin{subequations}
    \begin{align}
        \frac{\partial c^{\p}_{m,i}}{\partial t}
        &=
        k^{\mathrm{sc}}_{m,i} \left( c^{\p}_{0,i} - c^{\p}_{m,i} \right)
        - \frac{ 1 - \varepsilon^{\mathrm{sc}}_{m}}{\varepsilon^{\mathrm{sc}}_{m}}
        \sum_{k=1}^{N^{\mathrm{b}}_i} f^{\mathrm{bind}}_{m,i,k}(\vec{c}^{\p}_{m}, \vec{c}^{\s}_{m}),
        \\
        \frac{\partial c^{\s}_{m,i,k}}{\partial t}
        &=
        f^{\mathrm{bind}}_{m,i,k}(\vec{c}^{\p}_{m}, \vec{c}^{\s}_{m}).
    \end{align}
\end{subequations}

The boundary conditions are given on $(0, T^{\mathrm{end}})$ by
\begin{subequations}
\label{eq:model_particle_side_cavities_BC}
\begin{align}
    - \varepsilon^{\mathrm{mp}} \left. \left( D^{\p}_{i} \frac{\partial c^{\p}_{0,i}}{\partial r} \right) \right|_{r=0}
    &= 0,
    \label{eq:model_particle_1D_side_cavities_liquid_0}
    \\
	-\left( 1 - \varepsilon^{\mathrm{mp}} \right) \left. \left( D^{\s}_{i,k} \frac{\partial c^{\s}_{0, i, k}}{\partial r} \right) \right|_{r=0}
    &= 0,
    \label{eq:model_particle_side_cavities_BC_solid_0}\\
    d^{\mathrm{mp}} \varepsilon^{\mathrm{mp}} \left(D^{\p}_{i} \left. \frac{\partial c^{\p}_{0, i}}{\partial r} \right)\right|_{r = R^{\mathrm{p}}}
    &= 
    k^{\mathrm{f}}_{i} \left( c^{\b}_i - \left. c^{\p}_{0,i} \right|_{r = R^{\mathrm{p}}} \right)
    \label{eq:model_particle_side_cavities_BC_liquid_Rp}\\
    \left(1 - \varepsilon^{\mathrm{mp}} \right) \left( \left. D^{\s}_{i, k} \frac{\partial c^{\s}_{0, i, k}}{\partial r} \right) \right|_{r = R^{\mathrm{p}}}
    &= 0. \label{eq:model_particle_side_cavities_BC_solid_Rp}
\end{align}
\end{subequations}
In \cref{eq:model_particle_side_cavities_BC_liquid_Rp}, the factor $d^{\mathrm{mp}} \varepsilon^{\mathrm{mp}}$ is the part of the particle porosity belonging to the main pore network, so that $k^{\mathrm{f}}_{i}$ retains its usual meaning per unit of total external particle surface.
The model is complemented by the bulk concentration $c^{\b}_{i}$ (or externally supplied data).
The bulk transport equations of Section~\ref{sec:bulk_domain_transport} apply unchanged, with the surface concentration taken in the main pore network, $\left. c^{\p}_{j,0,i}\right|_{r = R^{\mathrm{p}}_{j}}$.
The initial conditions at $t = 0$ are given for the entire pore phase $m\in\{0,\dots,N^{\mathrm{sc}}\}$ by
\begin{subequations}
\begin{alignat}{2}
    \left. c^{\p}_{m,i} \right|_{t = 0} &= c^{\p,\mathrm{init}}_{m,i} & & \qquad\text{in } (0, R^{\mathrm{p}}),\\
    \left. c^{\s}_{m,i,k} \right|_{t = 0} &= c^{\s, \mathrm{init}}_{m,i,k} & & \qquad\text{in } (0, R^{\mathrm{p}}).
\end{alignat}\label{eq:model-particle_side_cavities-ic}
\end{subequations}
\end{model}

\begin{remark}[(PCb)]
The spatial dimension of the model can be reduced by assuming radial homogeneity, as in \cref{model:particle_0D}. In that case, \cref{eq:model_particle_side_cavities_liquid,eq:model_particle_side_cavities_solid} are replaced by

\begin{subequations}
\begin{align}
    \frac{\partial c^{\p}_{0, i}}{\partial t}
    &=
    \frac{a^{\mathrm{p}}}{d^{\mathrm{mp}} \varepsilon^{\mathrm{mp}}} k^{\mathrm{f}}_{i} \left( c^{\b}_i - c^{\p}_{0,i} \right)
    - \frac{1 - \varepsilon^{\mathrm{mp}}}{\varepsilon^{\mathrm{mp}}} \sum_{k=1}^{N^{\mathrm{b}}_i} f^{\mathrm{bind}}_{0,i,k}\mleft( \vec{c}_0^{\p}, \vec{c}^{\s}_0 \mright)
     - \sum_{m=1}^{N^{\mathrm{sc}}} \frac{d^{\mathrm{sc}}_{m} \varepsilon^{\mathrm{sc}}_m}{d^{\mathrm{mp}} \varepsilon^{\mathrm{mp}}} k^{\mathrm{sc}}_{m,i} \left( c^{\p}_{0,i} - c^{\p}_{m,i} \right), \label{eq:model_particle_side_cavities_homogeneous_liquid}
    \\
    \frac{\partial c^{\s}_{0, i, k}}{\partial t}
    &=
    f^{\mathrm{bind}}_{0,i,k}\mleft( \vec{c}^{\p}_0, \vec{c}^{\s}_0 \mright),
    \label{eq:model_particle_side_cavities_homogeneous_solid}
\end{align}
\end{subequations}
where $a^{\mathrm{p}}$ denotes the surface-to-volume ratio defined in \cref{eq:model-particle0D-surface-to-volume}.
The film diffusion term follows from averaging \cref{eq:model_particle_side_cavities_liquid} over the particle volume and applying \cref{eq:model_particle_side_cavities_BC_liquid_Rp}; consequently, \cref{eq:model_particle_side_cavities_BC} is dropped.
\end{remark}

\begin{remark}[(PCc)]
The cavity exchange coefficient $k^\mathrm{sc}_{m,i}$, associated with the entrance of side-cavity type $m$, can depend on the concentrations of all solutes in that type, thereby limiting the access of component $i$ to that type.
Furthermore, \cref{model:particle_side_cavities} allows different binding models in each side-cavity type.
This flexibility is particularly important when different portions of the pore phase have restricted accessibility due to steric constraints and exhibit different ligand densities or surface chemistries.
In this manner, the model simultaneously yet independently captures the effects of size exclusion and binding.
\end{remark}

\subsection{Binding and Reaction Models}
\label{sec:chemistry_models}

\subsubsection{Binding Models}
\label{sec:models-binding}

The primary source of variability in chromatographic models often lies in the description of binding once column geometry and mass transport parameters have been fixed.
In the preceding sections, arbitrary binding models were intentionally incorporated into the particle models in order to decouple mass transport from binding behavior.
This separation allows transport descriptions to be combined flexibly with different binding models and constitutes an important design principle of the modeling architecture.
In particular, it facilitates the exploration and comparison of alternative binding descriptions when analyzing experimental data.

Within this framework, binding is represented through functions $f^{\mathrm{bind}}_{i,k}\left( \vec{c}^{\p}, \vec{c}^{\s} \right)$ defined for each component $i$ and bound state $k$.
Depending on the formulation, these functions represent either equilibrium relationships or kinetic source terms (see \cref{sec:binding_equilibrium_kinetics}).
Here, kinetic binding refers to binding governed by finite-rate kinetics, whereas rapid-equilibrium binding is described by algebraic equilibrium constraints.
The underlying binding mechanisms can vary widely between systems and components, and the corresponding models are often motivated by thermodynamic considerations~\citep{Mollerup2008}.
For example, binding models can account for finite solid-phase capacity, competitive interactions between components, and the influence of solution conditions or mobile-phase modifiers.
A large number of binding models and variants have been proposed in the literature~\mbox{(see, e.g., the review in}~\citealp{mozaffari_majd_adsorption_2022}).
From the perspective of modeling architecture, these formulations can be expressed uniformly through the functions $f^{\mathrm{bind}}_{i,k}$ introduced above.

We introduce the linear model, the simplest binding model, using the notation established for the hierarchical transport models.
It is applicable to very low concentrations, i.e., for dilute systems under analytical conditions \cite[Sec.~3.1.2]{Guiochon2006}.
The net binding rate is given by
\begin{equation}
	f^{\mathrm{bind}}_{i}\mleft(\vec{c}^{\p}, \vec{c}^{\s}\mright) =
	k^{\mathrm{a}}_{i} c^{\p}_i - k^{\mathrm{d}}_{i} c^{\s}_i
	\qquad \forall\  i \in \{1, \dots, N^\mathrm{c} \}.
	\label{eq:model-bnd-linear}
\end{equation}

Here, $k^{\mathrm{a}}_{i} \geq 0$ denotes the adsorption rate constant and $k^{\mathrm{d}}_{i} \geq 0$ the desorption rate constant.
Each component binds independently; that is, there are no competitive effects.
The binding capacity is unbounded.
Alternatively, the model can be parameterized in terms of the equilibrium constant
$k^{\mathrm{eq}}_{i} = k^{\mathrm{a}}_{i} / k^{\mathrm{d}}_{i}$:
\begin{equation}
	f^{\mathrm{bind}}_{i}\mleft(\vec{c}^{\p}, \vec{c}^{\s}\mright) =
	k^{\mathrm{d}}_{i} \left( k^{\mathrm{eq}}_{i} c^{\p}_i - c^{\s}_i \right)
	\qquad \forall\  i \in \{1, \dots, N^\mathrm{c} \}.
	\label{eq:model-bnd-linear-keq}
\end{equation}

In Section~\ref{supp:sec:binding_models}, we present a selection of binding models and their formulation in our notation while referring the reader to the literature for comprehensive discussions of the underlying assumptions.

\subsubsection{Reaction Models}
\label{sec:models-reaction}

The integration of chemical reactions may be desired for several reasons.
Some processes couple separation and reaction within a single unit operation, as in chromatographic reactors \citep{Russo2019,Lode2001}.
Other intentional applications include on-column refolding \citep{Freydell2010} and racemization \citep{Nimmig2013}, while unintended reactions such as denaturation and aggregation may accompany chromatographic separation as well \citep{Ojala2014,Ojala2015}.
Chemical reactions also govern buffer equilibria \citep{Choy2014,Diedrich2019}.

We adapt the approach proposed by \cite{Diedrich2019} and integrate it into the hierarchical modeling framework developed in this work.
A detailed discussion of reactions and reaction kinetics can be found in the textbook by \cite{Toth2018}.
We first consider kinetic reactions, that is, reactions governed by finite-rate kinetics.
Rapid-equilibrium reactions are assumed to have infinitely fast kinetics and therefore require a reformulation of the model in terms of conserved moieties to avoid overdetermined systems.
This reformulation is discussed in Section~\ref{supp:sec:models-reaction-rapidequi}.

We consider reactions involving $N^{\mathrm{c}}$ components of the form 
\begin{equation}
    \label{eq:model-rctn-reaction-single}
	\sum_{i=1}^{N^{\mathrm{c}}} a_i c_i \to \sum_{i=1}^{N^{\mathrm{c}}} b_i c_i,
\end{equation}
where the coefficients $a_i, b_i \in \Integers$, $i \in \{ 1, \dots, N^{\mathrm{c}} \}$, specify the stoichiometric proportions of the participating components.
They are collected in the reactant and product complex vectors $\vec{a} \in \Integers^{N^{\mathrm{c}}}$ and $\vec{b} \in \Integers^{N^{\mathrm{c}}}$, respectively.
The net change in the concentration vector $\vec{c}$ is governed by the \gls{ode}
\begin{equation}
	\frac{\mathrm{d}\vec{c}}{\mathrm{d}t} = \left( \vec{b} - \vec{a} \right) r\mleft( \vec{c} \mright), \label{eq:model-rctn-ode-single}
\end{equation}
where $r\colon \Reals^{N^{\mathrm{c}}} \to \Reals$ denotes the reaction rate function.
The reaction rate function is not restricted to mass-action kinetics but may also represent more specific rate laws, such as Michaelis–Menten or Monod kinetics.
The notation is independent of the underlying rate model, allowing different reaction rate laws to be combined, building a reaction network.

In a reaction network, we collect the net stoichiometric coefficient vectors of all reactions in the columns of the stoichiometric matrix $\mat{S}$ of size $N^{\mathrm{c}} \times N^{\mathrm{r}}$, whose entries are given by
\[
S_{ij} = b_i^{(j)} - a_i^{(j)}, 
\qquad i=1,\dots,N^{\mathrm{c}},\; j=1,\dots,N^{\mathrm{r}}.
\]
The corresponding reaction rate functions are collected in the vector-valued function $\vec{r}\colon \Reals^{N^{\mathrm{c}}} \to \Reals^{N^{\mathrm{r}}}$.
The change in the concentration vector $\vec{c}$ due to the reaction system is then governed by the \gls{ode}
\begin{equation}
	\frac{\mathrm{d}\vec{c}}{\mathrm{d}t} = \mat{S} \vec{r}\mleft( \vec{c} \mright). \label{eq:model-rctn-ode}
\end{equation}

Real systems cannot always be represented accurately by integer stoichiometric coefficients, and our framework supports arbitrary real-valued coefficients.
The direction of a reaction is determined by the sign of its rate function, whereas the arrow in \cref{eq:model-rctn-reaction-single} only specifies which components are consumed or produced, that is, the sign pattern of the corresponding column of the stoichiometric matrix $\mat{S}$.
Moreover, a reaction system cannot be described by $\mat{S}$ alone; for example, $2A \to A + B$ and $A \to B$ share the same stoichiometric matrix $\mat{S} = (-1,1)^{\T}$ yet represent different reactions. 
This difference is captured by their rate functions, which may depend on the reactant and product complex vectors.

In this section, we illustrate the integration of the reaction source terms $\mat{S}\vec{r}(\vec{c})$ into \cref{model:1D_continuum_bulk_transport_model,model:particle_0D}, while the corresponding integration for the remaining bulk and particle transport models is presented in Section~\ref{supp:sec:reactions}.
Including these source terms in the mass balances yields the following equations for all components $i\in\{1,\dots,N^\mathrm{c}\}$ and bound states $k\in\{1,\dots,N^\mathrm{b}\}$:
\begin{subequations}
 \label{eq:reactions_1D_bulk_0D_particle}
\begin{align}
    \varepsilon^{\mathrm{b}} \frac{\partial c^{\b}_i}{\partial t}
    &= - u \frac{\partial (\varepsilon^{\mathrm{b}} c^{\b}_i)}{\partial z}
    + \frac{\partial}{\partial z} \left( \varepsilon^{\mathrm{b}} D^{\mathrm{ax}}_{i} \frac{\partial c^{\b}_i}{\partial z} \right)
    \nonumber \\
    &- \left(1 - \varepsilon^{\mathrm{b}} \right) \sum_{j=1}^{N^{\mathrm{p}}} \frac{3 d_j}{R^{\mathrm{p}}_{j}} k^{\mathrm{f}}_{j,i} \left(c^{\b}_i - \left. c^{\p}_{j,i} \right|_{r = R^{\mathrm{p}}_{j}} \right)
    \nonumber \\
    &+ \varepsilon^\mathrm{b} f^\mathrm{\b,reac}_{i}\mleft( \vec{c}^{\b} \mright),
    \\
    \frac{\partial c^{\p}_{i}}{\partial t}
    &=
    \frac{a^{\mathrm{p}}}{\varepsilon^{\p}} k^{\mathrm{f}}_{i} \left( c^{\b}_i - c^{\p}_{i} \right)
    -\frac{1 - \varepsilon^{\p}}{\varepsilon^{\p}} \sum_{k=1}^{N^{\mathrm{b}}_i} f^{\mathrm{bind}}_{i,k}\mleft( \vec{c}^{\p}, \vec{c}^{\s} \mright)
     + f^\mathrm{\p,reac}_{i}\mleft( \vec{c}^{\p} \mright)
    \\
    \frac{\partial c^{\s}_{i,k}}{\partial t}
    &=
    f^{\mathrm{bind}}_{i,k}\mleft( \vec{c}^{\p}, \vec{c}^{\s} \mright) + f^\mathrm{\s,reac}_{i,k}\mleft( \vec{c}^{\s} \mright).
\end{align}
\end{subequations}
The reaction source functions are given by 
\begin{equation}
    \vec{f}^\mathrm{\kappa,reac}\mleft( \vec{c}^{\kappa} \mright) = \mat{S}^\kappa \vec{r}^\mathrm{\kappa}\mleft( \vec{c}^{\kappa} \mright) 
\end{equation}
where $\kappa \in \{\b,\p,\s\}$ and $N^{\mathrm{r},\kappa}$ denotes the number of reactions in phase $\kappa$.
The stoichiometric matrices are $\mat{S}^\b,\mat{S}^\p\in\Reals^{N^\mathrm{c}\times N^{\mathrm{r},\kappa}}$ and $\mat{S}^\s\in\Reals^{N^\mathrm{totb}\times N^{\mathrm{r},\s}}$ with $N^\mathrm{totb} = \sum_{i=1}^{N^\mathrm{c}} N^\mathrm{b}_i$, so that $\vec{f}^\mathrm{\b,reac},\vec{f}^\mathrm{\p,reac}\in\Reals^{N^\mathrm{c}}$ and $\vec{f}^\mathrm{\s,reac}\in\Reals^{N^\mathrm{totb}}$.

\subsection{Rapid-Equilibrium Approximations}
\label{sec:rapid-equilibria}

Within the same model, mass transfer processes can span a wide range of time scales.
For example, film diffusion is often several orders of magnitude faster than convective mass transport.
Every physical process has finite kinetics; however, explicitly resolving these kinetics is unnecessary if the processes can be treated as instantaneous relative to the characteristic time scales of the system.
From a mathematical perspective, rate-limited and instantaneous processes lead to structurally different classes of equations.
From a numerical perspective, assuming rapid equilibrium is often preferable since explicitly resolving kinetics across multiple time scales typically introduces stiffness and increases the computational cost of solving the resulting differential equation systems.
In the following, we introduce model variants that treat film diffusion, binding, and reactions under rapid-equilibrium assumptions.

\subsubsection{Film Diffusion in Rapid Equilibrium (Model PN)}
\label{sec:rapid-equilibrium_film_diffusion}
We derive the rapid-equilibrium film diffusion variants of \cref{model:1D_continuum_bulk_transport_model} combined with either \cref{model:particle_1D} or \cref{model:particle_0D}.
As interfacial mass transfer couples the transport models in the bulk and particle domains through their boundary conditions, they cannot be treated independently here.
We still denote the resulting model by PN to preserve the combinatorial hierarchy introduced in \cref{sec:taxonomy}.

\begin{assumption}[(PN)]\label{asmpt:rapid-equilibrium_film_diffusion}
    In addition to~\cref{asmpt:model-1D,asmpt:model-1D-particle}, we assume that film diffusion is infinitely fast; that is, we consider the limit $k^{\mathrm{f}}_{j,i} \to \infty$.
\end{assumption}
This assumption can be applied independently to individual particle types $j$ or components $i$.
For readability, however, we apply it here to all particle types and components.

First, we apply \cref{asmpt:rapid-equilibrium_film_diffusion} to the combined \cref{model:1D_continuum_bulk_transport_model,model:particle_1D}.
Before taking the limit $k^{\mathrm{f}}_{j,i} \to \infty$, we substitute the boundary condition \cref{eq:model_particle_1D_BC_liquid_Rp} into the bulk-domain \gls{pde} \eqref{eq:model-1d-bulk-pde}, which yields on $(0, T^{\mathrm{end}}) \times (0, L^\mathrm{b})$:
\begin{subequations}
    \label{eq:model-1d-bulk-pde-infkf-and-cp=cb}
    \begin{multline}\allowdisplaybreaks[0]
    \label{eq:model-1d-bulk-pde-infkf}
    	\varepsilon^{\mathrm{b}} \frac{\partial c^{\b}_i}{\partial t}
        = -u \frac{\partial \left( \varepsilon^{\mathrm{b}} c^{\b}_i \right)}{\partial z} + \frac{\partial}{\partial z} \left( \varepsilon^{\mathrm{b}} D^{\mathrm{ax}}_{i} \frac{\partial c^{\b}_i}{\partial z} \right)
    	- \left(1 - \varepsilon^{\mathrm{b}} \right) \sum_{j=1}^{N^{\mathrm{p}}} \frac{3 d_j}{R^{\mathrm{p}}_{j}} \varepsilon^{\p}_{j} \left. \left( D^{\p}_{j,i} \frac{\partial c^{\p}_{j,i}}{\partial r} \right) \right|_{r = R^{\mathrm{p}}_{j}}.
    \end{multline}
    Dividing the boundary condition \cref{eq:model_particle_1D_BC_liquid_Rp} by $k^{\mathrm{f}}_{j,i}$ and taking the limit $k^{\mathrm{f}}_{j,i} \to \infty$ yields
    \begin{equation*}
    	\left( c^{\b}_i - \left. c^{\p}_{j,i} \right|_{r = R^{\mathrm{p}}_{j}} \right)
        = \frac{\varepsilon^{\p}_{j}}{k^{\mathrm{f}}_{j,i}} \left( \left. D^{\p}_{j,i} \frac{\partial c^{\p}_{j,i}}{\partial r} \right) \right|_{r = R^{\mathrm{p}}_{j}} \to 0,
    \end{equation*}
    which implies continuity of concentration at $r = R^{\mathrm{p}}_{j}$ on $(0,T^{\mathrm{end}}] \times (0, L^\mathrm{b})$:
    \begin{equation}
        c^{\b}_i = \left. c^{\p}_{j,i} \right|_{r = R^{\mathrm{p}}_{j}}.
        \label{eq:model-1d-particle-bulk-bc-liquid-infkf}
    \end{equation}
\end{subequations}

In summary, the bulk-domain \gls{pde} \eqref{eq:model-1d-bulk-pde} is replaced by \cref{eq:model-1d-bulk-pde-infkf} and the Robin boundary condition \eqref{eq:model_particle_1D_BC_liquid_Rp} is replaced by the Dirichlet boundary condition \eqref{eq:model-1d-particle-bulk-bc-liquid-infkf}.

Alternatively, we apply \cref{asmpt:rapid-equilibrium_film_diffusion} to the combined \cref{model:1D_continuum_bulk_transport_model,model:particle_0D}.
In this case, we use the particle liquid-phase \gls{ode} \eqref{eq:model_particle_0D_liquid} to replace the film-diffusion source term in the bulk-domain \gls{pde} \eqref{eq:model-1d-bulk-pde}, yielding
\begin{alignat}{1}
	\varepsilon^{\mathrm{b}} \frac{\partial c^{\b}_i}{\partial t}
    = &- u \frac{\partial \left( \varepsilon^{\mathrm{b}} c^{\b}_i \right)}{\partial z} + \frac{\partial}{\partial z} \left( \varepsilon^{\mathrm{b}} D^{\mathrm{ax}}_{i} \frac{\partial c^{\b}_i}{\partial z} \right)
    \nonumber \\
    &- \left(1 - \varepsilon^{\mathrm{b}} \right) \sum_{j=1}^{N^{\mathrm{p}}} d_j \chi_j \left( \varepsilon^{\p}_{j} \frac{\partial c^{\p}_{j,i}}{\partial t}
    + \left( 1- \varepsilon^{\p}_{j} \right) \sum_{k=1}^{N^{\mathrm{bind}}_{j,i}} f^{\mathrm{bind}}_{j,i,k}\mleft( \vec{c}^{\p}_j, \vec{c}^{\s}_j \mright) \right). \label{eq:model-lrm-bulk-infkf}
\end{alignat}
Here, $N^{\mathrm{bind}}_{j,i}$ denotes the number of bound states of component $i$ in particle type $j$, that is, the particle-type-resolved counterpart of $N^{\mathrm{b}}_{i}$.
The factor
\begin{equation}
    \chi_j \coloneqq \frac{3}{R^{\mathrm{p}}_{j} a^{\mathrm{p}}_{j}} = 1 - \left( \frac{R^{\mathrm{c}}_{j}}{R^{\mathrm{p}}_{j}} \right)^3 \label{eq:model-lrm-accessible-volume-fraction}
\end{equation}
is the accessible volume fraction of particle type $j$, which arises because $c^{\p}_{j,i}$ is averaged over the accessible particle volume, whereas $d_j$ refers to the total particle volume.
It equals one for particles without a solid core.
Then, we rearrange \cref{eq:model_particle_0D_liquid}, divide by $k^{\mathrm{f}}_{j,i}$, and take the limit $k^{\mathrm{f}}_{j,i} \to \infty$:
\begin{equation*}
    \frac{a^{\mathrm{p}}_{j}}{\varepsilon^{\p}_{j}} \left( c^{\b}_i - c^{\p}_{j,i} \right)
    =
    \frac{1}{k^{\mathrm{f}}_{j,i}} \left[ \frac{\partial c^{\p}_{j,i}}{\partial t} + \frac{1 - \varepsilon^{\p}_{j}}{\varepsilon^{\p}_{j}} \sum_{k=1}^{N^{\mathrm{bind}}_{j,i}} f^{\mathrm{bind}}_{j,i,k}\mleft( \vec{c}^{\p}_j, \vec{c}^{\s}_j \mright) \right] \to 0,
\end{equation*}
which implies continuity of concentration, $c^{\b}_i = c^{\p}_{j,i}$, for \cref{model:particle_0D} as well.
This eliminates $c^{\p}_{j,i}$ as an independent variable, and we replace all occurrences by $c^{\b}_i$ and substitute \eqref{eq:model_particle_0D_solid} to obtain
\begin{align*}
    \varepsilon^{\mathrm{b}} \frac{\partial c^{\b}_i}{\partial t}
    + \left(1 - \varepsilon^{\mathrm{b}} \right) \sum_{j=1}^{N^{\mathrm{p}}} d_j \chi_j \varepsilon^{\p}_{j} \frac{\partial c^{\b}_{i}}{\partial t}
    = &- u \frac{\partial \left( \varepsilon^{\mathrm{b}} c^{\b}_i \right)}{\partial z} + \frac{\partial}{\partial z} \left( \varepsilon^{\mathrm{b}} D^{\mathrm{ax}}_{i} \frac{\partial c^{\b}_i}{\partial z} \right)
    \\
    &- \left(1 - \varepsilon^{\mathrm{b}} \right) \sum_{j=1}^{N^{\mathrm{p}}} d_j \chi_j \left( 1- \varepsilon^{\p}_{j} \right) \sum_{k=1}^{N^{\mathrm{bind}}_{j,i}} \frac{\partial c^{\s}_{j,i,k}}{\partial t}.
\end{align*}

For a single particle type, the bulk and particle porosities $\varepsilon^{\mathrm{b}}$ and $\varepsilon^{\p}$ can be combined into a total porosity, and the accessible solid phase into an adsorbent volume fraction,
\begin{equation}
\varepsilon^{\mathrm{t}} \coloneqq \varepsilon^{\mathrm{b}} + (1-\varepsilon^{\mathrm{b}}) \chi \varepsilon^{\p},
\qquad
\varepsilon^{\s} \coloneqq (1-\varepsilon^{\mathrm{b}}) \chi \left( 1 - \varepsilon^{\p} \right),
\end{equation}
with the corresponding total liquid domain $\Omega^{\l} \coloneqq \Omega^{\b} \cup \Omega^{\p}$ and  liquid-phase concentration
\begin{align}
    c^{\l} \coloneqq \frac{\varepsilon^{\mathrm{b}}}{\varepsilon^{\mathrm{t}}} c^{\b} + \frac{1 - \varepsilon^{\mathrm{b}}}{\varepsilon^{\mathrm{t}}} \chi \varepsilon^{\p} c^{\p}
    \stackrel{\eqref{eq:model-1d-particle-bulk-bc-liquid-infkf}}{=} c^\b.
\end{align}
The column volume is thus partitioned into liquid, adsorbent, and solid cores, $\varepsilon^{\mathrm{t}} + \varepsilon^{\s} + (1-\varepsilon^{\mathrm{b}})(1-\chi) = 1$.
For particles without a solid core, $\chi = 1$ and the adsorbent occupies the entire non-liquid volume, $\varepsilon^{\s} = 1 - \varepsilon^{\mathrm{t}}$.
For axially uniform porosities, we introduce an apparent velocity $\tilde{u}$ and an apparent axial dispersion coefficient $\tilde{D}^{\mathrm{ax}}_{i}$ \mbox{(see also}~\citealp{KATSOULAS2023464345}),
\begin{equation}
\label{eq:model-lrm-reparameterization}
    \tilde{u} \coloneqq \frac{\varepsilon^{\mathrm{b}}}{\varepsilon^{\mathrm{t}}} u, \qquad \tilde{D}^{\mathrm{ax}}_{i} \coloneqq \frac{\varepsilon^{\mathrm{b}}}{\varepsilon^{\mathrm{t}}}
    D^{\mathrm{ax}}_{i}.
\end{equation}
With this reparameterization, we obtain a specific model variant formulated solely in terms of liquid- and solid-phase concentrations $c^{\l}_i$ and $c^{\s}_{i,k}$.

\begin{model}{(PN)}[Lumped Rate Model Without Pores]\label{model:lumped_rate_model_without_pores}
Consider \cref{model:1D_continuum_bulk_transport_model} combined with \cref{model:particle_0D}. Using the apparent velocity, apparent axial dispersion coefficient, total porosity, adsorbent volume fraction,  and liquid-phase concentrations as defined above, mass transfer in the homogenized domain $(0, T^{\mathrm{end}}) \times (0, L^\mathrm{b})$ is governed by the following convection-diffusion-reaction equations:
    \begin{subequations}
    \begin{align}
        \frac{\partial c^{\l}_i}{\partial t} + \frac{\varepsilon^{\mathrm{s}}}{\varepsilon^{\mathrm{t}}} \sum_{k=1}^{N^{\mathrm{b}}_i} \frac{\partial c^{\s}_{i,k}}{\partial t}
        &= - \tilde{u} \frac{\partial c^{\l}_i}{\partial z} + \tilde{D}^{\mathrm{ax}}_{i} \frac{\partial^2 c^{\l}_i}{\partial z^2}, \label{eq:model-lrm-rdm-classic-bulk-pde} \\
        \frac{\partial c^{\s}_{i,k}}{\partial t} &= f^{\mathrm{bind}}_{i,k}\mleft( \vec{c}^{\l}, \vec{c}^{\s} \mright) \label{eq:model-lrm-rdm-classic-particlesolid}
    \end{align}
    \end{subequations}
    with Danckwerts boundary conditions
    \begin{equation*}
        \tilde{u} c^{\mathrm{in}}_{i} = \left.\left( \tilde{u} c^{\l}_i - \tilde{D}^{\mathrm{ax}}_{i} \frac{\partial c^{\l}_i}{\partial z} \right)\right|_{z=0}, \qquad
        0 = - \tilde{D}^{\mathrm{ax}}_{i} \left. \frac{\partial c^{\l}_i}{\partial z} \right|_{z=L^\mathrm{b}} \qquad\text{on } (0, T^{\mathrm{end}}).
    \end{equation*}
    The initial conditions at $t = 0$ are given by
    \begin{equation}
        \left. c^{\l}_i \right|_{t = 0} = c^{\l,\mathrm{init}}_{i}, \qquad \left. c^{\s}_{i,k} \right|_{t = 0} = c^{\s,\mathrm{init}}_{i,k}  \qquad\text{in } (0, L^\mathrm{b}). \label{eq:model-lrm-rdm-classic-ic}
    \end{equation}
\end{model}

\Needspace{6\baselineskip}
\begin{remark}[(PN)]
In contrast to \cref{model:particle_0D}, where the particle pore phase is represented as a compartment without spatial resolution, \cref{model:lumped_rate_model_without_pores} neglects the explicit representation of the particle pore phase through homogenization with the bulk domain.
The apparent velocity and axial dispersion coefficients implicitly account for the effects of mass transfer into and out of the particle pore phase.
Analogous model variants with film diffusion in rapid equilibrium can be derived by combining the bulk \cref{model:3D_continuum_bulk_transport_model}, \cref{model:2D_continuum_bulk_transport_model} or \cref{model:tank} with the particle \cref{model:particle_1D} or \cref{model:particle_0D}.
\end{remark}

\subsubsection{Binding in Rapid Equilibrium}
\label{sec:binding_equilibrium_kinetics}
The binding source function $f^{\mathrm{bind}}_{i,k}( \vec{c}^{\p}, \vec{c}^{\s})$ can be expressed as the difference between adsorption and desorption rates:
\begin{equation*}
	f^{\mathrm{bind}}_{i,k}\mleft( \vec{c}^{\p}, \vec{c}^{\s} \mright) = k^{\mathrm{ads}}_{i} r^{\mathrm{ads}}_{i}\mleft( \vec{c}^{\p}, \vec{c}^{\s} \mright) - k^{\mathrm{des}}_{i} r^{\mathrm{des}}_{i}\mleft( \vec{c}^{\p}, \vec{c}^{\s} \mright),
\end{equation*}
with rate functions $r^{\mathrm{ads}}_{i} \geq 0$, $r^{\mathrm{des}}_{i} \geq 0$, and rate constants $k^{\mathrm{ads}}_{i} \geq 0$, $k^{\mathrm{des}}_{i} \geq 0$.
This formulation generalizes the linear binding model introduced in \cref{sec:models-binding}, allowing the rate functions to account for saturation effects, competition between multiple components, and other physicochemical effects.
The source function can be reparameterized in terms of an equilibrium constant $k^{\mathrm{eq}}_{i} = k^{\mathrm{ads}}_{i} / k^{\mathrm{des}}_{i}$ and a kinetic constant $k^{\mathrm{kin}}_{i} = k^{\mathrm{des}}_{i}$:
\begin{equation}
\label{eq:binding_kinetics_reparameterized}
	f^{\mathrm{bind}}_{i,k}\mleft( \vec{c}^{\p}, \vec{c}^{\s} \mright) = k^{\mathrm{kin}}_{i} \left(k^{\mathrm{eq}}_{i} r^{\mathrm{ads}}_{i}\mleft( \vec{c}^{\p}, \vec{c}^{\s} \mright) - r^{\mathrm{des}}_{i}\mleft( \vec{c}^{\p}, \vec{c}^{\s} \mright) \right).
\end{equation}

This reparameterization motivates an alternative representation of binding kinetics based on a \gls{ldf} approximation.
Rather than expressing the binding rate through separate adsorption and desorption processes, the \gls{ldf} approach assumes that the rate is proportional to the deviation from equilibrium.
Two non-equivalent variants are commonly used: in one the binding rate is proportional to the deviation of the solid-phase concentration from its equilibrium value, Eq.~\eqref{eq:ldf_solid_variant}, while the other uses the corresponding deviation of the liquid-phase concentration from its equilibrium value, Eq.~\eqref{eq:ldf_liquid_variant}.
\begin{subequations}
\label{eq:linear_driving_force_approximation}
\begin{align}
\label{eq:ldf_solid_variant}
    f^{\mathrm{bind}}_{i,k}
    &=
    k^{\mathrm{kin}}_{i}
    \left(
        c^{\s,*}_{i}(\vec{c}^{\p})
        -
        c^{\s}_{i}
    \right),
    \\
\label{eq:ldf_liquid_variant}
    f^{\mathrm{bind}}_{i,k}
    &=
    k^{\mathrm{kin}}_{i}
    \left(
        c^{\p}_{i}
        -
        c^{\p,*}_{i}(\vec{c}^{\s})
    \right),
\end{align}
\end{subequations}
where $c^{\s,*}_{i}$ and $c^{\p,*}_{i}$ are determined by the equilibrium isotherm.
The \gls{ldf} formulation replaces the original adsorption and desorption rate constants by the equilibrium constant $k^{\mathrm{eq}}_{i}$ and an effective kinetic constant $k^{\mathrm{kin}}_{i}$.

Both the native kinetic formulation~\eqref{eq:binding_kinetics_reparameterized} and the \gls{ldf} approximations~\eqref{eq:linear_driving_force_approximation} recover the same equilibrium model in the limit of infinitely fast binding kinetics.
The limit $k^{\mathrm{kin}}_{i} \to \infty$ is referred to as the rapid-equilibrium assumption \cite[Sec.~2.2.1]{Ingalls2013} or the quasi-equilibrium approximation \cite[Sec.~6.3.3]{Klipp2016}.
The derivation of model variants with binding in rapid equilibrium is demonstrated for \cref{model:particle_1D}; an analogous procedure can be applied to \cref{model:particle_0D}, as detailed in Section~\ref{supp:sec:model_particle_0D-req}.
We first eliminate the binding source function $f^{\mathrm{bind}}_{i,k}(\vec{c}^{\p}, \vec{c}^{\s})$ from \cref{eq:model_particle_1D_liquid,eq:model_particle_1D_solid} by taking a weighted sum of these equations for all bound states, which yields the mass balance for $c^{\p}_i + \frac{1-\varepsilon^{\p}}{\varepsilon^{\p}} \sum_{k=1}^{N^{\mathrm{b}}_{i}} c^{\s}_{i,k}$
\begin{equation*}
    \frac{\partial c^{\p}_i}{\partial t} + \frac{1-\varepsilon^{\p}}{\varepsilon^{\p}} \sum_{k=1}^{N^{\mathrm{b}}_{i}} \frac{\partial c^{\s}_{i,k}}{\partial t}
    =
    \frac{1}{r^2} \frac{\partial }{\partial r} \left( r^2 D_{i}^{\p} \frac{\partial c^{\p}_{i}}{\partial r} \right)
    + \frac{1-\varepsilon^{\p}}{\varepsilon^{\p}} \sum_{k=1}^{N^{\mathrm{b}}_{i}}
    \frac{1}{r^2} \frac{\partial }{\partial r} \left( r^2 D_{i}^{\s} \frac{\partial c^{\s}_{i,k}}{\partial r} \right).
\end{equation*}

The linear combination $c^{\p}_i + \frac{1-\varepsilon^{\p}}{\varepsilon^{\p}} \sum_{k=1}^{N^{\mathrm{b}}_{i}} c^{\s}_{i,k}$ is invariant under the binding process and is hence called a conserved moiety.
The corresponding mass balance is underdetermined with respect to the original liquid- and solid-phase concentrations. 
The remaining degrees of freedom are determined by taking the limit $k^{\mathrm{kin}}_{i} \to \infty$ of the binding source function for each bound state.
Specifically, we divide the solid-phase mass balance \cref{eq:model_particle_1D_solid} by $k^{\mathrm{kin}}_{i}$,
\begin{equation*}
    \frac{1}{k^{\mathrm{kin}}_{i}} \frac{\partial c^{\s}_{i,k}}{\partial t}
    =
    \frac{1}{k^{\mathrm{kin}}_{i}}
    \frac{1}{r^2} \frac{\partial }{\partial r} \left( r^2 D_{i}^{\s} \frac{\partial c^{\s}_{i,k}}{\partial r} \right)
    + \frac{1}{k^{\mathrm{kin}}_{i}} f^{\mathrm{bind}}_{i,k}\mleft( \vec{c}^{\p}, \vec{c}^{\s} \mright),
\end{equation*}
and take the limit $k^{\mathrm{kin}}_{i} \to \infty$:
\begin{alignat*}{2}
    & & 0 &= \lim\limits_{k^{\mathrm{kin}}_{i} \to \infty} \frac{1}{k^{\mathrm{kin}}_{i}} f^{\mathrm{bind}}_{i,k}\mleft( \vec{c}^{\p}, \vec{c}^{\s} \mright)
    \\
    & & 0 &= \frac{1 - \varepsilon^{\p}}{\varepsilon^{\p}} \lim\limits_{k^{\mathrm{kin}}_{i} \to \infty} \frac{k^{\mathrm{kin}}_{i}}{k^{\mathrm{kin}}_{i}} \left( k^{\mathrm{eq}}_{i} r^{\mathrm{ads}}_{i}\mleft( \vec{c}^{\p}, \vec{c}^{\s} \mright) - r^{\mathrm{des}}_{i}\mleft( \vec{c}^{\p}, \vec{c}^{\s} \mright) \right)
    \\
    & \Leftrightarrow&\qquad 0 &= k^{\mathrm{eq}}_{i} r^{\mathrm{ads}}_{i}\mleft( \vec{c}^{\p}, \vec{c}^{\s} \mright) - r^{\mathrm{des}}_{i}\mleft( \vec{c}^{\p}, \vec{c}^{\s} \mright).
\end{alignat*}
This is equivalent to
\begin{equation*}
	f^{\mathrm{bind}}_{i,k}\mleft( \vec{c}^{\p}, \vec{c}^{\s} \mright) = 0.
\end{equation*}

Consequently, the structure of the model changes as one \gls{pde} is converted to a purely algebraic equation. For the components with binding in rapid equilibrium, the particle-domain \gls{pde} \eqref{eq:model_particle_1D} is reformulated as a \gls{pdae}:
\begin{subequations}
\label{eq:model_particle_1D-req}
\begin{alignat}{2}
    \frac{\partial c^{\p}_i}{\partial t} + \sum_{k=1}^{N^{\mathrm{b}}_{i}} \frac{1 - \varepsilon^{\p}}{\varepsilon^{\p}} \frac{\partial c^{\s}_{i,k}}{\partial t}
    &=
    \frac{1}{r^2} \frac{\partial }{\partial r} \left( r^2 D_{i}^{\p} \frac{\partial c^{\p}_{i}}{\partial r} \right)
    + \frac{1 - \varepsilon^{\p}}{\varepsilon^{\p}} \sum_{k=1}^{N^{\mathrm{b}}_{i}}
    \frac{1}{r^2} \frac{\partial }{\partial r} \left( r^2 D_{i}^{\s} \frac{\partial c^{\s}_{i,k}}{\partial r} \right),
    \label{eq:model_particle_1D_liquid-req} \\
	0 &= f^{\mathrm{bind}}_{i,k}\mleft( \vec{c}^{\p}, \vec{c}^{\s} \mright) \quad \forall k. \label{eq:model_particle_1D_solid-req}
\end{alignat}
\end{subequations}

Finally, we determine the boundary conditions for the conserved moiety.
They are obtained by taking the same weighted sum of the boundary conditions \cref{eq:model_particle_1D_BC_liquid_Rp,eq:model_particle_1D_BC_solid_Rp} at $r = R^{\mathrm{p}}$ and \cref{eq:model_particle_1D_BC_liquid_0,eq:model_particle_1D_BC_solid_0} at $r = R^{\mathrm{c}}$.
\begin{subequations}
\label{eq:model_particle_1D_BC_rapidequi}
\begin{align}
    \left.\left( \varepsilon^{\p} D^{\p}_{i} \frac{\partial c^{\p}_{i}}{\partial r} + \left( 1 - \varepsilon^{\p} \right) \sum_{k=1}^{N^{\mathrm{b}}_{i}} D^{\s}_{i,k} \frac{\partial c^{\s}_{i,k}}{\partial r} \right)\right|_{r = R^{\mathrm{p}}}
    &=
    k^{\mathrm{f}}_{i} \left( c^{\b}_i - \left. c^{\p}_{i} \right|_{r = R^{\mathrm{p}}} \right), \label{eq:model_particle_1D_BC_Rp_rapidequi}
    \\
    -\left.\left( \varepsilon^{\p} D^{\p}_{i} \frac{\partial c^{\p}_{i}}{\partial r} + \left( 1 - \varepsilon^{\p} \right) \sum_{k=1}^{N^{\mathrm{b}}_{i}} D^{\s}_{i,k} \frac{\partial c^{\s}_{i,k}}{\partial r} \right) \right|_{r = R^{\mathrm{c}}}
    &= 0. \label{eq:model_particle_1D_BC_0_rapidequi}
\end{align}
\end{subequations}
Note that the outer boundary condition allows surface diffusion to contribute to the mass transfer between the bulk and particle domains.

Moreover, the purely algebraic nature of \cref{eq:model_particle_1D_solid-req} reduces the numerical stiffness of the system by removing the fast binding time scale from the governing equations.
This reformulation, however, requires consistent initial data $\vec{c}^{\p,\mathrm{init}}$ and $\vec{c}^{\s,\mathrm{init}}$ that satisfy the algebraic constraints
\begin{equation*}
    \vec{f}_{\mathrm{bind}}\mleft( \vec{c}^{\p,\mathrm{init}}, \vec{c}^{\s,\mathrm{init}} \mright) = \vec{0} \qquad \text{in } \Omega^{\mathrm{par}}.
\end{equation*}

Some binding models contain components with much faster binding kinetics than others, such as salt ions.
In that case, we define subsets $\mset{N}_{i}^{\mathrm{b},\mathrm{req}}, \mset{N}_{i}^{\mathrm{b},\mathrm{kin}} \subset$ $\{1,\dots, N^{\mathrm{b}}_{i}\}, \mset{N}_{i}^{\mathrm{b},\mathrm{req}} \cup \mset{N}_{i}^{\mathrm{b},\mathrm{kin}} \equiv \{1,\dots, N^{\mathrm{b}}_{i}\}$ representing bound states with rapid-equilibrium binding and kinetic binding, respectively.
\cref{eq:model_particle_1D-req} and \cref{eq:model_particle_1D_BC_rapidequi} are accordingly modified, as detailed in Section~\ref{supp:sec:mixed_kinetic_and_equilibrium_components}.

Note that model variants with film diffusion and binding in rapid equilibrium are not fully orthogonal.
The resulting model equations depend on the order in which these modifications are applied, as detailed in Section~\ref{supp:sec:model-reqBindFilmDiff-notorthogonal}.
The appropriate formulation depends on whether surface diffusion is assumed to contribute to molecular access to the particle domain.

\subsubsection{Reactions in Rapid Equilibrium}
\label{sec:models-reaction-rapidequi}

For reactions in rapid equilibrium, we apply the same time-scale separation as for binding processes.
The conserved moieties evolve dynamically, while the reactions are assumed to remain in instantaneous equilibrium.
For binding in rapid equilibrium, the conserved moieties are directly obtained as linear combinations of the liquid- and solid-phase concentrations.
Reaction networks, in contrast, require an additional elimination procedure to identify the conserved moieties.
For a stoichiometric matrix $\mat{S}$, a conserved moiety is a linear combination of concentrations $\vec{m}^{\T} \vec{c}$, $\vec{m} \in \Reals^{N^{\mathrm{c}}}$, that is structurally invariant under the reaction system.
Equivalently, the coefficient vector $\vec{m}$ lies in the left null space of $\mat{S}$, that is, $\vec{m}^{\T} \mat{S} = \vec{0}$.
The coefficient vectors of the conserved moieties are collected as rows of a conservation matrix $\mat{M}$.
Consequently, $\mat{M}$ satisfies $\mat{M} \mat{S} = \mat{0}$.
Note that the rapid-equilibrium assumption can be selectively applied to individual reactions in a reaction network.

Finally, the original \gls{pde} system is transformed into a \gls{pdae} system in which the transport equations are reformulated in terms of the conserved moieties and supplemented with equilibrium constraints to close the remaining degrees of freedom.
The general workflow for reactions in rapid equilibrium is as follows:
\begin{enumerate}
\item Select the reactions that are assumed to be in rapid equilibrium.
\item Determine the components involved in these reactions from the stoichiometric matrix.
\item Compute the conserved moieties associated with the rapid-equilibrium reactions from their stoichiometric matrix.
\item Construct the conservation matrix from the unaffected components and the conserved moieties.
\item Left-multiply the governing equations, boundary conditions, and initial conditions by the conservation matrix to obtain the reduced system.
\item Add algebraic constraints enforcing instantaneous equilibrium for all rapid-equilibrium reactions.
\end{enumerate}
A more detailed description, including examples that illustrate the construction of conservation matrices and the resulting reformulation of the governing equations, is provided in Supplementary~\ref{supp:sec:models-reaction-rapidequi}.

\section{Model Taxonomy and Literature Comparison}
\label{sec:relation_existing_models}

In this section, we first propose a unified taxonomy for the model family that results from arbitrary combinations of the modular model components described in the preceding sections.
We then discuss models from the literature within the logic of our framework to showcase its utility.

\subsection{Unified Taxonomy}
\label{sec:taxonomy}

The notation and terminology for the modeled phenomena and their associated parameters introduced throughout this work are designed to maintain consistency across the various models while clearly distinguishing quantities that represent conceptual differences.
Where possible, we followed established conventions from the literature.
For convenience, a consolidated overview of symbols and notation is provided in the Supplementary Table~\ref{supp:table:notation_termonology}, including notable differences from the conventions used in the literature.

Previous model classifications, such as those of \citet[][Figure~6.3]{Schmidt-Traub2020} and \citet[][Figure~6.2]{Seidel-Morgenstern2020-Chap6}, categorize models according to the number and type of band-broadening effects. In contrast, the present taxonomy is based on two structural criteria: 1) the physical domains represented as distinct entities, and 2) the dimensionality of the corresponding spatial domains.
Consequently, a model class is defined by the combination of bulk and particle transport models.
A combinatorial overview of bulk and particle models according to this taxonomy is provided in \cref{table:taxonomy_primary}, alongside the governing equations that define the respective model class.
We distinguish between zero-dimensional singleton models without spatial resolution, and column models with one- to three-dimensional spatial representations.
A tank without particles corresponds to the classical \gls{str}, which includes \gls{bstr} and \gls{cstr} models, while a column without particles reduces to a \gls{pfr}, which includes \gls{ipfr} and \gls{dpfr} models.
For both tanks and columns with particles, three alternative particle modeling approaches are distinguished.
The \gls{grm} accounts for finite film diffusion as well as finite pore and surface diffusion and correspondingly resolves the radial coordinate of the particle.
In the \gls{lrmp}, film diffusion is finite, but the particle domain is treated as a singleton without radial concentration gradients.
The \gls{lrm} further enforces equilibrium between bulk and particle pore concentrations, thereby considering the total liquid volume as a single entity.
Further subdivision into secondary model classes, e.g., based on specific particle properties such as \gls{psd}, core-shell particles, or non-spherical geometry, is provided in Table~\ref{supp:table:taxonomy_secondary}.
We deliberately omit binding equilibrium assumptions from the model classification itself \mbox{(as done by e.g.}~\citealp{Schmidt-Traub2020}), since within our modular formulation, rapid-equilibrium binding may be specified independently for each component.
Instead, binding models may be added as a suffix to the primary model name, e.g. \gls{grm} with Langmuir binding.
Further details are then specified through the parameterization of the resulting model.
The complete set of governing equations for all model classes defined by this taxonomy can be generated automatically using the \texttt{CADET-Equations} Python package \citep{breuer_2026_18339287}, as described in \cref{sec:cadet-equations}.

\begin{table}[!htbp]
\caption{Reduced-order model classification. Primary model classification is based on 1) which domains are considered separately and 2) the spatially resolved dimensions.
Further model variants are classified in Table~\ref{supp:table:taxonomy_secondary}.
Corresponding model equations can be generated with \texttt{CADET-Equations} \citep{breuer_2026_18339287}}
\centering
\renewcommand{\arraystretch}{1.2} %

\begin{tabular}{x{2.3cm}|x{2cm}x{2.15cm}x{2cm}x{2cm}}
\toprule
\textbf{Bulk Model} & \multicolumn{4}{c}{\textbf{Particle Model}} \\
 & No Particle & No Pores (\cref{model:lumped_rate_model_without_pores}) & 0D Singleton (\cref{model:particle_0D}) & 1D Continuum (\cref{model:particle_1D}) \\
\midrule
0D Singleton (\cref{model:tank}) & \gls{str} & 0D \gls{lrm} & 0D LRMP & 0D \gls{grm} \\
1D Continuum (\cref{model:1D_continuum_bulk_transport_model}) & \gls{pfr} & \gls{lrm} & \gls{lrmp} & \gls{grm} \\
2D Continuum (\cref{model:2D_continuum_bulk_transport_model}) & 2D \gls{pfr} & 2D \gls{lrm} & 2D \gls{lrmp} & 2D \gls{grm} \\
3D Continuum (\cref{model:3D_continuum_bulk_transport_model}) & 3D \gls{pfr} & 3D \gls{lrm} & 3D \gls{lrmp} & 3D \gls{grm} \\
\bottomrule
\end{tabular}

\label{table:taxonomy_primary}
\end{table}

\subsection{Comparison with Established Models}

We discuss how prominent packed-bed chromatography models from the literature can be represented within our proposed framework.
\Cref{table:literature_comp} summarizes a selection of widely adopted models and maps them to the model classification introduced in \cref{sec:taxonomy}.
This comparison highlights the considerable ambiguity present in the field and demonstrates the descriptive power of the taxonomy proposed here.

Some model names emphasize a single modeled phenomenon.
The ``Lumped Kinetic Model'', for instance, is used in some works to describe kinetic binding, while in other works it refers to mass transfer limitations due to film diffusion.
Similarly, the ``Linear Driving Force Model'' denotes a specific binding model, which, in our framework, can be combined with any of the transport models presented.
Another example is the (Lumped) Pore Diffusion Model which
in some studies represents a lumped film diffusion parameter consistent with an \gls{lrmp}, whereas in others, it describes a single-rate model with radial particle resolution (pore diffusion) while neglecting axial dispersion and film diffusion.
The latter case emphasizes the need for a modular perspective on bulk and particle transport.
Conversely, models that appear distinct based on their names are often closely related or even equivalent. For example, both the ``Lumped Finite Adsorption Rate Model'' and the ``Lumped Kinetic Model'' describe an \gls{lrm} with kinetic binding.

\begin{table}[!htbp]
\caption{Mapping of prominent chromatographic models from the literature to the taxonomy of the unified modeling framework presented in this work.}
\centering
\small
\renewcommand{\arraystretch}{1.2}

\begin{tabular}{p{0.47\textwidth} p{0.47\textwidth}}
\toprule
\textbf{Literature} & \textbf{Model class (\cref{table:taxonomy_primary})} \\
\hline
 & \cref{model:tank}
 \\
Continuous Stirred Tank\textsuperscript{1}
& $N^p=0$, $Q^\mathrm{filter}=0$
\\
Pore Diffusion Model\textsuperscript{12}, Single Particle Model\textsuperscript{9,10,11} (no explicit name)
& $N^p = 1$, $Q^\mathrm{filter}=0$
\\
\hline
 & \cref{model:1D_continuum_bulk_transport_model}
 \\
Ideal Plug Flow\textsuperscript{1}
& $N^p=0$, $D^\mathrm{ax}=0$ \\
Dispersive Plug Flow\textsuperscript{1}
& \\
\hline
 & \cref{model:1D_continuum_bulk_transport_model,model:lumped_rate_model_without_pores} -- \gls{lrm}
\\
Ideal Model\textsuperscript{1,2}
& $D^\mathrm{ax} = 0$, equilibrium binding
\\
Thomas Model\textsuperscript{1,6}
& $D^\mathrm{ax}=0$, kinetic binding
\\
Equilibrium Dispersive Model\textsuperscript{1,2,4}
& equilibrium binding
\\
Lumped Finite Adsorption Rate Model\textsuperscript{1}
& reparameterization~\eqref{eq:model-lrm-reparameterization} for $\varepsilon^\mathrm{p}$, kinetic Langmuir binding~\eqref{supp:sec:models-binding-langmuir}
\\
Lumped Kinetic Model\textsuperscript{2}
& kinetic binding
\\
Linear Driving Force Model\textsuperscript{7}
& $D^\mathrm{ax}=0$, linear driving force binding~\eqref{eq:linear_driving_force_approximation}
\\
Transport Model with solid or liquid phase driving force\textsuperscript{7}
& %
linear driving force binding~\eqref{eq:linear_driving_force_approximation}
\\
\hline
& \cref{model:1D_continuum_bulk_transport_model,model:particle_0D} -- \gls{lrmp}
\\
Lumped Pore Diffusion Model\textsuperscript{2,4}
& equilibrium binding
\\
Transport Dispersive Model\textsuperscript{1}
& equilibrium binding
\\
Lumped Kinetic Model\textsuperscript{1} %
& equilibrium binding
\\
\hline
& \cref{model:1D_continuum_bulk_transport_model,model:particle_1D} variant~\eqref{eq:model-1d-bulk-pde-infkf-and-cp=cb}
\\
Pore Diffusion Model\textsuperscript{7}
& $D^\mathrm{ax} = 0$, equilibrium binding
\\
\hline
 & \cref{model:1D_continuum_bulk_transport_model,model:particle_1D} -- \gls{grm}
 \\
General Rate Model (GRM)\textsuperscript{1,2,3}
& 
\\
\gls{grm} with cored beads\textsuperscript{3}
& $R^\mathrm{c} > 0$
\\
Versatile Reaction-Separation Model\textsuperscript{8}
&
kinetic reactions
\\
\gls{grm} with \gls{psd}\textsuperscript{5}
&
$N^\p \geq 1$, $R^\p_i \neq R^\p_j$
\\
\hline
 & \cref{model:2D_continuum_bulk_transport_model,model:particle_1D} -- 2D \gls{grm}
 \\
2D \gls{grm}\textsuperscript{13}
& spatially uniform parameters
\\
2D \gls{grm} with Core–Shell Particles\textsuperscript{14}
& $R^\mathrm{c}>0$, equilibrium binding
\\
2D \gls{grm} with \gls{psd}\textsuperscript{5}
& $N^\p\geq 1$, $R^\p_i \neq R^\p_j$, piecewise constant parameters
\\
\hline
High-Definition Model\textsuperscript{15}
& \cref{model:column-highres3D}
\\
\bottomrule
\end{tabular}

{\footnotesize\noindent
\textsuperscript{1} \cite{Schmidt-Traub2020};
\textsuperscript{2} \cite{Guiochon2006};
\textsuperscript{3} \cite{Gu2015};
\textsuperscript{4} \cite{KATSOULAS2023464345};
\textsuperscript{5} \cite{RAO2025134409};
\textsuperscript{6} \cite{GOLSHANSHIRAZI19921};
\textsuperscript{7} \cite{CARTA2026466781};
\textsuperscript{8} \cite{whitley_effects_1991};
\textsuperscript{9} \cite{Susanto2007};
\textsuperscript{10} \cite{GALLANT2004189};
\textsuperscript{11} \cite{Teske2006};
\textsuperscript{12} \cite{Traylor2014};
\textsuperscript{13} \cite{Qamar2017};
\textsuperscript{14} \cite{Brhane_2dGRM_2019};
\textsuperscript{15} \cite{RAO2023108355};
}

\label{table:literature_comp}
\end{table}

\clearpage
\section{CADET-Equations}
\label{sec:cadet-equations}

\begin{figure}[!htbp]
    \centering
    \includegraphics[width=1.0\linewidth]{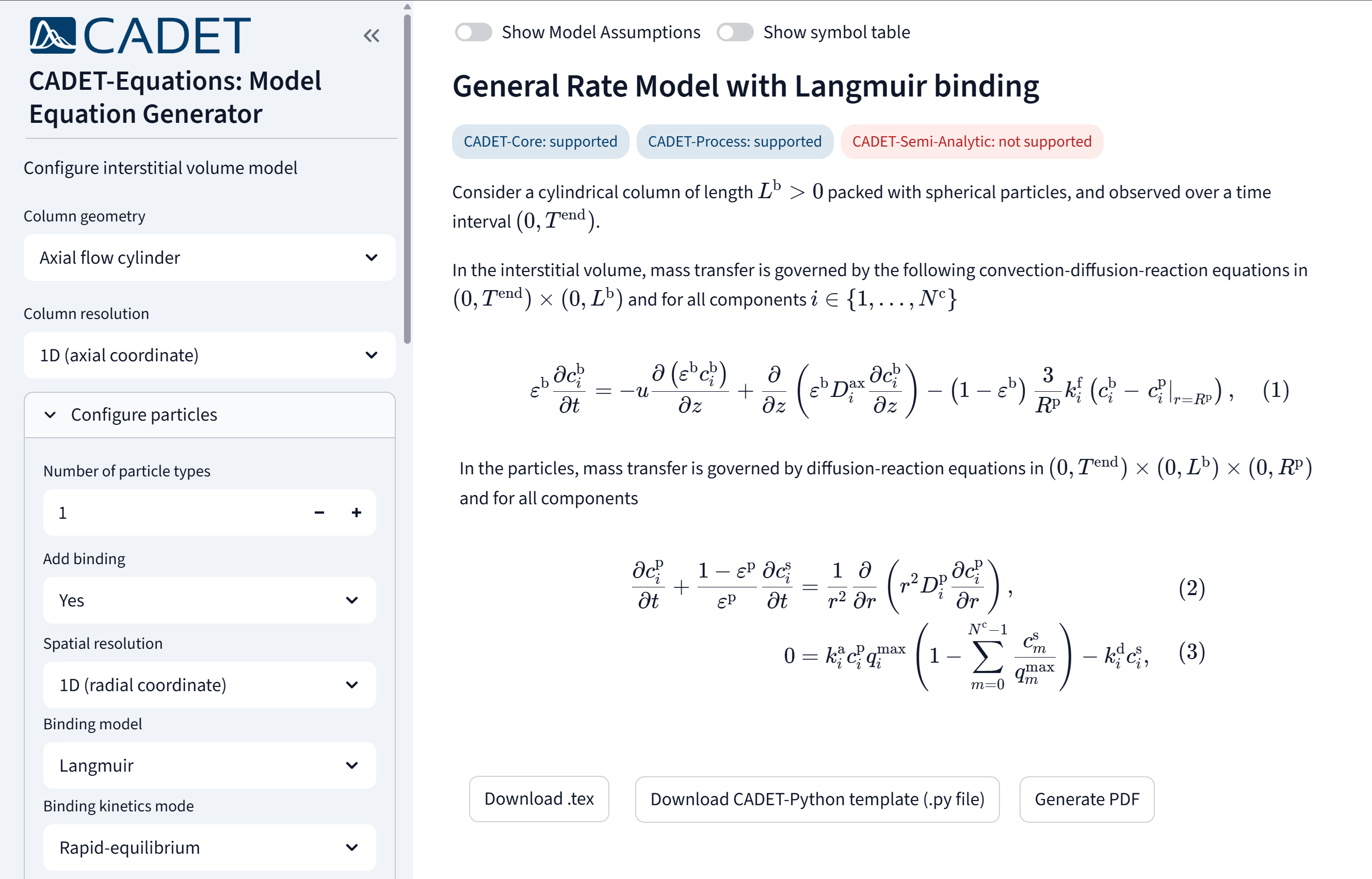}
    \caption{Screenshot of the \texttt{CADET-Equations} graphical user interface. The displayed configuration is a \gls{grm} with rapid-equilibrium Langmuir binding. The left panel allows users to select and combine all model variants presented in this work, including column geometries, binding models, and transport model variants. Model assumptions and parameter descriptions can be displayed on demand using the button at the top. At the bottom, download buttons provide \texttt{.tex} files containing the corresponding equations and \texttt{.py} template files for generating simulation input in the CADET format.}
    \label{fig:cadet-equations}
\end{figure}

To make the unified modeling framework readily accessible and facilitate consistent translation between model specification, mathematical formulation, and simulation, we developed \texttt{CADET-Equations}.
This application allows users to interactively assemble the models introduced in this work by selecting their individual components through a \gls{gui}, as illustrated in \cref{fig:cadet-equations}.
The corresponding governing equations are automatically generated in the standardized notation used throughout this work and can be exported as \LaTeX{} files. The application also displays model assumptions and parameter descriptions, making the basis of individual modeling choices explicit.

Furthermore, \texttt{CADET-Equations} generates Python template files defining the corresponding \texttt{CADET-Core} simulation input. Conversely, users can upload their simulation files to reconstruct and visualize the associated governing equations, thereby improving the transparency and reproducibility of simulation studies.
The application also indicates which model variants are supported by the CADET software stack and which numerical solution methods are available.

\clearpage

\section{Conclusion}
\label{sec:conclusion}

This work introduces a unified, modular, and composable framework for mechanistic modeling of packed-bed chromatography, bringing the diverse family of existing models into a single coherent structure.
Starting from a three-dimensional high-resolution model, we systematically derived a hierarchy of reduced-order models through successive model order reduction, thereby clarifying the mathematical relationships between all model variants.
The four orthogonal components of the framework, bulk transport, intraparticle transport, binding, and reaction, form independently meaningful modules that can be combined systematically to construct consistent model variants.

Existing chromatography models are recovered as particular instances of the framework, while previously unexplored model combinations arise naturally through its composable structure.
The framework therefore provides a unified mathematical description of the mechanistic packed-bed chromatography model family.
By establishing a common taxonomy, notation, and derivation strategy, the framework facilitates communication, comparison, implementation, and extension of chromatography models.
The modular structure of the framework enables systematic extension to increasingly sophisticated physical models while maintaining mathematical consistency across software implementations.

To facilitate the use of the composable modeling framework, we developed \texttt{CADET-Equations}, an interactive web application that assembles chromatographic models from user-selected components and automatically generates the corresponding governing equations in a standardized mathematical notation.
Integrated into the CADET documentation (\url{https://cadet.github.io}), it also serves as an educational resource by enabling users to explore and experiment with the composable modeling framework. In addition, it enhances the transparency and reproducibility of CADET simulations by allowing users to upload a simulation file and automatically generate the corresponding governing equations.

The same design principles are increasingly reflected in the CADET software ecosystem, including the ongoing redesign of \texttt{CADET-Core} toward a more general and composable architecture.

More broadly, we envision the framework serving as a common foundation for future developments in mechanistic chromatography modeling, numerical methods, parameter estimation, process optimization, and digital-twin applications.
By combining mathematical rigor with modular software implementation, it provides a scalable basis for the next generation of computational tools in bioseparation engineering.

Beyond the examples presented here, the composable structure of the framework facilitates the integration of increasingly sophisticated physical models. As a next step, future work will focus on implementing the side-cavity model proposed in \cref{sec:component_specific_pore_accessibility}, developing efficient numerical solution methods, and validating the model against experimental data for simultaneous size-exclusion and binding.

\section*{Declaration of Competing Interest}
The authors declare that they have no known competing financial interests or personal relationships that could have appeared to influence the work reported in this paper.

\section*{CRediT authorship contribution statement}
\textbf{Jan M. Breuer}: Conceptualization, Formal Analysis, Investigation, Methodology, Software, Visualization, Writing - Original Draft.
\textbf{Samuel Leweke}: Conceptualization, Formal Analysis, Methodology, Visualization, Writing - Original Draft.
\textbf{Johannes Schm\"{o}lder}: Writing - Review \& Editing.
\textbf{Antonia Berger}: Methodology, Software.
\textbf{Gregor J. Gassner}: Resources, Supervision.
\textbf{Eric von Lieres}: Conceptualization, Methodology, Project Administration, Resources, Supervision, Writing - Original Draft, Writing - Review \& Editing.

\section*{Acknowledgements}
Jan M. Breuer has received funding by the Deutsche Forschungsgemeinschaft (DFG, German Research Foundation) – 548805630 (grant no. GA 2160/8-1).
Antonia Berger has received funding by the Bundesministerium für Forschung, Technologie und Raumfahrt (BMFTR) - Graduate Cluster AUFBRUCH (grant no. 031B1170).

\bibliographystyle{model5-names}
\bibliography{literature}

  \clearpage

  \let\definition\suppdefinition     \let\enddefinition\endsuppdefinition
  \let\assumption\suppassumption     \let\endassumption\endsuppassumption
  \let\remark\suppremark             \let\endremark\endsuppremark

  \setcounter{section}{0}
  \setcounter{figure}{0}
  \setcounter{table}{0}
  \setcounter{equation}{0}
  \renewcommand{\thesection}{S\arabic{section}}
  \renewcommand{\thesubsection}{\thesection.\arabic{subsection}}
  \renewcommand{\thesubsubsection}{\thesubsection.\arabic{subsubsection}}
  \renewcommand{\thefigure}{S\arabic{figure}}
  \renewcommand{\thetable}{S\arabic{table}}
  \renewcommand{\theequation}{S\arabic{equation}}

  \phantomsection
  \addcontentsline{toc}{section}{Supplementary Material}
  \begin{center}
    \large\bfseries Supplementary Material
  \end{center}
  \vspace{1em}

\section{Vector Calculus}
\label{supp:sec:vectorcalculus}

We consider theorems and identities involving the linear differential operators $\nabla$, $\divOp$, and $\Delta$.
The results are taken from \cite{Borisenko1979}.

\subsection{Cartesian Coordinates}
\label{supp:sec:vectorcalculus-cartesian}

Consider a scalar-valued function $f$ and a vector-valued function $\vec{g}$ with
\begin{align*}
	f\colon (x, y, z) &\mapsto f(x, y, z) \in \Reals, \\ 
	\vec{g}\colon (x, y, z) &\mapsto g_{x}(x, y, z) \vec{e}_{x} + g_{y}(x, y, z) \vec{e}_{y} + g_{z}(x, y, z) \vec{e}_{z} \in \Reals^3.
\end{align*}
The differential operators are given by
\begin{align*}
	\nabla f &= \frac{\partial f}{\partial x} \vec{e}_{x} + \frac{\partial f}{\partial y} \vec{e}_{y} + \frac{\partial f}{\partial z} \vec{e}_{z}, \\ 
	\nabla \cdot \vec{g} &= \frac{\partial g_{x}}{\partial x} + \frac{\partial g_{y}}{\partial y} + \frac{\partial g_z}{\partial z}, \\
	\Delta f &= \frac{\partial^2 f}{\partial x^2} + \frac{\partial^2 f}{\partial y^2} + \frac{\partial^2 f}{\partial z^2}.
\end{align*}

\subsection{Cylindrical Coordinates}
\label{supp:sec:vectorcalculus-cylindrical}

The coordinate transformation is given by
\begin{equation*}
	\begin{pmatrix}
		x \\ y \\ z
	\end{pmatrix} = \begin{pmatrix}
		\rho \cos(\varphi) \\ \rho \sin(\varphi) \\ z
	\end{pmatrix}, \qquad \rho \geq 0, \, \varphi \in [0, 2\pi), \, z \in \Reals.
\end{equation*}
The determinant of the transform's Jacobian (i.e., the differential volume element) is $\rho$.
The corresponding unit vectors $\vec{e}_{\rho}$, $\vec{e}_{\varphi}$, $\vec{e}_{z}$ are pairwise orthogonal.

Consider a scalar-valued function $f$ and a vector-valued function $\vec{g}$ with
\begin{align*}
	f\colon (\rho, \varphi, z) &\mapsto f(\rho, \varphi, z) \in \Reals, \\
	\vec{g}\colon (\rho, \varphi, z) &\mapsto g_{\rho}(\rho, \varphi, z) \vec{e}_{\rho} + g_{\varphi}(\rho, \varphi, z) \vec{e}_{\varphi} + g_{z}(\rho, \varphi, z) \vec{e}_{z} \in \Reals^3.
\end{align*}
The differential operators transform to
\begin{subequations}
\begin{align}
	\nabla f &= \frac{\partial f}{\partial \rho} \vec{e}_{\rho} + \frac{1}{\rho} \frac{\partial f}{\partial \varphi} \vec{e}_{\varphi} + \frac{\partial f}{\partial z} \vec{e}_{z}, \\ 
	\nabla \cdot \vec{g} &= \frac{1}{\rho} \frac{\partial \left( \rho g_{\rho} \right)}{\partial \rho} + \frac{1}{\rho} \frac{\partial g_{\varphi}}{\partial \varphi} + \frac{\partial g_z}{\partial z}, \\
	\Delta f &= \frac{1}{\rho} \frac{\partial}{\partial \rho} \left( \rho  \frac{\partial f }{\partial \rho} \right) + \frac{1}{\rho^2} \frac{\partial^2 f}{\partial \varphi^2} + \frac{\partial^2 f}{\partial z^2}. \label{supp:eq:appx-math-cylinder-laplacian}
\end{align}
\end{subequations}

\subsection{Spherical Coordinates}
\label{supp:sec:vectorcalculus-spherical}

The coordinate transformation is given by
\begin{equation*}
	\begin{pmatrix}
		x \\ y \\ z
	\end{pmatrix} = \begin{pmatrix}
		r \sin(\vartheta) \cos(\varphi) \\ r \sin(\vartheta) \sin(\varphi) \\ r \cos(\vartheta)
	\end{pmatrix}, \qquad r \geq 0, \, \vartheta \in [0, \pi), \, \varphi \in [0, 2\pi).
\end{equation*}
The determinant of the transform's Jacobian (i.e., the differential volume element) is $r^2 \sin(\vartheta)$.
The corresponding unit vectors $\vec{e}_{r}$, $\vec{e}_{\vartheta}$, $\vec{e}_{\varphi}$ are pairwise orthogonal.

Consider a scalar-valued function $f$ and a vector-valued function $\vec{g}$ with
\begin{align*}
	f\colon (r, \vartheta, \varphi) &\mapsto f(r, \vartheta, \varphi) \in \Reals, \\ 
	\vec{g}\colon (r, \vartheta, \varphi) &\mapsto g_{r}(r, \vartheta, \varphi) \vec{e}_{r} + g_{\varphi}(r, \vartheta, \varphi) \vec{e}_{\varphi} + g_{\vartheta}(r, \vartheta, \varphi) \vec{e}_{\vartheta} \in \Reals^3.
\end{align*}
The differential operators transform to
\begin{align*}
	\nabla f &= \frac{\partial f}{\partial r} \vec{e}_{r} + \frac{1}{r} \frac{\partial f}{\partial \vartheta} \vec{e}_{\vartheta} + \frac{1}{r \sin(\vartheta)} \frac{\partial f}{\partial \varphi} \vec{e}_{\varphi}, \\ 
	\nabla \cdot \vec{g} &= \frac{1}{r^2} \frac{\partial \left( r^2 g_{r} \right)}{\partial r} + \frac{1}{r \sin(\vartheta)} \frac{\partial \left( \sin(\vartheta) g_{\vartheta} \right)}{\partial \vartheta} + \frac{1}{r \sin(\vartheta)} \frac{\partial g_\varphi}{\partial \varphi}, \\
	\Delta f &= \frac{1}{r^2} \frac{\partial}{\partial r} \left( r^2  \frac{\partial f }{\partial r} \right) + \frac{1}{r^2 \sin(\vartheta)} \frac{\partial}{\partial \vartheta} \left( \sin(\vartheta) \frac{\partial f}{\partial \vartheta} \right) + \frac{1}{r^2 \sin^2(\vartheta)} \frac{\partial^2 f}{\partial \varphi^2}.
\end{align*}

\subsection{Identities}

Consider a sufficiently smooth scalar-valued function $f$ and a sufficiently smooth vector-valued function $\vec{g}$.
Regardless of the coordinate system, we have
\begin{align*}
	\div{\nabla f} &= \Delta f, \\
	\div{f \vec{g}} &= f \left(\nabla \cdot \vec{g}\right) + \left(\nabla f \right) \cdot \vec{g}.
\end{align*}

\subsection{Divergence Theorem}

\begin{theorem}[Divergence Theorem, \citealp{Borisenko1979}]\label{supp:thm:GaussDivergence}
	Given an open domain $\Omega \subset \Reals^3$ bounded by a piecewise smooth closed surface $\Gamma \coloneqq \partial \Omega$ and a continuously differentiable vector field $\vec{g}\colon \closure{\Omega} \to \Reals^3$, then
	\begin{equation*}
		\int_{\Omega} \nabla \cdot \vec{g}(\vec{x}) \,\mathrm{d}\vec{x} = \int_{\Gamma} \vec{g}(\vec{x}) \cdot \vec{n}(\vec{x}) \,\mathrm{d}\omega(\vec{x}),
	\end{equation*}
	where $\omega$ is the surface measure and $\vec{n}(\vec{x})$ denotes the outward-facing unit normal.
\end{theorem}

\section{Volume Averages}
\label{supp:sec:volaverages}

Consider a two-phase system with phases $\gamma$ (e.g., fluid in the interstitial volume) and $\kappa$ (e.g., particles in the packed bed).

\begin{definition}[Volume Averages, \citealp{Whitaker1999}]\label{supp:def:VolumeAverages}
	For a simply connected open volume $V \subset \Reals^3$ bounded by a piecewise smooth surface $\partial V$, the \emph{phase average} of an integrable quantity $f\colon \Reals^3 \to \Reals$ is defined by
	\begin{equation*}
		\avg{f} \coloneqq \frac{1}{\vol{V}} \int_{V_\gamma} f(\vec{x}) \,\mathrm{d}\vec{x},
	\end{equation*}
	where $V_\gamma \subset V$ denotes the volume of the $\gamma$ phase contained in $V$.
	The \emph{intrinsic average} of $f$ is defined by
	\begin{equation*}
		\avg{f}^\gamma \coloneqq \frac{1}{\vol{V_\gamma}} \int_{V_\gamma} f(\vec{x}) \,\mathrm{d}\vec{x}.
	\end{equation*}

	The volume fraction $\varepsilon$ is defined by $\varepsilon \coloneqq \vol{V_\gamma} / \vol{V}$.
	This relates $\avg{f}$ to $\avg{f}^\gamma$ via
	\begin{equation*}
		\avg{f} = \varepsilon \avg{f}^\gamma.
	\end{equation*}

	The averages of vector-valued quantities $\vec{g}\colon \Reals^3 \to \Reals^3$ are defined component-wise.
\end{definition}

\begin{theorem}[Spatial Averaging Theorem, \citealp{Howes1985,Whitaker1999}]\label{supp:thm:SpatialAveraging}
	Consider a simply connected open volume $V_0 \subset \Reals^3$ bounded by a piecewise smooth surface $\partial V_0$ that contains a representative amount of both phases $\gamma$ and $\kappa$.
	Let $\vec{y}_0 \in V_0$ be the centroid of $V_0$ and define the mapping $\vec{y} \mapsto V(\vec{y})$ by translating $V_0$:
	\begin{equation*}
		V(\vec{y}) \coloneqq \left\{ \vec{y} - \vec{y}_0 + \vec{x} : \vec{x} \in V_0 \right\}.
	\end{equation*}

	Let $f$ be a scalar-valued continuously differentiable quantity and assume that $f$ and $\nabla f$ are both integrable in an open set that strictly contains $V_0$.
	Consequently, the volume averages
	\begin{equation*}
		\avg{f}(\vec{y}) \coloneqq \frac{1}{\vol{V(\vec{y})}} \int_{V_\gamma(\vec{y})} f(\vec{x}) \,\mathrm{d}\vec{x}
	\end{equation*}
	are well-defined for $\vec{y}$ in a sufficiently small neighborhood of $\vec{y}_0$.
	Then, for $\vec{y}$ in a sufficiently small neighborhood of $\vec{y}_0$, the mapping $\vec{y} \mapsto \avg{f}(\vec{y})$ is differentiable and it holds that
	\begin{equation*}
		\avg{\nabla f}(\vec{y}) = \nabla_{\vec{y}} \avg{f}(\vec{y}) + \frac{1}{\vol{V(\vec{y})}} \int_{\partial V_{\gamma\kappa}(\vec{y})} f(\vec{x}) \vec{n}(\vec{x}) \,\mathrm{d}\omega(\vec{x}),
	\end{equation*}
	where $\partial V_{\gamma\kappa}(\vec{y})$ denotes the interface between the $\gamma$ and $\kappa$ phase and $\vec{n}$ is the outward-facing unit normal from $\gamma$ into $\kappa$.

	Analogously, let $\vec{g}$ be a vector-valued continuously differentiable quantity and assume that $\vec{g}$ and $\nabla \cdot \vec{g}$ are both integrable in an open set that strictly contains $V_0$.
	Then, for $\vec{y}$ in a sufficiently small neighborhood of $\vec{y}_0$, the mapping $\vec{y} \mapsto \avg{\vec{g}}(\vec{y})$ is differentiable and it holds that
	\begin{equation*}
		\avg{\nabla \cdot \vec{g}}(\vec{y}) = \nabla_{\vec{y}} \cdot \avg{\vec{g}}(\vec{y}) + \frac{1}{\vol{V(\vec{y})}} \int_{\partial V_{\gamma\kappa}(\vec{y})} \vec{g}(\vec{x}) \cdot \vec{n}(\vec{x}) \,\mathrm{d}\omega(\vec{x}).
	\end{equation*}
\end{theorem}

\section{Modeling Components}
\label{supp:sec:Modeling}

This section contains extensions of the modeling framework and their incorporation into the framework that were too extensive or too specific for the main document.

\subsection{Binding Models}
\label{supp:sec:binding_models}

In Section~\ref{sec:a_hierarchical_family_of_transport_models}, we incorporated arbitrary binding models as modular components of the chromatography model. 
To provide a structured overview, binding models can be grouped according to their mechanistic and mathematical foundations:

\begin{description}

\item[\textbf{Mass Action Law Models}]
These models describe adsorption and desorption through mass-action laws.

\begin{itemize}

\item \textbf{Linear}~\citep{Guiochon2006}  
The linear isotherm assumes independent binding of each component with no competitive interactions.
Adsorption and desorption are proportional to the liquid- and solid-phase concentrations, respectively, and the solid phase has no saturation limit, which makes the model appropriate for dilute or analytical conditions.

\item \textbf{Langmuir}~\citep{Langmuir1916}  
The Langmuir model introduces a finite number of binding sites on the stationary phase.
The adsorption rate therefore decreases as the remaining binding capacity is depleted, and different components compete for the same sites.

\item \textbf{Steric Mass Action (SMA)}%
~\citep{Brooks1992,Diedrich2017,Huuk2017}  
Ion-exchange models describe how charged species interact with charged ligands on the resin.
The SMA model accounts for the number of ionic binding sites occupied or sterically shielded by each adsorbed molecule. It enforces electroneutrality by coupling protein adsorption and desorption to counter-ion exchange, making it particularly suitable for modeling biomolecule separations under varying ionic strength, such as salt-gradient elution.

\end{itemize}

\item[\textbf{Mobile Phase Modifier Models}]
These models account for the influence of mobile-phase components (e.g., salt concentration, pH, or solvent composition) on binding behavior. In these models, mobile-phase modifiers influence binding through concentration-dependent model parameters rather than participating explicitly in the binding mechanism.

\begin{itemize}

\item \textbf{Mobile Phase Modulator (MPM)}~\citep{Melander1989,Karlsson2004}  
MPM models describe how mobile-phase modifiers alter adsorption equilibria, typically by modifying effective interaction strengths between solutes and the stationary phase.
Parameters of these binding models are often expressed as functions of modifier concentrations.

\item \textbf{Hydrophobic Interaction Chromatography (HIC)}%
~\citep{Wang2016,Melander1989,Jaepel2025}  
HIC models capture the dependence of hydrophobic interactions on both salt type and concentration.
Kosmotropic salts generally strengthen these interactions and promote binding, whereas chaotropic salts can weaken them and facilitate elution. With kosmotropic salts, decreasing the salt concentration reduces adsorption and drives elution.

\end{itemize}

\item[\textbf{Colloidal Models}]
These models describe adsorption and desorption processes in systems where colloidal interactions dominate, such as large biomolecules or particles \citep{Xu2009,Briskot2019}.
The Colloidal Particle Adsorption (CPA) model \citep{Briskot2021} provides a mechanistic description of protein adsorption based on electrostatic interactions, accounting for pH-dependent protein charges and ionic strength.
Proteins or particles are treated as interacting colloidal entities.
Adsorption and desorption are influenced by electrostatic and steric interactions between particles and previously adsorbed species, which can lead to cooperative or hindered adsorption behavior.

\item[\textbf{Empirical Models}]
Phenomenological models designed to reproduce experimental observations without explicitly representing the underlying mechanisms.

\begin{itemize}

\item \textbf{Freundlich}~\citep{freundlich192}  
The Freundlich isotherm describes nonlinear adsorption without an explicit saturation limit.
It is often used as a simple empirical model for heterogeneous surfaces.

\item \textbf{Sips}~\citep{sips1948}  
The Sips model combines features of the Freundlich and Langmuir isotherms, capturing heterogeneous adsorption behavior while recovering Langmuir-type saturation at high concentrations.

\end{itemize}

\item[\textbf{Data-Driven and Hybrid Models}]
Approaches that use statistical or machine-learning methods to capture complex binding behavior, optionally incorporating physical constraints.

\begin{itemize}

\item \textbf{Artificial Neural Networks (ANN)}  
Neural-network models learn nonlinear relationships between process conditions and adsorption behavior directly from data. Physics-Informed Neural Networks (PINNs) additionally incorporate physical constraints into the training process.

\item \textbf{Gaussian Process Regression}  
Gaussian processes provide a probabilistic, nonparametric description of adsorption behavior. They also quantify predictive uncertainty, which can support uncertainty quantification in chromatography models.

\end{itemize}

\end{description}

Next, we present a selection of commonly used binding models using our notation. 
For the sake of brevity, and because each particle type $j$ may be described by a different binding model, we omit the particle-type index $j$ throughout this section.

\subsubsection{Multi-Component Langmuir Model}
\label{supp:sec:models-binding-langmuir}

The multi-component Langmuir binding model \cite[Sec.~6.2.2.6]{Seidel-Morgenstern2020-Chap6} adds competitive effects and a saturation term, which accounts for limited resin capacity.
\begin{equation}
	f^{\mathrm{bind}}_{i}\mleft(\vec{c}^{\p}, \vec{c}^{\s}\mright) = k^{\mathrm{a}}_{i} c^{\s,\mathrm{max}}_{i}\left( 1 - \sum_{n = 1}^{N^{\mathrm{c}}} \frac{c^{\s}_{n}}{c^{\s,\mathrm{max}}_{n}} \right) c^{\p}_i - k^{\mathrm{d}}_{i} c^{\s}_i \qquad \forall\  i \in \{1, \dots, N^{\mathrm{c}} \}. \label{supp:eq:model-bnd-langmuir}
\end{equation}
In addition to the adsorption rate constant $k^{\mathrm{a}}_{i} \geq 0$ and the desorption rate constant $k^{\mathrm{d}}_{i} \geq 0$, each component $i$ is equipped with a capacity $c^{\s,\mathrm{max}}_{i} > 0$.
All components compete for the same binding sites as indicated by the sum in \cref{supp:eq:model-bnd-langmuir}. 

Once the capacity is reached for some component $i$ (i.e., $c^{\s}_i = c^{\s,\mathrm{max}}_{i}$ for $N^{\mathrm{c}} = 1$), the net rate of this component cannot be positive, that is, $f^{\mathrm{bind}}_{i}(\vec{c}^{\p}, \vec{c}^{\s}) \leq 0$.
This prevents further binding of the component to the fully saturated resin and shows that $c^{\s}_i \in [0, c^{\s,\mathrm{max}}_{i}]$ for each component $i \in \{1, \dots, N^{\mathrm{c}} \}$.

\subsubsection{Steric Mass Action Model}
\label{supp:sec:models-binding-sma}

\cite{Brooks1992} introduced the steric mass action (SMA) model, which is widely used for modeling ion exchange chromatography.
Ion exchange chromatography exploits the different charge distributions of the molecules to steer the binding process by controlling the salt concentration in the mobile phase.
The SMA model is based on the assumption that the resin has a finite number of ionic binding sites $\Lambda > 0$, which is the ionic capacity of the solid phase.
These binding sites are initially occupied by (monovalent) salt ions.
Each molecule of component $i$ displaces $\nu_i > 0$ (characteristic charge) bound salt ions upon adsorption.
Additionally, the molecule may shield $\sigma_i \geq 0$ (steric factor) binding sites due to its size and shape, which are not available for binding of other molecules.

We choose component $i=1$ as salt and formulate the balance equation for the binding sites
\begin{equation}
	\Lambda = c^{\s}_1 + \sum_{n=2}^{N^{\mathrm{c}}} \nu_{n} c^{\s}_{n}, \label{supp:eq:model-bnd-salt}
\end{equation}
which represents electro-neutrality and is used to calculate the number of available binding sites (or displaceable salt ions on the surface) $\overbar{c}^{\s}_1$:
\begin{equation*}
	\Lambda = \overbar{c}^{\s}_1 + \sum_{n = 2}^{N^{\mathrm{c}}} \left( \nu_{n} + \sigma_{n} \right) c^{\s}_{n} \qquad \Leftrightarrow \qquad \overbar{c}^{\s}_1 = \Lambda - \sum_{n = 2}^{N^{\mathrm{c}}} \left( \nu_{n} + \sigma_{n} \right) c^{\s}_{n} = c^{\s}_1 - \sum_{n = 2}^{N^{\mathrm{c}}} \sigma_{n} c^{\s}_{n}.
\end{equation*}
The binding rate for the protein components $i \in \{ 2, \dots, N^{\mathrm{c}}\}$ follows from the mass action law:
\begin{equation}
	f^{\mathrm{bind}}_{i}\mleft(\vec{c}^{\p}, \vec{c}^{\s}\mright) = k^{\mathrm{a}}_{i} \left( \overbar{c}^{\s}_1 \right)^{\nu_i} c^{\p}_i - k^{\mathrm{d}}_{i} \left( c^{\p}_1 \right)^{\nu_i} c^{\s}_i \qquad \forall\  i \in \{2, \dots, N^{\mathrm{c}} \}. \label{supp:eq:model-bnd-sma-pre}
\end{equation}
The protein components $i \geq 2$ are coupled by the electro-neutrality condition \cref{supp:eq:model-bnd-salt} and, thus, compete for the available binding sites $\overbar{c}^{\s}_1$.
Due to the finite capacity $\Lambda < \infty$, the solid-phase protein concentrations are bounded, that is $c^{\s}_1 \in [0, \Lambda]$ and $c^{\s}_i \in [0, \Lambda / (\nu_i + \sigma_i)]$ for $i \geq 2$.

\Cref{supp:eq:model-bnd-salt} treats the binding of the salt ions as being in rapid equilibrium, assuming that it occurs much faster than the other binding and mass transfer processes.
To incorporate the model into the framework developed above, we impose this equilibrium constraint on component $i=1$ (salt) and define
\begin{equation*}
	f^{\mathrm{bind}}_{i}\mleft(\vec{c}^{\p}, \vec{c}^{\s}\mright) = \begin{cases}
		\Lambda - \left( c^{\s}_1 + \sum_{n=2}^{N^{\mathrm{c}}} \nu_{n} c^{\s}_{n} \right) & \text{for } i=1, \\
		k^{\mathrm{a}}_{i} \left( \overbar{c}^{\s}_1 \right)^{\nu_i} c^{\p}_i - k^{\mathrm{d}}_{i} \left( c^{\p}_1 \right)^{\nu_i} c^{\s}_i & \text{for } i \in \{2, \dots, N^{\mathrm{c}} \}.
	\end{cases}
\end{equation*}

\begin{remark}
	The extension to polyvalent salt ions $\nu_1 > 1$ is straightforward:
	\begin{equation*}
		f^{\mathrm{bind}}_{i}\mleft(\vec{c}^{\p}, \vec{c}^{\s}\mright) = \begin{cases}
			\Lambda - \left( \nu_1 c^{\s}_1 + \sum_{n=2}^{N^{\mathrm{c}}} \nu_{n} c^{\s}_{n} \right) & \text{for } i=1, \\
			k^{\mathrm{a}}_{i} \left( \overbar{c}^{\s}_1 \right)^{\nu_i / \nu_1} c^{\p}_i - k^{\mathrm{d}}_{i} \left( c^{\p}_1 \right)^{\nu_i / \nu_1} c^{\s}_i & \text{for } i \in \{2, \dots, N^{\mathrm{c}} \}.
		\end{cases}
	\end{equation*}
\end{remark}

\subsubsection{Multi-Component Bi-Langmuir Model}
\label{supp:sec:models-binding-bilangmuir}

\cite{Graham1953} extended the Langmuir isotherm (for one component) to a second type of binding sites.
The two types of binding sites are independent and modeled using different sets of parameters.
In order to describe the solid-phase concentration on the second type of binding sites, we need to introduce additional solid-phase unknowns, such that
\begin{equation*}
	\vec{c}^{\s} = \left(c^{\s}_{1,1}, c^{\s}_{1,2}, c^{\s}_{2,1}, c^{\s}_{2,2}, \dots, c^{\s}_{N^{\mathrm{c}},1}, c^{\s}_{N^{\mathrm{c}},2} \right).
\end{equation*}
That is, we have the solid-phase concentrations $c^{\s}_{i,k}$, where $k \in \{1,2\}$ denotes the binding site type.
We can equivalently say that each component $i$ has $N^{\mathrm{b}}_{i} = 2$ bound states, which correspond to the two types of binding sites.

For each type of binding site, the net binding rates are given by the usual multi-component Langmuir rates from \cref{supp:eq:model-bnd-langmuir} in \cref{supp:sec:models-binding-langmuir}:
\begin{equation*}
	f^{\mathrm{bind}}_{i,k}\mleft(\vec{c}^{\p}, \vec{c}^{\s}\mright) = k^{\mathrm{a}}_{i,k}\, c^{\s,\mathrm{max}}_{i,k}\left( 1 - \sum_{n = 1}^{N^{\mathrm{c}}} \frac{c^{\s}_{n,k}}{c^{\s,\mathrm{max}}_{n,k}} \right) c^{\p}_i - k^{\mathrm{d}}_{i,k} \, c^{\s}_{i,k} \quad \forall\  i \in \{1, \dots, N^{\mathrm{c}} \}, \, k \in \{1,2\}.
\end{equation*}
The total change in concentration of component $i$ is obtained by summing $f^{\mathrm{bind}}_{i,k}\mleft(\vec{c}^{\p}, \vec{c}^{\s}\mright)$ over the binding site types $k = 1, \dots, N^{\mathrm{b}}_{i}$.
Note that this model easily generalizes to more than $k = 2$ types of binding sites.

\subsubsection{Multi-State Steric Mass Action Model}
\label{supp:sec:models-binding-mssma}

The multi-state SMA model extends the SMA model from \cref{supp:sec:models-binding-sma} to multiple bound states.
This model is introduced by \cite{Diedrich2017} to allow molecules of the same component to bind with different strengths (i.e., in different bound states).
It applies the fundamental principle of the spreading model \citep{Yang2003,Lundstrom1985}, which allows for multiple bound states in a Langmuir-type model, to the SMA model and allows each component to transition between different bound states.

Multiple bound states can, for example, arise due to multiple binding orientations \citep{Ghosh2013}, which are modeled by different characteristic charges $\nu$ and shielding factors $\sigma$.
\cite{Diedrich2017} used the multi-state SMA model to analyze experiments with tentacle resins.
The proposed mechanism is that the tentacle folds around the macromolecule, thereby transferring the molecule to progressively more strongly bound states.

To formulate the model, we first introduce the number of bound states $N^{\mathrm{b}}_{i} \geq 1$ per component $i$.
As in the standard SMA model, we reserve component $i = 1$ for salt and set $N^{\mathrm{b}}_{1} = 1$.
The solid-phase concentration vector is extended to
\begin{equation*}
	\vec{c}^{\s} = \left(c^{\s}_{1,1}, \dots, c^{\s}_{1, N^{\mathrm{b}}_{1}}, c^{\s}_{2,1}, \dots, c^{\s}_{2, N^{\mathrm{b}}_{2}}, \dots, c^{\s}_{N^{\mathrm{c}},1}, \dots, c^{\s}_{N^{\mathrm{c}}, N^{\mathrm{b}}_{N^{\mathrm{c}}}} \right).
\end{equation*}
We assume that the bound states are ordered from weakly to strongly bound states, that is,
\begin{equation*}
  \nu_{i,k} \leq \nu_{i,k+1} \qquad \forall\  i \in \{2, \dots, N^{\mathrm{c}}\}, \, k \in \{1, \dots, N^{\mathrm{b}}_{i} - 1 \},
\end{equation*}
where $\nu_{i,k} > 0$ is the number of salt ions displaced by component $i$ in bound state $k$.
Similarly, we denote by $\sigma_{i,k} \geq 0$ the number of shielded binding sites in bound state $k$.

As for the standard SMA model, we assume that the salt component is in rapid equilibrium and define the electroneutrality equation
\begin{equation}
    \Lambda =  c^{\s}_1 + \sum_{n=2}^{N^{\mathrm{c}}} \sum_{k=1}^{N^{\mathrm{b}}_{n}} \nu_{n,k} c^{\s}_{n,k} . \label{supp:eq:model-bnd-mssma-salt}
\end{equation}
We need to account for the additional bound states in order to obtain the amount of free binding sites
\begin{equation*}
	\overbar{c}^{\s}_1 = \Lambda - \sum_{n=2}^{N^{\mathrm{c}}} \sum_{k=1}^{N^{\mathrm{b}}_{n}} \left( \nu_{n,k} + \sigma_{n,k} \right) c^{\s}_{n,k} = c^{\s}_1 - \sum_{n=2}^{N^{\mathrm{c}}} \sum_{k=1}^{N^{\mathrm{b}}_{n}} \sigma_{n,k} c^{\s}_{n,k}.
\end{equation*}
We can now define the net binding rate for the components $i \in \{2, \dots, N^{\mathrm{c}}\}$ and binding states $k \in \{1, \dots, N^{\mathrm{b}}_{i}\}$.
In doing so, we need to account for the necessary additional free binding sites when a component transitions to a more strongly bound state.
Vice versa, salt ions from the mobile phase occupy the binding sites released by the transition to a more weakly bound state.
Using the mass action law (see \cref{supp:sec:models-reaction-massaction}), these considerations result in
\begin{multline}
	f^{\mathrm{bind}}_{i,k}\mleft(\vec{c}^{\p}, \vec{c}^{\s}\mright) = k^{\mathrm{a}}_{i,k} \left(\overbar{c}^{\s}_1\right)^{\nu_{i,k}} c^{\p}_{i} - k^{\mathrm{d}}_{i,k} \left(c^{\p}_1\right)^{\nu_{i,k}} c^{\s}_{i,k} \\
	- \sum_{n = 1}^{k-1} \underbrace{k^{(i)}_{\mathrm{e},k,n}\: \left(c^{\p}_1\right)^{\left(\nu_{i,k} - \nu_{i,n}\right)}\: c^{\s}_{i,k}}_{\text{to more weakly bound state}} - \sum_{n = k+1}^{N^{\mathrm{b}}_{i}} \underbrace{k^{(i)}_{\mathrm{e},k,n}\: \left(\overbar{c}^{\s}_1\right)^{\left(\nu_{i,n} - \nu_{i,k}\right)}\: c^{\s}_{i,k}}_{\text{to more strongly bound state}} \\
	+ \sum_{n = 1}^{k-1} \underbrace{k^{(i)}_{\mathrm{e},n,k}\: \left(\overbar{c}^{\s}_1\right)^{\left(\nu_{i,k} - \nu_{i,n}\right)}\: c^{\s}_{i,n}}_{\text{from more weakly bound  state}} + \sum_{n = k+1}^{N^{\mathrm{b}}_{i}} \underbrace{k^{(i)}_{\mathrm{e},n,k}\: \left(c^{\p}_1\right)^{\left(\nu_{i,n} - \nu_{i,k}\right)}\: c^{\s}_{i,n}}_{\text{from more strongly bound state}}, \label{supp:eq:model-bnd-mssma-prot}
\end{multline}
where $k^{(i)}_{\mathrm{e},k,n} \geq 0$ denotes the exchange rate constant of component $i$ from bound state $k$ to bound state $n$, $k^{\mathrm{a}}_{i,k} \geq 0$ is the adsorption rate constant to bound state $k$ of component $i$, and $k^{\mathrm{d}}_{i,k} \geq 0$ is the desorption rate constant from bound state $k$ of component $i$.

The multi-state SMA model is different from the models considered before in that it includes transitions between bound states, which leads to reaction-type terms.
These reactions do not fit into the framework developed above (i.e., they do not match the form of \eqref{supp:eq:model-hrm-jbind-rapidequi}) when rapid equilibrium is assumed.
However, the reaction terms in \cref{supp:eq:model-bnd-mssma-prot} follow the mass action law and depend explicitly on the liquid- and solid-phase salt concentrations, reflecting counter-ion exchange during transitions between bound states.
In \cref{supp:sec:models-reaction-rapidequimssma}, this insight will enable us to derive the rapid-equilibrium variant of this model, which is not covered by \citet{Diedrich2017}.

\begin{remark}
	Similarly to the SMA model, the extension to polyvalent salt ions $\nu_1 > 1$ is straightforward:
	\begin{align*}
		 \Lambda =& \ \nu_1 c^{\s}_1 + \sum_{n=2}^{N^{\mathrm{c}}} \sum_{k=1}^{N^{\mathrm{b}}_{n}} \nu_{n,k} c^{\s}_{n,k} , \\
		f^{\mathrm{bind}}_{i,k}\mleft(\vec{c}^{\p}, \vec{c}^{\s}\mright) =
		&\begin{multlined}[t]
			k^{\mathrm{a}}_{i,k} \left(\overbar{c}^{\s}_1\right)^{\nu_{i,k} / \nu_1} c^{\p}_{i} - k^{\mathrm{d}}_{i,k} \left(c^{\p}_1\right)^{\nu_{i,k} / \nu_1} c^{\s}_{i,k} \\
			- \sum_{n = 1}^{k-1} \underbrace{k^{(i)}_{\mathrm{e},k,n}\: \left(c^{\p}_1\right)^{\left(\nu_{i,k} - \nu_{i,n}\right) / \nu_1}\: c^{\s}_{i,k}}_{\text{to more weakly bound state}} - \sum_{n = k+1}^{N^{\mathrm{b}}_{i}} \underbrace{k^{(i)}_{\mathrm{e},k,n}\: \left(\overbar{c}^{\s}_1\right)^{\left(\nu_{i,n} - \nu_{i,k}\right) / \nu_1}\: c^{\s}_{i,k}}_{\text{to more strongly bound  state}} \\
			+ \sum_{n = 1}^{k-1} \underbrace{k^{(i)}_{\mathrm{e},n,k}\: \left(\overbar{c}^{\s}_1\right)^{\left(\nu_{i,k} - \nu_{i,n}\right) / \nu_1}\: c^{\s}_{i,n}}_{\text{from more weakly bound  state}} + \sum_{n = k+1}^{N^{\mathrm{b}}_{i}} \underbrace{k^{(i)}_{\mathrm{e},n,k}\: \left(c^{\p}_1\right)^{\left(\nu_{i,n} - \nu_{i,k}\right) / \nu_1}\: c^{\s}_{i,n}}_{\text{from more strongly bound  state}}.
		\end{multlined}
	\end{align*}
\end{remark}

\subsection{Rapid-Equilibrium Binding and Non-Limiting Film Diffusion}
\label{supp:sec:model-reqBindFilmDiff-notorthogonal}

The combination of rapid-equilibrium binding (Section~\ref{sec:binding_equilibrium_kinetics}) and non-limiting film diffusion (Section~\ref{sec:rapid-equilibrium_film_diffusion}) is not orthogonal when surface diffusion is present in the model.
Consider the axially resolved bulk domain from Section~\ref{model:1D_continuum_bulk_transport_model} combined with a radially resolved particle, Eq.~\eqref{eq:model_particle_1D}, including surface diffusion.
Starting with non-limiting film diffusion and then applying rapid-equilibrium binding, we arrive at the bulk domain PDE
\begin{equation}
    \varepsilon^{\mathrm{b}} \frac{\partial c^{\b}_i}{\partial t}
    =
    -u \frac{\partial \left( \varepsilon^{\mathrm{b}} c^{\b}_i \right)}{\partial z} + \frac{\partial}{\partial z} \left( \varepsilon^{\mathrm{b}} D^{\mathrm{ax}}_{i} \frac{\partial c^{\b}_i}{\partial z} \right)
    - \left(1 - \varepsilon^{\mathrm{b}} \right) \sum_{j=1}^{N^{\p}} \frac{3 d_j}{R^{\p}_{j}} \varepsilon^{\p}_{j} \left.\left( D^{\p}_{j,i} \frac{\partial c^{\p}_{j,i}}{\partial r} \right) \right|_{r = R^{\p}_{j}}.
    \label{supp:eq:model-1d-bulk-filmdiffrapidequi}
\end{equation}
Conversely, letting $k^{\mathrm{f}}_{j,i} \to \infty$ in the rapid-equilibrium binding variant yields
\begin{equation}
    \varepsilon^{\mathrm{b}} \frac{\partial c^{\b}_i}{\partial t}
    =
    -u \frac{\partial \left( \varepsilon^{\mathrm{b}} c^{\b}_i \right)}{\partial z} + \frac{\partial}{\partial z} \left( \varepsilon^{\mathrm{b}} D^{\mathrm{ax}}_{i} \frac{\partial c^{\b}_i}{\partial z} \right)
    - \left(1 - \varepsilon^{\mathrm{b}} \right) \sum_{j=1}^{N^{\p}} \left. \frac{3 d_j}{R^{\p}_{j}} \left( \varepsilon^{\p}_{j} D^{\p}_{j,i} \frac{\partial c^{\p}_{j,i}}{\partial r} + \left( 1 - \varepsilon^{\p}_{j} \right) D^{\s}_{j,i} \frac{\partial c^{\s}_{j,i}}{\partial r} \right) \right|_{r = R^{\p}_{j}},
    \label{supp:eq:model-1d-bulk-rapidequifilmdiff}
\end{equation}
which also includes a surface diffusion flux term due to Eq.~\eqref{eq:model_particle_1D_BC_Rp_rapidequi}.
The remaining equations are the same for both orders:
\begingroup\allowdisplaybreaks
\begin{align*}
    c^{\b}_i
    &=
    \left. c^{\p}_{j,i} \right|_{r = R^{\p}_{j}},
    \\
    \frac{\partial c^{\p}_{j,i}}{\partial t} + \frac{1 - \varepsilon^{\p}_{j}}{\varepsilon^{\p}_{j}} \frac{\partial c^{\s}_{j,i}}{\partial t}
    &=
    \frac{1}{r^2} \frac{\partial }{\partial r} \left( r^2 D_{j,i}^{\p} \frac{\partial c^{\p}_{j,i}}{\partial r} \right) + \frac{1 - \varepsilon^{\p}_{j}}{\varepsilon^{\p}_{j}} \frac{1}{r^2} \frac{\partial }{\partial r} \left( r^2 D_{j,i}^{\s} \frac{\partial c^{\s}_{j,i}}{\partial r} \right), \\
    0
    &=
    f^{\mathrm{bind}}_{j,i,k}\mleft( \vec{c}^{\p}_j, \vec{c}^{\s}_j \mright),
    \\
    0 
    &= 
    -\left.\left(\varepsilon^{\p}_{j} D^{\p}_{j,i} \frac{\partial c^{\p}_{j,i}}{\partial r} + \left( 1 - \varepsilon^{\p}_{j} \right) D^{\s}_{j,i} \frac{\partial c^{\s}_{j,i}}{\partial r} \right) \right|_{r=R^{\mathrm{c}}}.
\end{align*}\endgroup

Clearly, \cref{supp:eq:model-1d-bulk-filmdiffrapidequi} and \cref{supp:eq:model-1d-bulk-rapidequifilmdiff} are only equivalent for $D^{\s}_{j,i} = 0$ or $\varepsilon^{\p}_{j} = 1$.
Whereas the first variant is more in line with the assumption that molecules cannot enter the particle by means of surface diffusion, which is also expressed in the original boundary condition \eqref{eq:model_particle_1D_BC_solid_Rp}, the second one emphasizes that only the conserved moiety $\varepsilon^{\p}_{j} c^{\p}_{j,i} + (1-\varepsilon^{\p}_{j}) c^{\s}_{j,i}$ is transported.
Thus, the choice between the two versions hinges on whether surface diffusion is assumed to be important for the molecules to access the particle pores.

\subsection{Binding Equilibrium}
\label{supp:sec:adsorption_equilibrium_kinetics}

\subsubsection{Homogeneous Particles}
\label{supp:sec:model_particle_0D-req}

Following the same procedure from Section~\ref{sec:binding_equilibrium_kinetics}, but for the homogeneous particle Model~\ref{model:particle_0D}, we add Eq.~(\ref{eq:model_particle_0D_liquid}) and Eq.~(\ref{eq:model_particle_0D_solid}) together (for all bound states), which yields the mass balance for the conserved quantity $c^{\p}_i + \frac{1-\varepsilon^{\p}}{\varepsilon^{\p}} \sum_{k=1}^{N^{\mathrm{b}}_{i}} c^{\s}_{i,k}$
\begin{equation*}
    \frac{\partial c^{\p}_i}{\partial t} + \frac{1-\varepsilon^{\p}}{\varepsilon^{\p}} \sum_{k=1}^{N^{\mathrm{b}}_{i}} \frac{\partial c^{\s}_{i,k}}{\partial t}
    =
    \frac{a^{\mathrm{p}}}{\varepsilon^{\p}} k^{\mathrm{f}}_{i} \left( c^{\b}_i - c^{\p}_{i} \right).
\end{equation*}
The combined quantity $c^{\p}_i + \frac{1-\varepsilon^{\p}}{\varepsilon^{\p}} \sum_{k=1}^{N^{\mathrm{b}}_{i}} c^{\s}_{i,k}$ is conserved with respect to binding and changes only due to transport.
The ratio between liquid-phase concentration $c^{\p}_i$ and solid-phase concentrations $c^{\s}_{i,k}$ is then determined by a vanishing net binding rate.
To this end, we divide the solid-phase mass balance Eq.~\eqref{eq:model_particle_0D_solid} by $k^{\mathrm{kin}}_{i}$ and take the limit $k^{\mathrm{kin}}_{i} \to \infty$:
\begin{alignat*}{2}
    & & 0 &= \lim\limits_{k^{\mathrm{kin}}_{i} \to \infty} \frac{1}{k^{\mathrm{kin}}_{i}} f^{\mathrm{bind}}_{i,k}\mleft( \vec{c}^{\p}, \vec{c}^{\s} \mright)
    \\
    & & 0 &=
    \lim\limits_{k^{\mathrm{kin}}_{i} \to \infty} \frac{k^{\mathrm{kin}}_{i}}{k^{\mathrm{kin}}_{i}} \left( k^{\mathrm{eq}}_{i} r^{\mathrm{ads}}_{i}\mleft( \vec{c}^{\p}, \vec{c}^{\s} \mright) - r^{\mathrm{des}}_{i}\mleft( \vec{c}^{\p}, \vec{c}^{\s} \mright) \right)
    \\
    & \Leftrightarrow&\qquad 0 &= k^{\mathrm{eq}}_{i} r^{\mathrm{ads}}_{i}\mleft( \vec{c}^{\p}, \vec{c}^{\s} \mright) - r^{\mathrm{des}}_{i}\mleft( \vec{c}^{\p}, \vec{c}^{\s} \mright),
\end{alignat*}
which is equivalent to a vanishing net rate
\begin{equation*}
	f^{\mathrm{bind}}_{i,k}\mleft( \vec{c}^{\p}, \vec{c}^{\s} \mright) = 0
\end{equation*}
if we take any finite non-zero value for $k^{\mathrm{kin}}_{i}$.
For the components with rapid-equilibrium binding, the particle differential system, Eq.~\eqref{eq:model_particle_0D}, becomes the differential-algebraic system
\begin{subequations}
\label{supp:eq:model_particle_0D-req}
\begin{alignat}{2}
    \frac{\partial c^{\p}_i}{\partial t} + \sum_{k=1}^{N^{\mathrm{b}}_{i}} \frac{1 - \varepsilon^{\p}}{\varepsilon^{\p}} \frac{\partial c^{\s}_{i,k}}{\partial t}
    &=
    \frac{a^{\mathrm{p}}}{\varepsilon^{\p}} k^{\mathrm{f}}_{i} \left( c^{\b}_i - c^{\p}_{i} \right),
    \label{supp:eq:model_particle_0D_liquid-req} \\
	0 &= f^{\mathrm{bind}}_{i,k}\mleft( \vec{c}^{\p}, \vec{c}^{\s} \mright). \label{supp:eq:model_particle_0D_solid-req}
\end{alignat}
\end{subequations}
Components with kinetic binding and components with rapid-equilibrium binding can be combined as described in Section~\ref{supp:sec:mixed_kinetic_and_equilibrium_components}.

\subsubsection{Combined Kinetic and Rapid-Equilibrium Binding}
\label{supp:sec:mixed_kinetic_and_equilibrium_components}

In the case of binding models having components that bind much faster than the others (e.g., salt counter-ions in ion-exchange models), the model can have subsets $\mset{N}_{i}^{\mathrm{b},\mathrm{req}}, \mset{N}_{i}^{\mathrm{b},\mathrm{kin}} \subset$ ($\{1,\dots, N^{\mathrm{b}}_{i}\}, \mset{N}_{i}^{\mathrm{b},\mathrm{req}} \cup \mset{N}_{i}^{\mathrm{b},\mathrm{kin}} \equiv \{1,\dots, N^{\mathrm{b}}_{i}\}$) representing bound states with rapid-equilibrium binding and kinetic binding, respectively.
In that case, Eq.~\eqref{eq:model_particle_1D_liquid-req} changes for all affected components $i \in \left\{ i\in\{1,\dots,N^{\mathrm{c}}\} \colon \lvert \mset{N}_{i}^{\mathrm{b},\mathrm{req}} \rvert > 0 \right\} =: \mset{N}^{\mathrm{c},\mathrm{req}}$ to
\begin{subequations}
\label{supp:eq:model_particle_1D-combined-req-kin}
\begin{alignat}{2}
    \begin{split}
        \frac{\partial c^{\p}_i}{\partial t} + \frac{1 - \varepsilon^{\p}}{\varepsilon^{\p}} \sum_{k\in\mset{N}_{i}^{\mathrm{b},\mathrm{req}}} \frac{\partial c^{\s}_{i,k}}{\partial t}
        &= 
        \frac{1}{r^2} \frac{\partial }{\partial r} \left( r^2 D_{i}^{\p} \frac{\partial c^{\p}_{i}}{\partial r} \right) \\
        &- \frac{1 - \varepsilon^{\p}}{\varepsilon^{\p}} \sum_{k\in\mset{N}_{i}^{\mathrm{b},\mathrm{kin}}} f^{\mathrm{bind}}_{i,k}\mleft( \vec{c}^{\p}, \vec{c}^{\s} \mright) \\
        &+ \frac{1 - \varepsilon^{\p}}{\varepsilon^{\p}} \sum_{k\in \mset{N}_{i}^{\mathrm{b},\mathrm{req}}} \frac{1}{r^2} \frac{\partial }{\partial r} \left( r^2 D_{i}^{\s} \frac{\partial c^{\s}_{i,k}}{\partial r} \right).
    \end{split} \label{supp:eq:model_particle_1D_liquid-combined-req-kin}
\end{alignat}
\begin{alignat}{2}
\intertext{For all bound states $k\in \mset{N}_{i}^{\mathrm{b},\mathrm{kin}}$ with kinetic binding, Eq.~\eqref{eq:model_particle_1D_solid-req} becomes}
    \frac{\partial c_{i,k}^{\s}}{\partial t}
    &=
    \frac{1}{r^2} \frac{\partial }{\partial r} \left( r^2 D_{i}^{\s} \frac{\partial c^{\s}_{i,k}}{\partial r} \right)
    + f^{\mathrm{bind}}_{i,k}\mleft( \vec{c}^{\p}, \vec{c}^{\s} \mright),
    \label{supp:eq:model_particle_1D_solid-combined-kin} \\
\intertext{and for all bound states $k\in \mset{N}_{i}^{\mathrm{b},\mathrm{req}}$ with rapid-equilibrium binding}
    0 &= f^{\mathrm{bind}}_{i,k}\mleft( \vec{c}^{\p}, \vec{c}^{\s} \mright).
\label{supp:eq:model_particle_1D_solid-combined-req}
\end{alignat}
\end{subequations}
Similarly, the boundary conditions, Eq.~\eqref{eq:model_particle_1D_BC_rapidequi}, are modified by summing up the respective equations:
\begingroup\allowdisplaybreaks[0]
\begin{subequations} 
\label{supp:eq:model_particle_1D_BC_solid-combined-req-kin}
\begin{gather}
    \! \left.\Biggl[ \varepsilon^{\p} D^{\p}_{i} \frac{\partial c^{\p}_{i}}{\partial r} + \left( 1 - \varepsilon^{\p} \right)\! \sum_{k \in \mset{N}_{i}^{\mathrm{b},\mathrm{req}}}\!\! D_{i,k}^{\s} \frac{\partial c^{\s}_{i,k}}{\partial r} \Biggr]\right|_{r = R^{\mathrm{p}}} \hspace*{-10pt}
    =
    k^{\mathrm{f}}_{i} \left( c^{\b}_i - \left. c^{\p}_{i} \right|_{r = R^{\mathrm{p}}}\! \right), \label{supp:eq:model_particle_1D_BC_Rp_solid-combined-req-kin}
    \\
    0 = -\left.\Biggl[\varepsilon^{\p} D^{\p}_{i} \frac{\partial c^{\p}_{i}}{\partial r} + \left( 1 - \varepsilon^{\p} \right) \sum_{k \in \mset{N}_{i}^{\mathrm{b},\mathrm{req}}} D_{i,k}^{\s} \frac{\partial c^{\s}_{i,k}}{\partial r} \Biggr] \right|_{r = R^{\mathrm{c}}}. \label{supp:eq:model_particle_1D_BC_0_solid-combined-req-kin}
\end{gather}
\end{subequations}
\endgroup

Again, we gain consistency conditions
\begin{equation*}
f^{\mathrm{bind}}_{i,k}\mleft( \vec{c}^{\p,\mathrm{init}}, \vec{c}^{\s,\mathrm{init}} \mright) = 0 \qquad \text{in } (0, L^\mathrm{b}) \times (R^{\mathrm{c}}, R^{\mathrm{p}})
\end{equation*}
for the initial data of the algebraic unknowns $i \in \mset{N}^{\mathrm{c},\mathrm{req}}$ and $k \in \mset{N}_{i}^{\mathrm{b},\mathrm{req}}$.

\subsection{Reactions}
\label{supp:sec:reactions}

This section extends the transport models introduced in Section~\ref{sec:a_hierarchical_family_of_transport_models} to include reaction models. Reactions in rapid equilibrium require additional consideration to prevent an overdetermined system by reformulating the equations in terms of \emph{conserved moieties}, as described in \cref{supp:sec:models-reaction-rapidequi}.

\subsubsection{Kinetic Reactions}
\label{supp:sec:models-reaction-dynamic}

We consider reactions involving $N^{\mathrm{c}}$ components of the form 
\begin{equation}
    \label{supp:eq:model-rctn-reaction-single}
	\sum_{i=1}^{N^{\mathrm{c}}} a_i c_i \to \sum_{i=1}^{N^{\mathrm{c}}} b_i c_i,
\end{equation}
where the coefficients $a_i, b_i \in \Integers$, $i \in \{ 1, \dots, N^{\mathrm{c}} \}$, specify the stoichiometric proportions of the participating components.
They are collected in the reactant and product complex vectors $\vec{a} \in \Integers^{N^{\mathrm{c}}}$ and $\vec{b} \in \Integers^{N^{\mathrm{c}}}$, respectively.
The net change in the concentration vector $\vec{c}$ is governed by the ODE
\begin{equation}
	\frac{\mathrm{d}\vec{c}}{\mathrm{d}t} = \left( \vec{b} - \vec{a} \right) r\mleft( \vec{c} \mright), \label{supp:eq:model-rctn-ode-single}
\end{equation}
where $r\colon \Reals^{N^{\mathrm{c}}} \to \Reals$ denotes the reaction rate function.

We extend this single reaction to a system of $N^{\mathrm{r}}$ reactions.
We collect the net stoichiometric coefficients $\vec{b} - \vec{a}$ in the columns of the stoichiometric matrix $\mat{S}$ of size $N^{\mathrm{c}} \times N^{\mathrm{r}}$.
The rate functions $r$ are summarized in a vector-valued rate function $\vec{r}\colon \Reals^{N^{\mathrm{c}}} \to \Reals^{N^{\mathrm{r}}}$.
The change in concentration $\vec{c}$ due to the reaction system is then given by the ODE
\begin{equation}
	\frac{\mathrm{d}\vec{c}}{\mathrm{d}t} = \mat{S} \vec{r}\mleft( \vec{c} \mright). \label{supp:eq:model-rctn-ode}
\end{equation}

\begin{remark}
	We observe three points:\vspace*{-0.5\baselineskip}
	\begin{itemize}\setlength{\itemsep}{0\baselineskip}
		\item The constraint of integer stoichiometric coefficients is not always satisfied in reality.
			Our framework does not depend on this constraint and supports arbitrary real-valued coefficients.
		\item The direction of a reaction is determined by the sign of its rate function.
			The arrow in \cref{supp:eq:model-rctn-reaction-single} only determines which components are consumed or produced, that is, it determines the sign of the column in $\mat{S}$ that corresponds to this reaction.
		\item We cannot reduce the description of a reaction system to its stoichiometry $\mat{S}$.
			The reactions $2A \to A + B$ and $A \to B$ have the same stoichiometry $\mat{S} = (-1, 1)^{\T}$, but are very different.
			The difference is encoded in their rate functions, which can make use of the reactant and product complex vectors.
	\end{itemize}
\end{remark}

\begin{example}
    \label{supp:example:reaction_incorporation}
	Consider the system of reactions $A + B \to C$ and $2B + 1C \to 4A$ of three hypothetical components $A$, $B$, and $C$.
	The corresponding $3\times 2$ stoichiometric matrix $\mat{S}$ is given by
	\begin{equation*}
		\mat{S} = \begin{pmatrix*}[r]
			-1 & 4 \\ -1 & -2 \\ 1 & -1
		\end{pmatrix*}.
	\end{equation*}
	Using the rate functions $r_1, r_2\colon \Reals^3 \to \Reals$, the kinetics of this reaction system is governed by
	\begin{equation*}
		\frac{\mathrm{d}}{\mathrm{d}t} \begin{pmatrix}
			c_A \\ c_B \\ c_C
		\end{pmatrix} = \frac{\mathrm{d}\vec{c}}{\mathrm{d}t} = \mat{S} \vec{r}(\vec{c}) = \begin{pmatrix}
			-r_1(\vec{c}) + 4 r_2(\vec{c}) \\
			-r_1(\vec{c}) - 2 r_2(\vec{c}) \\
			\phantom{-}r_1(\vec{c}) - \phantom{1} r_2(\vec{c}) \\
		\end{pmatrix}.	
	\end{equation*}	
\end{example}

We now extend the model family by introducing abstract reaction terms of the form \cref{supp:eq:model-rctn-ode}.
Since the method of adding the terms is always the same, we only show the procedure on the two-dimensional Model~\ref{model:2D_continuum_bulk_transport_model}, which is one of the most complex models discussed in this work.
To this end, we first shorten the equations by collecting the transport terms in linear differential operators.
First, the bulk mass transfer PDE~\eqref{eq:model-2d-bulk-pde} is changed into
\begin{equation}
	\varepsilon^{\mathrm{b}} \frac{\partial \vec{c}^{\b}}{\partial t} = \vecop{B}_{\mathrm{2d}}\vec{c}^{\b} - \left(1 - \varepsilon^{\mathrm{b}} \right) \sum_{j=1}^{N^{\mathrm{p}}} \frac{3 d_j}{R^{\mathrm{p}}_{j}} \mat{K}^{\mathrm{f}}_{j} \left(\vec{c}^{\b} - \left. \vec{c}_j^{\p} \right|_{r = R^{\mathrm{p}}_{j}} \right), \label{supp:eq:model-rctn-2d-bulk} %
\end{equation}
where $\mat{K}^{\mathrm{f}}_{j} \coloneqq \diag{k^{\mathrm{f}}_{j,1}, \dots, k^{\mathrm{f}}_{j,N^{\mathrm{c}}}}$ is the diagonal matrix of film diffusion coefficients of particle type $j$, we have collected all concentrations $c^{\b}_i$ in the vector $\vec{c}^{\b}$ and represented convection and dispersion by the diagonal operator $\vecop{B}_{\mathrm{2d}}$, which is defined by its action
\begin{equation*}
	\vecop{B}_{\mathrm{2d}}\vec{c}^{\b} = \begin{pmatrix}
		-u \frac{\partial \left( \varepsilon^{\mathrm{b}} c^{\b}_1 \right)}{\partial z} + \frac{\partial}{\partial z} \left( \varepsilon^{\mathrm{b}} D^{\mathrm{ax}}_{1} \frac{\partial c^{\b}_1}{\partial z} \right) + \frac{1}{\rho} \frac{\partial}{\partial \rho} \left( \rho \varepsilon^{\mathrm{b}} D^{\mathrm{rad}}_{1}  \frac{\partial c^{\b}_1}{\partial \rho} \right) \\
		\vdots \\ 
		-u \frac{\partial \left( \varepsilon^{\mathrm{b}} c^{\b}_{N^{\mathrm{c}}} \right)}{\partial z} + \frac{\partial}{\partial z} \left( \varepsilon^{\mathrm{b}} D^{\mathrm{ax}}_{N^{\mathrm{c}}} \frac{\partial c^{\b}_{N^{\mathrm{c}}}}{\partial z} \right) + \frac{1}{\rho} \frac{\partial}{\partial \rho} \left( \rho \varepsilon^{\mathrm{b}} D^{\mathrm{rad}}_{N^{\mathrm{c}}}  \frac{\partial c^{\b}_{N^{\mathrm{c}}}}{\partial \rho} \right)
	\end{pmatrix}.
\end{equation*}
Second, the PDEs for the particles \eqref{eq:model_particle_1D_liquid}-\eqref{eq:model_particle_1D_solid} are shortened to
\begin{equation}
	\mat{E}_j \frac{\partial}{\partial t} \begin{pmatrix} \vec{c}^{\p}_j \\ \vec{c}^{\s}_j \end{pmatrix} = \mat{E}_j \vecop{P}_{\mathrm{2d},j} \begin{pmatrix} \vec{c}^{\p}_j \\ \vec{c}^{\s}_j \end{pmatrix} + \left( 1 - \varepsilon^{\p}_{j} \right) \mat{S}^{\mathrm{bind}}_{j} \vec{f}^{\mathrm{bind}}_{j}\mleft( \vec{c}^{\p}_j, \vec{c}^{\s}_j \mright), \label{supp:eq:model-rctn-2d-par} %
\end{equation}
where $\mat{E}_j$ is a diagonal matrix of phase volume fractions and $\vecop{P}_{\mathrm{2d},j}$ is a diagonal diffusion operator:
\begingroup\allowdisplaybreaks
\begin{align*}
	\mat{E}_j &= \diag{\varepsilon^{\p}_{j} \mat{1}_{N^{\mathrm{c}}\times 1}, (1-\varepsilon^{\p}_{j}) \mat{1}_{N^{\mathrm{totb}}_{j}\times 1} } \in \Reals^{\left( N^{\mathrm{c}} + N^{\mathrm{totb}}_{j} \right) \times \left( N^{\mathrm{c}} + N^{\mathrm{totb}}_{j} \right)}, \\
	\vecop{P}_{\mathrm{2d},j} \begin{pmatrix} \vec{c}^{\p}_j \\ \vec{c}^{\s}_j \end{pmatrix} &= \frac{1}{r^2} \frac{\partial }{\partial r} \begin{pmatrix}
		r^2 D_{j,1}^{\p} \frac{\partial c^{\p}_{j,1}}{\partial r} \\
		\vdots \\
		r^2 D_{j,N^{\mathrm{c}}}^{\p} \frac{\partial c^{\p}_{j,N^{\mathrm{c}}}}{\partial r} \\
		r^2 D_{j,1,1}^{\s} \frac{\partial c^{\s}_{j,1,1}}{\partial r} \\
		\vdots \\
		r^2 D_{j,N^{\mathrm{c}},N^{\mathrm{b}}_{j,N^{\mathrm{c}}}}^{\s} \frac{\partial c^{\s}_{j,N^{\mathrm{c}},N^{\mathrm{b}}_{j,N^{\mathrm{c}}}}}{\partial r}
	\end{pmatrix}.
\end{align*}\endgroup
The matrix $\mat{S}^{\mathrm{bind}}_{j}$ is defined by
\begingroup
\allowdisplaybreaks
\begin{align*}
	\mat{S}^{\mathrm{bind}}_{j} &= \begin{pmatrix}
		-\mat{S}^{\mathrm{bind}}_{\p,j} \\ \diag{\mat{1}_{N^{\mathrm{totb}}_{j}\times 1}}
	\end{pmatrix} \in \Reals^{\left(N^{\mathrm{c}} + N^{\mathrm{totb}}_{j}\right) \times N^{\mathrm{totb}}_{j}}, \\
	\mat{S}^{\mathrm{bind}}_{\p,j} &= \begin{pmatrix}
		\mat{1}_{1\times N^{\mathrm{b}}_{j,1}} & & & \\
		& \mat{1}_{1\times N^{\mathrm{b}}_{j,2}} & & \\
		& & \ddots & \\
		& & & \mat{1}_{1\times N^{\mathrm{b}}_{j,N^{\mathrm{c}}}}
	\end{pmatrix} \in \Reals^{N^{\mathrm{c}} \times N^{\mathrm{totb}}_{j}}. 
\end{align*}
\endgroup

\begin{remark}\label{rem:model-rctn-bindingisrxn}
	Removing the transport operator $\vecop{P}_{\mathrm{2d},j}$ in \cref{supp:eq:model-rctn-2d-par}, we obtain
	\begin{equation*}
		\mat{E}_j \frac{\partial}{\partial t} \begin{pmatrix} \vec{c}^{\p}_j \\ \vec{c}^{\s}_j \end{pmatrix} = \left( 1 - \varepsilon^{\p}_{j} \right) \mat{S}^{\mathrm{bind}}_{j} \vec{f}^{\mathrm{bind}}_{j}\mleft( \vec{c}^{\p}_j, \vec{c}^{\s}_j \mright).
	\end{equation*}
	Multiplying the phase-specific concentrations by $\mat{E}_j$ expresses them per unit particle volume. In these variables, the binding equations take the standard form of a reaction system, with $\mat{S}^{\mathrm{bind}}_{j}$ as the stoichiometric matrix and $\vec{f}^{\mathrm{bind}}_{j}$ as the reaction-rate vector.
	Binding can therefore be interpreted as a cross-phase reaction governed by the chosen binding rate law.
\end{remark}

In the following example, we illustrate how the model equations are recovered from the condensed equations developed above.

\begin{example}
    \label{supp:example:reaction_recovery}
	Consider a model with two components ($N^{\mathrm{c}} = 2$), one particle type ($N^{\mathrm{p}} = 1$), one bound state for component one ($N^{\mathrm{b}}_{1,1} = 1$), and two bound states for component two ($N^{\mathrm{b}}_{1,2} = 2$).
	Consequently, we have the vectors
	\begin{equation*}
		\left( \vec{c}^{\b} \right)^{\T} = \left( c^{\b}_1, c^{\b}_2 \right), 
		\qquad \left( \vec{c}^{\p} \right)^{\T} = \left( \vec{c}^{\p}_1 \right)^{\T} = \left( c^{\p}_{1,1}, c^{\p}_{1,2} \right), \qquad
		\left( \vec{c}^{\s} \right)^{\T} = \left( \vec{c}^{\s}_1 \right)^{\T} = \left( c^{\p}_{1,1,1}, c^{\p}_{1,2,1}, c^{\p}_{1,2,2} \right).
	\end{equation*}
	The bulk domain PDE \eqref{supp:eq:model-rctn-2d-bulk} expands to
	\begin{multline*}
			\varepsilon^{\mathrm{b}} \frac{\partial \vec{c}^{\b}}{\partial t} %
			= -u \frac{\partial \left( \varepsilon^{\mathrm{b}} \vec{c^{\b}} \right)}{\partial z} + \frac{\partial}{\partial z} \left( \varepsilon^{\mathrm{b}} \mat{D}_{\mathrm{ax}} \frac{\partial \vec{c^{\b}}}{\partial z} \right) + \frac{1}{\rho} \frac{\partial}{\partial \rho} \left( \rho \varepsilon^{\mathrm{b}} \mat{D}_{\mathrm{rad}} \frac{\partial \vec{c^{\b}}}{\partial \rho} \right) \\
			- \left(1 - \varepsilon^{\mathrm{b}} \right) \sum_{j=1}^{N^{\mathrm{p}}} \frac{3 d_j}{R^{\mathrm{p}}_{j}} \mat{K}^{\mathrm{f}}_{j} \left(\vec{c}^{\b} - \left. \vec{c}_j^{\p} \right|_{r = R^{\mathrm{p}}_{j}} \right),
	\end{multline*}
	where $\mat{D}_{\mathrm{ax}}$ and $\mat{D}_{\mathrm{rad}}$ are diagonal matrices consisting of dispersion coefficients.
	Clearly, this is equivalent to Eq.~\eqref{eq:model-2d-bulk-pde} for each component $i$ of Model~\ref{model:2D_continuum_bulk_transport_model}.

	In order to expand the particle PDE \eqref{supp:eq:model-rctn-2d-par}, we first state the matrix $\mat{E}_j$ and the matrix $\mat{S}^{\mathrm{bind}}_{1}$ :
	\begin{equation*}
		\mat{E}_j = \begin{pmatrix}
			\varepsilon^{\mathrm{p}}_{1} & & & & \\
			& \varepsilon^{\mathrm{p}}_{1} & & & \\
			& & 1 - \varepsilon^{\mathrm{p}}_{1} & & \\
			& & & 1 - \varepsilon^{\mathrm{p}}_{1} & \\
			& & & & 1 - \varepsilon^{\mathrm{p}}_{1}
		\end{pmatrix}, \qquad 
		\mat{S}^{\mathrm{bind}}_{1} = \begin{pmatrix*}[r]
			-1 & & \\
			& -1 & -1 \\
			1 & & \\
			& 1 & \\
			& & 1
		\end{pmatrix*}.
	\end{equation*}
	The vector-valued binding rate function $\vec{f}^{\mathrm{bind}}_{1}$ has $N_{\mathrm{totb},1} = N^{\mathrm{b}}_{1,1} + N^{\mathrm{b}}_{1,2} = 3$ entries:
	\begin{equation*}
		\vec{f}^{\mathrm{bind}}_{1} = \left(
			f^{\mathrm{bind}}_{1,1,1} , f^{\mathrm{bind}}_{1,2,1} , f^{\mathrm{bind}}_{1,2,2} \right)^{\T}.
	\end{equation*}
	We decompose \cref{supp:eq:model-rctn-2d-par} into the first two rows, which contain the PDEs for the particle liquid phase, and the last three rows, which contain the particle solid phase.
	The liquid-phase portion expands to
	\begin{equation*}
		\varepsilon^{\mathrm{p}}_{1} \frac{\partial \vec{c}^{\p}_1}{\partial t} = \varepsilon^{\mathrm{p}}_{1} \frac{1}{r^2} \frac{\partial }{\partial r} \left( r^2 \mat{D}_{1}^{\p} \frac{\partial \vec{c}^{\p}_{1}}{\partial r} \right) - \left( 1 - \varepsilon^{\p}_{j} \right) \begin{pmatrix}
			f^{\mathrm{bind}}_{1,1,1}\mleft( \vec{c}^{\p}_1, \vec{c}^{\s}_1 \mright) \\
			f^{\mathrm{bind}}_{1,2,1}\mleft( \vec{c}^{\p}_1, \vec{c}^{\s}_1 \mright) + f^{\mathrm{bind}}_{1,2,2}\mleft( \vec{c}^{\p}_1, \vec{c}^{\s}_1 \mright)
		\end{pmatrix},
	\end{equation*}
	where $\mat{D}_{1}^{\p}$ is a diagonal matrix with pore diffusion coefficients.
	Rewriting this as
	\begin{equation*}
		\varepsilon^{\mathrm{p}}_{1} \frac{\partial c^{\p}_{1,i}}{\partial t} = \varepsilon^{\mathrm{p}}_{1} \frac{1}{r^2} \frac{\partial }{\partial r} \left( r^2 D_{1,i}^{\p} \frac{\partial c^{\p}_{1,i}}{\partial r} \right) - \left( 1 - \varepsilon^{\p}_{j} \right) \sum_{k=1}^{N^{\mathrm{b}}_{1,i}} f^{\mathrm{bind}}_{1,i,k}\mleft( \vec{c}^{\p}_1, \vec{c}^{\s}_1 \mright) \qquad \forall\  i \in \{1,2\},
	\end{equation*}
	we see that this part is equivalent to \eqref{eq:model_particle_1D_liquid}.
	Using the diagonal surface diffusion coefficient matrix $\mat{D}_{1}^{\s}$, the solid-phase portion reads
	\begin{equation*}
		\left(1-\varepsilon^{\mathrm{p}}_{1}\right) \frac{\partial \vec{c}^{\s}_1}{\partial t} = \left(1-\varepsilon^{\mathrm{p}}_{1}\right) \frac{1}{r^2} \frac{\partial }{\partial r} \left( r^2 \mat{D}_{1}^{\s} \frac{\partial \vec{c}^{\s}_{1}}{\partial r} \right) + \left( 1 - \varepsilon^{\p}_{j} \right) \vec{f}^{\mathrm{bind}}_{1}\mleft( \vec{c}^{\p}_1, \vec{c}^{\s}_1 \mright),
	\end{equation*}
	which is equivalent to
	\begin{equation*}
		\left(1-\varepsilon^{\mathrm{p}}_{1}\right) \frac{\partial c^{\s}_{1,i,k}}{\partial t} = \left(1-\varepsilon^{\mathrm{p}}_{1}\right) \frac{1}{r^2} \frac{\partial }{\partial r} \left( r^2 D_{1,i,k}^{\s} \frac{\partial c^{\s}_{1,i,k}}{\partial r} \right) + \left( 1 - \varepsilon^{\p}_{j} \right) f^{\mathrm{bind}}_{1,i,k}\mleft( \vec{c}^{\p}_1, \vec{c}^{\s}_1 \mright)
	\end{equation*}
	for all $i \in \{1, 2\}$ and $k \in \{1, \dots N^{\mathrm{b}}_{1,i} \}$.
	This recovers Eq.~\eqref{eq:model_particle_1D_solid}.
\end{example}

Having prepared the model equations, we simply add reaction terms of the form $\mat{S}\vec{r}(\vec{c})$ to the mass balances.
For the bulk domain mass balance, we obtain 
\begin{equation}
	\varepsilon^{\mathrm{b}} \frac{\partial \vec{c}^{\b}}{\partial t} = \vecop{B}_{\mathrm{2d}}\vec{c}^{\b} - \left(1 - \varepsilon^{\mathrm{b}} \right) \sum_{j=1}^{N^{\mathrm{p}}} \frac{3 d_j}{R^{\mathrm{p}}_{j}} \mat{K}^{\mathrm{f}}_{j} \left(\vec{c}^{\b} - \left. \vec{c}_j^{\p} \right|_{r = R^{\mathrm{p}}_{j}} \right) + \varepsilon^{\mathrm{b}} \mat{S}^{\b} \vec{r}^{\b}\mleft( \vec{c}^{\b} \mright) \label{supp:eq:model-rctn-2d-bulk-short}
\end{equation}
with $N^{\mathrm{c}} \times N^{\mathrm{r},\b}$ matrix $\mat{S}^{\b}$ and rate function $\vec{r}^{\b}\colon \Reals^{N^{\mathrm{c}}} \times \Reals^{N^{\mathrm{r},\b}}$.
For the particles, we get
\begin{equation}
	\mat{E}_j \frac{\partial}{\partial t} \begin{pmatrix} \vec{c}^{\p}_j \\ \vec{c}^{\s}_j \end{pmatrix} = \mat{E}_j \vecop{P}_{\mathrm{2d},j} \begin{pmatrix} \vec{c}^{\p}_j \\ \vec{c}^{\s}_j \end{pmatrix} + \mat{S}^{\ps}_j \vec{r}^{\ps}_j\mleft( \vec{c}^{\p}_j, \vec{c}^{\s}_j \mright), \label{supp:eq:model-rctn-2d-particle-short}
\end{equation}
where the $(N^{\mathrm{c}} + N^{\mathrm{totb}}_{j}) \times N^{\mathrm{r}, \ps}_{j}$ stoichiometric matrix $\mat{S}^{\ps}_j$ and the rate function $\vec{r}^{\ps}_j\colon \Reals^{N^{\mathrm{c}}} \times \Reals^{N^{\mathrm{totb}}_{j}} \to \Reals^{N^{\mathrm{r}, \ps}_{j}}$ are given by
\begin{align*}
	\mat{S}^{\ps}_j &= \left( \begin{array}{c|c|cc}
		& & \varepsilon^{\p}_{j} \mat{S}_j^{\p} & \mat{0} \\ 
		\raisebox{.5\normalbaselineskip}[0pt][0pt]{$\left( 1 - \varepsilon^{\p}_{j} \right) \mat{S}^{\mathrm{bind}}_{j}$} & \raisebox{.5\normalbaselineskip}[0pt][0pt]{$\left( 1 - \varepsilon^{\p}_{j} \right) \mat{S}^{\crossPhase}_j$} & \mat{0} & \left( 1 - \varepsilon^{\p}_{j} \right) \mat{S}_j^{\s} \end{array} \right), \\
	\vec{r}^{\ps}_j\mleft( \vec{c}^{\p}_j, \vec{c}^{\s}_j \mright) &= \begin{pmatrix} \vec{f}^{\mathrm{bind}}_{j}\mleft( \vec{c}^{\p}_j, \vec{c}^{\s}_j \mright) \\ \vec{r}^{\crossPhase}_{j}\mleft( \vec{c}^{\p}_j, \vec{c}^{\s}_j \mright) \\ \vec{r}_j^{\p}\mleft( \vec{c}^{\p}\mright) \\ \vec{r}_j^{\s}\mleft(\vec{c}^{\s} \mright) \end{pmatrix}.
\end{align*}
Here, we have combined binding reactions described by $\mat{S}^{\mathrm{bind}}_{j}\vec{f}^{\mathrm{bind}}_{j}( \vec{c}^{\p}_j, \vec{c}^{\s}_j )$, other cross-phase reactions $\mat{S}^{\crossPhase}_j \vec{r}^{\crossPhase}_{j}( \vec{c}^{\p}_j, \vec{c}^{\s}_j )$, liquid-phase reactions $\mat{S}^{\p}_j \vec{r}^{\p}_{j}( \vec{c}^{\p}_j )$, and solid-phase reactions $\mat{S}^{\s}_j \vec{r}^{\s}_{j}( \vec{c}^{\s}_j )$.
This results in a total of
\begin{equation*}
	N^{\mathrm{r}, \ps}_{j} = N^{\mathrm{totb}}_{j} + N_{\mathrm{r},j}^{\crossPhase} + N_{\mathrm{r},j}^{\p} + N_{\mathrm{r},j}^{\s}
\end{equation*}
reactions.
Whereas liquid- and solid-phase reaction rates are multiplied with their respective porosity factor, the cross-phase reaction rates, which include binding, are computed \emph{per unit of solid-phase domain}.
That is, they are multiplied by the solid-phase porosity factor $1 - \varepsilon^{\p}_{j}$.
As pointed out in Remark~\ref{rem:model-rctn-bindingisrxn}, binding is treated as a cross-phase reaction.

\begin{remark}
	Since boundary and initial conditions are not changed by the introduction of the reaction terms, the other models are modified analogously.
	This also holds for model variants such as rapid-equilibrium film diffusion.
\end{remark}

\subsubsection{Reactions in Rapid Equilibrium}
\label{supp:sec:models-reaction-rapidequi}

Similar to binding, some reactions may happen on a much faster time scale than the other processes.
This is the case for buffer reactions, for example, but can also apply to binding, which in this framework can be treated as just another reaction.
Hence, we want to apply the rapid-equilibrium assumption to selected reactions.
In Section~\ref{sec:binding_equilibrium_kinetics} on rapid-equilibrium model variants, we have already seen the procedure to exploit the rapid-equilibrium assumption.
We form the conservation relation that eliminates the fast reaction terms from the equations, replace the corresponding mass balance PDEs by the PDE for the conserved moiety, and fix the remaining unknowns by imposing the vanishing rate condition.

Before we apply the same procedure to this setting, we examine conserved moieties more closely and their relation to the stoichiometry.

\begin{example}
	Consider the reaction $A + B \to C$ of three components $A$, $B$, and $C$.
	The corresponding $3\times 1$ stoichiometric matrix $\mat{S}$ is given by
	\begin{equation*}
		\mat{S} = (-1, -1, 1)^{\T}.
	\end{equation*}
	We observe that $c^{\mathrm{A}} + c^{\mathrm{C}}$ and $c^{\mathrm{B}} + c^{\mathrm{C}}$ are constant.
	That is,
	\begin{align}
	\begin{split}
		\frac{\mathrm{d}\left( c^{\mathrm{A}} + c^{\mathrm{C}} \right)}{\mathrm{d}t} &= \frac{\mathrm{d}c^{\mathrm{A}}}{\mathrm{d}t} + \frac{\mathrm{d}c^{\mathrm{C}}}{\mathrm{d}t} = -r_1\mleft( \vec{c} \mright) + r_1\mleft( \vec{c} \mright) = 0, \\
		\frac{\mathrm{d}\left( c^{\mathrm{B}} + c^{\mathrm{C}} \right)}{\mathrm{d}t} &= \frac{\mathrm{d}c^{\mathrm{B}}}{\mathrm{d}t} + \frac{\mathrm{d}c^{\mathrm{C}}}{\mathrm{d}t} = -r_1\mleft( \vec{c} \mright) + r_1\mleft( \vec{c} \mright) = 0
	\end{split}\label{supp:eq:model-rctn-consmoi-split}
	\end{align}
	regardless of the reaction rate $r_1\colon \Reals^3 \to \Reals$.
	We capture these linear combinations of concentrations in a $2 \times 3$ matrix
	\begin{equation*}
		\mat{M} = \begin{pmatrix}
			1 & 0 & 1 \\ 0 & 1 & 1
		\end{pmatrix}.
	\end{equation*}
	Using this matrix and the fact that $\mat{M} \mat{S} = \mat{0}$, we can write \cref{supp:eq:model-rctn-consmoi-split} in a short form:
	\begin{equation*}
		\frac{\mathrm{d}\left(\mat{M}\vec{c} \right)}{\mathrm{d}t} = \mat{M} \frac{\mathrm{d}\vec{c}}{\mathrm{d}t} = \mat{M} \mat{S} \vec{r}\mleft( \vec{c} \mright) = \vec{0}.
	\end{equation*}
\end{example}

The example illustrated how conserved moieties can be identified as combinations of concentrations that remain invariant under the reactions.
We have already used conserved moieties when deriving the model equations for the binding rapid-equilibrium assumption.
Hence, we formalize this observation in the following definition.

\begin{definition}[Conserved Moieties]\label{supp:def:model-rctn-consmoiety} %
	For a system of reactions described by the stoichiometry $\mat{S}$, a conserved moiety is a linear combination of concentrations $\vec{m}^{\T} \vec{c}$, $\vec{m} \in \Reals^{N^{\mathrm{c}}}$, that is (structurally) invariant under reactions.
	This means, $\vec{m}$ is in the left null space of $\mat{S}$ (or, equivalently, in the kernel of $\mat{S}^{\T}$), that is, $\vec{m}^{\T} \mat{S} = \vec{0}$.

	The coefficient vectors of the conserved moieties are collected as rows of a conservation matrix $\mat{M}$.
	Consequently, $\mat{M}$ satisfies $\mat{M} \mat{S} = \mat{0}$.
\end{definition}

\Needspace{5\baselineskip}
\begin{remark}
	If no reactions are present in the system, each component is a conserved moiety.
	That is, $\mat{M} = \mat{I}$.
\end{remark}

Note that we have not required $\vec{m}$ to be nonnegative, which would be reasonable for the conserved moiety in order to be physical.
Another advantage of nonnegative coefficients $\vec{m}$ is that positive definite discretizations of the diffusion problem remain positive definite.
This, however, may come at the price of increased condition numbers of $\mat{M}$.

Definition~\ref{supp:def:model-rctn-consmoiety} also hints at a method for automatically determining the conservation matrix $\mat{M}$ given a stoichiometry $\mat{S}$.
We simply compute a basis of $\operatorname{ker}\left(\mat{S}^{\T}\right)$ and collect the basis vectors in the rows of $\mat{M}$.
The former can be achieved by means of QR or singular value decomposition, for example.

We can use Definition~\ref{supp:def:model-rctn-consmoiety} if we want to consider all reactions in rapid equilibrium.
For a subset of reactions, we need a refined definition of a conserved moiety, which is illustrated in the following example. \enlargethispage{1\baselineskip}

\begin{example}\label{ex:model-rctn-unwantedmoieties}
	Consider the reaction $A \to C$ in a three component system $A$, $B$, and $C$.
	The stoichiometric matrix is given by
	\begin{equation*}
		\mat{S} = (-1, 0, 1)^{\T}.
	\end{equation*}
	According to Definition~\ref{supp:def:model-rctn-consmoiety}, the conservation matrix is
	\begin{equation*}
		\mat{M} = \begin{pmatrix}
			1 & 0 & 1 \\ 0 & 1 & 0
		\end{pmatrix}.
	\end{equation*}
	We, therefore, have the conserved moieties $A+C$ and $B$.
\end{example}

We eventually want to eliminate the reaction terms of the reactions in rapid equilibrium.
Example~\ref{ex:model-rctn-unwantedmoieties} shows that, using Definition~\ref{supp:def:model-rctn-consmoiety}, we might get additional conserved moieties that do not eliminate the terms.
However, we observe that the desired conservation relation $(1,0,1)$ has the same support as the stoichiometric coefficients $(-1, 0, 1)^{\T}$, whereas the additional relation $(0,1,0)$ has a different support.
We exploit this in the next definition.

\begin{definition}[Eliminating Conserved Moieties]\label{supp:def:model-rctn-elimconsmoiety}
	For a single reaction described by the stoichiometric vector $\vec{s}$, the eliminating conserved moiety is a linear combination of concentrations $\vec{m}^{\T} \vec{c}$, $\vec{m} \in \Reals^{N^{\mathrm{c}}}$, that satisfies
	\begin{equation*}
		\vec{m}^{\T} \vec{s} = 0, \qquad \supp{\vec{m}} = \supp{\vec{s}}.
	\end{equation*}
	The coefficient vectors of the eliminating conserved moieties of a system of reactions described by the $N^{\mathrm{c}} \times N^{\mathrm{r}}$ stoichiometry $\mat{S}$ form the rows of the $N_{\mathrm{m}} \times N^{\mathrm{c}}$ matrix $\mat{M}$ that satisfies
	\begin{equation}
		\mat{M} \mat{S} = \mat{0}, \qquad \supp{\vec{e}_i^{\T} \mat{M}} \subset \bigcup_{j=1}^{N^{\mathrm{r}}} \supp{\mat{S}\vec{e}_j} \qquad \forall\  i = 1, \dots, N_{\mathrm{m}}. \label{supp:eq:model-rctn-elimconsmoiety-sys}
	\end{equation}
\end{definition}
The last part of condition \cref{supp:eq:model-rctn-elimconsmoiety-sys} of Definition~\ref{supp:def:model-rctn-elimconsmoiety} requires that the support of each row of $\mat{M}$ is contained in the combined support of the columns of $\mat{S}$.

The eliminating conserved moieties can also be computed as a basis of the left null space.
We simply remove all rows that only contain zeros from $\mat{S}$, compute the conservation relations for the modified stoichiometry, and insert zero columns for the removed rows in $\mat{M}$.

\pagebreak[4]
Before we apply the procedure outlined above, we first left-multiply \cref{supp:eq:model-rctn-2d-bulk-short,supp:eq:model-rctn-2d-particle-short} by conservation matrices $\mat{\tilde{M}}^{\b}$ and $\mat{\tilde{M}}^{\ps}_j$, which we will later use to add the equations together in order to eliminate reaction terms:
\begin{subequations}
\begin{align}
	\varepsilon^{\mathrm{b}} \mat{\tilde{M}}^{\b} \frac{\partial \vec{c}^{\b}}{\partial t} &= \mat{\tilde{M}}^{\b} \vecop{B}_{\mathrm{2d}}\vec{c}^{\b} - \left(1 - \varepsilon^{\mathrm{b}} \right) \sum_{j=1}^{N^{\mathrm{p}}} \frac{3 d_j}{R^{\mathrm{p}}_{j}} \mat{\tilde{M}}^{\b} \mat{K}^{\mathrm{f}}_{j} \left(\vec{c}^{\b} - \left. \vec{c}_j^{\p} \right|_{r = R^{\mathrm{p}}_{j}} \right)
     + \varepsilon^{\mathrm{b}} \mat{\tilde{M}}^{\b} \mat{S}^{\b} \vec{r}^{\b}\mleft( \vec{c}^{\b} \mright), \\
	\mat{\tilde{M}}^{\ps}_j \mat{E}_j \frac{\partial}{\partial t} \begin{pmatrix} \vec{c}^{\p}_j \\ \vec{c}^{\s}_j \end{pmatrix} &= \mat{\tilde{M}}^{\ps}_j \mat{E}_j \vecop{P}_{\mathrm{2d},j} \begin{pmatrix} \vec{c}^{\p}_j \\ \vec{c}^{\s}_j \end{pmatrix} + \mat{\tilde{M}}^{\ps}_j \mat{S}^{\ps}_j \vec{r}^{\ps}_j\mleft( \vec{c}^{\p}_j, \vec{c}^{\s}_j \mright).
\end{align}\label{supp:eq:model-rctn-2d-pre-rapidequi}%
\end{subequations}%

We still need to determine the conservation matrices $\mat{\tilde{M}}^{\b}$ and $\mat{\tilde{M}}^{\ps}_j$.
To this end, let $\mset{R}_{\mathrm{re}}^{\b} \subset \{1, \dots, N^{\mathrm{r}, \b}\}$ and $\mset{R}_{\mathrm{re},j}^{\ps} \subset \{1, \dots, N^{\mathrm{r}, \ps}_{j}\}$ be index sets of reactions in rapid equilibrium.
The reaction index sets are accompanied by component index sets $\mset{C}_{\mathrm{re}}^{\b} \subset \{1, \dots, N^{\mathrm{c}}\}$ and $\mset{C}_{\mathrm{re},j}^{\ps} \subset \{1, \dots, N^{\mathrm{c}} + N^{\mathrm{totb}}_{j}\}$ defined by
\begin{equation}
	\mset{C}_{\mathrm{re}}^{\b} \coloneqq \bigcup_{i \in \mset{R}_{\mathrm{re}}^{\b}} \supp{\mat{S}^{\b} \vec{e}_i}, \qquad \mset{C}_{\mathrm{re},j}^{\ps} \coloneqq \bigcup_{i \in \mset{R}_{\mathrm{re},j}^{\ps}} \supp{\mat{S}^{\ps}_j \vec{e}_i}. \label{supp:eq:model-rctn-compinds}
\end{equation}
These contain the indices of the components participating in reactions that are in rapid equilibrium.
Consequently, their complements $\mset{C}_{\mathrm{dyn}}^{\b} \coloneqq \{1, \dots, N^{\mathrm{c}}\} \setminus \mset{C}_{\mathrm{re}}^{\b}$ and $\mset{C}_{\mathrm{dyn},j}^{\ps} \coloneqq \{1, \dots, N^{\mathrm{c}} + N^{\mathrm{totb}}_{j}\} \setminus \mset{C}_{\mathrm{re},j}^{\ps}$ contain the indices of the remaining components.
Denote by $\mat{M}^{\b}$ the eliminating conserved moieties corresponding to the stoichiometric matrix $\submat{\mat{S}^{\b}}{\all}{\mset{R}_{\mathrm{re}}^{\b}}$ consisting of the columns $\mset{R}_{\mathrm{re}}^{\b}$ of $\mat{S}^{\b}$.
Similarly, $\mat{M}^{\ps}_j$ denotes the eliminating conserved moieties corresponding to the stoichiometric matrix $\submat{\mat{S}_j^{\ps}}{\all}{\mset{R}_{\mathrm{re},j}^{\ps}}$ consisting of the columns $\mset{R}_{\mathrm{re},j}^{\ps}$ of $\mat{S}^{\ps}_j$.
Using these preparations, we define
\begin{equation}
	\mat{\tilde{M}}^{\b} \coloneqq \begin{pmatrix}
		\submat{\mat{I}_{N^{\mathrm{c}}}}{\mset{C}_{\mathrm{dyn}}^{\b}}{\all} \\
		\mat{M}^{\b}
	\end{pmatrix}, \qquad
	\mat{\tilde{M}}^{\ps}_j \coloneqq \begin{pmatrix}
		\submat{\mat{I}_{N^{\mathrm{c}} + N^{\mathrm{totb}}_{j}}}{\mset{C}_{\mathrm{dyn},j}^{\ps}}{\all} \\
		\mat{M}^{\ps}_j
	\end{pmatrix}. \label{supp:eq:model-rctn-consrelmat}
\end{equation}
The first rows of the conservation matrices contain the standard basis vectors corresponding to the components with indices $\mset{C}_{\mathrm{dyn}}^{\b}$ and $\mset{C}_{\mathrm{dyn},j}^{\ps}$, respectively.
These components do not participate in reactions that are in rapid equilibrium.
The lower part consists of the respective eliminating conservation relations.

\begin{remark}
	If all reactions are kinetic, we have $\mset{C}_{\mathrm{re}}^{\b} = \mset{C}_{\mathrm{re},j}^{\ps} = \emptyset$ and, therefore, empty matrices $\mat{M}^{\b}$ and $\mat{M}^{\ps}_j$.
	The conservation matrices are the full identity matrices of the respective size.
\end{remark}

By means of the conservation matrices $\mat{\tilde{M}}^{\b}$ and $\mat{\tilde{M}}^{\ps}_j$, we have eliminated the rapid-equilibrium reaction terms from the differential equations.
Since there are fewer eliminating conservation relations (i.e., rows of $\mat{M}^{\b}$ and $\mat{M}^{\ps}_j$) than $\card*{\mset{C}_{\mathrm{re}}^{\b}}$ and $\card*{\mset{C}_{\mathrm{re},j}^{\ps}}$, respectively, we need to impose vanishing rate conditions for the reactions in rapid equilibrium, which fixes the remaining unknowns.
For the bulk domain, we obtain
\begin{subequations}
\begin{align}
    \varepsilon^{\mathrm{b}} \mat{\tilde{M}}^{\b} \frac{\partial \vec{c}^{\b}}{\partial t} &= \mat{\tilde{M}}^{\b} \vecop{B}_{\mathrm{2d}}\vec{c}^{\b} - \left(1 - \varepsilon^{\mathrm{b}} \right) \sum_{j=1}^{N^{\mathrm{p}}} \frac{3 d_j}{R^{\mathrm{p}}_{j}} \mat{\tilde{M}}^{\b} \mat{K}^{\mathrm{f}}_{j} \left(\vec{c}^{\b} - \left. \vec{c}_j^{\p} \right|_{r = R^{\mathrm{p}}_{j}} \right) + \varepsilon^{\mathrm{b}} \mat{\tilde{M}}^{\b} \mat{S}^{\b} \vec{r}^{\b}\mleft( \vec{c}^{\b} \mright), \label{supp:eq:model-rctn-2d-bulk-rapidequi-mt} \\
	0 &= \vec{e}_i^{\T} \vec{r}^{\b}\mleft( \vec{c}^{\b} \mright) \qquad \forall\  i \in \mset{R}_{\mathrm{re}}^{\b} \label{supp:eq:model-rctn-2d-bulk-rapidequi-rate}
\end{align}\label{supp:eq:model-rctn-2d-bulk-rapidequi}%
\end{subequations}%
and for the particles of type $j$, we get
\begin{subequations}
\begin{align}
	\mat{\tilde{M}}^{\ps}_j \mat{E}_j \frac{\partial}{\partial t} \begin{pmatrix} \vec{c}^{\p}_j \\ \vec{c}^{\s}_j \end{pmatrix} &= \mat{\tilde{M}}^{\ps}_j \mat{E}_j \vecop{P}_{\mathrm{2d},j} \begin{pmatrix} \vec{c}^{\p}_j \\ \vec{c}^{\s}_j \end{pmatrix} + \mat{\tilde{M}}^{\ps}_j \mat{S}^{\ps}_j \vec{r}^{\ps}_j\mleft( \vec{c}^{\p}_j, \vec{c}^{\s}_j \mright), \\
	0 &= \vec{e}_i^{\T} \vec{r}^{\ps}_j\mleft( \vec{c}^{\p}_j, \vec{c}^{\s}_j \mright) \qquad \forall\  i \in \mset{R}_{\mathrm{re},j}^{\ps}. \label{supp:eq:model-rctn-2d-par-rapidequi-rate}
\end{align}\label{supp:eq:model-rctn-2d-par-rapidequi}%
\end{subequations}%
Note that the vanishing rate conditions \cref{supp:eq:model-rctn-2d-bulk-rapidequi-rate,supp:eq:model-rctn-2d-par-rapidequi-rate} are removed if their corresponding index sets $\mset{R}_{\mathrm{re}}^{\b}$ and $\mset{R}_{\mathrm{re},j}^{\ps}$ are empty.

Since we have transitioned from concentrations to (eliminating) conserved moieties as differential unknowns, at least for some concentrations, we need to adapt boundary and initial conditions to this new setting.
This is achieved by multiplying the respective equations by $\mat{\tilde{M}}$.
For the bulk boundary conditions~\eqref{eq:model-2d-bulk-bc}, we get 
\begin{subequations}
\begin{alignat}{2}
	u \mat{\tilde{M}}^{\b} \vec{c}^{\mathrm{in}} &= \left.\mat{\tilde{M}}^{\b} \left( u \vec{c}^{\b} - \mat{D}_{\mathrm{ax}} \frac{\partial \vec{c}^{\b}}{\partial z} \right)\right|_{z=0} & &\qquad\text{on } (0, T^{\mathrm{end}}) \times (0, \Rho^{\mathrm{c}}), \\
	\vec{0} &= \left. - \mat{\tilde{M}}^{\b} \mat{D}_{\mathrm{ax}} \frac{\partial \vec{c}^{\b}}{\partial z} \right|_{z=L} & &\qquad\text{on } (0, T^{\mathrm{end}}) \times (0, \Rho^{\mathrm{c}}), \\
	\vec{0} &= \left.  - \mat{\tilde{M}}^{\b} \mat{D}_{\mathrm{rad}} \frac{\partial \vec{c}^{\b}}{\partial \rho} \right|_{\rho=\Rho^{\mathrm{c}}} & &\qquad\text{on } (0, T^{\mathrm{end}}) \times (0, L^\mathrm{b}), \\
	\vec{0} &= \left. - \mat{\tilde{M}}^{\b} \mat{D}_{\mathrm{rad}} \frac{\partial \vec{c}^{\b}}{\partial \rho} \right|_{\rho=0} & &\qquad\text{on } (0, T^{\mathrm{end}}) \times (0, L^\mathrm{b}),
\end{alignat}\label{supp:eq:model-rctn-2d-rapidequi-bulk-bc}%
\end{subequations}%
where $\mat{D}_{\mathrm{ax}}$ and $\mat{D}_{\mathrm{rad}}$ are diagonal matrices consisting of dispersion coefficients.
The particle boundary conditions~\eqref{eq:model_particle_1D_BC} on $(0,T^{\mathrm{end}}] \times (0, \Rho^{\mathrm{c}}) \times (0, L^\mathrm{b})$ are modified to
\begin{subequations}
\begin{align}
	\left. \left( 1 - \varepsilon^{\mathrm{b}} \right) \mat{\tilde{M}}^{\ps}_j \mat{E}_j \mat{D}^{\ps}_j \frac{\partial}{\partial r} \begin{pmatrix}
		\vec{c}^{\p}_{j} \\ \vec{c}^{\s}_{j}
	\end{pmatrix} \right|_{r = R^{\mathrm{p}}_{j}} &= \left( 1 - \varepsilon^{\mathrm{b}} \right) \mat{\tilde{M}}^{\ps}_j \begin{pmatrix}
		\mat{K}_{\mathrm{f},j} \left( \vec{c}^{\b} - \left. \vec{c}^{\p}_{j} \right|_{r = R^{\mathrm{p}}_{j}} \right) \\ \vec{0}
	\end{pmatrix}, \label{supp:eq:model-rctn-2d-rapidequi-particle-bc-outer} \\
	\left. -\mat{\tilde{M}}^{\ps}_j \mat{E}_j \mat{D}^{\ps}_j \frac{\partial}{\partial r} \begin{pmatrix}
		\vec{c}^{\p}_{j} \\ \vec{c}^{\s}_{j}
	\end{pmatrix} \right|_{r = R^\mathrm{c}_j} &= \vec{0},
\end{align}\label{supp:eq:model-rctn-2d-rapidequi-particle-bc}%
\end{subequations}%
where $\mat{D}^{\ps}_j$ and $\mat{K}_{\mathrm{f},j}$ are diagonal matrices collecting the diffusion and film diffusion coefficients, respectively.
The initial conditions~\eqref{eq:model-2d-bulk-ic} change to
\begin{subequations}
\begin{alignat}{2}
    \mat{\tilde{M}}^{\b} \left. \vec{c}^{\b} \right|_{t = 0} &= \mat{\tilde{M}}^{\b} \vec{c}^{\b,\mathrm{init}} & & \qquad\text{in } (0, \Rho^{\mathrm{c}}) \times (0, L^\mathrm{b}), \\
    \mat{\tilde{M}}^{\ps}_j \left.
    \begin{pmatrix}
      \vec{c}^{\p}_{j} \\ \vec{c}^{\s}_{j}
    \end{pmatrix} \right|_{t = 0} &= \mat{\tilde{M}}^{\ps}_j
    \begin{pmatrix}
        \vec{c}^{\p,\mathrm{init}}_{j} \\ \vec{c}^{\s,\mathrm{init}}_{j}
    \end{pmatrix}  & & \qquad\text{in } (0, \Rho^{\mathrm{c}}) \times (0, L^\mathrm{b}) \times (R^{\mathrm{c}}_{j}, R^{\mathrm{p}}_{j})
\end{alignat}\label{supp:eq:model-rctn-2d-rapidequi-bulk-ic}%
\end{subequations}%
and we gain consistency conditions on the initial data
\begin{subequations}
\begin{alignat}{2}
	\vec{e}_i^{\T} \vec{r}^{\b}\mleft( \vec{c}^{\b,\mathrm{init}} \mright) &= 0 & &\qquad \forall\  i \in \mset{R}_{\mathrm{re}}^{\b}, \\
	\vec{e}_i^{\T} \vec{r}^{\ps}_j\mleft( \vec{c}^{\p,\mathrm{init}}_{j}, \vec{c}^{\s,\mathrm{init}}_{j} \mright) &= 0 & &\qquad \forall\  i \in \mset{R}_{\mathrm{re},j}^{\ps}.
\end{alignat}\label{supp:eq:model-rctn-2d-rapidequi-consinit}%
\end{subequations}%

\vspace{0.2cm}\noindent\textit{Rapid-Equilibrium Film Diffusion}\vspace{0.1cm}

Having the general framework in place, we proceed by investigating the rapid-equilibrium film diffusion variant discussed in Section~\ref{sec:rapid-equilibrium_film_diffusion}.
Mass transfer between bulk and particle domains (i.e., film diffusion) can, in general, also be described as a cross-phase reaction.
However, this formulation would be slightly more complicated since the coupling is between bulk domain PDE and particle domain boundary condition.
We therefore retain the existing formulation and treat film diffusion explicitly.

We apply the same procedure as in Section~\ref{sec:rapid-equilibrium_film_diffusion}.
First, we eliminate the film diffusion term, which transforms \cref{supp:eq:model-rctn-2d-bulk-rapidequi-mt} into
\begin{equation*}
	\begin{split}
		\varepsilon^{\mathrm{b}} \mat{\tilde{M}}^{\b} \frac{\partial \vec{c}^{\b}}{\partial t} &= \mat{\tilde{M}}^{\b} \vecop{B}_{\mathrm{2d}}\vec{c}^{\b} - \left(1 - \varepsilon^{\mathrm{b}} \right) \sum_{j=1}^{N^{\mathrm{p}}} \frac{3 d_j}{R^{\mathrm{p}}_{j}} \varepsilon^{\p}_{j} \mat{\tilde{M}}^{\b} \mat{D}^{\p}_{j} \left. \frac{\partial \vec{c}^{\p}_{j}}{\partial r} \right|_{r = R^{\mathrm{p}}_{j}} + \varepsilon^{\mathrm{b}} \mat{\tilde{M}}^{\b} \mat{S}^{\b} \vec{r}^{\b}\mleft( \vec{c}^{\b} \mright),
	\end{split}
\end{equation*}
where $\mat{D}^{\p}_{j}$ is a diagonal matrix with diffusion coefficients.
Second, boundary condition \cref{supp:eq:model-rctn-2d-rapidequi-particle-bc-outer} is replaced by a mix of Dirichlet and Neumann conditions:
\begin{subequations}\label{supp:eq:model-rctn-rapidequi-infkf-bc}%
\begin{align}
	\left( 1 - \varepsilon^{\mathrm{b}} \right)& \mat{\tilde{M}}^{\ps}_j \begin{pmatrix}
		\left.\vec{c}^{\p}_{j}\right|_{r=R^{\mathrm{p}}_{j}} \\ \vec{0}_{N^{\mathrm{totb}}_{j}\times 1}
	\end{pmatrix} = \left( 1 - \varepsilon^{\mathrm{b}} \right) \mat{\tilde{M}}^{\ps}_j \begin{pmatrix}
		\vec{c}^{\b} \\ \vec{0}_{N^{\mathrm{totb}}_{j}\times 1}
	\end{pmatrix}, \label{supp:eq:model-rctn-rapidequi-infkf-liquid} \\
	\left(1 - \varepsilon^{\mathrm{b}} \right)& \left(1 - \varepsilon^{\p}_{j} \right) \submat{\mat{\tilde{M}}^{\ps}_j}{\mset{M}_j^{\s}}{\all} \begin{pmatrix}
		\vec{0}_{N^{\mathrm{c}}\times 1} \\
		\mat{D}^{\s}_{j} \left. \frac{\partial \vec{c}^{\s}_{j}}{\partial r} \right|_{r = R^{\mathrm{p}}_{j}}
	\end{pmatrix} = \vec{0}_{\card*{\mset{C}_{\mathrm{dyn},j}^{\ps}}\times 1}, \label{supp:eq:model-rctn-rapidequi-infkf-solid}
\end{align} 
\end{subequations}%
where $\mset{M}_j^{\s}$ is the index set of pure solid-phase conservation relations:
\begin{equation*}
	\mset{M}_j^{\s} \coloneqq \left\{ i \in \left\{1, \dots, \numrows{\mat{\tilde{M}}^{\ps}_j} \right\} : \vec{e}_i^{\T} \mat{\tilde{M}}^{\ps}_j \vec{e}_k = 0 \text{ for } k = 1, \dots, N^{\mathrm{c}} \right\}.
\end{equation*}
In other words, $\mset{M}_j^{\s}$ contains the indices of the rows of $\mat{\tilde{M}}^{\ps}_j$ whose first $N^{\mathrm{c}}$ entries are zero.
Due to the construction of $\mset{M}_j^{\s}$, \cref{supp:eq:model-rctn-rapidequi-infkf-bc} results in $\numrows{\mat{\tilde{M}}^{\ps}_j}$ conditions.
The Dirichlet condition \cref{supp:eq:model-rctn-rapidequi-infkf-liquid} is similar to~\eqref{eq:model-1d-particle-bulk-bc-liquid-infkf} and imposes the conditions for the liquid-phase concentrations of the conservation relations.
The remaining liquid-phase concentrations are determined by vanishing reaction rate conditions.
The Neumann condition \cref{supp:eq:model-rctn-rapidequi-infkf-solid} copies~\eqref{eq:model_particle_1D_BC_solid_Rp} for the solid-phase concentrations that are not part of reactions in rapid equilibrium.
Again, the remaining solid-phase concentrations are determined by the other concentrations due to the rapid-equilibrium reactions.

\vspace{0.2cm}\noindent\textit{Constraints for Well-Posed Reaction-Systems}\vspace{0.1cm}

The presented framework can yield an overdetermined system, which may have no solution unless the additional constraints are compatible.
For obtaining a non-overdetermined system, it is sufficient that
\begin{subequations}
\begin{align}
	N^{\mathrm{c}} &= \card{\mset{C}_{\mathrm{dyn}}^{\b}} + \rank{\mat{M}^{\b}} + \card{\mset{R}_{\mathrm{re}}^{\b}}, \\
	N^{\mathrm{c}} + N^{\mathrm{totb}}_{j} &= \card{\mset{C}_{\mathrm{dyn},j}^{\ps}} + \rank{\mat{M}^{\ps}_j} + \card{\mset{R}_{\mathrm{re},j}^{\ps}},
\end{align}\label{supp:eq:model-rctn-rapidequi-correctnumeq}%
\end{subequations}%
where the rank of the eliminating conservation relation matrices is given by their respective number of rows.
This is illustrated with the following example.

\begin{example}
	Consider the cyclic reaction system $A \to B \to C \to A$, which consists of three chemical components and three reactions.
	Its stoichiometry is given by
	\begin{equation*}
		\mat{S} = \begin{pmatrix*}[r]
			-1 & 0 & 1 \\ 1 & -1 & 0 \\ 0 & 1 & -1
		\end{pmatrix*}.
	\end{equation*}
	Suppose all reactions are in rapid equilibrium, that is, $\mset{R}_{\mathrm{re}} = \{1, 2, 3\}$.
	Using \cref{supp:eq:model-rctn-compinds}, the component indices are given by
	\begin{equation*}
		\mset{C}_{\mathrm{re}} = \bigcup_{i \in \mset{R}_{\mathrm{re}}} \supp{\mat{S} \vec{e}_i}
		= \{1,2 \} \cup \{ 2,3\} \cup \{ 1,3\} = \{1,2,3\} \qquad \Rightarrow \qquad \mset{C}_{\mathrm{dyn}} = \emptyset.
	\end{equation*}
	Next, we compute the eliminating conservation relations of $\mat{S}$:
	\begin{equation*}
		\operatorname{ker} \mleft(\mat{S}^{\T}\mright) = \operatorname{ker} \begin{pmatrix*}[r]
			-1 & 1 & 0 \\ 0 & -1 & 1 \\ 1 & 0 & -1
		\end{pmatrix*} = \spann{\begin{pmatrix}
			1 \\ 1 \\ 1
		\end{pmatrix}} \qquad \Rightarrow \qquad \mat{M} = \begin{pmatrix}
			1 & 1 & 1
		\end{pmatrix}.
	\end{equation*}
	By \cref{supp:eq:model-rctn-consrelmat}, this results in $\mat{\tilde{M}} = \mat{M}$ and we obtain the system
	\begin{equation*}
		\mat{\tilde{M}}\frac{\mathrm{d}\vec{c}}{\mathrm{d}t} = \mat{\tilde{M}} \mat{S} \vec{r}\mleft( \vec{c} \mright), \qquad \vec{r}\mleft( \vec{c} \mright) = \vec{0}
	\end{equation*}
	with initial conditions
	\begin{equation*}
		\left.\mat{\tilde{M}}\vec{c}\right|_{t=0} = \mat{\tilde{M}} \vec{c}^{\mathrm{init}}, \qquad \vec{r}\mleft( \vec{c}^{\mathrm{init}} \mright) = \vec{0}.
	\end{equation*}
	The system consists of one differential and three algebraic equations for three unknowns.
\end{example}

The example shows that cycles in the reaction network can result in overdetermined systems.
In the example, the concentration of $B$ is determined by the concentration of $A$.
Adding the link between $C$ and $A$, hence, requires compatibility as $C$ is already determined by $A$ through $B$.
Cycles can introduce additional equilibrium constraints whose compatibility must be checked.
A cyclic reaction network may nevertheless admit a consistent equilibrium model if the additional constraints are redundant.

The well-posedness also depends on the form of the rate functions.
In general, a non-overdetermined reaction system is well-posed if the rate functions corresponding to rapid equilibrium reactions, are uniquely solvable with respect to $\subvec{\vec{c}}{\mset{C}_{\mathrm{re}}}$.
This must hold for every physical admissible input $\subvec{\vec{c}}{\mset{C}_{\mathrm{dyn}}}$, subjected to its conserved constrains.
That is, there is a mapping $\subvec{\vec{c}}{\mset{C}_{\mathrm{dyn}}} \mapsto \subvec{\vec{c}}{\mset{C}_{\mathrm{re}}}(\subvec{\vec{c}}{\mset{C}_{\mathrm{dyn}}})$ that satisfies
\begin{equation*}
	\vec{e}_i^{\T} \vec{r}\mleft( \subvec{\vec{c}}{\mset{C}_{\mathrm{dyn}}}, \subvec{\vec{c}}{\mset{C}_{\mathrm{re}}}\mleft( \subvec{\vec{c}}{\mset{C}_{\mathrm{dyn}}} \mright) \mright)= 0\qquad \forall\  i \in \mset{R}_{\mathrm{re}}.
\end{equation*}
In addition, if the rate functions are compatible, an overdetermined system can also be solved uniquely since some equations are redundant in this case.

\subsubsection{Rapid-Equilibrium Multi-State SMA}
\label{supp:sec:models-reaction-rapidequimssma}

The multi-state SMA model introduced in \cref{supp:sec:models-binding-mssma} can formally be interpreted as combining the standard SMA model of \cref{supp:sec:models-binding-sma} with reactions in the solid phase.
These solid-phase reactions describe transitions between bound states of a component.
By casting the model into the framework developed in this section, we easily obtain its rapid-equilibrium variant.

We remove the kinetic reaction terms from \cref{supp:eq:model-bnd-mssma-prot} and obtain the standard SMA binding rate as in \cref{supp:eq:model-bnd-sma-pre}:
\begin{equation*}
    f^{\mathrm{bind}}_{i,k}\mleft(\vec{c}^{\p}, \vec{c}^{\s}\mright)
    =
    k^{\mathrm{a}}_{i,k}(\overbar{c}^{\s}_1)^{\nu_{i,k}} \, c^{\p}_i - k^{\mathrm{d}}_{i,k}(c^{\p}_1)^{\nu_{i,k}} \, c^{\s}_{i,k}
\end{equation*}
The exchanges between different bound states of a component are instead formulated as solid-phase reactions.
To this end, we consider the reaction graph for the bound states of each component $i$.
Since there is no interaction between the bound states of different components, we can consider each component separately.

The reaction graph for the bound states of component $i$ is essentially specified by the rate constant matrix $\mat{K}^{(i)}_{\mathrm{e}}$ that contains the coefficients $k^{(i)}_{\mathrm{e},k,b}$ for all indices $k,b$.
A connection between two bound states $k$ and $b$ exists if $k^{(i)}_{\mathrm{e},k,b} \neq 0$ or $k^{(i)}_{\mathrm{e},b,k} \neq 0$.
From \cref{supp:sec:models-reaction-rapidequi}, we know that an acyclic graph is, in general, necessary for a well-posed non-overdetermined system.
This imposes a constraint on the rate constant matrices $\mat{K}^{(i)}_{\mathrm{e}}$ that determines whether a rapid-equilibrium variant is possible.

Assuming an acyclic graph, we translate the graph connectivity of the bound states into their stoichiometric counterpart by adding columns to $\mat{S}^{\s}_j$.
For this purpose, we fix the nominal direction of the reaction from weakly bound state $k$ to strongly bound state $b > k$.
Since only two bound states are connected at a time, the column's non-zero entries comprise exactly one $-1$ at the weakly bound state $k$ and one $1$ entry at the strongly bound state $b$.
The corresponding entry in the rate function $\vec{r}^{\s}_{j}$ is given by
\begin{equation*}
	k^{(i)}_{\mathrm{e},k,b}\: \left(\overbar{c}^{\s}_1\right)^{\left(\nu_{i,b} - \nu_{i,k}\right)}\: c^{\s}_{i,k} - k^{(i)}_{\mathrm{e},b,k}\: \left(c^{\p}_1\right)^{\left(\nu_{i,b} - \nu_{i,k}\right)}\: c^{\s}_{i,b}.
\end{equation*}
\Needspace{3\baselineskip}
Finally, we add the created reactions to the index set $\mset{R}_{\mathrm{re},j}^{\ps}$ of reactions that are in rapid equilibrium.

\subsubsection{Mass-Action Rate Laws}
\label{supp:sec:models-reaction-massaction}

Except for the excursion to the multi-state SMA model above, we have avoided explicitly stating the rate function $\vec{r}$.
As we have seen, the actual form of $\vec{r}$ does not matter for constructing the rapid-equilibrium model variant.
Kinetic binding models can generally be interpreted as rate functions, providing a wide range of possible formulations for $\vec{r}$.

According to \citet[Sec.~4.1.2]{Klipp2016}, the law of mass action states that the reaction rate is proportional to the concentration of the reactants to the power of their molecularity.
The molecularity of a reactant is the number of molecules that enter the reaction (e.g., $a_i$ in \cref{supp:eq:model-rctn-reaction-single}).
For the general reaction \cref{supp:eq:model-rctn-reaction-single}, which is given by
\begin{equation*}
	\sum_{i=1}^{N^{\mathrm{c}}} a_i c_i \to \sum_{i=1}^{N^{\mathrm{c}}} b_i c_i,
\end{equation*}
the rate of the forward reaction postulated by the mass action law reads
\begin{equation*}
	k \prod_{i=1}^{N^{\mathrm{c}}} \left(c_i\right)^{a_i},
\end{equation*}
where $k > 0$ is the proportionality constant.
Note that we use the convention $0^0 = 1$.
Thus, the net rate of the (reversible) reaction \cref{supp:eq:model-rctn-reaction-single} is given by
\begin{equation}
	k_{\mathrm{fwd}} \prod_{i=1}^{N^{\mathrm{c}}} \left(c_i\right)^{a_i} - k_{\mathrm{bwd}} \prod_{i=1}^{N^{\mathrm{c}}} \left(c_i\right)^{b_i}, \label{supp:eq:model-rctn-massaction-netrate}
\end{equation}
where $k_{\mathrm{fwd}} \geq 0$ and $k_{\mathrm{bwd}} \geq 0$ are proportionality constants.
We model irreversible reactions by setting one of these constants to zero.

\subsubsection{Modified Mass-Action Rate Law}

In order to provide more flexibility, we allow components that are not directly involved in a reaction to have an accelerating or decelerating effect.
This is achieved by permitting arbitrary exponents in \cref{supp:eq:model-rctn-massaction-netrate}, such that each entry of $\vec{r}^{\b}$ is given by
\begin{equation*}
	r_{\kappa}^{\b}\mleft(\vec{c}^{\b}\mright) = k^{\b}_{\mathrm{fwd},\kappa} \prod_{i=1}^{N^{\mathrm{c}}} \left(c_i^{\b}\right)^{e^{\b}_{\mathrm{fwd},\kappa,i}} - k^{\b}_{\mathrm{bwd},\kappa} \prod_{i=1}^{N^{\mathrm{c}}} \left(c_i^{\b}\right)^{e^{\b}_{\mathrm{bwd},\kappa,i}}.
\end{equation*}
For the particle domain reactions, we additionally allow components in the solid phase to accelerate or decelerate a reaction in the liquid phase and vice versa.
This is 
accomplished by rate functions of the form
\begin{multline*}
	r^{\ps}_{j,\kappa}\mleft( \vec{c}^{\p}_j, \vec{c}^{\s}_j \mright) = k^{\ps}_{\mathrm{fwd},\kappa} \prod_{i=1}^{N^{\mathrm{c}}} \left(c_{j,i}\right)^{e^{\p}_{\mathrm{fwd},j,\kappa,i}} \prod_{k=1}^{N^{\mathrm{totb}}_{j}} \left(c_{j,k}^{\s}\right)^{e^{\s}_{\mathrm{fwd},j,\kappa,k}} \\
	- k^{\ps}_{\mathrm{bwd},\kappa} \prod_{i=1}^{N^{\mathrm{c}}} \left(c_{j,i}^{\p}\right)^{e^{\p}_{\mathrm{bwd},j,\kappa,i}} \prod_{k=1}^{N^{\mathrm{totb}}_{j}} \left(c_{j,k}^{\s}\right)^{e^{\s}_{\mathrm{bwd},j,\kappa,k}},
\end{multline*}
where the index $k$ runs through the solid-phase vector $\vec{c}^{\s}_j$ ordered as
\begin{equation}
    \label{supp:eq:model-bnd-solidordering}
	\vec{c}^{\s}_j = \left(c^{\s}_{j,1,1}, \dots, c^{\s}_{j,1, N^{\mathrm{b}}_{j,1}}, c^{\s}_{j,2,1}, \dots, c^{\s}_{j,2, N^{\mathrm{b}}_{j,2}}, \dots, c^{\s}_{j,N^{\mathrm{c}},1}, \dots, c^{\s}_{j,N^{\mathrm{c}}, N^{\mathrm{b}}_{j,N^{\mathrm{c}}}} \right) \in \Reals^{N^{\mathrm{totb}}_{j}}.
\end{equation}

For simple reactions where the stoichiometric coefficients $\vec{a}$ and $\vec{b}$ have disjoint support $\supp{\vec{a}} \cap \supp{\vec{b}} = \emptyset$, we can infer the exponents from the stoichiometric matrix $\mat{S}$:
\begin{equation*}
	e_{\mathrm{fwd},\kappa,i} = \max\mleft\{ -S_{i,\kappa}, 0 \mright\}, \qquad e_{\mathrm{bwd},\kappa,i} = \max\mleft\{ S_{i,\kappa}, 0 \mright\}.
\end{equation*}
Note that we cannot infer the exponents of the modifier components because they do not show up in the stoichiometry as they are neither consumed nor produced.

\subsection{High-Resolution Model}
\label{supp:sec:hrm}

The High-Resolution Model presented in this work is based on the following customary assumptions, which also serve as the starting point for deriving reduced-order models.
\begin{assumption}[(HR)]\label{supp:asmpt:model-hrm}
We assume that \citep{Gu2015,Guiochon2006,Seidel-Morgenstern2020-Chap6}
\begin{multicols}{2}
\begin{itemize}\setlength{\itemsep}{0\baselineskip}
    \item the packed bed is confined within a rigid, perfectly cylindrical column of radius $\Rho^\mathrm{c} > 0$, with fluid entering and exiting through the inlet and outlet faces;
    \item fluid density and viscosity are constant in time and space (in particular, the fluid is incompressible);
    \item fluid flow is laminar;
    \item fluid inside the particles is stagnant (i.e., no convective flow in the particle pores);
    \item particles are perfectly spherical, rigid, and immobile;
    \item particles are homogeneous and have uniform porosity, and the intraparticle pores are equally accessible to all components;
    \item diffusion is Fickian and independent of component concentrations;
    \item solute concentrations are sufficiently low such that changes in solution volume caused by solute transport and binding can be neglected;
    \item solvent is not adsorbed.
\end{itemize}
\end{multicols}
\end{assumption}

A detailed discussion of most points in Assumption~\ref{asmpt:model-hrm} is provided in Section~2.1.2 of \cite{Guiochon2006}.
Hence, we will refrain from repeating the full arguments here and instead highlight a few selected aspects:

\begin{itemize}\setlength{\itemsep}{0\baselineskip}
    \item Chromatography systems typically operate at low Reynolds numbers ($\mathrm{Re} < 1$) \citep[Section~6.1]{Nicoud2015}, which justifies the assumption of laminar flow.
    \item Fluid viscosity is not always constant in real-world applications.
    Variations may arise from switching between feeds of different viscosities during operation or from highly concentrated solute regions formed during step elution within the column.
    \item Although temperature changes due to binding are negligible in most cases \citep{Yun1997}, some systems employ temperature-responsive particles, enabling binding behavior to be controlled through the temperature distribution within the column, e.g. using heating and cooling jackets \citep{Muller2013}.
    Temperature can also become significant when chemical reactions are involved \citep{Sainio2007}.
    \item Chromatography particles are not perfectly spherical \cite[Section~3.1]{Gu2015}.
    Bed compression, whether mechanical or flow-induced, can cause slight deformation.
    However, simulations by \cite{Dorn2016} indicate that these effects are minor.
     \item Assuming uniformly porous and homogeneous particles allows for a continuum representation in which mobile and stationary phases coexist at every point within the particle. This enables a macroscopic treatment of binding, where the solute distribution between phases is described using binding functions, often defined by adsorption isotherms, and the ratio of pore and solid domains.
    \item Surface diffusion, which describes diffusion of adsorbed solute along the internal solid surfaces of the porous particle, is known to depend on solute concentrations \cite[Section~6.2.3]{Carta2010}.
    Pore diffusion may also vary with concentration, particularly when highly viscous or concentrated feeds are processed \citep{Guiochon2006}.
\end{itemize}

\subsubsection{Model Variants}
\label{supp:sec:models-hrm-variants}

Imposing additional assumptions on parameters or time scales, we can slightly reduce the complexity of the model.
In the following, we demonstrate two common simplifications, similar to the model variants introduced for the continuum models in Section~\ref{sec:bulk_domain_transport}.

\Needspace{6\baselineskip}
\vspace{0.2cm}\noindent\textit{Non-Limiting Film Diffusion}\vspace{0.1cm}
\newline
Model~\ref{model:column-highres3D} admits several modifications.
As mentioned in Remark~\ref{rem:model-highres}, we can additionally assume that the film around the particles does not limit mass transfer.
That is, we take the limit $k^{\mathrm{f}}_{i} \to \infty$, which changes the boundary conditions~\eqref{eq:model-hrm-mtcoupling}.
First, we add \eqref{eq:model-hrm-bulkparticlebc} to \eqref{eq:model-hrm-particlebulkbc} and obtain on the boundary $\Gamma^{\mathrm{par}}$ the flux continuity condition
\begin{alignat}{2}
	& & \left( \vec{u} c^{\b}_i - D^{\b}_i \nabla c^{\b}_i \right) \cdot \vec{n}^{\b} -\varepsilon^{\mathrm{p}} D^{\p}_i \nabla c^{\p}_i \cdot \vec{n}^{\p} &= k^{\mathrm{f}}_{i} \left( c^{\b}_i - c^{\p}_i \right) + k^{\mathrm{f}}_{i} \left( c^{\p}_i - c^{\b}_i \right) = 0 \nonumber \\
	&\Leftrightarrow\qquad & \left( \vec{u} c^{\b}_i - D^{\b}_i \nabla c^{\b}_i \right) \cdot \vec{n}^{\b} &= \varepsilon^{\mathrm{p}} D^{\p}_i \nabla c^{\p}_i \cdot \vec{n}^{\p}. \label{supp:eq:model-hrm-mt-kfvar-fluxcont}
\end{alignat}
Second, we divide \eqref{eq:model-hrm-mtcoupling} by $k^{\mathrm{f}}_{i}$,
\begin{alignat*}{4}
	& & \left( \vec{u} c^{\b}_i - D^{\b}_i \nabla c^{\b}_i \right) \cdot \vec{n}^{\b} &= k^{\mathrm{f}}_{i} \left( c^{\b}_i - c^{\p}_i \right)
	& &\qquad\wedge\qquad & -\varepsilon^{\mathrm{p}} D^{\p}_i \nabla c^{\p}_i \cdot \vec{n}^{\p} &= k^{\mathrm{f}}_{i} \left( c^{\p}_i - c^{\b}_i \right) \\
	&\Leftrightarrow\qquad & \frac{1}{k^{\mathrm{f}}_{i}} \left( \vec{u} c^{\b}_i - D^{\b}_i \nabla c^{\b}_i \right) \cdot \vec{n}^{\b} &= c^{\b}_i - c^{\p}_i
	& &\qquad\wedge\qquad & -\varepsilon^{\mathrm{p}} \frac{1}{k^{\mathrm{f}}_{i}} D^{\p}_i \nabla c^{\p}_i \cdot \vec{n}^{\p} &= c^{\p}_i - c^{\b}_i,
\end{alignat*}
and take the limit $k^{\mathrm{f}}_{i} \to \infty$ on both sides to arrive at
\begin{equation}
	0 = c^{\b}_i - c^{\p}_i 
	 \qquad\wedge\qquad  0 = c^{\p}_i - c^{\b}_i, \label{supp:eq:model-hrm-mt-kfvar-ccont}
\end{equation}
which is the continuity of the concentrations $c^{\p}_i = c^{\b}_i$ at the boundary $\Gamma^{\mathrm{par}}$.
We only need one equation to replace Eq.~\eqref{eq:model-hrm-mtcoupling}.
Both \cref{supp:eq:model-hrm-mt-kfvar-fluxcont} and \cref{supp:eq:model-hrm-mt-kfvar-ccont} are possible and produce equal results.
However, we favor the flux continuity condition \cref{supp:eq:model-hrm-mt-kfvar-fluxcont}, since such expressions naturally appear in boundary integrals.
If in addition the initial data $c^{\b,\mathrm{init}}_{i}$ and $c^{\p,\mathrm{init}}_{i}$ are equal on $\Gamma^{\mathrm{par}}$, then \cref{supp:eq:model-hrm-mt-kfvar-ccont} holds for the state variables throughout time.
Since we have effectively removed film diffusion, we could proceed to remove the stagnant film of fluid around the particles.
This is achieved by replacing the no-slip condition~\eqref{eq:model-hrm-noslipbc} on $\Gamma^{\mathrm{par}}$ by a vanishing normal flow condition $\vec{u} \cdot \vec{n} = 0$.

\begin{remark}
	We have assumed that $k^{\mathrm{f}}_{i} = \infty$ for all components $i$.
	Clearly, we can choose to maintain a finite film diffusion coefficient $k^{\mathrm{f}}_{i} < \infty$ for some components $i$ while setting $k^{\mathrm{f}}_{i} = \infty$ for the rest.
\end{remark}

\vspace{0.2cm}\noindent\textit{Rapid-Equilibrium Binding}\vspace{0.1cm}
\newline
Another variant arises on the assumption that adsorption and desorption of some (or all) components occur on a much faster time scale than the other mass transfer processes (e.g., convection, diffusion).
This is called the \emph{rapid-equilibrium assumption} \citep[Sec.~2.2.1]{Ingalls2013} (or the \emph{quasi-equilibrium approximation} \citep[Sec.~6.3.3]{Klipp2016}).
In general, the net binding rate $f^{\mathrm{bind}}_{i}( \vec{c}^{\p}, \vec{c}^{\s})$ decomposes into an adsorption rate $r^{\mathrm{ads}}_{i} \geq 0$, a desorption rate $r^{\mathrm{des}}_{i} \geq 0$, an equilibrium constant $k^{\mathrm{eq}}_{i} \geq 0$, and a kinetic constant $k^{\mathrm{kin}}_{i} \geq 0$:
\begin{equation*}
	f^{\mathrm{bind}}_{i}\mleft( \vec{c}^{\p}, \vec{c}^{\s} \mright) = k^{\mathrm{kin}}_{i} \left(k^{\mathrm{eq}}_{i} r^{\mathrm{ads}}_{i}\mleft( \vec{c}^{\p}, \vec{c}^{\s} \mright) - r^{\mathrm{des}}_{i}\mleft( \vec{c}^{\p}, \vec{c}^{\s} \mright) \right).
\end{equation*}
Another equivalent parameterization is given by
\begin{equation}
	f^{\mathrm{bind}}_{i}\mleft( \vec{c}^{\p}, \vec{c}^{\s} \mright) = k^{\mathrm{ads}}_{i} r^{\mathrm{ads}}_{i}\mleft( \vec{c}^{\p}, \vec{c}^{\s} \mright) - k^{\mathrm{des}}_{i} r^{\mathrm{des}}_{i}\mleft( \vec{c}^{\p}, \vec{c}^{\s} \mright), \label{supp:eq:model-hrm-jbind-rapidequi}
\end{equation}
where $k^{\mathrm{kin}}_{i} = k^{\mathrm{des}}_{i}$ and $k^{\mathrm{eq}}_{i} = k^{\mathrm{ads}}_{i} / k^{\mathrm{des}}_{i}$.
In rapid equilibrium, binding kinetics is assumed to be infinitely fast, that is, we let the kinetic constant $k^{\mathrm{kin}}_{i} \to \infty$.
Alternatively, we let $k^{\mathrm{ads}}_{i}, k^{\mathrm{des}}_{i} \to \infty$ such that their ratio $k^{\mathrm{ads}}_{i} / k^{\mathrm{des}}_{i} = k^{\mathrm{eq}}_{i}$ remains constant.

\Needspace{5\baselineskip}
\begin{remark}
	Since both the liquid and solid phases are required for binding to make sense, we assume $\varepsilon^{\mathrm{p}} \in (0,1)$ for the derivation of this model variant.
	The resulting model can be used with $\varepsilon^{\mathrm{p}} \in [0,1]$, though.
\end{remark}

This assumption changes the structure of Eq.~\eqref{eq:model-hrm-particle} as one partial differential equation is converted to a purely algebraic equation.
Similarly to the variant without film diffusion, we first eliminate the net binding rate $f^{\mathrm{bind}}_{i}(\vec{c}^{\p}, \vec{c}^{\s})$ from Eqs.~\eqref{eq:model-hrm-particleliquid} and~\eqref{eq:model-hrm-particlesolid} by adding them together, which yields the mass balance for the conserved quantity $\varepsilon^{\mathrm{p}} c^{\p}_i + (1-\varepsilon^{\mathrm{p}}) c^{\s}_i$:
\begin{equation*}
	\varepsilon^{\mathrm{p}} \frac{\partial c^{\p}_i}{\partial t} + \left( 1 - \varepsilon^{\mathrm{p}} \right) \frac{\partial c^{\s}_i}{\partial t} = \varepsilon^{\mathrm{p}} \div{D^{\p}_i \nabla c^{\p}_i} + \left( 1 - \varepsilon^{\mathrm{p}} \right) \div{D^{\s}_i \nabla c^{\s}_i}.
\end{equation*}
This shows that the quantity $\varepsilon^{\mathrm{p}} c^{\p}_i + (1-\varepsilon^{\mathrm{p}}) c^{\s}_i$ is conserved with respect to binding, that is, it is invariant under binding and only changes due to transport.
Such quantities are also called \emph{conserved moieties} (see Definition~\ref{supp:def:model-rctn-consmoiety}) and will become important in Section~\ref{sec:models-reaction} when reaction terms are added to the models.

The ratio between liquid-phase concentration $c^{\p}_i$ and solid-phase concentration $c^{\s}_i$ is determined by a vanishing net binding rate.
This can be seen by dividing the solid-phase mass balance Eq.~\eqref{eq:model-hrm-particlesolid} by $k^{\mathrm{kin}}_{i}$,
\begin{equation*}
	\frac{1 - \varepsilon^{\mathrm{p}}}{k^{\mathrm{kin}}_{i}} \frac{\partial c^{\s}_i}{\partial t} = \frac{1 - \varepsilon^{\mathrm{p}}}{k^{\mathrm{kin}}_{i}} \div{D^{\s}_i \nabla c^{\s}_i} + \frac{1 - \varepsilon^{\mathrm{p}}}{k^{\mathrm{kin}}_{i}} f^{\mathrm{bind}}_{i}\mleft( \vec{c}^{\p}, \vec{c}^{\s} \mright),
\end{equation*}
and taking the limit $k^{\mathrm{kin}}_{i} \to \infty$:
\begin{alignat*}{2}
	& & 0 &= \lim\limits_{k^{\mathrm{kin}}_{i} \to \infty} \frac{1 - \varepsilon^{\mathrm{p}}}{k^{\mathrm{kin}}_{i}} f^{\mathrm{bind}}_{i}\mleft( \vec{c}^{\p}, \vec{c}^{\s} \mright) \\
	& & 0 &= \left( 1 - \varepsilon^{\mathrm{p}} \right) \lim\limits_{k^{\mathrm{kin}}_{i} \to \infty} \frac{k^{\mathrm{kin}}_{i}}{k^{\mathrm{kin}}_{i}} \left( k^{\mathrm{eq}}_{i} r^{\mathrm{ads}}_{i}\mleft( \vec{c}^{\p}, \vec{c}^{\s} \mright) - r^{\mathrm{des}}_{i}\mleft( \vec{c}^{\p}, \vec{c}^{\s} \mright) \right) \\
	& \Leftrightarrow&\qquad 0 &= k^{\mathrm{eq}}_{i} r^{\mathrm{ads}}_{i}\mleft( \vec{c}^{\p}, \vec{c}^{\s} \mright) - r^{\mathrm{des}}_{i}\mleft( \vec{c}^{\p}, \vec{c}^{\s} \mright),
\end{alignat*}
which is equivalent to a vanishing net rate
\begin{equation*}
	f^{\mathrm{bind}}_{i}\mleft( \vec{c}^{\p}, \vec{c}^{\s} \mright) = 0
\end{equation*}
if we take any finite non-zero value for $k^{\mathrm{kin}}_{i}$.
For the components $i$ in rapid equilibrium, the particle PDE \eqref{eq:model-hrm-particle} is changed to the partial differential algebraic equation (PDAE)
\begin{subequations}
\begin{alignat}{2}
	\varepsilon^{\mathrm{p}} \frac{\partial c^{\p}_i}{\partial t} + \left( 1 - \varepsilon^{\mathrm{p}} \right) \frac{\partial c^{\s}_i}{\partial t} &= \varepsilon^{\mathrm{p}} \div{D^{\p}_i \nabla c^{\p}_i} + \left( 1 - \varepsilon^{\mathrm{p}} \right) \div{D^{\s}_i \nabla c^{\s}_i} & &\quad \text{in } (0, T^{\mathrm{end}}) \times \Omega^{\mathrm{par}}, \\
	0 &= \left( 1 - \varepsilon^{\mathrm{p}} \right) f^{\mathrm{bind}}_{i}\mleft( \vec{c}^{\p}, \vec{c}^{\s} \mright) & &\quad \text{in } (0, T^{\mathrm{end}}) \times \Omega^{\mathrm{par}}. \label{supp:eq:model-hrm-particlesolid-rapidequi} %
\end{alignat}
\end{subequations}
Note that we have kept the factor $1 - \varepsilon^{\mathrm{p}}$ in \cref{supp:eq:model-hrm-particlesolid-rapidequi} in order to later account for the case $\varepsilon^{\mathrm{p}} = 1$.
In addition, we replace the boundary conditions~\eqref{eq:model-hrm-particlebcsolid} and~\eqref{eq:model-hrm-particlebulkbc} with the single condition
\begin{equation}
	-\left(\varepsilon^{\mathrm{p}} D^{\p}_i \nabla c^{\p}_i + \left(1-\varepsilon^{\mathrm{p}}\right) D^{\s}_i \nabla c^{\s}_i \right) \cdot \vec{n}^{\p} = k^{\mathrm{f}}_{i} \left( c^{\p}_i - c^{\b}_i \right) \label{supp:eq:model-hrm-particlebc-rapidequi}
\end{equation}
for the conserved moiety, which also allows transport into the particle by surface diffusion.
In fact, we cannot impose boundary condition~\eqref{eq:model-hrm-particlebcsolid} in addition to \cref{supp:eq:model-hrm-particlebc-rapidequi} since the solid-phase concentration $\vec{c}^{\s}$ is fully determined by the liquid-phase concentration $\vec{c}^{\p}$ via \cref{supp:eq:model-hrm-particlesolid-rapidequi} and the system would be over-determined.

Under the rapid-equilibrium assumption, \cref{supp:eq:model-hrm-particlesolid-rapidequi} is a purely algebraic equation.
This reduces the stiffness of the system since it effectively removes the faster time scale of the binding process from the system.
The resulting algebraic constraints require consistent initial data $c^{\p,\mathrm{init}}_{i}$ and $c^{\s,\mathrm{init}}_{i}$ satisfying
\begin{equation*}
	\vec{f}^{\mathrm{bind}}\mleft( \vec{c}_{\mathrm{init}}^{\p}, \vec{c}_{\mathrm{init}}^{\s} \mright) = \vec{0} \qquad \text{in } \Omega^{\mathrm{par}}.
\end{equation*}

Note that we can choose a subset of components whose binding processes are assumed to be in equilibrium.
This is important for binding models having components that bind much faster than the others, such as salt counter-ions in ion-exchange models.

\Needspace{6\baselineskip}
\begin{remark}
	Both model variants, non-limiting film diffusion and binding in rapid equilibrium, can be easily combined since the model changes are orthogonal.
	If combined, the boundary condition \cref{supp:eq:model-hrm-mt-kfvar-fluxcont} is modified as detailed in the derivation of the rapid-equilibrium variant to
	\begin{equation*}
		\left( \vec{u} c^{\b}_i - D^{\b}_i \nabla c^{\b}_i \right) \cdot \vec{n}^{\b} = \left(\varepsilon^{\mathrm{p}} D^{\p}_i \nabla c^{\p}_i + \left(1-\varepsilon^{\mathrm{p}}\right) D^{\s}_i \nabla c^{\s}_i \right) \cdot \vec{n}^{\p}.
	\end{equation*}
\end{remark}

\section{Taxonomy}

Further (secondary) model classification that complements the primary classification from Table~\ref{table:taxonomy_primary} is provided in \cref{supp:table:taxonomy_secondary}.

The notation and terminology for the modeled phenomena and their associated parameters introduced throughout this work are designed to maintain consistency across the various models while clearly distinguishing quantities that represent conceptual differences.
For example, indices are consistently placed in subscripts, whereas descriptive labels are given in superscripts.
Where possible, we followed established conventions from the literature.
For convenience, a consolidated overview is provided in Table~\ref{supp:table:notation_termonology}, including notable differences from the conventions used in the literature.

\begin{table}[!htbp]
\caption{Further model classification complementing Table~\ref{table:taxonomy_primary}. Note that these model variants can be combined. The governing equations of the model classes can be generated using the Python package CADET-Equations \citep{breuer_2026_18339287}.}
\centering
\renewcommand{\arraystretch}{1.2} %

\begin{tabular}{x{2.2cm}|x{2cm}x{2cm}x{2cm}x{2cm}}
\toprule
Model class (Table~\ref{table:taxonomy_primary}) & \multicolumn{2}{c}{$N^{\p}>1$} & & \\
 & $R^{\p}_{j} \neq R^{\p}_k$ & $f^{\mathrm{bind}}_{j} \not\equiv f^{\mathrm{bind}}_{k}$ & $R^{\mathrm{c}}>0$ & $f^{\mathrm{react}} \not\equiv 0$ \\
\midrule
GRM & GRM with PSD & GRM with PTD & GRM with core-shell particles & reactive GRM \\
\bottomrule
\end{tabular}

\label{supp:table:taxonomy_secondary}
\end{table}

\begin{longtable}{p{0.09\textwidth}|p{0.47\textwidth}|p{0.36\textwidth}}

\caption{Notation and terminology of model parameters, variables and other definitions.}

\label{supp:table:notation_termonology}\\

\toprule
\textbf{Symbol} & \textbf{Description} & \textbf{Literature} \\
\midrule
\endfirsthead
\multicolumn{3}{l}{\textbf{Table~\thetable} \textit{(continued)}} \\[3pt]
\toprule
\textbf{Symbol} & \textbf{Description} & \textbf{Literature} \\
\midrule
\endhead
$c^{\mathrm{b}}$
& Bulk liquid-phase concentration
& \emph{Mobile phase concentration}\textsuperscript{1} $c$
\\
$c^{\mathrm{\ell}}$
& Total liquid-phase concentration
& 
Usually, $c$ is reused\textsuperscript{1}
\\
$c^{\mathrm{p}}$
& Particle liquid-phase concentration
& $c_\mathrm{p}$ concentration of the solute in the particle pores\textsuperscript{1} %
\\
$c^{\mathrm{s}}$
& Solid-phase concentration
& Commonly denoted as $q$, \emph{solid phase loading}\textsuperscript{1}
\\
$D^\mathrm{ang}$
& Angular dispersion
& --
\\
$D^\mathrm{ax}$
& Axial dispersion
& $D_\mathrm{ax}$ common\textsuperscript{1}
\\
$\tilde{D}^\mathrm{ax}$
& Apparent axial dispersion
& $D_\text{app}$ \emph{apparent dispersion}\textsuperscript{1}
\\
$D^\mathrm{rad}$
& Radial dispersion
& $D^\rho$ common\textsuperscript{5}
\\
$D^\p$
& Pore diffusion
& $D_\text{pore}$ common\textsuperscript{1}
\\
$D^\s$
& Surface diffusion
& $D_\text{solid}$ common\textsuperscript{1}
\\
$k^\mathrm{f}$
& Film diffusion %
& $k_\mathrm{film}$ \emph{film mass transfer}\textsuperscript{1}. Also interpreted as (lumped) film and pore diffusion\textsuperscript{1,4} $k_\mathrm{eff}$ or $k_\mathrm{e}$
\\
$k^{sc}$
& Cavity exchange coefficient
& --
\\
$L^\mathrm{b}$
& Column or bed length
& $L_\mathrm{c}$ common\textsuperscript{1}
\\
$N^\mathrm{b}$
& Number of bound states per component
& --
\\
$N^\mathrm{c}$
& Number of components
& $N_\mathrm{comp}$ common\textsuperscript{1}
\\
$N^\mathrm{p}$
& Number of particle types
& --
\\
$T^\mathrm{end}$
& Modeled end time
& $T_\text{end}$ common\textsuperscript{1}
\\
$\Rho^\mathrm{c}$
& Column radius
& $D_c$ diameter of the column\textsuperscript{1}
\\
$\Rho^\mathrm{0}$
& Column radius at flow directional coordinate $0$
& --
\\
$\Rho^\mathrm{L^b}$
& Column radius at flow directional coordinate $L^\mathrm{b}$
& --
\\
$R^\mathrm{p}$
& Radius of the particle
& $r_\mathrm{p}$ particle radius\textsuperscript{1}
\\
$R^\mathrm{c}$
& Radius of the impenetrable particle core
& $R_\text{core}$ may be used\textsuperscript{6}
\\
$a^\mathrm{p}$
& Ratio of outer particle surface to accessible particle volume
& --
\\
$u$
& Interstitial velocity
& $u_\mathrm{int,e}$ interstitial external velocity\textsuperscript{1}
\\
$\tilde{u}$
& Effective velocity in the Lumped Rate Model without Pores, see Model~\ref{model:lumped_rate_model_without_pores}
& $u^\text{hypo}_\text{int,t}$ hypothetical effective velocity\textsuperscript{1}
\\
$V^\mathrm{b}$
& Interstitial or bulk volume
& $V_\text{int}$ interstitial volume\textsuperscript{1}
\\
$V^{\ell}$
& Combined liquid volume
& $V_\text{m}$ overall liquid volume\textsuperscript{1}
\\
$V^\mathrm{par}$
& Particle volume
& $V_\text{ads}$ volume of the stationary phase within a column\textsuperscript{1}
\\
$V^\mathrm{p}$
& Volume of the particle pores
& $V_\text{pore}$
\\
$V^\mathrm{s}$
& Solid-phase volume
& $V_\text{solid}$ volume of the stationary phase within a column\textsuperscript{1}
\\
$V^\mathrm{tot}$
& Total column volume
& $V_\text{c}$ total column volume\textsuperscript{1}
\\
$\varepsilon^{\mathrm{b}}$
& Interstitial or \emph{bulk porosity}
& Also referred to as \emph{external porosity}\textsuperscript{1,4}, denoted $\epsilon_e$
\\
$\varepsilon^{\mathrm{p}}_j$
& \emph{Particle porosity} of particle type $j$
& Also referred to as \emph{internal porosity}\textsuperscript{1,4}, denoted $\epsilon_i$
\\
$\varepsilon^{\mathrm{t}}$
& Total porosity $\varepsilon^{\mathrm{b}} + (1-\varepsilon^{\mathrm{b}}) \sum_{j=1}^{N^{\mathrm{p}}} d_j \varepsilon^{\mathrm{p}}_j$
& $\varepsilon^{\mathrm{t}}$ total porosity\textsuperscript{1}
\\
$\Omega^\mathrm{b}$
& Bulk spatial domain
& --
\\
$\Omega^\mathrm{par}$
& Particle spatial domain
& --
\\
$\Omega^\mathrm{mp}$
& Spatial domain of the main pore in Model~\ref{model:particle_side_cavities}
& --
\\
$\Omega^\mathrm{sc}$
& Spatial domain of a side cavity in Model~\ref{model:particle_side_cavities}
& --
\\
$t$
& time coordinate
&  $t$ common\textsuperscript{1}
\\
$r$
& radial particle coordinate
&  $r$ common\textsuperscript{1}
\\
$z$
& axial cylinder/frustum coordinate
& $z$ common\textsuperscript{1}
\\
$\rho$
& radial cylinder coordinate
& $\rho$ common\textsuperscript{5}
\\
$\varphi$
& angular cylinder coordinate
& --
\\
\bottomrule

\end{longtable}

{\footnotesize\noindent
\textsuperscript{1} \cite{Schmidt-Traub2020};
\textsuperscript{2} \cite{Guiochon2006};
\textsuperscript{3} \cite{Gu2015};
\textsuperscript{4} \cite{KATSOULAS2023464345};
\textsuperscript{5} \cite{Qamar2017};
\textsuperscript{6} \cite{Brhane_2dGRM_2019}.
}

\end{document}